\documentclass{aa}  
\usepackage{graphicx}
\usepackage{float} %to be able to apply [H] option for figure placement
\usepackage{txfonts}
\usepackage{grffile} %checks for image file names containing more than 1 dots (multidot)
\usepackage{booktabs}
\usepackage{amsmath}
\usepackage[colorlinks=true, allcolors=blue]{hyperref}
\bibpunct{(}{)}{;}{a}{}{,}
\graphicspath{{SED/}{matisse_observability/}{0_model_fit_10um7/}}

\begin{document} 
	
	\title{VLTI/MIDI atlas of disks around low- and intermediate-mass young stellar objects}
	%Low-mass young stellar object disks from mid-infrared interferometry
	\titlerunning{VLTI/MIDI atlas of disks around low- and intermediate-mass YSOs}
	
	% \subtitle{I. Overviewing the $\kappa$-mechanism}
	
	\author{J. Varga\inst{1} \and
		P. \'Abrah\'am\inst{1}\and
		L. Chen\inst{1} \and
		Th. Ratzka\inst{2} \and
		K. \'E. Gab\'anyi\inst{1} \and  
		\'A. K\'osp\'al\inst{1,3} \and  
		A. Matter\inst{4} \and
		R. van Boekel\inst{3} \and  
		Th. Henning\inst{3} \and
		W. Jaffe\inst{5} \and
		A. Juh\'asz\inst{6} \and
		B. Lopez\inst{4} \and
		J. Menu\inst{7} \and
		A. Mo\'or\inst{1} \and
		L. Mosoni\inst{1,8} \and
		N. Sipos\inst{1}
	}
	\authorrunning{J. Varga et al.}
	
	\institute{Konkoly Observatory, Research Centre for Astronomy and Earth Sciences, Hungarian Academy of Sciences, Konkoly Thege Mikl\'os \'ut 15-17., H-1121 Budapest, Hungary \\
		\email{varga.jozsef@csfk.mta.hu}
		\and
		Institute for Physics/IGAM, NAWI Graz, University of Graz, Universitätsplatz 5/II, 8010, Graz, Austria              
		\and
		Max Planck Institute for Astronomy, K\"{o}nigstuhl 17, D-69117 Heidelberg, Germany
		\and
		Laboratoire Lagrange, Universit\'e C\^ote d'Azur, Observatoire de la C\^ote d'Azur, CNRS, Boulevard de l'Observatoire, CS 34229, 06304 Nice Cedex 4, France
		\and
		Leiden Observatory, Leiden University, Niels Bohrweg 2, 2333 CA Leiden, The Netherlands
		\and
		Institute of Astronomy, Madingley Road, Cambridge CB3 OHA, UK
		\and         
		Instituut voor Sterrenkunde, KU Leuven, Celestijnenlaan 200D, 3001, Leuven, Belgium        
		\and
		Park of Stars in Zselic, 064/2 hrsz., H-7477 Zselickisfalud,  Hungary        
		%\and
		%Institute for Astronomy, ETH Z\"{u}rich, Wolfgang-Pauli-Strasse 27, 8093 Z\"{u}rich, Switzerland
		%\thanks{The university of heaven temporarily does not accept e-mails}
	}
	
	\date{Received; accepted}
	
	% \abstract{}{}{}{}{} 
	% 5 {} token are mandatory
	
	\abstract
	{Protoplanetary disks show large diversity regarding their morphology and dust composition. With mid-infrared interferometry the thermal emission of disks can be spatially resolved, and the distribution and properties of the dust within can be studied. 
	}
	{Our aim is to perform a statistical analysis on a large sample of $82$ disks around low- and intermediate-mass young stars, based on mid-infrared interferometric observations. We intend to study the distribution of disk sizes, variability, and the silicate dust mineralogy.
	}
	{Archival mid-infrared interferometric data from the MIDI instrument on the Very Large Telescope Interferometer are homogeneously reduced and calibrated. Geometric disk models are used to fit the observations to get spatial information about the disks. An automatic spectral decomposition pipeline is applied to analyze the shape of the silicate feature. 
	}
	{We present the resulting data products in the form of an atlas, containing $N$ band correlated and total spectra, visibilities, and differential phases. The majority of our data can be well fitted with a continuous disk model, except for a few objects, where a gapped model gives a better match. From the mid-infrared size--luminosity relation we find that disks around T Tauri stars are generally colder and more extended with respect to the stellar luminosity than disks around Herbig Ae stars. 
		We find that in the innermost part of the disks ($r \lesssim 1$~au) the silicate feature is generally weaker than in the outer parts, suggesting that in the inner parts the dust is substantially more processed. We analyze stellar multiplicity and find that in two systems (AB Aur and HD 72106)  data suggest a new companion or asymmetric inner disk structure. We make predictions for the observability of our objects with the upcoming Multi-AperTure mid-Infrared SpectroScopic Experiment (MATISSE) instrument, supporting the practical preparations of future MATISSE observations of T Tauri stars.
	}  
	{}
	
	
	\maketitle
	
	\section{Introduction}
	\label{sec:intro}
	
	\begin{figure}
		\centering
		\includegraphics[width=0.99\linewidth]{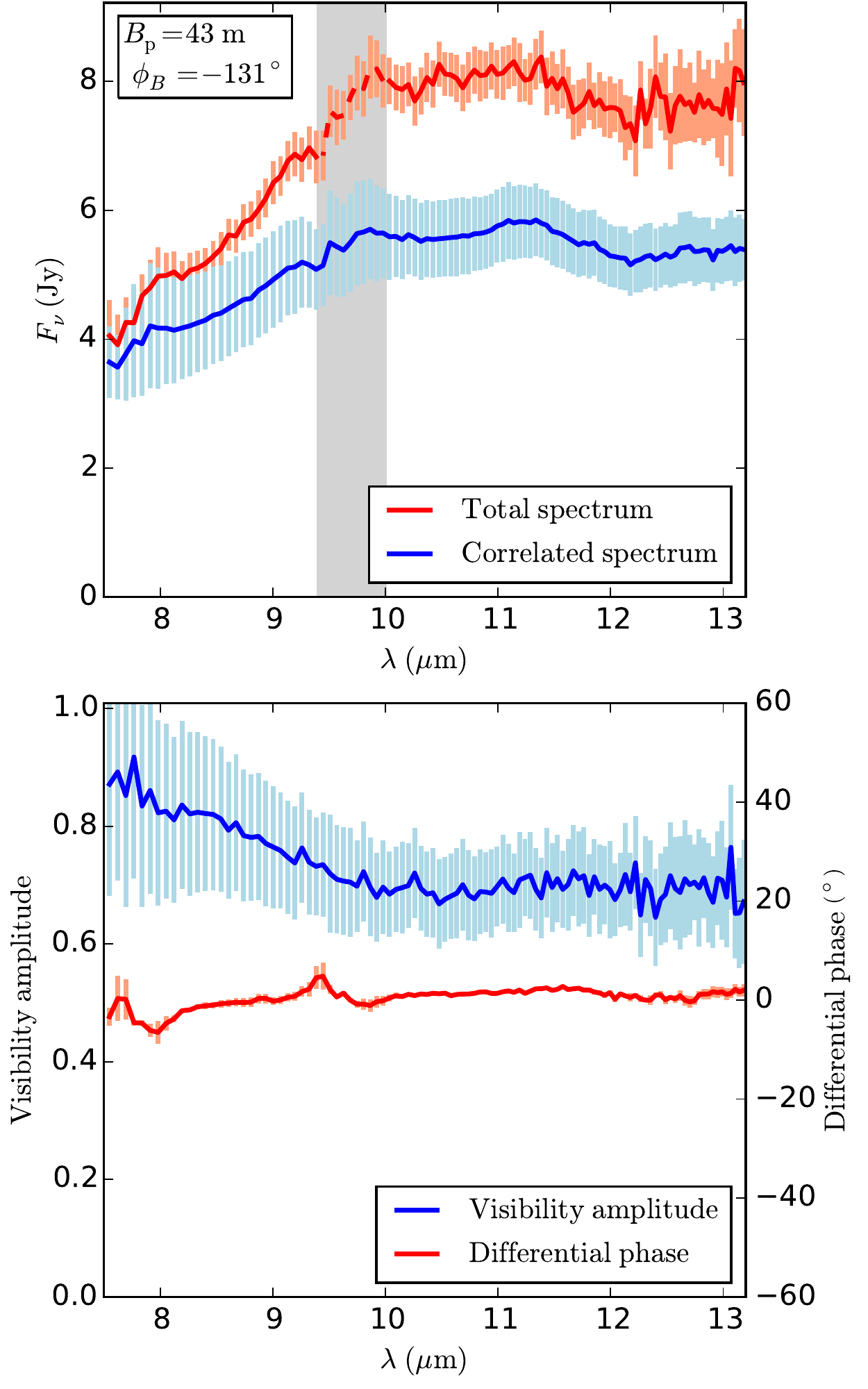}
		\caption{Calibrated MIDI spectra of T Tau N taken on 2004 October 31. Top panel: Total (red) and correlated (blue) spectrum. Gray shading indicates the region affected by telluric ozone absorption. Bottom panel: Visibility amplitude (blue) and differential phase (red).} 
		\label{fig:spectra}
	\end{figure}
	
	Disks around pre-main sequence stars are the places where planetary systems are born. Circumstellar disks, consisting of $\sim 99\%$ gas and $\sim 1\%$ dust, are formed when an interstellar cloud core gravitationally collapses into a protostar. Initially the disk is embedded in the collapsing envelope, then along with the central star it becomes visible as the core disperses. Meanwhile the disk material is being accreted onto the central star, while angular momentum is transported outwards. Later, as the accretion rate decreases, the disk becomes passive, as the main source of heating is radiation by the central star.
	In parallel, the gas and dust content in the disk undergo substantial physical and chemical processing: dust grains coagulate and grow, volatile material freezes out to the grains, complex chemical reactions occur on the surface of the dust grains. Due to the radial temperature gradient, different processes work at different regions in the disk. Planet formation also shapes the disk structure by depleting material by accretion, inducing asymmetries, and carving gaps into the disk. After approximately ten million years most of the gas content of the disk disperses, mainly by photoevaporation driven by the central star, leaving a remnant debris disk behind (\citealp[for dust dissipation see, e.g.,][]{Haisch2001,Hernandez2007}, for gas dissipation we refer to \citealp{Fedele2010}). Photoevaporation is not always the main mechanism of disk dissipation, as disk winds and accretion can also play significant roles \citep{Ercolano2017}. Disks are thought to evolve mostly steadily, interrupted by short-lived (a few hundred years), recurring outbursts, when the accretion rate is much higher than normal. General reviews on circumstellar disks can be found in, for example, \citet{Williams2011}, \citet{Dullemond2010}, and \citet{Sicilia-Aguilar2016}.
	
	Circumstellar disks have quite diverse morphology and their appearance is largely dependent on the wavelength in which one chooses to observe them. In the optical and in the near-infrared (near-IR) one can observe the disk by the scattered stellar light. From the mid-infrared (mid-IR) to the millimeter wavelengths disks are seen by their thermal emission. Generally, disks do not have a definite outer edge, and the spatial distribution of the gas (along with the small grains) can be largely different from the distribution of the larger grains. Recent theoretical work \citep{Birnstiel2014} and observations \citep{Andrews2012,deGregorio-Monsalvo2013}, however, indicate that at millimeter wavelengths there may be a well defined disk edge at some hundreds of au radii.
	As an example, millimeter continuum imaging of HD 169142, \citep{Fedele2017} complemented with near-IR scattered light observations \citep{Pohl2017}, show a sharp outer disk edge at $83$~au, indicating radial drift of large grains. Sharp features in the density distribution can be also present at the outer edge of the dead-zone \citep{Flock2015,Isella2016}.
	At mid-IR wavelengths the disk emission originates in the inner regions, where the temperature is sufficiently high. 
	To characterize the size of the emitting region at these wavelengths, it is common to define the half-light radius, encircling $50\%$ of the integrated emission. Apparent disk sizes are in the range of $0.1 - 100$~au, depending on the wavelength, because at larger wavelengths the emitting region is also larger \citep{Williams2011}. At the typical distance of the nearby star forming regions ($\sim 150$~pc) the angular size of disks ranges from $1$~mas to $1\arcsec$. Therefore, to spatially resolve their radiation, sub-arcsecond scale observations are needed. Such high resolution could be achieved with several approaches, for example, using space telescopes (to get rid of the atmospheric seeing), or using ground-based telescopes with adaptive optics, speckle imaging, and interferometry. 
	
	Long-baseline infrared interferometry offers the possibility to achieve the angular resolution required to resolve the innermost regions ($0.1 - 10$~au) of the circumstellar disks, where planets form. Observations with the recently decommissioned Mid-infrared Interferometric Instrument \citep[MIDI,][]{midi} at the Very Large Telescope Interferometer (VLTI) in Chile provided a wealth of information about the structure of the protoplanetary disks and the spatial distribution of the dust species therein \citep[e.g.,][]{Boekel_silicate,Menu2015}. \citet{midi_reduc_Leinert} determined the mid-infrared sizes of the disks around seven Herbig Ae/Be stars. A number of studies provided a detailed analysis of individual Herbig Ae stars, like \citet{Preibisch2006} on HR 5999, \citet{diFolco2009} on AB Aur, \citet{Benisty2010} on HD 100546, \citet{Matter2014} on HD 139614, \citet{Jamialahmadi2015,Jamialahmadi2018} on MWC480, \citet{Matter2014,Matter2016c} on HD 139614, and \citet{Kreplin2016} on UX Ori. \citet{Fedele2008} presented a study of three intermediate-mass young stellar objects (YSOs) (HD 101412, HD 135344 B, and HD 179218), and suggested that these systems may form an evolutionary sequence from an earlier flared geometry to a flat disk.
	
	Further studies, based on MIDI data, focused on the geometry and structure of individual low-mass sources. Some studies applied simple modeling (either geometric models or the model of \citealp{Chiang1997}) to characterize the observed disks, like \citet{Abraham2006} on V1647 Ori, \citet{Quanz2006} on FU Ori, \citet{Ratzka2007} and \citet{Akeson2011} on TW Hya, \citet{Roccatagliata2011} on Haro 6-10, and \citet{Vural2012} on S CrA N.
	Radiative transfer modeling became a popular tool to interpret MIDI data, because it can give physically more realistic results, although an adequate dataset and a large amount of fine-tuning are needed for a proper analysis. Examples are \citet{Schegerer2008} on RY Tau, \citet{Schegerer2009} on a sample consisting of DR Tau, RU Lup, S CrA N, S CrA S, HD 72106, HBC 639, and GW Ori, \citet{Juhasz2012} on EX Lup, \citet{Mosoni2013} on V1647 Ori, \citet{Schegerer2013} on HD 142666, AS 205 N, and AS 205 S, \citet{Menu2014} on TW Hya, \citet{Scicluna2016} on VV CrA, and \citet{Brunngraber2016} on DR Tau. \citet{Varga2017} applied both geometric and radiative transfer modeling to study the variable silicate emission of DG Tau. The presence of gaps in the inner disks of several objects were also revealed \citep{Ratzka2007,Schegerer2009,Schegerer2013}. The structure of the transitional disk of TW Hya has long been a puzzle, as the results were inconsistent when using data obtained at different wavelengths \citep{Calvet2002,Ratzka2007,Andrews2012}. Recently, \citet{Menu2014} found a suitable solution, with an inner disk radius of $0.3-0.5$~au, using a large set of multi-wavelength interferometric observations. 
	
	Young multiple stellar systems were also valuable targets for MIDI observations. By determining disk geometries and orientations, cloud fragmentation and binary formation theories can be tested \citep{Ratzka2009,Roccatagliata2011,Scicluna2016}. Tidal truncation of disks was observed in the hierarchical triple system T Tau by \citet{Ratzka2008}. The authors presented a detailed study and determined the relative positions of the components of the southern binary T Tau Sab.
	
	MIDI data are also suitable for spectral analysis in the $8-13~\mu$m wavelength range. This spectral region is usually dominated by the silicate feature, emitted by silicate dust grains. The shape of this feature can be used as a tracer for grain growth and chemical processing. MIDI observations provided the first evidence that the spatial distribution of the dust species and also their properties are not homogeneous in protoplanetary disks; \citet{Boekel_silicate} found that the dust content of the inner disks of three Herbig Ae stars are highly crystallized, while the outer disks show a much lower crystalline fraction. Signs of dust evolution were later observed in low-mass systems as well \citep{Schegerer2008,Schegerer2009,Ratzka2009}. \citet{Varga2017}, however, found that the outer disk of DG Tau shows a crystalline silicate emission feature, while the inner disk shows amorphous absorption. 
	
	Multi-epoch interferometric sequences can be used to explore changes in the disk structure, enabling us to study the temporal physical processes and the dynamics of the inner disk \citep{Abraham2006}. Young eruptive stars are ideal candidates for such studies. Six months after the 2008 eruption of EX Lupi \citep{Abraham2009}, \citet{Juhasz2012} observed that the crystallinity in the silicate feature decreased, suggesting fast radial transport of crystals, for example, by stellar or disk wind. \citet{Mosoni2013} reported changes of the interferometric visibilities of V1647 Ori, indicating structural changes during the outburst. Mid-IR interferometric variability can be also detected in non-eruptive systems.  \citet{Brunngraber2016} observed changes in the size of the mid-IR emitting region of DR Tau. Recently, \citet{Varga2017} revealed that the large variations in the amplitude of the silicate feature in the spectrum of DG Tau originate from a disk region outside $1-3$~au radius. 
	
	Recently, two general atlases of homogeneously reduced MIDI observations of young stellar objects (YSOs) were published. \citet{Boley2013} conducted a survey  on a sample of $24$ intermediate- and high-mass YSOs. \citet{Menu2015} analyzed a sample of $64$ disks around intermediate-mass young stars, searching for links between the structure and evolutionary status of the disks. They found evidence that a fraction of group II (flat) Herbig disks also possess gaps, and they propose a new evolutionary scenario for Herbig Ae/Be disks, as an alternative to earlier schemes, like \citet{Waelkens1994}, \citet{Meeus2001}, and \citet{Maaskant2013}.

	Continuing this line of research, here we present a study on a sample of disks around low- and intermediate-mass pre-main sequence stars, consisting of $82$ objects observed with MIDI. We perform a homogeneous data processing of the MIDI data, allowing statistical analysis of the sample, in a similar manner to \citet{Menu2015}. The resulting data products (e.g., interferometric spectra, visibilities, size estimates), presented in the form of an atlas, are made publicly available for further analysis. We apply interferometric model fitting to determine disk sizes and explore the mid-IR size-luminosity relation. We also study the mid-IR variability of the disks and  perform a spectral analysis on the shape of the silicate feature. This work supports the preparation of science observations with the next generation mid-IR interferometric instrument, Multi-AperTure mid-Infrared SpectroScopic Experiment (MATISSE), the successor of MIDI at the VLTI \citep{Lopez2014_MATISSE,Matter2016a}.
	
	The structure of the paper is as follows. In Sect.~\ref{sec:obs_datared} we describe the sample, the observations, and the data reduction. In Sect.~\ref{sec:res} we show our results: the published atlas and the interferometric modeling. In Sect.~\ref{sec:discuss}, based on our findings, we study the multiple stellar systems, the mid-IR size-luminosity relation, the variability of the sources, and the silicate dust mineralogy. We also analyze the observability of our sources with MATISSE. Finally, in Sect.~\ref{sec:summ}, we summarize our results.

	%__________________________________________________________________
	
	\section{Observations and data reduction}
	\label{sec:obs_datared}
	
	\subsection{Sample description}
	\label{sec:sample}

	\begin{table*}
		\centering
		\caption{Overview of the sample. The distance ($d$) and spectral type are from literature, while the stellar luminosity ($L_\star$) and optical extinction ($A_V$) are from our SED fits. The coordinates were taken from the ESO archive. For FU Orionis-type young eruptive stars no photospheric spectral type can be obtained.}\label{tab:sample}
		{\tiny\begin{tabular}{rlr@{ }c@{ }lr@{ }c@{ }llr@{$\pm$}lrr@{$\pm$}lr@{$\pm$}ll}
				\hline\hline
				\#& Name & \multicolumn{3}{c}{RA (J2000)} & \multicolumn{3}{c}{Dec (J2000)} & Type & \multicolumn{2}{c}{$d$} & Spectral & \multicolumn{2}{c}{$L_\star$} & \multicolumn{2}{c}{$A_V$} & References\\
				& &\multicolumn{3}{c}{(h m s)}& \multicolumn{3}{c}{($^\circ$ $\arcmin$ $\arcsec$)} & &\multicolumn{2}{c}{(pc)}& type & \multicolumn{2}{c}{($L_\sun$)} & \multicolumn{2}{c}{ } & \\
				\hline
				1 & SVS 13A1 & $03$ & $29$ & $03.84$ & $+31$ & $16$ & $07.14$ & eruptive & \multicolumn{2}{c}{250} &  & \multicolumn{2}{c}{} & \multicolumn{2}{c}{} &a,1,2,3 \\ 
				2 & LkH$\alpha$ 330 & $03$ & $45$ & $48.34$ & $+32$ & $24$ & $12.10$ & TT & \multicolumn{2}{c}{250} & G2 & 8.0 & 1.2 & 2.2 & 0.2 & a,1,2,3 \\ 
				3 & RY Tau & $04$ & $21$ & $57.43$ & $+28$ & $26$ & $35.34$ & TT/HAe & 177 & 27 & K1 & 18.8 & 5.0 & 2.1 & 0.4 & b,1,2,3 \\ 
				4 & T Tau N & $04$ & $21$ & $59.47$ & $+19$ & $32$ & $06.76$ & TT & 139 & 6 & K0 & 9.5 & 2.3 & 1.5 & 0.4 & b,4,2,3 \\ 
				5 & T Tau S & $04$ & $21$ & $59.47$ & $+19$ & $32$ & $04.99$ & TT & 139 & 6 & K7 & \multicolumn{2}{c}{11\tablefootmark{a}} & \multicolumn{2}{c}{15\tablefootmark{a}} & b,5,4 \\ 
				6 & DG Tau & $04$ & $27$ & $04.70$ & $+26$ & $06$ & $16.67$ & TT & \multicolumn{2}{c}{140} & K6 & 1.98 & 0.39 & 1.8 & 0.4 & a,6,7,8 \\ 
				7 & Haro 6-10N & $04$ & $29$ & $23.72$ & $+24$ & $33$ & $00.61$ & TT & \multicolumn{2}{c}{140} & K5 & 0.23 & 0.04 & \multicolumn{2}{c}{13} & a,9,10 \\ 
				8 & Haro 6-10S & $04$ & $29$ & $23.72$ & $+24$ & $33$ & $00.61$ & TT & \multicolumn{2}{c}{140} & K5 & 5.42 & 0.28 & \multicolumn{2}{c}{9.1} & a,9,10,11 \\ 
				9 & HBC 393 & $04$ & $31$ & $34.08$ & $+18$ & $08$ & $04.88$ & eruptive & \multicolumn{2}{c}{140} &  & \multicolumn{2}{c}{} & \multicolumn{2}{c}{} &c,12,11 \\ 
				10 & HL Tau & $04$ & $31$ & $38.46$ & $+18$ & $13$ & $56.86$ & TT & \multicolumn{2}{c}{140} & K5 & 0.47 & 0.09 & 2.6 & 0.4 & a,1,13,14,15 \\ 
				11 & GG Tau Aab & $04$ & $32$ & $30.38$ & $+17$ & $31$ & $40.37$ & TT & \multicolumn{2}{c}{140} & M0 & 1.16 & 0.07 & 0.8 & 0.1 & d,16,3 \\ 
				12 & LkCa 15 & $04$ & $39$ & $17.77$ & $+22$ & $21$ & $02.45$ & TT & \multicolumn{2}{c}{140} & K5 & 1.09 & 0.11 & 1.4 & 0.1 & d,17,11,3 \\ 
				13 & DR Tau & $04$ & $47$ & $06.01$ & $+16$ & $58$ & $42.02$ & TT & 207 & 13 & K5 & 4.00 & 0.62 & 1.5 & 0.3 & b,1,18,19 \\ 
				14 & UY Aur B & $04$ & $51$ & $47.36$ & $+30$ & $47$ & $12.52$ & TT & 151 & 9 & M2 & 0.24 & 0.05 & 2.0 & 0.2 & b,16,20,21,22 \\ 
				15 & UY Aur A & $04$ & $51$ & $47.38$ & $+30$ & $47$ & $12.73$ & TT & 151 & 9 & M0 & 0.67 & 0.17 & 0.46 & 0.08 & b,16,20,22 \\ 
				16 & GM Aur & $04$ & $55$ & $11.04$ & $+30$ & $21$ & $59.22$ & TT & \multicolumn{2}{c}{140} & K3 & 0.67 & 0.05 & 0.35 & 0.08 & a,1,3,23 \\ 
				17 & AB Aur & $04$ & $55$ & $45.83$ & $+30$ & $33$ & $04.39$ & HAe & 153 & 10 & A0 & 64.2 & 6.0 & 0.64 & 0.02 & b,1 \\ 
				18 & SU Aur & $04$ & $55$ & $59.38$ & $+30$ & $34$ & $00.95$ & TT/HAe & 142 & 12 & G2 & 9.0 & 1.8 & 1.0 & 0.2 & b,1,2,3 \\ 
				19 & MWC 480 & $04$ & $58$ & $46.25$ & $+29$ & $50$ & $36.82$ & HAe & 142 & 7 & A5 & 21.7 & 4.6 & 0.35 & 0.09 & b,1 \\ 
				20 & UX Ori & $05$ & $04$ & $30.02$ & $-03$ & $47$ & $13.74$ & HAe & 346 & 34 & A4 & 9.9 & 1.1 & 0.36 & 0.06 & b,1,24 \\ 
				21 & CO Ori & $05$ & $27$ & $38.33$ & $+11$ & $25$ & $39.25$ & HAe & 435 & 93 & F7 & 62 & 12 & 2.1 & 0.2 & b,25 \\ 
				22 & GW Ori & $05$ & $29$ & $08.39$ & $+11$ & $52$ & $12.68$ & TT/HAe & 469 & 102 & G5 & 77 & 20 & 1.3 & 0.3 & b,26,27 \\ 
				23 & MWC 758 & $05$ & $30$ & $27.60$ & $+25$ & $19$ & $57.14$ & HAe & 151 & 9 & A8 & 19.2 & 5.7 & 0.7 & 0.2 & b,1,28 \\ 
				24 & NY Ori & $05$ & $35$ & $35.89$ & $-05$ & $12$ & $24.98$ & eruptive & \multicolumn{2}{c}{414} & G6 & 4.4 & 1.0 & 1.6 & 0.3 & e,29 \\ 
				25 & CQ Tau & $05$ & $35$ & $58.53$ & $+24$ & $44$ & $53.20$ & HAe & 160 & 7 & F5 & 6.80 & 0.89 & 1.11 & 0.10 & b,30 \\ 
				26 & V1247 Ori & $05$ & $38$ & $05.29$ & $-01$ & $15$ & $21.53$ & HAe & 320 & 27 & A7 & 14.9 & 3.4 & 0.4 & 0.2 & b,24 \\ 
				27 & V883 Ori & $05$ & $38$ & $18.10$ & $-07$ & $02$ & $26.34$ & eruptive & \multicolumn{2}{c}{460} &  & \multicolumn{2}{c}{} & \multicolumn{2}{c}{} &f,31,2 \\ 
				28 & MWC 120 & $05$ & $41$ & $02.28$ & $-02$ & $42$ & $59.98$ & HAe & 420 & 49 & B9 & 225.6 & 7.8 & 0.30 & 0.05 & b,30,3 \\ 
				29 & FU Ori & $05$ & $45$ & $22.42$ & $+09$ & $04$ & $13.30$ & eruptive & 353 & 66 &  & \multicolumn{2}{c}{} & \multicolumn{2}{c}{} &b,32,2,7 \\ 
				30 & V1647 Ori & $05$ & $46$ & $13.22$ & $+00$ & $06$ & $04.00$ & eruptive & \multicolumn{2}{c}{400} &  & \multicolumn{2}{c}{} & \multicolumn{2}{c}{} &e,33,34,35 \\ 
				31 & V900 Mon & $06$ & $57$ & $22.22$ & $-08$ & $23$ & $17.20$ & eruptive & 1100 & 120 &  & \multicolumn{2}{c}{} & \multicolumn{2}{c}{} &g,36 \\ 
				32 & Z CMa & $07$ & $03$ & $43.19$ & $-11$ & $33$ & $06.19$ & eruptive & \multicolumn{2}{c}{1050} &  & \multicolumn{2}{c}{} & \multicolumn{2}{c}{} &h,30,24 \\ 
				33 & V646 Pup & $07$ & $50$ & $35.62$ & $-33$ & $06$ & $23.08$ & eruptive & 1800 & 360 &  & \multicolumn{2}{c}{} & \multicolumn{2}{c}{} &i,37,2,38 \\ 
				34 & HD 72106 & $08$ & $29$ & $34.75$ & $-38$ & $36$ & $20.16$ & HAe & 288 & 119 & A0 & 38 & 12 & 0.1 & 0.1 & j,30,24 \\ 
				35 & CR Cha & $10$ & $59$ & $07.20$ & $-77$ & $01$ & $40.69$ & TT & 188 & 8 & K4 & 4.91 & 0.97 & 1.5 & 0.4 & b,6,2 \\ 
				36 & TW Hya & $11$ & $01$ & $51.92$ & $-34$ & $42$ & $16.99$ & TT & 59.5 & 0.9 & K6 & 0.40 & 0.05 & 0.3 & 0.2 & b,1 \\ 
				37 & DI Cha & $11$ & $07$ & $20.81$ & $-77$ & $38$ & $07.33$ & TT & 198 & 9 & G2 & 16.6 & 5.2 & 2.5 & 0.4 & b,6,39,40 \\ 
				38 & Glass I & $11$ & $08$ & $15.47$ & $-77$ & $33$ & $53.82$ & TT & 179 & 16 & K7 & 3.14 & 0.53 & 1.5 & 0.3 & k,41,3,2 \\ 
				39 & Sz 32 & $11$ & $09$ & $53.49$ & $-76$ & $34$ & $25.61$ & TT & 179 & 16 & K5 & 13.3 & 5.3 & 9 & 1 & k,6,2,3 \\ 
				40 & WW Cha & $11$ & $10$ & $00.44$ & $-76$ & $34$ & $58.33$ & TT & 179 & 16 & K5 & 6.2 & 1.5 & 3.9 & 0.5 & k,6,39 \\ 
				41 & CV Cha & $11$ & $12$ & $27.97$ & $-76$ & $44$ & $22.60$ & TT & 199 & 9 & G9 & 7.9 & 2.1 & 1.7 & 0.4 & b,1 \\ 
				42 & DX Cha & $12$ & $00$ & $05.10$ & $-78$ & $11$ & $34.55$ & HAe & 104 & 3 & A4 & 47 & 11 & 0.7 & 0.1 & b,30,24 \\ 
				43 & DK Cha & $12$ & $53$ & $16.94$ & $-77$ & $07$ & $10.63$ & embHAe & 181 & 13 & F0 & 71 & 28 & 12.0 & 0.5 & k,42,2,3 \\ 
				44 & HD 135344B & $15$ & $15$ & $48.45$ & $-37$ & $09$ & $17.10$ & HAe & 156 & 11 & F8 & 10.7 & 1.9 & 0.31 & 0.08 & b,1,43,44,24 \\ 
				45 & HD 139614 & $15$ & $40$ & $46.38$ & $-42$ & $29$ & $54.78$ & HAe & 131 & 5 & A7 & 9.2 & 2.8 & 0.3 & 0.2 & b,45,24 \\ 
				46 & HD 141569 & $15$ & $49$ & $57.77$ & $-03$ & $55$ & $16.32$ & HAe & 111 & 5 & A0 & 23.0 & 8.3 & 0.3 & 0.2 & b,1 \\ 
				47 & HD 142527 & $15$ & $56$ & $41.91$ & $-42$ & $19$ & $23.99$ & HAe & 156 & 6 & F6 & 21.5 & 3.6 & 0.8 & 0.2 & b,1,24,28 \\ 
				48 & RU Lup & $15$ & $56$ & $42.17$ & $-37$ & $49$ & $16.21$ & TT & 169 & 9 & K7 & 1.89 & 0.22 & 0.2 & 0.2 & b,1,46 \\ 
				49 & HD 143006 & $15$ & $58$ & $36.90$ & $-22$ & $57$ & $14.72$ & TT & 166 & 10 & G8 & 5.46 & 0.55 & 1.16 & 0.04 & b,1,2,28 \\ 
				50 & EX Lup & $16$ & $03$ & $05.50$ & $-40$ & $18$ & $25.06$ & eruptive & \multicolumn{2}{c}{140} & M0 & 0.47 & 0.06 & 0.2 & 0.2 & a,1,38,2 \\ 
				51 & HD 144432 & $16$ & $06$ & $57.97$ & $-27$ & $43$ & $09.73$ & HAe & 253 & 95 & F0 & 43 & 10 & 0.3 & 0.2 & j,1 \\ 
				52 & V856 Sco & $16$ & $08$ & $34.32$ & $-39$ & $06$ & $18.36$ & HAe & 208 & 38 & A7 & 95 & 18 & 0.6 & 0.1 & j,45,28 \\ 
				53 & AS 205 N & $16$ & $11$ & $31.34$ & $-18$ & $38$ & $25.91$ & TT & \multicolumn{2}{c}{160} & K5 & 5.7 & 1.3 & 2.6 & 0.5 & a,1,47,48,49 \\ 
				54 & AS 205 S & $16$ & $11$ & $31.34$ & $-18$ & $38$ & $25.91$ & TT & \multicolumn{2}{c}{160} & M3 & 2.00 & 0.29 & 2.5 & 0.4 & a,1,47,48,49 \\ 
				55 & DoAr 20 & $16$ & $25$ & $56.08$ & $-24$ & $20$ & $47.29$ & TT & \multicolumn{2}{c}{120} & K0 & 1.30 & 0.33 & 2.1 & 0.4 & a,1,2 \\ 
				56 & V2246 Oph & $16$ & $26$ & $03.07$ & $-24$ & $23$ & $35.70$ & TT & \multicolumn{2}{c}{122} & K0 & 8.4 & 2.7 & 5.0 & 0.6 & a,1 \\ 
				57 & HBC 639 & $16$ & $26$ & $23.35$ & $-24$ & $21$ & $01.98$ & TT & \multicolumn{2}{c}{150} & K0 & 5.5 & 1.6 & 5.2 & 0.6 & d,41 \\ 
				58 & DoAr 25 & $16$ & $26$ & $23.65$ & $-24$ & $43$ & $13.62$ & TT & \multicolumn{2}{c}{150} & K5 & 2.03 & 0.47 & 3.3 & 0.4 & d,50,2,3 \\ 
				59 & Elias 24 & $16$ & $26$ & $24.21$ & $-24$ & $16$ & $14.88$ & TT & \multicolumn{2}{c}{150} & K6 & 5.54 & 0.34 & \multicolumn{2}{c}{9.3} & d,51,2,38 \\ 
				60 & SR 24N & $16$ & $26$ & $58.48$ & $-24$ & $45$ & $33.91$ & TT & \multicolumn{2}{c}{140} & M0 & 2.17 & 0.02 & \multicolumn{2}{c}{7.5} & l,52,19,47 \\ 
				61 & SR 24S & $16$ & $26$ & $58.51$ & $-24$ & $45$ & $35.39$ & TT & \multicolumn{2}{c}{140} & K2 & 2.05 & 0.77 & 4.9 & 0.9 & l,19 \\ 
				
				\hline
			\end{tabular}
		}
	\end{table*}
	
	\addtocounter{table}{-1}
	\begin{table*}
		\centering
		\caption{continued.}
		{\tiny\begin{tabular}{rlr@{ }c@{ }lr@{ }c@{ }llr@{$\pm$}lrr@{$\pm$}lr@{$\pm$}ll}
				\hline\hline
				\#& Name & \multicolumn{3}{c}{RA (J2000)} & \multicolumn{3}{c}{Dec (J2000)} & Type & \multicolumn{2}{c}{$d$} & Spectral & \multicolumn{2}{c}{$L_\star$} & \multicolumn{2}{c}{$A_V$} & References\\
				& &\multicolumn{3}{c}{(h m s)}& \multicolumn{3}{c}{($^\circ$ $\arcmin$ $\arcsec$)} & &\multicolumn{2}{c}{(pc)}& type & \multicolumn{2}{c}{($L_\sun$)} & \multicolumn{2}{c}{ } & \\
				\hline
				62 & Elias 29 & $16$ & $27$ & $09.48$ & $-24$ & $37$ & $18.05$ & TT & \multicolumn{2}{c}{120} &  & \multicolumn{2}{c}{41\tablefootmark{a}} & \multicolumn{2}{c}{9.8\tablefootmark{a}} & a,50 \\ 
				63 & SR 21A & $16$ & $27$ & $10.35$ & $-24$ & $19$ & $12.18$ & TT/HAe & \multicolumn{2}{c}{150} & F4 & 15.2 & 5.8 & 6.3 & 0.5 & d,6,2,3 \\ 
				64 & IRS 42 & $16$ & $27$ & $21.42$ & $-24$ & $41$ & $43.37$ & TT & \multicolumn{2}{c}{120} & K7 & \multicolumn{2}{c}{4.2\tablefootmark{a}} & \multicolumn{2}{c}{9.8\tablefootmark{a}} & m,50 \\ 
				65 & IRS 48 & $16$ & $27$ & $37.09$ & $-24$ & $30$ & $35.75$ & HAe & \multicolumn{2}{c}{120} & A0 & 12.2 & 4.7 & 10.2 & 0.3 & a,1,44,23,2,3 \\ 
				66 & V2129 Oph & $16$ & $27$ & $40.26$ & $-24$ & $22$ & $02.28$ & TT & \multicolumn{2}{c}{150} & K5 & 3.24 & 0.67 & 1.8 & 0.4 & d,6 \\ 
				67 & Haro 1-16 & $16$ & $31$ & $33.47$ & $-24$ & $27$ & $35.89$ & TT & \multicolumn{2}{c}{120} & K3 & 1.22 & 0.34 & 2.2 & 0.5 & a,1,3,14 \\ 
				68 & V346 Nor & $16$ & $32$ & $32.21$ & $-44$ & $55$ & $29.57$ & eruptive & \multicolumn{2}{c}{500} &  & \multicolumn{2}{c}{} & \multicolumn{2}{c}{} &n,12,20,2,3 \\ 
				69 & HD 150193 & $16$ & $40$ & $17.91$ & $-23$ & $53$ & $45.67$ & HAe & 145 & 6 & B9 & 91.8 & 1.1 & \multicolumn{2}{c}{2.0} & b,1,3 \\ 
				70 & AS 209 & $16$ & $49$ & $15.30$ & $-14$ & $22$ & $07.72$ & TT & 127 & 14 & K4 & 2.23 & 0.34 & 1.3 & 0.3 & b,1,2,3 \\ 
				71 & AK Sco & $16$ & $54$ & $44.78$ & $-36$ & $53$ & $18.49$ & HAe & 144 & 5 & F5 & 7.8 & 1.8 & 0.5 & 0.1 & b,26,24 \\ 
				72 & HD 163296 & $17$ & $56$ & $21.33$ & $-21$ & $57$ & $22.18$ & HAe & 122 & 15 & A1 & 34.3 & 7.1 & 0.33 & 0.08 & j,1 \\ 
				73 & HD 169142 & $18$ & $24$ & $29.85$ & $-29$ & $46$ & $49.04$ & HAe & 117 & 4 & A5 & 8.3 & 1.4 & 0.35 & 0.07 & b,1,44,28 \\ 
				74 & VV Ser & $18$ & $28$ & $47.87$ & $+00$ & $08$ & $40.13$ & HAe & \multicolumn{2}{c}{330} & A5 & 60 & 23 & 3.5 & 0.4 & o,1 \\ 
				75 & SVS20N & $18$ & $29$ & $57.76$ & $+01$ & $14$ & $07.19$ & TT & \multicolumn{2}{c}{415} & M5 & 13.75 & 0.56 & \multicolumn{2}{c}{19} & a,52,1,53,54 \\ 
				76 & SVS20S & $18$ & $29$ & $57.77$ & $+01$ & $14$ & $05.46$ & embHAe & \multicolumn{2}{c}{415} & A0 & 1799 & 682 & 22.4 & 0.2 & a,53,54 \\ 
				77 & S CrA S & $19$ & $01$ & $08.74$ & $-36$ & $57$ & $20.48$ & TT & \multicolumn{2}{c}{130} & M0 & 1.12 & 0.02 & \multicolumn{2}{c}{2.0} & a,55,56,57,58,49 \\ 
				78 & S CrA N & $19$ & $01$ & $08.75$ & $-36$ & $57$ & $19.98$ & TT & \multicolumn{2}{c}{130} & K3 & 2.51 & 0.29 & 1.99 & 0.07 & a,55,56,57,58,49 \\ 
				79 & T CrA & $19$ & $01$ & $58.60$ & $-36$ & $57$ & $51.34$ & HAe & \multicolumn{2}{c}{130} & F0 & 2.77 & 0.19 & \multicolumn{2}{c}{2.7} & a,1,14,2,38 \\ 
				80 & VV CrA NE & $19$ & $03$ & $06.72$ & $-37$ & $12$ & $49.21$ & TT & \multicolumn{2}{c}{130} & K7 & 3.16 & 0.24 & 10.2 & 0.3 & a,52,59,60 \\ 
				81 & VV CrA SW & $19$ & $03$ & $06.92$ & $-37$ & $12$ & $49.61$ & TT & \multicolumn{2}{c}{130} & M0 & 1.19 & 0.15 & 1.73 & 0.08 & a,1,60 \\ 
				82 & HD 179218 & $19$ & $11$ & $11.30$ & $+15$ & $47$ & $16.44$ & HAe & 293 & 31 & A0 & 280 & 22 & \multicolumn{2}{c}{1.3} & b,1,43 \\ 
				
				\hline
			\end{tabular}
			\tablefoot{\tablefoottext{a}{Luminosity and extinction values were taken from the literature.}}
			\tablebib{Distance references: (a) \citet{Salyk2013}; (b) \citet{GaiaDR1}; (c) \citet{Connelley2008}; (d) \citet{Mohanty2013}; (e) \citet{Audard2014}; (f) \citet{Molinari1993}; (g) \citet{Reipurth2012}; (h) \citet{Manoj2006}; (i) \citet{Reipurth2002}; (j) \citet{Hipparcos1997}; (k) \citet{Voirin2017}; (l) \citet{Beck2012}; (m) \citet{vanKempen2009}; (n) \citet{Lumsden2013}; (o) \citet{deLara1991}. Other references: (1) \citet{Salyk2013}; (2) \citet{Zacharias2004}; (3) \citet{Monet2003}; (4) \citet{Csepany2015}; (5) \citet{Koresko1997}; (6) \citet{Furlan2009}; (7) \citet{Droege2006}; (8) \citet{Kraus2011}; (9) \citet{Roccatagliata2011}; (10) \citet{Luhman2016}; (11) \citet{Ahn2012}; (12) \citet{Connelley2010}; (13) \citet{Ivanov2008}; (14) \citet{Zacharias2012}; (15) \citet{Abazajian2009}; (16) \citet{Kraus2009}; (17) \citet{Rebull2010}; (18) \citet{Nascimbeni2016}; (19) \citet{Henden2016}; (20) \citet{Skiff2014}; (21) \citet{WhiteGhez2001}; (22) \citet{Hioki2007}; (23) \citet{Fedorov2011}; (24) \citet{Fairlamb2015}; (25) \citet{Lazareff2017}; (26) \citet{Lopez-Martinez2015}; (27) \citet{Dolan2002}; (28) \citet{Bourges2014}; (29) \citet{DaRio2010}; (30) \citet{Chen2016}; (31) \citet{Fang2013}; (32) \citet{Pueyo2012}; (33) \citet{Acosta-Pulido2007}; (34) \citet{Abraham2006}; (35) \citet{Aspin2009}; (36) \citet{Reipurth2012}; (37) \citet{Samus2017}; (38) \citet{Denis2005}; (39) \citet{Roser2008}; (40) \citet{Luhman2004}; (41) \citet{Sartori2003}; (42) \citet{Spezzi2008}; (43) \citet{Fedele2008}; (44) \citet{Maaskant2013}; (45) \citet{Manoj2006}; (46) \citet{Alcala2014}; (47) \citet{Herbig1995}; (48) \citet{Flesch2016}; (49) \citet{Prato2003}; (50) \citet{Evans2009}; (51) \citet{Dunham2015}; (52) \citet{Dunham2013}; (53) \citet{Ciardi2005}; (54) \citet{Flewelling2016}; (55) \citet{Patten1998}; (56) \citet{Vural2012}; (57) \citet{Mason2001}; (58) \citet{Forbrich2007}; (59) \citet{Avilez2017}; (60) \citet{Scicluna2016}.}
		}
	\end{table*}

	We collected all available MIDI science observations on low-mass young stars from 2003 to 2015 from the European Southern Observatory (ESO) science archive, where all data are now publicly available. MIDI data, which were obtained with the Phase-Referenced Imaging and Micro-arcsecond Astrometry fringe sensor unit \citep[PRIMA-FSU, ][]{Muller2014} as external fringe tracker (approximately $100$ observations), have been excluded from the present study, because of special requirements for data reduction. In addition, we extended the sample with intermediate-mass stars, adopting an effective stellar photospheric temperature of $10000$~K as upper limit. The spectral types and effective temperatures of the candidate sources were checked in the literature. The Herbig Ae stars in our sample were also included in \citet{Menu2015}, but here we re-evaluated them to ensure consistency between the results of these complementary studies, and validate our modeling. Additionally, we included all observed young eruptive FU Orionis (FUor) or EX Lupi (EXor) type stars. Our sample contains 82 sources, of which 45 are T Tauri stars, 11 are young eruptive stars, and 26 are Herbig Ae stars. 
	
	The source list is shown in Table~\ref{tab:sample}, including the basic stellar parameters. Our modeling needs the stellar luminosity as input parameter. In order to determine the luminosities for each source homogeneously, we fitted the optical to near-IR spectral energy distributions (SEDs). Therefore  we collected distances (partly from the Gaia DR1 catalog\footnote{The Gaia DR1 catalog can be found at \url{http://gea.esac.esa.int/archive/}.} \citealp{GaiaDR1}, partly from the literature), optical ($U$, $B$, $V$, $R$, $I$), and near-IR ($J$, $H$, $K$)  photometric data from the literature (see Table~\ref{tab:sample} for references). For binaries we collected data for the individual components, when available. We fitted the SEDs with Kurucz-models \citep{Castelli2004} using the SED fitter code by \citet{Robitaille2007}. Fluxes $H$ and $K$ were used only as upper limits in the fits, because YSOs commonly have excess emission over the photosphere at these wavelengths. The derived stellar luminosities as well as extinction values are listed in Table~\ref{tab:sample}. We used the known spectral type and optical extinction as priors in the fitting. SEDs for each object along with the best-fit models are shown in Fig.~\ref{fig:stellar_SED} in the Appendix~\ref{app:sed}. We note that some of our sources are young eruptive stars (e.g., FU Ori, V1647 Ori, Z CMa) in outburst. In these cases the SED fit is related to the emission of the outbursting accretion disk, rather than to the stellar photosphere. For a few further objects (T Tau S, Elias 29, IRS 42) we did not find enough photometric data for the fitting, thus we use luminosities from the literature in these cases. There are two main reasons for this: either the object is deeply embedded, so it has very low optical fluxes, or it is a binary and measurements are spoiled by source confusion.
	Since the photometric data points are not simultaneously measured, the intrinsic variability of the sources may introduce uncertainties in the derived parameters. 
	\label{sec:sedfit}
	
	\subsection{Observations}
	\label{sec:obs}
	
	MIDI \citep{midi} is a Michelson-type interferometer with a half-reflecting plate optical recombiner. It combines signal from two telescopes of the Very Large Telescope (VLT) array, using either two 8.2 m unit telescopes (UT) or two 1.8 m auxiliary telescopes (AT). The provided data are spectrally resolved in the $7.5-13~\mu$m wavelength range. The spectral resolution $R$ is $30$ with a prism as a dispersing element, or $260$ with a grism. The majority of our data were observed with the HIGH-SENS mode, which is suitable for fainter objects (some brighter objects in the sample were also observed with SCI-PHOT mode). 
	
	The measurements were taken at various projected baseline lengths ($B_p$) and position angles ($\phi_B$). The projected baseline length ranges from $10$~m to $130$~m in our sample. Table~\ref{tab:obs} in  Appendix~\ref{app:obs} summarizes the dates, the baseline configurations, and the ESO Program IDs for the observations. We mark low quality datasets, flagged by the data reduction pipeline as unreliable, by the observation flag in the table. The majority of the data were observed with UTs and a smaller portion with ATs. Overall, we have $627$ interferometric measurements corresponding to our sample from $222$ individual nights. 
	
	\subsection{Data reduction} 
	\label{sec:datared}
	
	Data reduction was performed in a similar way to \citet{Menu2015}.
	We reduced the interferometric data using the Expert Work Station (EWS) package 2.0\footnote{The EWS package can be downloaded from: \url{home.strw.leidenuniv.nl/~nevec/MIDI/index.html}.} , which is a common tool for MIDI data processing. It uses a coherent linear averaging method to obtain correlated fluxes and visibilities \citep{midi_reduc1,faintMIDI}. EWS routines are called from a python environment developed by \citet{Menu2015}. Instead of visibility calibration the direct flux calibration method was applied, as described in \citet{faintMIDI}. In this scheme the usage of the less accurate total spectrum measurements can be avoided in the calibration of the correlated spectrum of the target. The higher uncertainty of the total spectrum is caused by the strong and variable instrumental and sky background. Background subtraction was performed by chopping, which can leave some residual signal. The correlated spectrum, however, is less affected by the background, because the incoherent background is efficiently canceled out by subtracting the two interferometric signals on the detector.
	
	The final product of the interferometric calibration is the calibrated correlated spectrum, which is computed by dividing the observed (raw) correlated spectrum of the target by $T_\mathrm{corr,\nu}$, the transfer function.  The transfer function, $T_\mathrm{corr,\nu}$, following \cite{Menu2015}, is expressed as
	\begin{equation}
	T_\mathrm{corr,\nu}=\frac{F^\mathrm{cal,raw}_\mathrm{corr,\nu}}{F^\mathrm{cal}_\mathrm{tot,\nu} V^\mathrm{cal}_\nu},
	\label{transfer_func}
	\end{equation}
	where $F^\mathrm{cal,raw}_\mathrm{corr,\nu}$ is the observed (raw) correlated spectrum of the calibrator, $F^\mathrm{cal}_\mathrm{tot,\nu}$ is the known total spectrum of the calibrator, and $V^\mathrm{cal}_\nu$ is the visibility of the calibrator (calculated from its known diameter). The transfer function has two components: the atmospheric transparency and the system response of the whole optical system. Following \citet{Menu2015}, we determined $T_\mathrm{corr,\nu}$ for each night as a time-dependent quantity by using all suitable calibrator observations. In addition, $T_\mathrm{corr,\nu}$ is dependent on the airmass, which we also take into account. Calibration of the total spectrum is performed similarly to that of the correlated spectrum, by applying $V^\mathrm{cal}_\nu \equiv 1$ in Eq. (\ref{transfer_func}). 
	
	The datasets after the whole reduction process consist of the calibrated total spectrum ($F_\mathrm{tot,\nu}$), and the calibrated correlated spectrum ($F_\mathrm{corr,\nu}$) of the target, both in the same wavelength range with the same spectral resolution. Visibility is defined as $V_\nu =  F_\mathrm{corr,\nu} / F_\mathrm{tot,\nu}$. The correlated spectrum, in the case of circularly or elliptically symmetric objects, can be interpreted as the spectrum of the inner region of an object. The size of the corresponding region is related to the fringe spacing or the full width at half maximum (FWHM) of a fringe ($\theta$), often taken as a measure for the resolution (see Table \ref{tab:obs}). One should, however, be aware that the baseline achieves this resolution only in one direction. Data in the $9.4-10.0~\mu$m wavelength range can be heavily affected by the atmospheric ozone absorption feature, increasing systematic errors and reducing the signal-to-noise ratio (SNR), especially in the total spectra. This should be taken into account during further data analysis.
	
	The error budget is part of the data reduction. As our sample consists of fainter objects than the sample of \citet{Menu2015}, generally we have somewhat lower SNR. The most significant error in HIGH-SENS mode is due to the atmospheric fluctuations, which can affect the total spectrum as much as $10-15\%$ in a time interval as short as a few minutes \citep{midi_reduc1}. Consequently, the visibility is uncertain at a similar level. This uncertainty can be constrained if there exists multiple calibrator observations close in time. Correlated spectra are affected by correlation losses in the interferometer, but still have considerably higher accuracy. The errors are mainly systematic in nature, caused by the uncertainty in the determination of the transfer function. The absolute level of both the correlated and total spectra can be determined less accurately, while the spectral shape is more reliable. Thus, when analyzing spectral shapes, systematic errors can be largely disregarded. Hence it is desirable to determine random and systematic errors separately. We calculate random uncertainties as the moving standard deviation of the detrended data (see Appendix~\ref{app:spec_class} for more details). \citet{Varga2017} found that the error bars provided by the EWS reduction pipeline are generally reasonable, and the typical total uncertainties (including random and systematic errors) are $\sim 6\%$ in the correlated spectra and $10-20\%$ in the total spectra (for a $\sim 4$~Jy bright source). We note that when there are bad atmospheric conditions the uncertainties can get much larger. As a rule of thumb the limiting correlated fluxes of MIDI with the UTs can reach values as low as $\sim 0.05-0.1$~Jy, while the limit for total fluxes is $\sim 0.2$~Jy (see, e.g., TW Hya). At low fluxes results become biased and a careful treatment is needed, as developed by \citet{faintMIDI} and also applied by \citet{Menu2015}. We used this approach in the data reduction for all sources in our sample. For more information about the calibration issues we refer to \citet{faintMIDI}.
	
	\section{Results}
	\label{sec:res}
	
	%%%%%%%%%%%%%%%%%%%%%%%%%%%%%%%%%%%%%%%%
	% main atlas figure: uv-plot, gaussian sizes, interferometric model fit
	\begin{figure}
		\centering
		% for a referee version, use a width of 0.5\columnwidth
		\includegraphics[width=0.72\columnwidth]{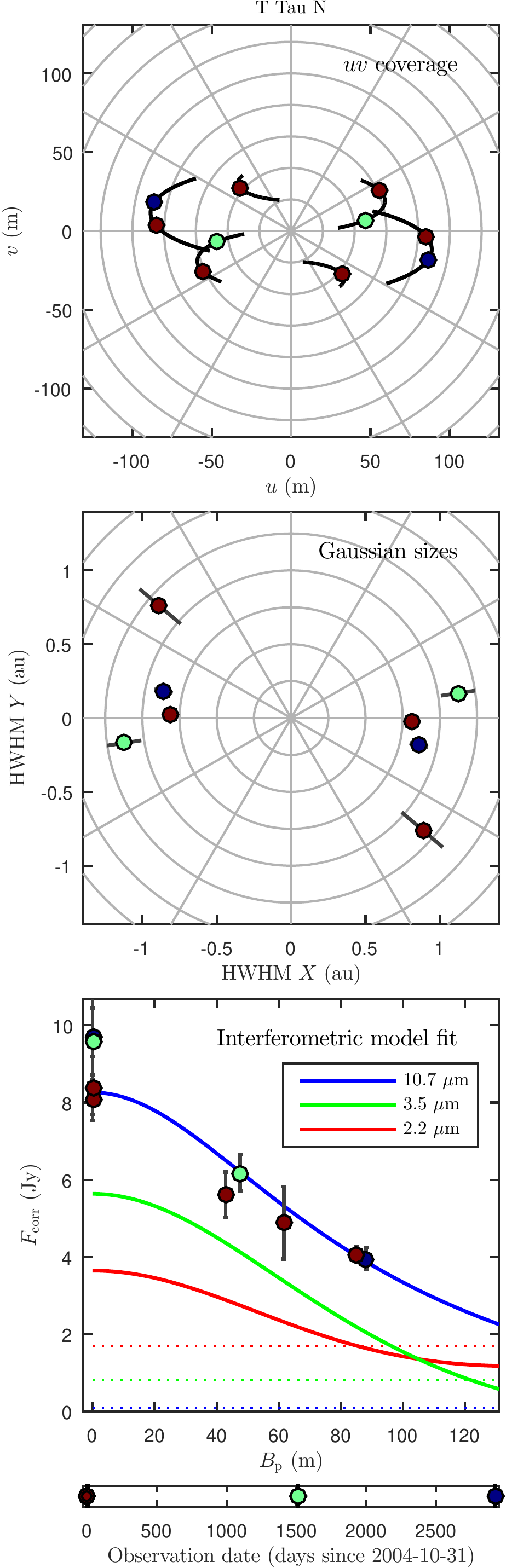}
		\caption{Figure outputs of the atlas for T Tau N as an example. Top: $uv$-coverage plot. Middle: Gaussian size diagram. The plot shows the physical sizes of the disk (HWHM, in au, see Eq.~\ref{eq:gaussian}) individually measured for each observation corresponding to different baseline angles. Bottom: Results of the interferometric modeling to the correlated and total fluxes (shown at $0$~m baseline) at $10.7~\mu$m as a function of baseline length. The blue solid line shows continuous model fit. Red and green lines show the correlated fluxes extrapolated from the continuous model to the K and L bands, respectively (see Sections \ref{sec:int_model} and \ref{sec:matisse} for more details). Dotted lines show the estimates for the unresolved stellar emission. Data points on each subplot are color-coded for observing date. For the other sources in the sample see Fig.~\ref{fig:atlas_fit_app} in Appendix~\ref{app:atlasfig}. 
		} 
		\label{fig:int_model}
	\end{figure}
	
	\subsection{The atlas}
	\label{sec:atlas}
	
	The source list is shown in Table~\ref{tab:sample} and the individual observations are given in Table~\ref{tab:obs}. The results of the basic data reduction, consisting of correlated spectra, total spectra, visibilities, and differential phases are presented in the online material. As an example, the data for T Tau N are plotted in Fig.~\ref{fig:spectra}. The error bars denote the total calibration uncertainty, including both the random and systematic errors. The wavelength region affected by the atmospheric ozone feature is highlighted as a gray stripe on the upper panel. Although the data products contain values between $5.2$ and $13.7\ \mu$m, due to atmospheric transmission the wavelengths below $7.5\ \mu$m and above $13\ \mu$m are generally not useful for scientific analysis. The $uv$-coverage of each source is shown in Fig.~\ref{fig:atlas_fit_app} in  Appendix \ref{app:atlasfig}. One example is shown in Fig.~\ref{fig:int_model}.
	
	For a few spectra, especially in the low flux ($\lesssim 1$~Jy) regime, the data quality was too low to perform a successful reduction, in some other cases the results are flagged by the pipeline as low quality. The main reason for these low quality results are the inadequate atmospheric conditions. We will not analyze these data further ($8\%$ of the interferometric measurements) in the paper. The observations have also been checked individually to ensure that the source designation corresponds to the actual pointing of the telescopes, which is especially important for binaries. The problematic datasets are listed in Table~\ref{tab:obs} for completeness, and are marked. The results presented here were carefully checked and are correct to the best of our knowledge. All these material and some other additional plots (e.g., SEDs and diagnostic plots) 
	can be found on the project website at \url{http://www.konkoly.hu/MIDI_atlas}. Reduced spectra (correlated and total spectra, visibilities, and differential phases) will be also made available at VizieR.\footnote{\url{http://vizier.u-strasbg.fr/}.} 
	
	\subsection{Gaussian sizes}
	
	In order to give a first impression of the spatial extent as well as azimuthal structure of the sources, we calculated the sizes for each observation assuming that the brightness distribution is Gaussian. Gaussian sizes at a specific wavelength ($10.7\ \mu$m in our case for easy comparison with the literature)\footnote{We used a $0.33\ \mu$m wide window around this wavelength over which we averaged the visibilities.} are calculated according to the formula
	\begin{equation}
	\mathrm{HWHM} = \frac{ \sqrt[]{\log(2)} } {\pi } \cdot \frac{\sqrt[]{-\log(V_{\nu})}}{B_\mathrm{p}/\lambda} \cdot 3600\cdot 180/\pi\cdot d,
	\label{eq:gaussian}
	\end{equation}
	where HWHM is the half width at half maximum of the real-space Gaussian brightness distribution, and $d$ is the distance in parsecs. The results are presented in polar plots in Fig.~\ref{fig:atlas_fit_app} for each source, and also in Fig.~\ref{fig:int_model} for T Tau N as an example. The radial range is au scale, and the azimuthal angle corresponds to the baseline position angle. Every point is plotted twice because of the $180\degr$ ambiguity of the position angle. The formal uncertainties of the sizes are overplotted. 
	
	A few objects in the sample have such a good baseline angle coverage that the disk inclination can be well constrained (e.g., GW Ori, EX Lup, RU Lup, V346 Nor). The disk of GW Ori has been resolved also in the millimeter with the Submillimeter Array \citep[SMA, ][]{Fang2017} and with the Atacama Large Millimeter Array \citep[ALMA, ][]{Czekala2017}, and the derived disk orientation agrees well with our MIDI data. The largest apparent size for a disk within the whole sample (averaged for different baselines and epochs) is $24$~au (V646 Pup), while the smallest is $0.4$~au (HBC 639) and the median is $1.3$~au. Apparent angular sizes range from $2.5$~mas (HBC 639) to $27$~mas (HD~179218), with a median of $7.5$~mas. For well-resolved disks the Gaussian model cannot provide a consistent size, hence the estimated sizes will depend on baseline length. Our results demonstrate that MIDI was able to resolve the planet-forming region in protoplanetary disks. 
	
	In some cases (e.g., DG Tau, EX Lup, HD 144132) signs of long-term temporal variability can be seen in the plots. This may indicate a rearrangement of the geometry of the inner disk. When the data indicate an elliptical shape, this may suggest an inclined circumstellar disk (GW Ori is a good example). A more circular distribution may correspond to a face-on disk or a spherical halo. 
	
	\subsection{Interferometric modeling}
	\label{sec:int_model}
	
	Following \citet{Menu2015} and \citet{Varga2017}, we modeled the interferometric data using a simple geometry to describe the brightness distribution of the disks. \citet{Menu2015} used a continuous disk model, starting from the sublimation radius out to a fixed outer radius of $300$~au. Typically the vast majority of $N$-band emission arises from the central few au for T Tauri stars and the central few ten au for the Herbig Ae stars. The aim of the model fitting is to measure the size of the mid-infrared emitting region of the disks. Because inner gaps are common features of circumstellar disks, for which also \citet{Menu2015} found indications in their study, here we will additionally try to reproduce the data with a second model where the inner radius is a free parameter. The visibilities and correlated spectra contain spatial information on the mid-IR brightness distribution of our objects. The visibilities suffer from large errors due to the total flux measurements (see Sect.~\ref{sec:datared}), therefore we use correlated fluxes for the fitting. The total fluxes are also taken into account as zero baseline correlated fluxes, but they had a lower weight in the fitting due to their large uncertainties.
	
	The geometry of our model is a thin, flat disk, beginning at the dust sublimation radius and extending to $R_\mathrm{out} = 300$~au, where the mid-IR radiation is already negligible. The disk emits blackbody radiation with a temperature decreasing as a power law
	\begin{equation}
	T\left(r\right) = T_\mathrm{sub}\left(\frac{r}{R_\mathrm{sub}}\right)^{-q},
	\end{equation}
	where $T_\mathrm{sub}$ is the dust sublimation temperature, fixed at $1500$~K. The sublimation radius ($R_\mathrm{sub}$) is calculated from the derived luminosity ($L_\star$, see Sect. \ref{sec:sedfit}) of the central star. We can express the total flux density of the object ($F_\mathrm{tot,\nu}$), which is measured by MIDI, as 
	\begin{equation}
	F_\mathrm{tot,\nu} = \int_{R_\mathrm{sub}}^{R_\mathrm{out}} 2 \pi r \left(1-\mathrm{e}^{-\tau}\right) B_\nu\left(T\left(r\right)\right) \mathrm{d}r,
	\end{equation} 
	where $B_\nu$ is the Planck function and $\tau$ is the constant optical depth, which is used here as a scaling parameter.
	The other observable we have is the correlated flux density ($F_\mathrm{corr,\nu}$) as a function of the projected baseline length ($B_{\mathrm{p}}$). We also take into account the stellar photospheric emission ($F_\mathrm{star,\nu}$) in our model, calculated from photometric SED-fits, as we discussed in Sect.~\ref{sec:sedfit}. The correlated flux density then can be expressed as 
	\begin{equation}
	F_\mathrm{corr,\nu}\left(B_{\mathrm{p}}\right) = \left| 
	\left( F_\mathrm{tot,\nu} - F_\mathrm{star,\nu} \right) V_\mathrm{disk,\nu}\left(B_{\mathrm{p}}\right) + F_\mathrm{star,\nu} \right|,
	\end{equation}
	where $V_\mathrm{disk,\nu}$ is the complex visibility function of the disk, related to the brightness distribution by means of a two-dimensional Fourier transform. Our model has two free parameters: $q$, which is the power-law exponent of the temperature profile, and $F_\mathrm{tot,\nu}$. 
	The exponent $q$ characterizes the concentration of the brightness distribution of the disk, thus it can be used as a measure of the disk size. Following \citet{midi_reduc_Leinert} and \citet{Menu2015} we define a half-light radius ($r_\mathrm{hl}$) as
	\begin{equation}
	\frac{F_\mathrm{tot,\nu}}{2} = 
	\int_{R_\mathrm{sub}}^{r_\mathrm{hl}} 2 \pi r I_\nu\left(r\right) \mathrm{d}r.
	\end{equation}
	We apply a maximum likelihood estimator to find the best-fit parameter values. The resulting fit parameters ($F_\mathrm{tot,\nu}$, $q$ and $r_\mathrm{hl}$), calculated at $\lambda = 10.7~\mu$m, are listed for each object in Table~\ref{tab:fit_param}. In one case (GG Tau Aab) the fitting failed, due to the complex geometry of the object (see Sect.~\ref{sec:binary}). For four additional objects (HBC 393, V2246 Oph, DoAr 25, SR 24N) we could not perform the modeling because of the lack of reliable data. 
	
	In the second model, which can better fit a gapped disk, the size of the inner radius is no longer fixed to the sublimation radius, but is a free parameter. Here we have the total flux fixed at the median of the total fluxes of the individual measurements. Therefore we have two free parameters: $q_\mathrm{gap}$, the power-law exponent of the temperature profile, and $R_\mathrm{in}$, the disk inner radius. We also calculate a half-light radius ($r_\mathrm{hl,\ gap}$) of the gapped disk in the same manner as in the continuous model. We list the resulting parameters ($q_\mathrm{gap}$, $R_\mathrm{in}$ and $r_\mathrm{hl,\ gap}$) of this model, calculated at $\lambda = 10.7~\mu$m, also in Table~\ref{tab:fit_param}. If the gap size could not be constrained from the data, we left the table entry empty. For a couple of sources fitting was not possible due to the small number of measurements.

	\begin{table*}
		\centering
		\caption{Results of the interferometric model fitting. The fit parameters $F_\mathrm{tot}$, $q$, and $r_\mathrm{hl}$ belong to the continuous disk model. The parameters $q_\mathrm{gap}$, $R_\mathrm{in}$, and $r_\mathrm{hl,\ gap}$ belong to the gapped disk model. We indicate $\chi^2$ values for both models. }\label{tab:fit_param}
		{\tiny\begin{tabular}{r l c c c c c c c c c}
				\hline\hline
				\# & Name &  $R_\mathrm{sub}$ & $F_\mathrm{tot}$ & $q$ & $r_\mathrm{hl}$ & $\chi^2$ & $q_\mathrm{gap}$ & $R_\mathrm{in}$ & $r_\mathrm{hl,\ gap}$ & $\chi^2_\mathrm{gap}$\\
				&   &  (au) & (Jy) &   & (au) & &   & (au) & (au) & \\
				\hline
				1 & SVS 13A1 & 0.066 & $4.4_{-0.5}^{+0.4}$ & $0.385_{-0.006}^{+0.009}$ & $5.1_{-0.8}^{+0.6}$ & $20.0$ & $0.686_{-0.040}^{+0.001}$ & $6.68_{-0.41}^{+0.06}$ & $7.03_{-0.18}^{+0.07}$ & $0.78$ \\ 
				2 & LkH$\alpha$ 330 & 0.20 & $0.48_{-0.03}^{+0.05}$ & $0.68_{-0.05}^{+0.04}$ & $0.9_{-0.1}^{+0.2}$ & $0.29$ & $1.12_{-0.55}^{+0.01}$ & $2.4_{-2.3}^{+0.5}$ & $2.5_{-1.3}^{+0.6}$ & $1.1$ \\ 
				3 & RY Tau & 0.30 & $8.8_{-0.5}^{+0.5}$ & $0.80_{-0.03}^{+0.04}$ & $0.94_{-0.08}^{+0.08}$ & $34.4$ &  &  &  &  \\ 
				4 & T Tau N & 0.21 & $8.3_{-0.2}^{+0.3}$ & $0.76_{-0.02}^{+0.02}$ & $0.74_{-0.04}^{+0.04}$ & $0.78$ &  &  &  &  \\ 
				5 & T Tau S & 0.23 & $2.8_{-2.0}^{+0.2}$ & $0.599_{-0.006}^{+2.401}$ & $1.65_{-1.39}^{+0.06}$ & $46.7$ & $0.96_{-0.07}^{+0.21}$ & $1.96_{-0.08}^{+1.78}$ & $2.19_{-0.02}^{+1.72}$ & $8.4$ \\ 
				6 & DG Tau & 0.097 & $4.1_{-0.2}^{+0.2}$ & $0.592_{-0.010}^{+0.010}$ & $0.73_{-0.05}^{+0.05}$ & $3.3$ &  &  &  &  \\ 
				7 & Haro 6-10N & 0.033 & $3.6_{-0.4}^{+0.3}$ & $0.385_{-0.005}^{+0.007}$ & $2.6_{-0.3}^{+0.3}$ & $6.8$ & $0.49_{-0.01}^{+0.04}$ & $1.07_{-0.02}^{+0.09}$ & $1.71_{-0.10}^{+0.09}$ & $0.47$ \\ 
				8 & Haro 6-10S & 0.16 & $1.4_{-0.2}^{+0.2}$ & $0.67_{-0.03}^{+0.04}$ & $0.8_{-0.1}^{+0.1}$ & $4.1$ &  &  &  &  \\ 
				10 & HL Tau & 0.047 & $4.4_{-0.5}^{+0.4}$ & $0.49_{-0.01}^{+0.02}$ & $0.9_{-0.1}^{+0.1}$ & $0.00$ &  &  &  &  \\ 
				12 & LkCa 15 & 0.072 & $0.33_{-0.09}^{+0.13}$ & $0.46_{-0.03}^{+0.06}$ & $1.7_{-0.8}^{+1.0}$ & $0.22$ &  &  &  &  \\ 
				13 & DR Tau & 0.14 & $2.02_{-0.06}^{+0.08}$ & $0.82_{-0.06}^{+0.09}$ & $0.41_{-0.07}^{+0.07}$ & $4.2$ &  &  &  &  \\ 
				14 & UY Aur B & 0.034 & $1.73_{-0.06}^{+0.20}$ & $0.56_{-0.05}^{+0.03}$ & $0.31_{-0.06}^{+0.18}$ & $0.00$ &  &  &  &  \\ 
				15 & UY Aur A & 0.057 & $2.2_{-0.2}^{+0.3}$ & $0.50_{-0.02}^{+0.02}$ & $0.9_{-0.1}^{+0.2}$ & $0.01$ &  &  &  &  \\ 
				16 & GM Aur & 0.057 & $0.60_{-0.17}^{+0.23}$ & $0.37_{-0.02}^{+0.02}$ & $6.6_{-2.3}^{+4.1}$ & $0.10$ &  &  &  &  \\ 
				17 & AB Aur & 0.55 & $15.5_{-0.6}^{+1.1}$ & $0.68_{-0.02}^{+0.01}$ & $2.6_{-0.2}^{+0.2}$ & $7.7$ &  &  &  &  \\ 
				18 & SU Aur & 0.21 & $3.6_{-0.9}^{+0.8}$ & $0.66_{-0.03}^{+0.04}$ & $1.1_{-0.2}^{+0.2}$ & $0.00$ &  &  &  &  \\ 
				19 & MWC 480 & 0.32 & $8.68_{-0.09}^{+1.13}$ & $1.88_{-1.04}^{+2.12}$ & $0.42_{-0.07}^{+0.49}$ & $0.55$ &  &  &  &  \\ 
				20 & UX Ori & 0.22 & $3.2_{-0.1}^{+0.1}$ & $0.483_{-0.005}^{+0.003}$ & $4.1_{-0.1}^{+0.2}$ & $12.1$ & $0.64_{-0.02}^{+0.05}$ & $2.42_{-0.09}^{+0.16}$ & $3.5_{-0.1}^{+0.1}$ & $4.0$ \\ 
				21 & CO Ori & 0.54 & $1.60_{-0.08}^{+0.16}$ & $0.82_{-0.05}^{+0.03}$ & $1.6_{-0.1}^{+0.2}$ & $3.3$ & $1.50_{-0.74}^{+0.40}$ & $1.8_{-1.2}^{+0.1}$ & $2.0_{-0.2}^{+0.2}$ & $5.0$ \\ 
				22 & GW Ori & 0.61 & $3.1_{-1.1}^{+1.1}$ & $0.75_{-0.05}^{+0.05}$ & $2.2_{-1.3}^{+1.3}$ & $27.8$ & $0.80_{-0.02}^{+0.04}$ & $6.81_{-0.09}^{+0.14}$ & $7.99_{-0.09}^{+0.07}$ & $35.6$ \\ 
				23 & MWC 758 & 0.30 & $6.0_{-0.4}^{+0.5}$ & $0.63_{-0.01}^{+0.01}$ & $1.8_{-0.1}^{+0.2}$ & $0.71$ & $0.590_{-0.008}^{+0.028}$ & $0.18_{-0.03}^{+0.13}$ & $2.2_{-0.2}^{+0.2}$ & $0.86$ \\ 
				24 & NY Ori & 0.14 & $0.45_{-0.19}^{+0.19}$ & $0.66_{-0.05}^{+0.05}$ & $0.7_{-1.3}^{+1.3}$ & $0.00$ &  &  &  &  \\ 
				25 & CQ Tau & 0.18 & $7.6_{-2.6}^{+2.8}$ & $0.51_{-0.11}^{+0.08}$ & $2.6_{-1.2}^{+7.5}$ & $0.00$ &  &  &  &  \\ 
				26 & V1247 Ori & 0.27 & $0.20_{-0.05}^{+0.19}$ & $0.67_{-0.16}^{+1.33}$ & $1.3_{-1.0}^{+2.5}$ & $1.3$ & $0.66_{-0.05}^{+0.27}$ & $5.8_{-1.9}^{+5.2}$ & $7.0_{-2.7}^{+4.9}$ & $0.09$ \\ 
				27 & V883 Ori & 2.3 & $25.1_{-1.3}^{+1.3}$ & $0.99_{-0.03}^{+0.03}$ & $4.9_{-0.2}^{+0.2}$ & $0.72$ & $0.83_{-0.04}^{+0.19}$ & $1.8_{-0.6}^{+1.5}$ & $6.1_{-0.5}^{+0.5}$ & $1.3$ \\ 
				28 & MWC 120 & 1.0 & $10.7_{-1.4}^{+1.6}$ & $0.88_{-0.09}^{+0.14}$ & $2.7_{-0.6}^{+0.7}$ & $0.00$ &  &  &  &  \\ 
				29 & FU Ori & 0.99 & $4.4_{-0.1}^{+0.2}$ & $1.37_{-0.11}^{+0.11}$ & $1.53_{-0.08}^{+0.10}$ & $0.82$ &  &  &  &  \\ 
				30 & V1647 Ori & 0.42 & $3.5_{-0.1}^{+0.1}$ & $0.577_{-0.007}^{+0.008}$ & $3.5_{-0.2}^{+0.2}$ & $23.8$ & $0.69_{-0.03}^{+0.28}$ & $2.0_{-0.2}^{+0.8}$ & $3.4_{-0.2}^{+0.2}$ & $18.0$ \\ 
				31 & V900 Mon & 1.4 & $3.8_{-0.6}^{+1.1}$ & $0.69_{-0.08}^{+0.09}$ & $6.4_{-1.8}^{+3.0}$ & $0.00$ &  &  &  &  \\ 
				32 & Z CMa & 3.9 & $93.4_{-2.8}^{+2.8}$ & $0.871_{-0.012}^{+0.009}$ & $10.3_{-0.2}^{+0.3}$ & $7.2$ &  &  &  &  \\ 
				33 & V646 Pup & 1.4 & $1.2_{-0.3}^{+0.6}$ & $0.49_{-0.04}^{+0.05}$ & $23.4_{-8.2}^{+14.1}$ & $0.00$ &  &  &  &  \\ 
				34 & HD 72106 & 0.42 & $2.0_{-0.1}^{+0.2}$ & $0.519_{-0.008}^{+0.007}$ & $5.6_{-0.4}^{+0.4}$ & $19.1$ & $0.80_{-0.03}^{+0.03}$ & $10.7_{-0.2}^{+0.2}$ & $11.62_{-0.05}^{+0.09}$ & $12.2$ \\ 
				35 & CR Cha & 0.15 & $1.22_{-0.07}^{+0.23}$ & $0.61_{-0.03}^{+0.01}$ & $1.03_{-0.07}^{+0.25}$ & $1.4$ & $1.25_{-0.11}^{+0.02}$ & $1.3_{-0.2}^{+0.3}$ & $1.4_{-0.2}^{+0.3}$ & $1.7$ \\ 
				36 & TW Hya & 0.043 & $0.69_{-0.03}^{+0.03}$ & $0.428_{-0.006}^{+0.004}$ & $1.7_{-0.1}^{+0.1}$ & $2.4$ & $0.54_{-0.02}^{+0.09}$ & $0.51_{-0.06}^{+0.16}$ & $0.97_{-0.07}^{+0.05}$ & $0.98$ \\ 
				37 & DI Cha & 0.28 & $1.9_{-0.2}^{+0.1}$ & $0.56_{-0.01}^{+0.01}$ & $2.6_{-0.2}^{+0.3}$ & $4.1$ & $0.80_{-0.07}^{+0.47}$ & $1.6_{-0.2}^{+0.4}$ & $2.13_{-0.05}^{+0.10}$ & $0.20$ \\ 
				38 & Glass I & 0.12 & $8.5_{-0.3}^{+0.3}$ & $0.564_{-0.010}^{+0.009}$ & $1.12_{-0.07}^{+0.09}$ & $2.6$ & $0.707_{-0.185}^{+0.005}$ & $2.2_{-1.0}^{+1.0}$ & $2.56_{-1.35}^{+0.02}$ & $4.3$ \\ 
				39 & Sz 32 & 0.25 & $2.9_{-0.3}^{+0.3}$ & $0.56_{-0.02}^{+0.02}$ & $2.5_{-0.3}^{+0.4}$ & $4.3$ &  &  &  &  \\ 
				40 & WW Cha & 0.17 & $5.4_{-0.2}^{+0.3}$ & $0.59_{-0.02}^{+0.02}$ & $1.3_{-0.1}^{+0.2}$ & $0.93$ &  &  &  &  \\ 
				41 & CV Cha & 0.19 & $2.3_{-0.1}^{+0.2}$ & $0.61_{-0.02}^{+0.02}$ & $1.3_{-0.1}^{+0.2}$ & $10.8$ &  &  &  &  \\ 
				42 & DX Cha & 0.47 & $13.6_{-0.5}^{+0.4}$ & $1.73_{-0.05}^{+0.07}$ & $0.642_{-0.011}^{+0.008}$ & $1.9$ & $1.22_{-0.15}^{+0.05}$ & $0.43_{-0.06}^{+0.03}$ & $0.77_{-0.01}^{+0.06}$ & $1.7$ \\ 
				43 & DK Cha & 0.58 & $27.2_{-1.9}^{+1.6}$ & $0.71_{-0.02}^{+0.02}$ & $2.5_{-0.2}^{+0.2}$ & $18.4$ & $1.479_{-0.378}^{+0.006}$ & $2.760_{-0.220}^{+0.000}$ & $2.91_{-0.04}^{+0.05}$ & $1.0$ \\ 
				44 & HD 135344B & 0.23 & $0.77_{-0.02}^{+0.03}$ & $1.35_{-0.16}^{+0.28}$ & $0.35_{-0.04}^{+0.04}$ & $5.2$ &  &  &  &  \\ 
				45 & HD 139614 & 0.21 & $3.76_{-0.08}^{+0.15}$ & $0.531_{-0.005}^{+0.003}$ & $2.48_{-0.06}^{+0.11}$ & $3.2$ & $0.73_{-0.02}^{+0.01}$ & $5.30_{-0.08}^{+0.04}$ & $5.96_{-0.06}^{+0.04}$ & $0.64$ \\ 
				46 & HD 141569 & 0.33 & $0.14_{-0.09}^{+0.14}$ & $0.55_{-0.37}^{+0.43}$ & $3.4_{-2.7}^{+185.6}$ & $0.02$ &  &  &  &  \\ 
				47 & HD 142527 & 0.32 & $12.8_{-0.4}^{+0.6}$ & $0.627_{-0.007}^{+0.006}$ & $1.95_{-0.07}^{+0.08}$ & $15.3$ & $0.73_{-0.02}^{+0.07}$ & $0.53_{-0.02}^{+1.45}$ & $1.45_{-0.03}^{+1.11}$ & $9.5$ \\ 
				48 & RU Lup & 0.095 & $2.60_{-0.08}^{+0.13}$ & $0.58_{-0.02}^{+0.01}$ & $0.77_{-0.05}^{+0.09}$ & $1.1$ & $0.62_{-0.07}^{+0.47}$ & $0.5_{-0.5}^{+0.5}$ & $1.05_{-0.09}^{+0.16}$ & $0.73$ \\ 
				49 & HD 143006 & 0.16 & $0.73_{-0.20}^{+0.15}$ & $0.55_{-0.03}^{+0.05}$ & $1.6_{-0.5}^{+0.5}$ & $0.00$ &  &  &  &  \\ 
				50 & EX Lup & 0.047 & $1.19_{-0.06}^{+0.06}$ & $0.497_{-0.009}^{+0.012}$ & $0.77_{-0.09}^{+0.08}$ & $58.7$ &  &  &  &  \\ 
				51 & HD 144432 & 0.45 & $9.3_{-0.5}^{+0.4}$ & $0.630_{-0.009}^{+0.009}$ & $2.7_{-0.1}^{+0.1}$ & $2.0$ & $0.63_{-0.02}^{+0.03}$ & $0.75_{-0.08}^{+0.08}$ & $3.0_{-0.2}^{+0.2}$ & $0.64$ \\ 
				52 & V856 Sco & 0.67 & $10.2_{-0.2}^{+0.3}$ & $1.67_{-0.09}^{+0.08}$ & $0.93_{-0.02}^{+0.03}$ & $4.7$ &  &  &  &  \\ 
				53 & AS 205 N & 0.16 & $5.7_{-0.3}^{+0.5}$ & $0.69_{-0.04}^{+0.05}$ & $0.8_{-0.1}^{+0.2}$ & $0.31$ &  &  &  &  \\ 
				54 & AS 205 S & 0.097 & $3.5_{-0.5}^{+0.8}$ & $0.55_{-0.03}^{+0.03}$ & $1.0_{-0.2}^{+0.3}$ & $0.26$ & $1.154_{-0.602}^{+0.007}$ & $0.9_{-0.8}^{+0.1}$ & $0.99_{-0.06}^{+0.33}$ & $0.20$ \\ 
				55 & DoAr 20 & 0.079 & $1.00_{-0.09}^{+0.13}$ & $0.64_{-0.03}^{+0.02}$ & $0.46_{-0.05}^{+0.08}$ & $0.99$ &  &  &  &  \\ 
				57 & HBC 639 & 0.16 & $1.91_{-0.10}^{+0.19}$ & $0.94_{-0.12}^{+0.13}$ & $0.37_{-0.06}^{+0.10}$ & $0.32$ & $1.48_{-0.67}^{+0.52}$ & $0.4_{-0.4}^{+0.2}$ & $0.5_{-0.2}^{+0.2}$ & $0.49$ \\ 
				59 & Elias 24 & 0.16 & $1.52_{-0.05}^{+0.17}$ & $0.692_{-0.025}^{+0.006}$ & $0.73_{-0.02}^{+0.08}$ & $4.8$ & $0.88_{-0.17}^{+0.57}$ & $0.8_{-0.2}^{+0.2}$ & $0.99_{-0.10}^{+0.07}$ & $3.7$ \\ 
				61 & SR 24S & 0.099 & $3.1_{-1.0}^{+0.8}$ & $0.53_{-0.02}^{+0.05}$ & $1.2_{-0.4}^{+0.3}$ & $0.00$ &  &  &  &  \\ 
				
				\hline
			\end{tabular}
		}
	\end{table*}
	
	\addtocounter{table}{-1}
	\begin{table*}
		\centering
		\caption{continued.}
		{\tiny\begin{tabular}{r l c c c c c c c c c}
				\hline\hline
				\# & Name &  $R_\mathrm{sub}$ & $F_\mathrm{tot}$ & $q$ & $r_\mathrm{hl}$ & $\chi^2$ & $q_\mathrm{gap}$ & $R_\mathrm{in}$ & $r_\mathrm{hl,\ gap}$ & $\chi^2_\mathrm{gap}$\\
				&   &  (au) & (Jy) &   & (au) & &   & (au) & (au) & \\
				\hline
				62 & Elias 29 & 0.44 & $16.3_{-5.6}^{+8.3}$ & $0.70_{-0.30}^{+0.19}$ & $1.9_{-0.8}^{+22.2}$ & $0.00$ &  &  &  &  \\ 
				63 & SR 21A & 0.27 & $2.0_{-0.9}^{+1.0}$ & $0.46_{-0.03}^{+0.04}$ & $7.0_{-2.4}^{+3.9}$ & $0.06$ & $0.59_{-0.03}^{+0.23}$ & $1.3_{-0.2}^{+21.7}$ & $3.1_{-0.9}^{+21.8}$ & $0.01$ \\ 
				64 & IRS 42 & 0.14 & $2.2_{-0.3}^{+0.3}$ & $0.66_{-0.02}^{+0.05}$ & $0.73_{-0.13}^{+0.10}$ & $0.04$ &  &  &  &  \\ 
				65 & IRS 48 & 0.24 & $1.5_{-0.6}^{+0.6}$ & $0.38_{-0.24}^{+0.03}$ & $19.6_{-8.8}^{+168.7}$ & $0.00$ &  &  &  &  \\ 
				66 & V2129 Oph & 0.12 & $1.8_{-0.5}^{+0.5}$ & $0.54_{-0.03}^{+0.05}$ & $1.4_{-0.4}^{+0.4}$ & $0.00$ &  &  &  &  \\ 
				67 & Haro 1-16 & 0.076 & $1.12_{-0.04}^{+0.12}$ & $0.65_{-0.04}^{+0.02}$ & $0.41_{-0.04}^{+0.11}$ & $0.94$ &  &  &  &  \\ 
				68 & V346 Nor & 0.29 & $3.5_{-0.1}^{+0.1}$ & $0.627_{-0.011}^{+0.008}$ & $1.76_{-0.08}^{+0.11}$ & $18.7$ &  &  &  &  \\ 
				69 & HD 150193 & 0.66 & $15.9_{-0.5}^{+0.3}$ & $0.767_{-0.006}^{+0.006}$ & $2.26_{-0.04}^{+0.04}$ & $23.1$ & $0.91_{-0.09}^{+0.02}$ & $7.846_{-3.008}^{+0.000}$ & $8.66_{-2.60}^{+0.09}$ & $25.8$ \\ 
				70 & AS 209 & 0.10 & $3.0_{-0.2}^{+0.1}$ & $0.65_{-0.01}^{+0.02}$ & $0.57_{-0.04}^{+0.04}$ & $1.4$ &  &  &  &  \\ 
				71 & AK Sco & 0.19 & $3.3_{-0.1}^{+0.2}$ & $0.65_{-0.01}^{+0.01}$ & $1.04_{-0.06}^{+0.08}$ & $1.7$ &  &  &  &  \\ 
				72 & HD 163296 & 0.40 & $19.2_{-0.6}^{+0.7}$ & $0.75_{-0.01}^{+0.01}$ & $1.44_{-0.05}^{+0.06}$ & $6.1$ &  &  &  &  \\ 
				73 & HD 169142 & 0.20 & $0.75_{-0.25}^{+0.26}$ & $0.40_{-0.05}^{+0.03}$ & $11.1_{-4.6}^{+15.5}$ & $0.09$ & $0.51_{-0.14}^{+0.11}$ & $16.8_{-13.3}^{+17.2}$ & $20.8_{-14.9}^{+15.3}$ & $0.07$ \\ 
				74 & VV Ser & 0.53 & $4.2_{-0.1}^{+0.3}$ & $0.89_{-0.03}^{+0.03}$ & $1.36_{-0.06}^{+0.09}$ & $0.17$ & $0.99_{-0.28}^{+0.97}$ & $1.0_{-0.7}^{+0.4}$ & $1.56_{-0.02}^{+0.40}$ & $0.20$ \\ 
				75 & SVS20N & 0.26 & $1.6_{-0.1}^{+0.1}$ & $0.54_{-0.01}^{+0.01}$ & $2.8_{-0.2}^{+0.3}$ & $19.2$ & $1.019_{-0.033}^{+0.002}$ & $4.37_{-0.07}^{+0.03}$ & $4.57_{-0.05}^{+0.05}$ & $10.2$ \\ 
				76 & SVS20S & 2.9 & $6.5_{-0.2}^{+0.3}$ & $1.29_{-0.01}^{+0.01}$ & $4.73_{-0.04}^{+0.04}$ & $75.3$ &  &  &  &  \\ 
				77 & S CrA S & 0.073 & $2.3_{-0.3}^{+0.4}$ & $0.61_{-0.02}^{+0.02}$ & $0.49_{-0.07}^{+0.07}$ & $0.00$ &  &  &  &  \\ 
				78 & S CrA N & 0.11 & $2.5_{-0.3}^{+0.3}$ & $0.54_{-0.02}^{+0.02}$ & $1.2_{-0.2}^{+0.2}$ & $21.9$ &  &  &  &  \\ 
				79 & T CrA & 0.11 & $6.0_{-1.0}^{+0.9}$ & $0.47_{-0.02}^{+0.02}$ & $2.5_{-0.6}^{+0.7}$ & $0.00$ &  &  &  &  \\ 
				80 & VV CrA NE & 0.12 & $5.0_{-0.3}^{+0.6}$ & $0.59_{-0.02}^{+0.02}$ & $0.92_{-0.09}^{+0.16}$ & $4.0$ &  &  &  &  \\ 
				81 & VV CrA SW & 0.075 & $14.5_{-0.6}^{+0.9}$ & $0.54_{-0.02}^{+0.01}$ & $0.84_{-0.07}^{+0.14}$ & $2.5$ & $0.53_{-0.02}^{+0.08}$ & $0.3_{-0.2}^{+0.3}$ & $1.1_{-0.1}^{+0.2}$ & $1.2$ \\ 
				82 & HD 179218 & 1.2 & $25.6_{-1.3}^{+1.5}$ & $0.535_{-0.005}^{+0.005}$ & $13.2_{-0.5}^{+0.6}$ & $31.0$ & $0.96_{-0.29}^{+0.06}$ & $10.4_{-7.0}^{+0.2}$ & $11.61_{-4.25}^{+0.04}$ & $16.3$ \\ 
				
				\hline
			\end{tabular}
		}
	\end{table*}

	As an example, in Fig.~\ref{fig:int_model} we plot the results of the interferometric model fitting. Overplotted on the data points are the best-fit model curves for the continuous model. The stellar contribution, which is unresolved for MIDI, is also plotted. To support the preparation of future MATISSE observations, we extrapolated our continuous disk model to the $K$ and $L$ bands. MATISSE will observe in the $L$, $M,$ and $N$ bands. %, with the possibility to use the fringe tracker of GRAVITY to track fringes in $K$. 
	The extrapolated model curves are also shown in Fig.~\ref{fig:int_model}. More details on the near and mid-IR observability of the sources can be found in Sect.~\ref{sec:matisse}. Information on the model fit plots for all sources can be found in Fig.~\ref{fig:atlas_fit_app} in Appendix~\ref{app:atlasfig}. Visual inspection of the plots suggests that the majority of our data can be well fitted by the continuous model, except for a few objects (e.g., Elias 24, DI Cha, DK Cha, Haro 6-10N, SVS20N, TW Hya) where the gapped model gives a better match to the data.
	
	%%%%%%%%%%%%%%%%%%%%%%%
	% ATLAS DATA PRODUCTS %
	%%%%%%%%%%%%%%%%%%%%%%%
	%appendix: 
	%list of observations table
	%table: result of the interf. modeling (q, hlr, etc.)
	%figure: uv-plots, gaussian size, model fits
	
	%vizier: 
	%source list table (info on binaries, but don't provide info on spectral shapes) 
	%table: result of the interf. modeling (q, hlr, etc.)
	%list of observations table, links to the spectrum data files, and links to the spectrum plots: corr, total, vis, phase
	%data files: corr, total, vis, phase - in fits
	
	%konkoly project website:
	%all atlas data
	%images
	%%%%%%%%%%%%%%%%%%%%%%%%%%
	
	%%%%%%%%%%%%%%%%%%%%%
	\section{Discussion}
	\label{sec:discuss}

	\subsection{Stellar multiplicity}
	\label{sec:binary}
	
	\begin{table*}
		\caption{Multiple stellar systems in the atlas.}             % title of Table
		\label{tab:binary}      % is used to refer this table in the text
		\centering      
		\tiny
		% used for centering table
		\begin{tabular}{r l l c c p{8.6cm} c}        % centered columns 
			
			\hline\hline                 % inserts double horizontal lines
			\# & Name & Type & Separation & Modulation of  & Comment & References\\
			& & & ($''$) & MIDI phases &  &\\% table heading 
			\hline                        % inserts single horizontal line
			1 & SVS 13A1 & eruptive &       0.3      &      no       &      Only one star (VLA 4B) in the close binary system has mid-IR emission associated with it.  & 1 \\
			4 & T Tau N & TT &      0.7      &      no       & Triple system: T Tau N and S is a wide binary, T Tau Sab is a close binary sytem. & 2 \\
			5 & T Tau S & TT &      0.1      &      yes      &      Sa, Sb: a close binary system itself. & 3 \\
			7 & Haro 6-10N & TT &   1.2      &      no       & Both components are embedded in a common envelope. The disks of the components are highly misaligned. & 4,5 \\
			8 & Haro 6-10S & TT &   1.2      &      no       & Both components are embedded in a common envelope. The disks of the components are highly misaligned. & 4,5 \\
			9 & HBC 393 & eruptive &        0.32     &      n/a      & Binary protostar, listed also as a FUor, (N, S components), both components have disks & 6 \\
			11 & GG Tau Aab & TT &  0.25     &      yes      &      Quadruple system: Aab -- Bab $10''$ separation, Aa -- Ab $0.25''$ separation, Ba -- Bb $1.48''$ separation, having a circumbinary disk. MIDI observed GG Tau Aab pair.  & 7,8 \\
			14 & UY Aur B & TT &    0.88     &      no       &      Both components have circumstellar disks, and a circumbinary ring  & 9 \\
			15 & UY Aur A & TT &    0.88     &      no       &      Both components have circumstellar disks, and a circumbinary ring  & 9 \\
			17 & AB Aur & HAe &              &      yes      & At the longest baselines phases show binary signal &  \\
			21 & CO Ori & HAe & 2 & no & The primary (CO Ori A) shows UX Ori-type variability. CO Ori B is a K2-K5 type star, which also has circumstellar material. & 10,11 \\
			22 & GW Ori & TT/HAe & 0.02 & yes & Triple system, containing a spectroscopic binary and a tertiary component at $\sim20$~mas from the primary. The components of the spectroscopic binary are separated by $\sim1$~au ($\sim 3.5$~mas), based on radial velocity measurements. At the longest baselines phases show weak modulation, probably caused by the tertiary component. & 12,13 \\
			29 & FU Ori & eruptive &        0.4      &      no       &  FU Ori N: main component, FU Ori S is embedded is ($A_V = 8$) and 5-6 mag fainter in J and H.  & 14,15 \\
			32 & Z CMa & eruptive &         0.1      &      yes      & The NW component is a Herbig Be star, which shows EXor-like outbursts, the SE component exhibits broad double-peaked optical absorption lines that are typical of a  FU Ori  object.  & 16,17 \\
			34 & HD 72106 & HAe &   0.8      &      yes      &      The primary (A) is a late B-type magnetic star, secondary (B) is a Herbig A. The weak modulation is possibly not the signal of the companion, but may be due to asymmetric disk structure. & 18,19 \\
			37 & DI Cha & TT &      0.2      &      no       &      It is a quadruple  system with a hierachical structure, consisting of two binaries: a G2/M6 pair (A and D, separation: $0.2''$) and a pair of two M5.5 dwarfs (B and C, separation: $0.06''$). Separation of A and B is $4.6''$. MIDI observed source A. Source D is possibly a brown dwarf.  & 20 \\
			38 & Glass I & TT &     2.4      &      no       &      Multiple system: Glass Ia is binary with K (WTTS) and M star components. Glass Ib is a G-type young star, which dominates the mid-IR emission of the system. MIDI observed Glass Ib. & 21,22 \\
			44 & HD 135344B & HAe &         21       &      no       &      Visual binary, the companion (HD135344 or HD135344A) is well separated. & 23,24 \\
			46 & HD 141569 & HAe &  7.5      &      no       &      Triple system in a comoving group, the primary (HD 141569A) was observed with MIDI. The companions (B and C, separation: $1.4''$) form a binary pair $7.5''$ from A. & 25 \\
			47 & HD 142527 & HAe &  0.088    &      yes      &      Faint companion HD 142527 B $13$~au from the primary (A). Modulation can be seen at the longest baselines.       & 26 \\
			51 & HD 144432 & HAe &  1.47     &      no       &      Hierarchical triple system. The components B and C are located $1.47''$ north of the primary  star (A).  B  and  C are  separated  by $0.1''$. MIDI observed component A.  & 27 \\
			53 & AS 205 N & TT &    1.3      &      no       &      This is the A (North) component.  & 28 \\
			54 & AS 205 S & TT &    1.3      & no & This is the B (South) component. & 28 \\
			57 & HBC 639 & TT &     2        &      no       &  Aka. DoAr 24E. The primary (DoAr 24E A) is a weak line T Tauri star. The secondary (DoAr 24E B) is an infrared companion, an active class I object, a binary itself. MIDI observed the primary component. & 28,29 \\
			60 & SR 24N & TT &      0.197    &      yes      &      North component of the triple system, itself a binary.   & 30 \\
			61 & SR 24S & TT &      5.2      &      no       & South component of the triple system, located $5.2''$ from SR 24N. & 31 \\
			63 & SR 21A & TT/HAe &  6.7      &      no       & SR 21A and SR 21B are classified as a binary, however it does not appear to be a bound, co-eval system. & 32,33 \\
			69 & HD 150193 & HAe &  1.1      &      yes      &  HD 150193A is physically associated with a K4 star (HD 150193B). The disk of component A (which was observed with MIDI) has an asymmetric structure due to the interaction with HD 150193B, where no disk was detected.  & 34,35 \\
			71 & AK Sco & HAe & 0.001 & no & Spectroscopic binary with a semimajor axis of $0.16$~au. It has a circumbinary disk. & 36 \\
			\hline
		\end{tabular}
	\end{table*}
	
	\addtocounter{table}{-1}
	\begin{table*}
		\centering
		\tiny
		\caption{continued.}
		\begin{tabular}{r l l c c p{8.6cm} c}        % centered columns 
			
			\hline\hline                 % inserts double horizontal lines
			\# & Name & Type & Separation & Modulation of  & Comment & References\\
			& & & ($''$) & MIDI phases &  &\\% table heading 
			\hline               
			75 & SVS20N & TT & 1.58 &       no       & Triple system: SVS20N and S is a wide binary, SVS20S is a close binary sytem. & 37,38 \\
			76 & SVS20S & embHAe & 0.32 &   no       & SVS20S is a close binary sytem itself. & 37,38 \\
			77 & S CrA S & TT &     1.4      &      no       &  & 31,39,40 \\
			78 & S CrA N & TT &     1.4      &      no       &  & 31,39,40 \\
			80 & VV CrA NE & TT &   2        &      no       &  & 31,40 \\
			81 & VV CrA SW & TT &   2        &      no       &        & 31,40 \\
			82 & HD 179218 & HAe &  2.5      &      yes      & The modulation is possibly not caused by the known companion.  & 41 \\
			
			\hline                                   %inserts single line
		\end{tabular}
		\tablebib{(1)~\citep{Fujiyoshi2015}; 
			(2) \citep{Dyck1982}; 
			(3) \citep{Ratzka2009}; 
			(4) \citep{Leinert1989}; 
			(5) \citep{Roccatagliata2011}; 
			(6) \citep{Rodriguez1998}; 
			(7) \citep{1991AA...250..407L}; 
			(8) \citep{2016AARv..24....5D}; 
			(9) \citep{Tang2014}; 
			(10) \citep{Correia2006}; 
			(11) \citep{Davies2018}; 
			(12) \citep{Berger2011}; 
			(13) \citep{Fang2014}; 
			(14) \citep{Wang2004}; 
			(15) \citep{Pueyo2012}; 
			(16) \citep{Hartmann1989}; 
			(17) \citep{Szeifert2010}; 
			(18) \citep{ESA1997}; 
			(19) \citep{Folsom2008}; 
			(20) \citep{Schmidt2013}; 
			(21) \citep{Chelli1988}; 
			(22) \citep{Kruger2013}; 
			(23) \citep{Coulson1995}; 
			(24) \citep{Dunkin1997}; 
			(25) \citep{2000ApJ...544..937W}; 
			(26) \citep{2012ApJ...753L..38B}; 
			(27) \citep{2011AA...535L...3M}; 
			(28) \citep{1993AJ....106.2005G}; 
			(29) \citep{2002AJ....124.1082K}; 
			(30) \citep{1995ApJ...443..625S}; 
			(31) \citep{1993AA...278...81R}; 
			(32) \citep{2003ApJ...591.1064B}; 
			(33) \citep{2003ApJ...584..853P}; 
			(34) \citep{2001IAUS..200..155B}; 
			(35) \citep{2014AA...568A..40G}; 
			(36) \citep{Janson2016}; 
			(37) \citep{1987AA...179..171E}; 
			(38) \citep{Duchene2007}; 
			(39) \citep{1985AA...153..278B}; 
			(40) \citep{Sicilia-Aguilar2011}; 
			(41) \citep{2007IAUS..240..250T}; 
		}
	\end{table*}
	
	About $30\%$ of the sources in our sample are known multiple systems. MIDI is sensitive to the binaries if the orientation of the components is not perpendicular to the projected baseline, they have large enough separation, and their flux ratio is not very small. If the components are widely separated, like in the case of T Tau N and S ($0.7''$), they can be individually observed. For close binaries, such as T Tau Sa and Sb ($0.1''$), both components are in the MIDI interferometric field of view, and cause a modulation in the interferometric signal. This, however, requires that the fluxes of the individual components are not too different from each other. As an example, the companion of FU Ori, which is $2.5$ mag fainter in the $N$ band than the primary, was not interferometrically detected, but could be seen on the MIDI acquisition images \citep{Quanz2006}.  In Table~\ref{tab:binary} we list all of the multiple sources that were known from the literature. Additionally, we checked all sources for binary signal in the MIDI data, and we also included in the table those where we found binary modulations at least in some measurements. 
	
	In six cases (T Tau S, GG Tau Aab, GW Ori, Z CMa, HD 142527, SR 24N) the known binary component was detected in the interferometric observables. We also identified two systems with modulations (AB Aur, HD 72106) where there was no companion known or the known companion is either too wide (separation $\gtrsim 250$ mas) or too faint (flux ratio $\lesssim 0.1$) to cause the observed signal. These modulations can be caused by a companion but can also be due to an asymmetric inner gap structure (e.g., for HD 142527 the correlated flux drops to zero, which would require equally bright components). The star HD 150193 has an asymmetric disk structure, presumably due to the secondary companion \citep[observed in $H$ band by][]{Fukagawa2003}. The star HD 179218 was studied by \citet{Fedele2008}, who concluded that the phase modulation is caused by asymmetric emission. The disk is inclined, and due to the flaring the far side appears brighter for the observer. The interpretation of these observations requires detailed studies that are beyond the scope of this atlas. The systems exhibiting significant modulations in the interferometric observables cannot be interpreted using our simple geometric model, and thus we exclude them from the further discussion (except Sect.~\ref{sec:var}) to allow an unbiased analysis. Our results in Table~\ref{tab:binary} suggest that MIDI was able to detect binaries with separations between $\sim$$50$~mas and $\sim$$250$~mas, depending on the baseline length and orientation.
	
	\subsection{Mid-infrared sizes}
	\label{sec:sizes}
	
	\begin{figure}
		\centering
		\includegraphics[width=\linewidth]{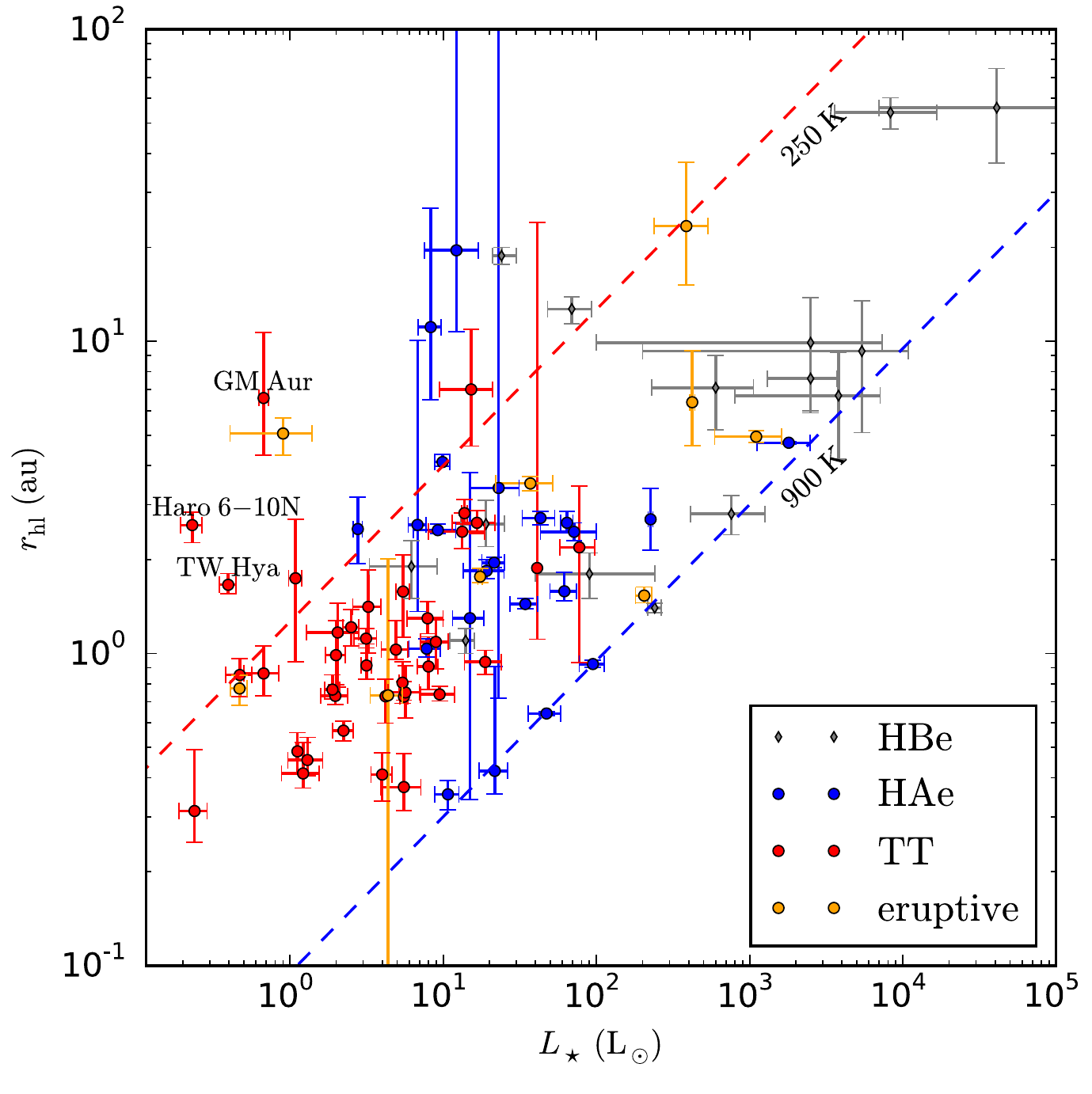}
		\caption{Mid-IR size versus stellar luminosity for our sample. Color of the symbols indicates the object type: Herbig Ae (blue), T Tauri (red), and  eruptive (orange). To extend the luminosity coverage, Herbig Be sources from \citet{Menu2015} are also plotted (gray diamonds). The radii that correspond to temperatures of $250$~K and $900$~K for optically thin gray dust are overplotted with red and blue dashed lines, respectively. The individually marked objects are discussed in Sect.~\ref{sec:sizes}.} 
		
		\label{fig:size_lum}
	\end{figure}
	
	\begin{figure}
		\centering
		\includegraphics[width=\linewidth]{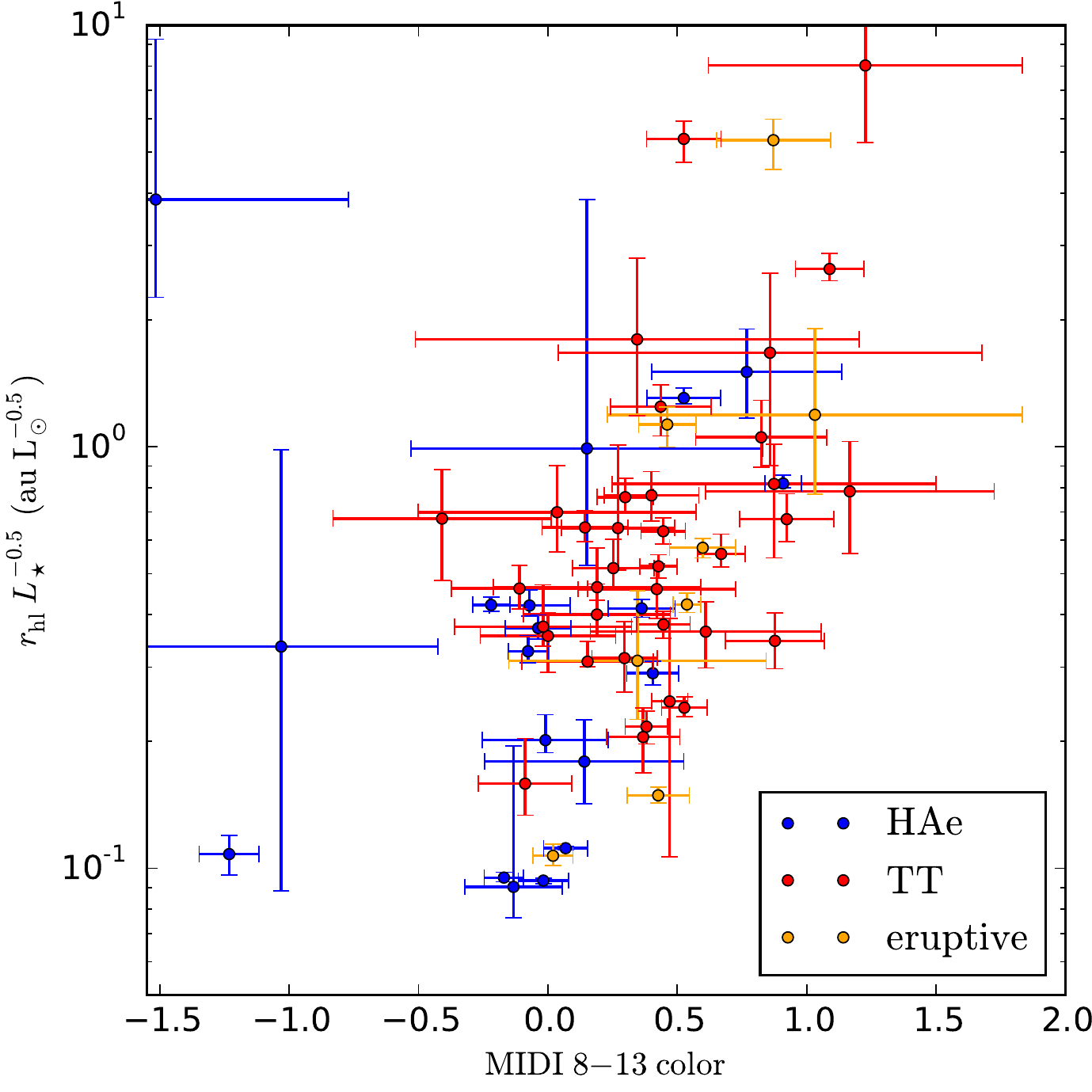}
		\caption{Normalized mid-IR size as function of MIDI $8-13$ color for our sample. Color of the symbols indicates the object type: Herbig Ae (blue), T Tauri (red), and  eruptive (orange).} 
		
		\label{fig:size_color_all}
	\end{figure}
	
	\begin{figure}
		\centering
		\includegraphics[width=\linewidth]{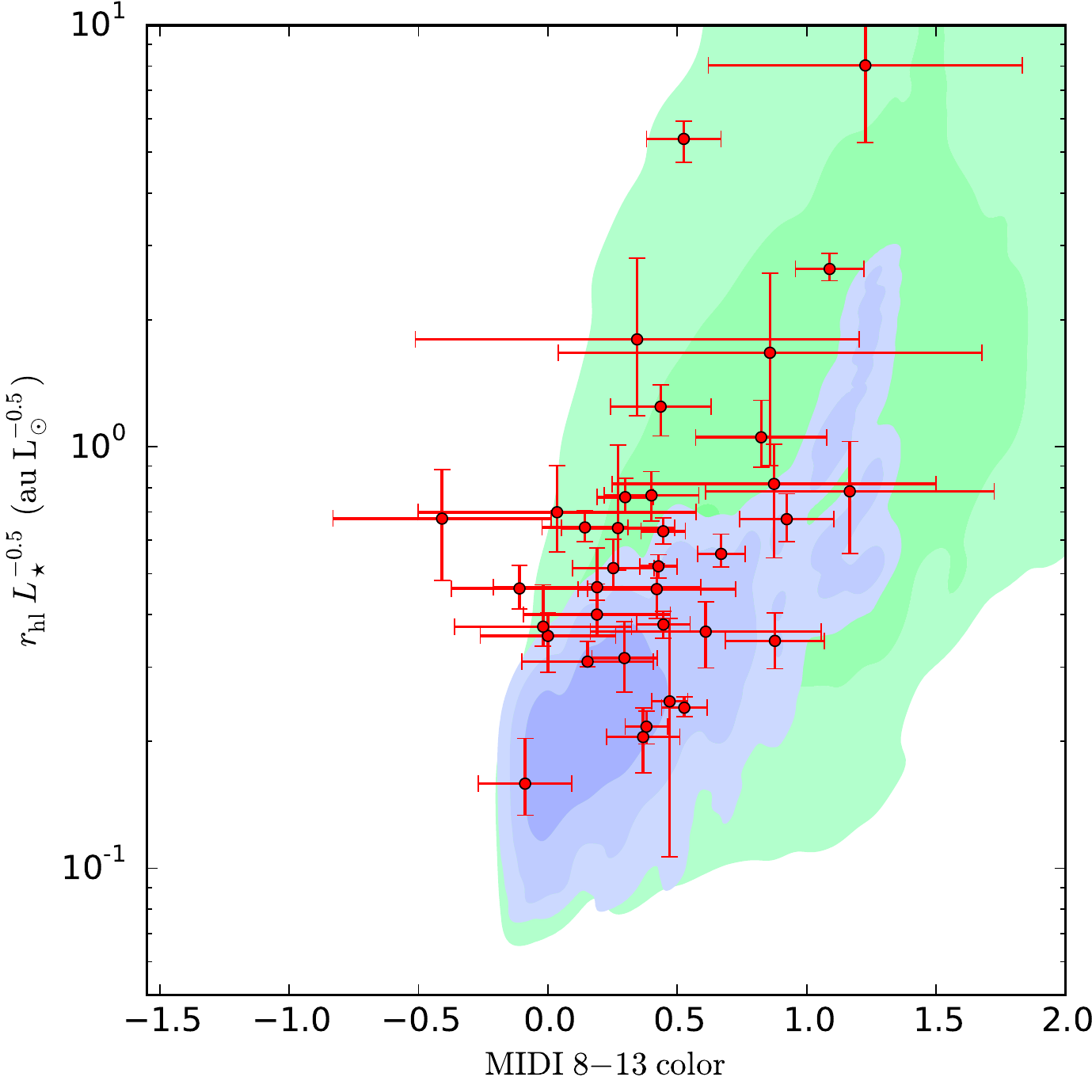}
		\caption{Same as Fig.~\ref{fig:size_color_all}, but only for the T Tauri stars. The shaded regions represent the distribution of radiative transfer models of disks with (green area) and without (blue area) gaps. Darker shades indicate higher density of models. See the text for more details.} 
		
		\label{fig:size_color_TT}
	\end{figure}
	
	\begin{table}
		\caption{Parameters used in our radiative transfer modeling. \newline
		}
		\label{tab:rt_param}
		\centering
		\begin{tabular}{l p{3.6cm} p{3.0cm} l}
			\toprule \toprule
			\multicolumn{4}{l}{\textit{System Parameters}}\\
			& {Distance} ($d$)                               & 60, 100, 140, 200, 400    &pc\\
			& {Inclination ($i$)}                    &0, 35, 70      &$^{\circ}$\\
			\multicolumn{4}{l}{\textit{Stellar Parameters}}\\
			& {Luminosity ($L_\star$) and\newline eff. temperature ($T_\mathrm{eff}$)}                             &(0.5 L$_\sun$ , 4000 K),\newline (1.0 L$_\sun$, 4300 K),\newline (2.0 L$_\sun$, 4800 K),\newline (5.0 L$_\sun$, 5400 K),\newline (10.0 L$_\sun$, 6000 K)     &  \\
			%&{Mass}  ($M_\star$)                           &1.0, ...     & M$_\sun$ \\
			%&\emph{Radius} ($R_{star}$)                           &  1.6     &$\rm{R}_{\odot}$\\
			%&\emph{Visual extinction} ($A_\mathrm{V}$)                    &  0.0      &\\
			%&\emph{Selective extinction coefficient} ($R_\mathrm{V}$)                    &  3.1      &\\
			\multicolumn{4}{l}{\textit{Halo}}\\
			&{Inner radius}    &$R_\mathrm{sub}$       & \\
			&{Outer radius}  &$1.3~R_\mathrm{sub}$  &\\
			\multicolumn{4}{l}{\textit{Dusty Disk}}\\
			&{Inner radii} ($R_\mathrm{in}^\mathrm{d}$)    & $R_\mathrm{sub}$, 0.5, 1, 2, 5      & au\\
			&{Outer radius} ($R_\mathrm{out}^\mathrm{d}$)   &100  &au\\
			&Inner scale height ($h_\mathrm{in}^\mathrm{d}$)               & 0.04, 0.07, 0.1     &\\
			&Outer scale height ($h_\mathrm{out}^\mathrm{d}$)               & 0.1, 0.15, 0.2     &\\
			%&Power-law index for scale height ($p_h$)               & 0.228    &\\
			&{Power-law index for surface density ($p_\Sigma$)}               & 0    &\\
			&optical depth integrated over the midplane              & $10^3$     &\\
			%&Mass ($M_\mathrm{disk}$)               & $3.4\times10^{-2}$     & M$_\sun$\\
			&{Gas-to-dust mass ratio} & $100$     & \\
			\multicolumn{4}{l}{\textit{Dust Grain Size Distribution}}\\
			&{Diameter range}  & $[10^{-2},~10^{3}]$ & $\mu$m \\
			&{Power-law index}  & $-3.5$ &  \\
			%&\emph{Dust species} & \multicolumn{2}{l}{\parbox[t]{3.6cm}{$50\%$ amorphous carbon,\\$50\%$ astronomical silicate}} \\
			&{Dust species:} & &\\
			& \hspace{0.3cm}{amorphous carbon} & 50 & $\%$\\
			& \hspace{0.3cm}{astronomical silicate} & 50 & $\%$\\
			%&Optical depth ($\tau$) & 2 & \\
			
			\bottomrule
		\end{tabular}
	\end{table}
	
	IR interferometry is ideally suited to determine the size of the emitting circumstellar region. While in the near-IR regime the size of the emitting area simply scales with the square root of the stellar luminosity \citep{Monnier2005}, in the mid-IR the situation is more complex, since the emitting region covers a broader disk area from the inner rim to the planet-forming region \citep{Monnier2009}. The temperature gradient in the disk is influenced by the height of the inner rim, the flaring of the disk, and by other structural peculiarities like gaps or rings \citep[e.g.,][]{Leinert2004}. Thus the mid-IR observations are indispensable to clarify the structure of the inner disks at scales of $1-10$~au, which cannot be done at other wavelengths. 
	
	To learn about the distribution of the mid-IR emission we will study the distribution of the fitted half-light radii ($r_\mathrm{hl}$), derived in Sect.~\ref{sec:int_model}, as function of other system parameters.
	Disks can have complex geometries, but our simple modeling approach has the advantage that we can characterize the disk sizes with only one parameter. The size of the emitting region is determined by the density and temperature distributions of the disk. %the luminosity of the star ($L_\star$), and 
	In our simple geometric models this is taken into account by the temperature gradient ($q$) and the inner radius ($R_\mathrm{sub}$ or $R_\mathrm{in}$). As our continuous model performs more robustly than the gapped model (Sect.~\ref{sec:int_model}), and it can be applied for all sources, in the following we will use the parameters of the continuous model ($q$, $r_\mathrm{hl}$) to analyze the size distribution of disks. 
	
	For continuous disks we expect that the size increases with the stellar luminosity. Disks with inner gaps have larger half-light radii than the continuous disks for the same stellar luminosity. The disk size--luminosity relation is well established for near-IR disk sizes \citep[e.g.,][]{Monnier2002,Eisner2007}, but for the mid-IR only a weak correlation was found \citep{Monnier2009,Menu2015}. The near-IR emission is dominated by dust near the sublimation radius, which is largely determined by the stellar luminosity, but the mid-IR emission generally comes from a much larger area. Thus, the specific disk geometry can have a much larger influence on the distribution of the mid-IR emission. Figure~\ref{fig:size_lum} shows $r_\mathrm{hl}$ as a function of stellar luminosity. More luminous sources from \citet{Menu2015} are also plotted in the figure for comparison and for extension of the distribution to higher luminosities. For comparison, the radii that correspond to temperatures of $250$~K and $900$~K for optically thin gray dust are overplotted on the figure. As expected, we see a correlation between $r_\mathrm{hl}$ and $L_\star$, as was observed earlier. The spread in $r_\mathrm{hl}$ may indicate a large variety of structure, for example, variance in the inner disk radii, flaring angle, and disk inclination.
	The T Tauri stars seem to be more concentrated within the range delimited by the $250$~K and $900$~K lines than the Herbig stars, as most T Tauris have $r_\mathrm{hl} \gtrsim 0.7$~au. It could mean that the low-mass sample is more homogeneous regarding disk structure. Moreover, the distribution of sizes of T Tauri disks are closer to the $250$~K line, that could mean disks around low-mass stars are generally colder and more extended with respect to the stellar luminosity than disks around Herbig stars. There are, however, several outliers, like TW Hya, GM Aur, and Haro 6-10N, which have a much larger size than expected from their luminosity, attributed to their inner gap. The lower boundary of the size distribution at low luminosities is set by the resolving power of the VLTI at mid-IR. We estimate that the practical resolution limit of MIDI is $r_\mathrm{hl} \approx 2$~mas, which is $0.3$~au at a typical $150$~pc distance of nearby star-forming regions. These disks are the smallest ones where the half-light radius could be reliably estimated. However, the actual resolution depends on not just the longest attainable baseline, but also on the $uv$-sampling and on the SNR of the measured flux. MATISSE therefore will probably have higher effective resolving power due to better sensitivity and baseline length coverage. 
	
	Following \citet{Menu2015}, in Fig.~\ref{fig:size_color_all} we plot the normalized size ($r_\mathrm{hl} L_\star^{-0.5}$) as a function of the $8-13\ \mu$m color.
	The color was calculated from the MIDI total fluxes as $-2.5 \log_{10} ( F_{\mathrm{tot,\nu,\,}8\mathrm{\,}\mu\mathrm{m}} / F_{\mathrm{tot,\nu,\,}13\mathrm{\,}\mu\mathrm{m}} )$.  The fluxes are error-weighted average values over a window $0.33~\mu$m wide, averaged over all individual measurements, using an error-weighted mean. By using normalized sizes we account for the influence of the stellar luminosity. If the disk structure of our objects were the same, all sources would lie on a horizontal line in the figure. The vertical spread therefore represents the distribution of the fitted temperature gradients ($q$). Five objects with very large uncertainties in their parameter values (error in the $8-13\ \mu$m color $> 1$ or error in $\log (r_\mathrm{hl} L_\star^{-0.5} )>0.7$ ) are not plotted on the figure.
	
	The distribution of the T Tauri stars in Fig.~\ref{fig:size_color_all} is very similar to that of the Herbig stars, presented in \citet{Menu2015}. However, the relative sizes of the T Tauri objects are more concentrated in the $0.3-0.7\mathrm{\ au\,L}^{-0.5}_\odot$ range. A second distinct group at $r_\mathrm{hl}L_\star^{-0.5} < 0.15\mathrm{\ au\,L}^{-0.5}_\odot$ can be observed for only Herbig stars, both in our sample and in the sample of \citet{Menu2015} as shown in their Fig.~4. Taking Fig.~\ref{fig:size_lum} into account, this lack of compact mid-IR emitting regions among T Tauri disks may be caused by observational limitations.The real abundance of disks around low-mass stars with $r_\mathrm{hl} \lesssim 0.3$~au is therefore an open question, which should be addressed with the new generation interferometric instruments GRAVITY \citep{Gravity2017} and MATISSE. 
	
	To interpret the distribution of the sources in the normalized size--color diagram we performed radiative transfer modeling using RADMC-3D \citep{Dullemond2012}, much the same way as in \citet{Menu2015}, but extending the parameter space of the models to lower luminosities. We use a population of models defined by a grid of parameters, listed in Table~\ref{tab:rt_param}. The model includes the central star, an optically thin spherical halo close to the star, and an optically thick flaring disk. The luminosity of the central star in our models ranges from $0.5$ to $10~L_\sun$, which is typical of T Tauri stars.\footnote{For a distribution of models of intermediate-mass stars we refer to \citet{Menu2015}.} The effective temperature of the star corresponds to the given luminosity, given by the location of typical pre-main sequence stars in the Hertzsprung-Russell diagram. The disk inner radius $R_\mathrm{in}^\mathrm{d}$ ranges from the sublimation radius to $5$~au. The models are calculated at $140$~pc distance. We distinguish between continuous disks ($R_\mathrm{in}^\mathrm{d} = R_\mathrm{sub}$) and disks with gaps ($R_\mathrm{in}^\mathrm{d} > R_\mathrm{sub}$). From the modeling we obtain SEDs and images, from which we calculate visibilities between $B_\mathrm{p} = 0.7$~m and $327$~m, with position angles sampled at $15^\circ$ intervals. Then we rescale the models at five selected distance values between $60$~pc and $400$~pc to cover the distance range present in the sample. Each visibility curve is then resampled using the real $B_\mathrm{p}$ distribution of the observed data. We have a set of $B_\mathrm{p}$ values for each of our $82$ objects, thus we produce $82$ samplings of a given modeled visibility curve. This ensures that the simulated data have statistically the same $uv$-configuration as the real data, to avoid biases in the fitting. Then we use the simulated SEDs and visibilities to calculate the color and the half-light radius in the same way as for the real data. 
	
	The results from the modeling are plotted in Fig.~\ref{fig:size_color_TT}. The figure shows the normalized mid-IR size as a function of the mid-IR color, like in Fig.~\ref{fig:size_color_all}. The distribution of the parameters derived by the model grid is shown as blue (continuous disks) and green (gapped disks) shaded regions. The blue region covers a much smaller area and overlaps with the green region. The locations of the regions mostly correspond to those in the modeling of intermediate-mass sources in \citet{Menu2015}. Parameter values for T Tauri stars from the MIDI data are overplotted. Most of our data points fall into the area defined by the models, with more sources occupying the area of the gapped models. It is interesting to note that the region with the highest density of continuous models is almost devoid of observed data points, indicating that the majority of the T Tauri stars in our sample may have gaps in their disks. A notable example is GM Aur, located at the top of Fig.~\ref{fig:size_color_TT}, which has the largest relative half-light radius in the sample. The star GM Aur has a transitional disk with a very large inner-hole radius of $20.5^{+1.0}_{-0.5}$~au \citep{Graefe2011}. MIDI observations of this source were taken at a single baseline, thus we are unable to constrain the gap radius. The half-light radius from the continuous model is $6.7^{+3.4}_{-2.6}$~au, the second highest value among T Tauri stars in the sample. It provides an excellent example to show that a very large half-light radius from the continuous modeling implies the presence of an inner hole. 
	
	\citet{Eisner2007} found that the lowest mass T Tauri disks (corresponding to $L_\star \lesssim 1~\mathrm{L}_\odot$) look oversized in the near-IR. They discuss several possible explanations for this, like viscous heating due to accretion, or a larger fraction of small dust grains  moving the sublimation radius outwards. Alternatively, these disks may be truncated by a different process, such as magnetospheric truncation, which is the scenario considered most likely by Eisner et al. However, \citet{Pinte2008} argue that for low-luminosity objects scattered starlight can make the disk appear to be more extended in near-IR than the size of the inner edge. Recently, \citet{Anthonioz2015} found that, taking into account scattered stellar light in the modeling, the inner disk radii of T Tauri stars measured from $H$ band interferometric data are consistent with the expected dust sublimation radii, although their data do not cover the $L_\star < 1~\mathrm{L}_\odot$ range.  We note that in the calculation of mid-IR sizes the scattered starlight is not an issue, because of its negligible contribution.
	
	If future observations do not find compact T Tauri objects, then it might indicate a general structural difference between Herbig and T Tauri disks. As a speculation, we propose the following scenario to explain this. As the inner rim of a dusty disk at the sublimation radius is thought to be extended vertically (puffed-up inner rim), it can cast shadows onto the outer parts of the disk \citep{Dullemond2001,Dullemond2004_flaring,vanBoekel2005_flaring}. Flat, self-shadowed disks appear to be compact, because only the inner-rim emits substantial mid-IR radiation. Puffed-up inner rims of T Tauri disks, however, are likely less pronounced than those of Herbig disks \citep{Dullemond2004_dust_settling,Dullemond2007}, causing less shadowing. Consequently, T Tauri disks can appear more spatially extended.
	
	\subsection{Variability}
	\label{sec:var}
	
	\begin{table*}
		\centering
		\caption{\label{tab:var} List of candidate objects for mid-IR variability.}
		\begin{tabular}{r l l l l l l l l l p{4cm}}
			\toprule
			\#&Name&$\chi^2_\mathrm{corr}$&$\chi^2_\mathrm{total}$&$\Delta t_\mathrm{max,\ corr}$&$\Delta t_\mathrm{max,\ total}$&$\sigma_\mathrm{rel,\ corr}$&$\sigma_\mathrm{rel,\ total}$&$N_\mathrm{corr}$&$N_\mathrm{total}$&Comment\\
			&&&&(yr)&(yr)\\
			\midrule
			50&EX Lup&90.4&16.5&4.9&4.9&1.5&1.1&20&17&Young eruptive star.\\
			5&T Tau S&76.1&7.6&8.2&8.2&0.7&0.5&4&3&T Tauri star, a close binary system. Known IR variable \citep{Ghez1991,Duchene2002}.\\
			69&HD 150193&40.3&2.8&6.9&6.9&7.0&0.1&13&11&Herbig Ae star. Interferometric spectra are also modulated by an apparent binary signal. A comparison of ISOPHOT-S, ISO-SWS, and UKIRT/CGS3 spectra indicate significant variations both in flux level and shape of the silicate feature \citep{kospal_atlas}.\\
			75&SVS20N&37.7&0.7&3.7&3.7&0.5&0.2&6&6&Embedded Herbig Ae star. A gapped model may also fit the data without the need for time variability.\\
			82&HD 179218&31.7&4.4&5.3&5.3&2.0&0.1&21&6&Herbig Ae star.\\
			30&V1647 Ori&30.9&2.6&0.9&0.6&0.7&0.4&9&3&Young eruptive star. MIDI observations were taken during its $2003-2006$ outburst.\\
			3&RY Tau&29.5&29.1&9.3&9.2&0.2&0.4&7&6&T Tauri star. \\
			78&S CrA N&26.6&4.2&1.1&1.1&0.9&0.3&6&4&T Tauri star. Mid-IR variability, with an amplitude of $\sim30\%$  was also observed with Spitzer IRS and ISOPHOT-S \citep{kospal_atlas}.\\
			68&V346 Nor&25.2&9.9&1.9&1.9&0.4&0.4&19&14&Young eruptive star.\\
			41&CV Cha&20.7&1.0&0.8&0.8&0.4&0.2&3&3&T Tauri star. Dispersion also can be due to disk inclination. Variability was marginally observed by \citet{kospal_atlas}.\\
			\bottomrule
		\end{tabular}
	\end{table*}
	
	Young stars commonly produce brightness variations, caused by various processes (variable accretion, stellar magnetic activity and rotation, disk turbulence, outflows) acting on different timescales \citep[for a review, see][]{Sicilia-Aguilar2016}. Variability is most pronounced at optical \citep{Rigon2017} and near-IR \citep{Eiroa2002,Rice2012} wavelengths, but it is still significant in the mid-IR \citep{Espaillat2011,kospal_atlas,Bary_spitzer,Varga2017}. \citet{kospal_atlas} studied ISOPHOT-S and Spitzer spectra of $47$ sources, and found that $79 \% $ of the sources varied by
	more than $0.1$~mag, while $43 \%$ changed more than $0.3$~mag. Timescales of mid-IR variability can be as short as a few days \citep{vanBoekel2010}. Moreover, some sources show enhanced variability in the mid-IR silicate feature \citep[see, e.g.,][on DG Tau]{Bary_spitzer,Varga2017}. With the future generation of IR interferometers, observing sources with the aim of image reconstruction will be a routine task. When planning such observations it is essential to know the timescale over which the source exhibits significant brightness changes. Large variations can make the image reconstruction unfeasible.
	
	For most sources in our sample there were no dedicated interferometric monitoring observations. Hence the distribution of observing epochs is not optimal for studying variability, that is, usually most of the observations were taken on a few consecutive nights. The longest time range for MIDI observations is approximately ten years, for instance in the cases of AB Aur, RY Tau, V856 Sco, and Z CMa. When there are observations with significant time differences, in most cases it means two or three really different epochs. Another issue is that of the $uv$-configuration: strictly one can directly compare correlated spectra from different epochs only if they were all measured with the same baseline length and position angle. This constraint would greatly reduce the number of comparable observations. Total spectra are always comparable, but have much larger uncertainty than the correlated spectra. As the cadences of observations are highly heterogeneous, automatic characterization of variability is hardly viable. 
	
	To tackle these issues we apply a simple statistical approach to study time variability. In this scheme we consider the best fitting continuous disk model (see Sect.~\ref{sec:int_model} and Fig.~\ref{fig:atlas_fit_app}) as a reference state of the disk, and we try to characterize the deviations from the model as variability. We also want to take into account the uncertainties of the observations to be able to tell the significance of the deviation from the model. Thus we calculate the reduced chi-square ($\chi^2_\mathrm{red}$) of the observed data points with respect to the model, as
	\begin{equation}
	\chi^2_\mathrm{red} = \frac{1}{N} \sum_{i=1}^N \frac{\left({D_i-M_i}\right)^2}{\sigma_{D_i}^2},
	\end{equation}
	where $N$ is the number of data points, $D_i$ are the observed fluxes, $M_i$ are the model fluxes, and $\sigma_{D_i}$ are the total uncertainties of the observed data, at $10.7~\mu$m. Chi-square values are calculated separately for the total and correlated fluxes, for all objects. We also calculate the largest time difference between the observations of the sources and the scatter of the data points normalized to the best-fit model ($\sigma_\mathrm{rel}$), that is, the standard deviation of $D_i / M_i$. The latter may be related to temporal brightness variations, but may also reflect that our continuous disk model does not account for fine structure, inclination, or gaps. As we know that the true intensity distributions of the disks are significantly different from that of this simple model, we should be cautious in the interpretation of the deviations from the model. 
	
	To identify candidates for mid-IR variability, we consider only the objects with the highest $\chi^2_\mathrm{red}$ values, and examine the sources by visual inspection. We selected a group of objects with $\chi^2_\mathrm{red} > 20$, as they clearly deviate from the rest of the sample. From this group we discarded those objects where the high $\chi^2_\mathrm{red}$ could be due to inclined disks (GW Ori, UX Ori) or gaps (DK Cha, SVS 13A1, HD 72106). The remaining sources with extreme $\chi^2_\mathrm{red}$ values are considered to be variable candidates and are shown in Table~\ref{tab:var}. The table lists the $\chi^2_\mathrm{red}$ values, the maximum time range of observations ($\Delta t_\mathrm{max}$), the standard deviation of the data points normalized to the best-fit model ($\sigma_\mathrm{rel}$), and the number of observations ($N$) for the correlated and total fluxes separately. Generally, $\chi^2_\mathrm{red}$ for correlated fluxes is larger because of the smaller uncertainties. The time span of the measurements ranges from seven months to nine years. Among these objects, there are known eruptive stars (EX Lup, V1647 Ori, and V346 Nor). The star EX Lup has one of the largest amplitudes of variability (a factor of approximately six), related to its historically largest eruption in 2008 \citep{Juhasz2012}. In several normal T Tauri and Herbig Ae systems in  Table~\ref{tab:var}, variability has already been reported in the literature, based on Infrared Space Observatory (ISO), Spitzer, and ground-based measurements (T Tau S, HD 150193, S CrA N, CV Cha), supporting our MIDI variability analysis.
	
	In contrast to correlated fluxes, total flux measurements are directly comparable without the need to involve modeling. We found that the characteristic amplitude of variations ($\sigma_\mathrm{rel}$) is $\sim$$0.3$~mag at $10.7~\mu$m, both for the sources in Table~\ref{tab:var} and also for the rest of the sample where multi-epoch MIDI observations are available. This is in agreement with previous results \citep{kospal_atlas}. The objects listed in Table~\ref{tab:var} can be considered as good targets for follow-up monitoring observations. Many other studies found that mid-IR variability on daily, weekly, and yearly timescales is ubiquitous \citep[see, e.g.,][and references therein]{kospal_atlas}. Thus, for future interferometric observations (e.g., with MATISSE), it is advisable to take into account these variability timescales, if image reconstruction is planned. On the other hand, the interferometric monitoring of the variability of eruptive stars  can lead to a better understanding of stellar activity and disk instabilities, especially with the efficient imaging and the multi-wavelength approach offered by MATISSE.
	
	\subsection{Dust mineralogy}
	\label{sec:sil_feat}
	
	\begin{table*}
		\caption{Disks showing peculiar spectral features (silicate in absorption or PAH).}             % title of Table
		\label{table:sil_abs}      % is used to refer this table in the text
		\centering                          % used for centering table
		\begin{tabular}{r l l l l p{4.5cm} p{4.5cm}}        % centered columns 
			\hline\hline                 % inserts double horizontal lines
			\# & Name & Object class & \multicolumn{2}{c}{Emission type}  & \multicolumn{2}{c}{Notes}\\
			& &  & correlated & total & correlated spectrum & total spectrum \\
			\hline
			1 & SVS 13A1 & eruptive & a & a & crystalline & crystalline \\
			2 & LkH$\alpha$ 330 & TT & wa,e & e & weak absorption only at longest baseline &  \\
			5 & T Tau S & TT & a & a &  &  \\
			6 & DG Tau & TT & a & e &  &  \\
			7 & Haro 6-10N & TT & a & a &  & The spectra are very similar to the ISM absorption profile \citep{Roccatagliata2011}.  \\
			8 & Haro 6-10S & TT & a & a &  & The spectra are very similar to the ISM absorption profile \citep{Roccatagliata2011}.\\
			10 & HL Tau & TT & a & a &  &  \\
			17 & AB Aur & HAe & a,e & e & binary signal can also be seen &  \\
			26 & V1247 Ori & HAe & wa & a?,PAH &  &  \\
			27 & V883 Ori & eruptive & a & a &  &  \\
			30 & V1647 Ori & eruptive & a,f & f & absorption at longest baseline &  \\
			32 & Z CMa & eruptive & a & a &  &  \\
			34 & HD 72106 & HAe & wa?,e & e & binary signal also can be seen at the longest baselines &  \\
			37 & DI Cha & TT & wa & e &  &  \\
			39 & Sz 32 & TT & a & wa &  &  \\
			43 & DK Cha & embHAe & a & a & crystalline &  \\
			44 & HD 135344B & HAe & wa,f & a,PAH? &  & may show weak PAH emission \\
			56 & V2246 Oph & TT & a & - &  & bad quality data \\
			57 & HBC 639 & TT & wa?,f & we,f & double peaked spectra, maybe a mix of emission and absorption &  \\
			58 & DoAr 25 & TT & wa & - &  & bad quality data \\
			62 & Elias 29 & TT & a & a &  &  \\
			63 & SR 21A & TT/HAe & a & PAH? & noisy spectra & noisy spectrum, may show PAH emission \\
			64 & IRS 42 & TT & a? & a? & double peaked spectra, maybe a mix of emission and absorption & double peaked spectra, maybe a mix of emission and absorption \\
			65 & IRS 48 & TT & a? & a?,PAH & noisy spectrum & maybe a combination of silicate absorption and PAH emission \\
			68 & V346 Nor & eruptive & a & a &  &  \\
			73 & HD 169142 & HAe & a?,f & PAH & noisy spectra &  \\
			75 & SVS20N & TT & a & a &  &  \\
			80 & VV CrA NE & TT & a & a &  &  \\
			81 & VV CrA SW & TT & a & a &  &  \\

			\hline                                   %inserts single line
		\end{tabular}
		\tablefoot{Emission type categories: a: absorption, wa: weak absorption, e: emission, we: weak emission, f: flat spectrum, PAH: PAH emission.
		}
	\end{table*}
	
	The $8-13~\mu$m spectrum of most circumstellar disks shows the prominent silicate feature, which contains precious information about the silicate dust. In passive, evolved systems the feature originates in the warm surface layer heated by the stellar radiation, appearing in emission. The shape of the feature contains information about the dust composition and grain size distribution. It is well known from theoretical studies that larger grains produce less pronounced silicate emission \citep[see][and references therein]{Natta2007}. Additionally, the silicate feature can show the signs of dust evolution: coagulation and crystallization \citep{vanBoekel2003,Boekel_silicate,vanBoekel2005}. 
	
	The majority of our sources show silicate emission in their spectra, but there are several objects showing peculiar features. One such group is the sources with silicate in absorption, as listed in Table~\ref{table:sil_abs}. Most of these sources have disks embedded in the remnant protostellar envelope, producing line-of-sight absorption. But other scenarios, such as edge-on viewing angle, thermal inversion in the disk due to high accretion, or a specific grain size distribution can also be the cause of absorption. The grain size distribution has an especially significant effect on the feature shape and depth. The majority of these objects show absorption characteristic of small amorphous silicate grains, similar to the interstellar medium (ISM), but in the case of SVS~13A1 we clearly see a crystalline absorption feature. Stars DK~Cha and V346 Nor also show some degree of crystallinity in their absorption spectra. The crystallinity may be linked to the eruptive processes in the disk \citep{Abraham2009}.
	
	\begin{figure*}
		\centering
		\includegraphics[width=0.49\textwidth]{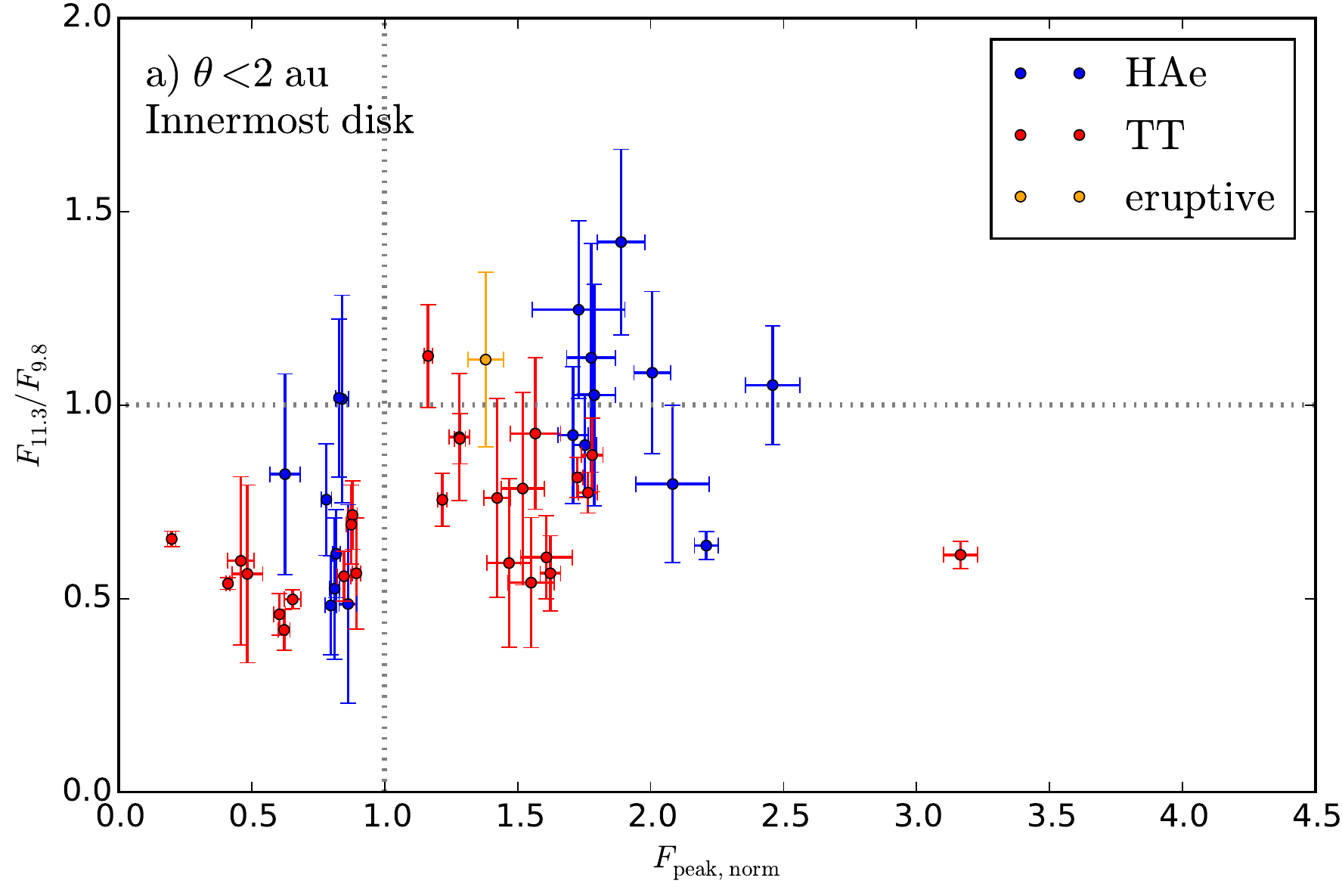}
		\includegraphics[width=0.49\textwidth]{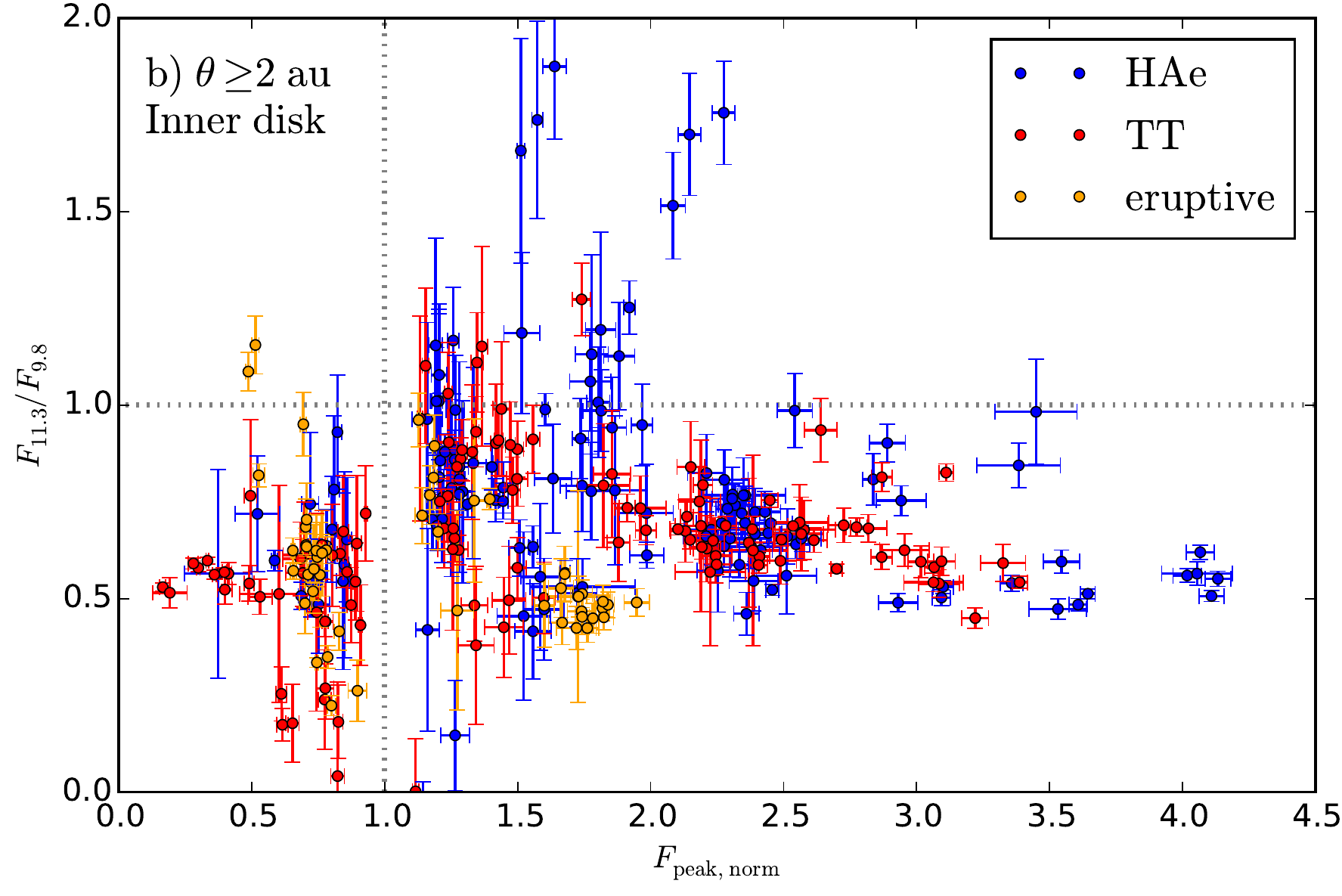}
		\caption{Ratio of the $11.3$ and the $9.8~\mu$m flux versus amplitude of the continuum-normalized silicate feature. The color of the symbols indicates the object type: Herbig Ae (blue), T Tauri (red), and  eruptive systems (orange). Panels a) and b) show the distribution of points extracted from correlated spectra (with physical resolutions smaller and larger than $2$~au, respectively). } 
		
		\label{fig:flux_ratio}
	\end{figure*}
	
	\begin{figure*}
		\centering
		\includegraphics[width=0.49\textwidth]{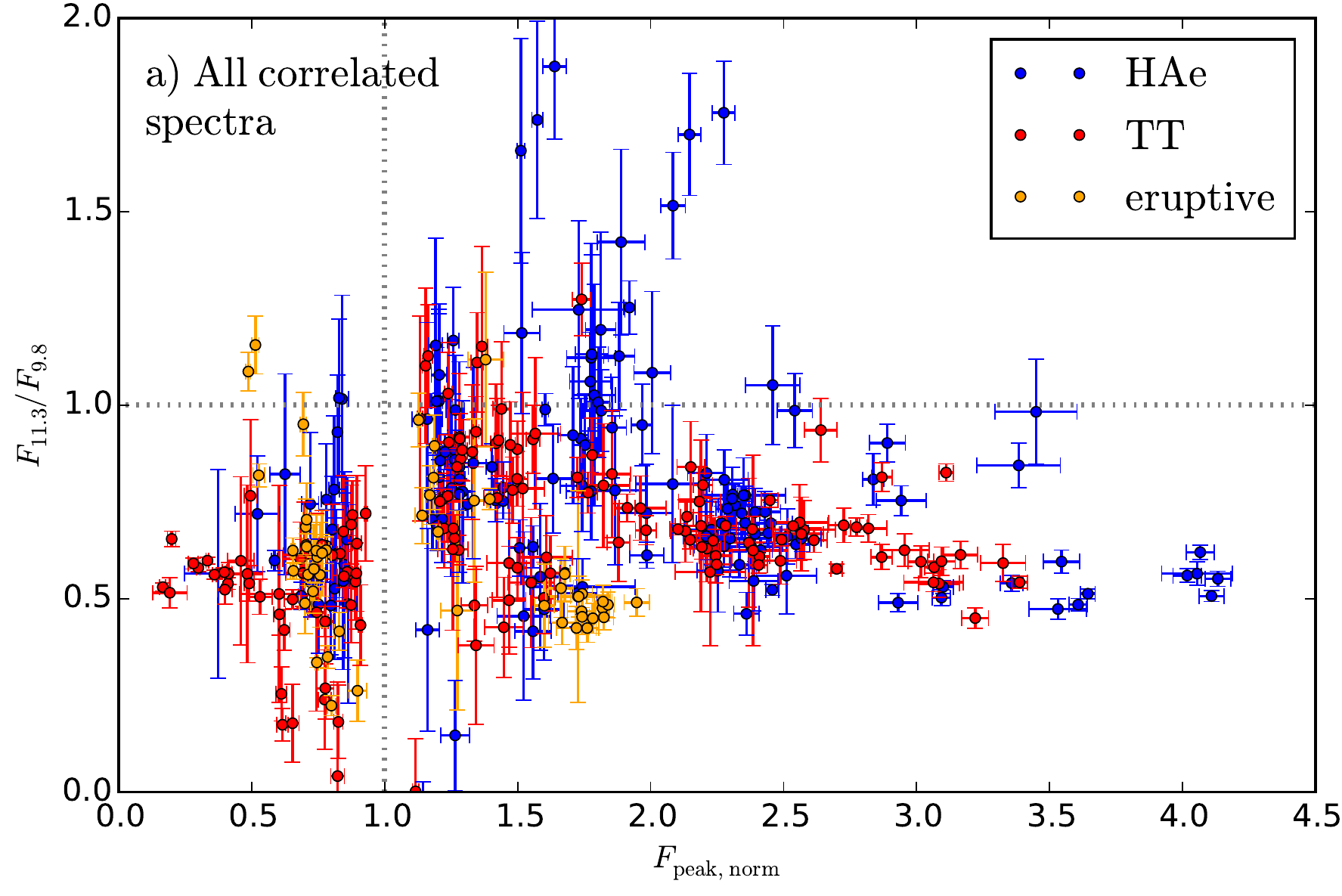}
		\includegraphics[width=0.49\textwidth]{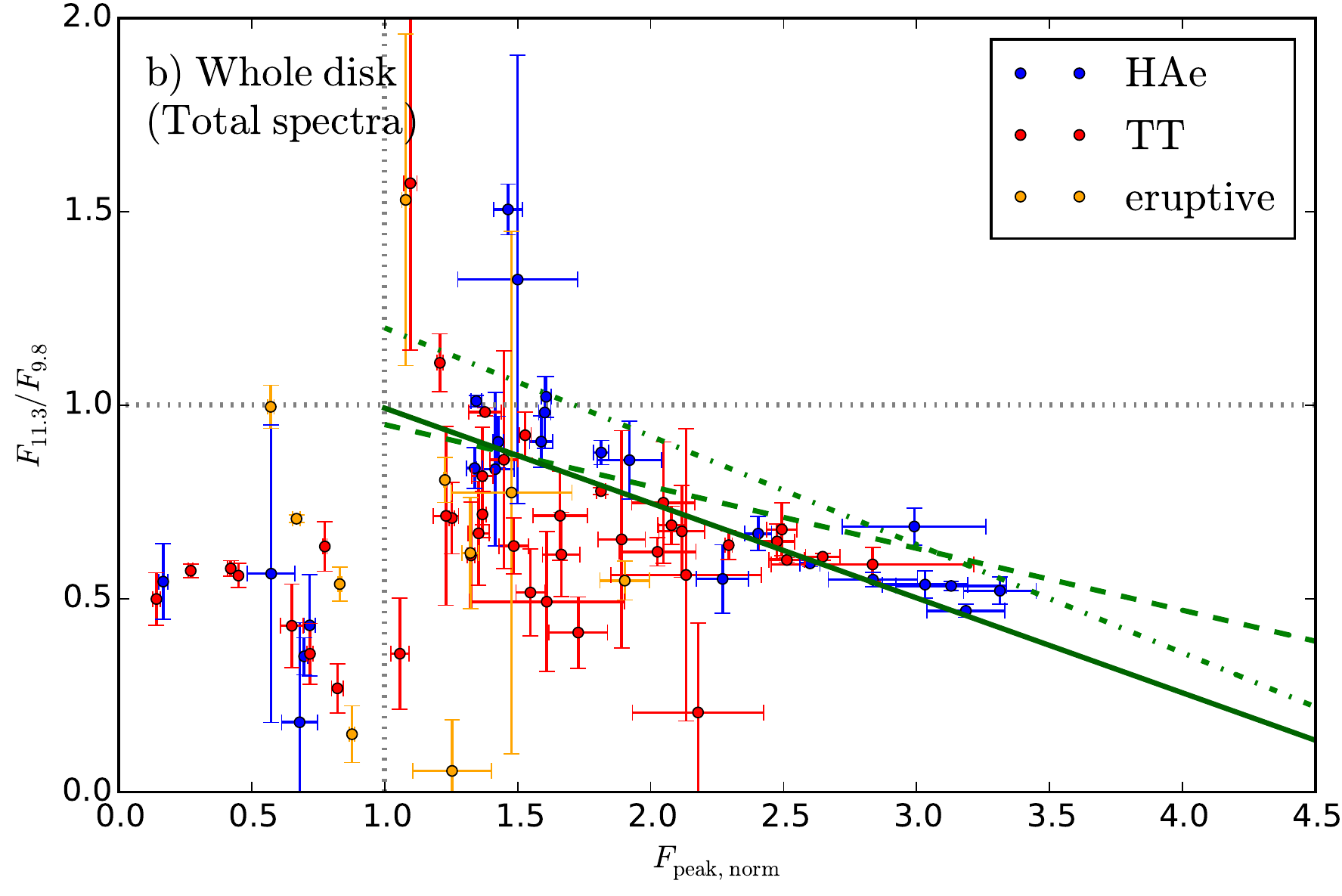}
		\caption{Panel a): same as Fig.~\ref{fig:flux_ratio}, but for all correlated spectra, representing all inner disk spectral shapes. On panel b) values derived from total spectra are shown, averaged for each individual object. The solid green line is our fit to the data, the dashed line is a fit from \citet{Przygodda2003} for T Tauri stars, and the dot-dashed line is a fit from \citet{vanBoekel2003} for Herbig Ae/Be stars. } 
		
		\label{fig:flux_ratio_all}
	\end{figure*}
	
	There are a few sources where the nature of the feature (whether it is in emission or absorption) differs between different baselines (including total spectra as zero baseline). These sources are of special interest, as was demonstrated by \citet{Varga2017} in the case of DG Tau. Here we identified six additional sources showing a switch between emission and absorption in separate regions in their disks: LkH$\alpha$ 330, AB Aur, V1647 Ori, HD~72106, DI Cha, and HBC 639. The first four seem to show absorption in the correlated spectra only at the longest baselines, while short baseline correlated spectra and total spectra show emission features. The boundary radius between the emission and absorption, defined by the resolution of the observation, can be placed at $r\sim 1.1,\ 0.7,\ 2.3,\ 2.6$~au, respectively for the four sources. In the case of AB Aur and HD~72106 the interferometric phases show binary signal at the longest baselines, hence the correlated fluxes are also modulated. The spectrum of DI Cha is more similar to that of DG Tau, showing weak absorption (or flat spectra) in the correlated spectra (inside $1.5-2$~au), and emission in the total spectra. The star HBC~639 is a peculiar case, as it shows almost flat spectra, but some of the correlated spectra can be a mix of absorption and emission features. 
	The low amplitude or absence of silicate emission in inner disks may be related to dust grain growth, but other processes (e.g., accretion heating or an inner absorbing halo) can also play a role, although the exact physical interpretation remains uncertain. Therefore this question should be addressed with the next generation of high angular resolution instruments.
	
	A couple of sources in our sample show polycyclic aromatic hydrocarbon (PAH) features in their spectra, as indicated in Table~\ref{table:sil_abs}. PAH emission can be frequently found in spectra of Herbig stars, but rarely among T Tauri stars and protostars \citep{Kamp2011}. PAHs are stochastically excited by incident ultraviolet photons from the central star. Generally, PAH emission originates from a large area, spanning tens of au around Herbig stars. As our sample mostly contains T Tauri stars, PAH emission is uncommon among our objects, observable in the total spectra of only five objects (V1247 Ori, HD 135344B, SR 21A, IRS 48, HD 169142). All of these have luminosities around $10~\mathrm{L}_\sun$, and can be classified as Herbig Ae stars. We do not observe PAHs in any of the correlated spectra, as expected. 
	
	%misc: dependence of parameters (hlr, color) on disk inclination: we shall not discuss it
	
	The $11.3/9.8\ \mu$m flux ratio ($F_{11.3}/F_{9.8}$) in the spectra of YSOs turned out to be a good proxy for grain growth and crystallinity \citep{vanBoekel2003,Przygodda2003}. For a spectrum of weakly processed, amorphous silicates, consisting of small grains, the flux ratio is significantly less than one, but for evolved dust with a high crystalline fraction and larger average grain size it is around or above one. It was observed that the flux ratio has a tight linear anti-correlation with the peak amplitude of the continuum-normalized silicate feature ($F_\mathrm{peak,\ norm}$), both for Herbig Ae/Be and T Tauri stars \citep{Bouwman2001,vanBoekel2003,Przygodda2003,Bouwman2008}. Up to now, this correlation was only delineated by using unresolved total fluxes. Now, for the first time, we are able to make a similar study of the flux ratios using our spatially resolved spectro-interferometric data. Thus, we perform an automatic characterization of all spectra in our interferometric atlas. Our pipeline groups the spectra by the following categories: emission feature, absorption feature, and weak feature (indicated in Table~\ref{tab:obs}). The last category stands for flat spectra and for those where the emission type cannot be determined due to low SNR. We also calculate $F_\mathrm{peak,\ norm}$ and $F_{11.3}/F_{9.8}$. The details of the data processing can be found in Appendix~\ref{app:spec_class}. 
	
	In order to plot the silicate amplitude versus flux ratio relation at different spatial scales, corresponding to different regions in the circumstellar disks, one can compare total spectra (representing the emission of the whole disk) with correlated spectra (dominated by the inner disk emission). Here we utilize an additional division of the correlated spectra, which results in two different zoom levels, defined by $\theta$, the physical resolution of the observations (indicated in Table~\ref{tab:obs}). For the division we use  $\theta = 2$~au, chosen to emphasize the differences between the groups, discussed later. Thus, we end up with three groups of spectra with different spatial scales: 1) correlated spectra with $\theta < 2$~au, where the emission is dominated by the innermost region of the disk; 2) correlated spectra with $\theta \geq 2$~au, where the emission comes from a larger, but still relatively compact region (inner $2-115$~au, depending on the baseline and distance, with a median $\theta$ of $3.5$~au); and 3) total spectra, containing the whole disk emission. The disk regions representing the groups are not disjointed, which means the $< 2$~au emission is included in the $\theta \geq 2$~au subsample, and both are included in the total spectra.
	
	One should be aware that the correlated fluxes are not proper spectra, but are always modulated by a spatial term that we do not know unless we know the true intensity distribution well (which usually is not the case). To the first approximation this spatial modulation acts as a slope, which is taken out by the normalization. The modulation also depends on the spectral shape of the silicate feature as was shown by \citet{vanBoekel2005_flaring}. Also, if visibilities get close to zero the modulation becomes enhanced, and one has to be particularly wary of this issue when interpreting correlated spectra.
	
	In Fig.~\ref{fig:flux_ratio} we plot the distribution of the $11.3/9.8\ \mu$m flux ratio versus $F_\mathrm{peak,\ norm}$ on two panels based on our processing from the correlated spectra. Panels a) and b) represent the $\theta < 2$~au and $\theta \geq 2$~au groups of correlated spectra, respectively. We excluded those points where the error in the flux ratio or in $F_\mathrm{peak,\ norm}$ is larger than $0.3$ or $0.2$, respectively, because in these cases the parameters are poorly constrained. From the points in panel a) and b) $45\%$ and $23\%$ were excluded by this criterion, respectively. The general distribution of the points did not change after this cut. We extended the parameter space to include objects with silicate absorption, shown in the region $F_\mathrm{peak,\ norm} < 1$. As silicate emission and absorption in circumstellar disks have typically different physical origins, the regions corresponding to absorption and emission should not be treated as a uniform parameter space. Nevertheless, comparing flux ratios of the two domains is perfectly viable with this extension.  The distribution of the points on panel a) differs greatly from that on panel b): we cannot see any tight correlation between the parameters, and there are barely any points with $F_\mathrm{peak,\ norm} > 2.5$. The lack of correlation could also be the consequence of the larger measurement errors on the flux ratio, but the cutoff in the feature amplitude is well justified by the uncertainties. These measurements can be significantly affected by the unknown spatial modulations, but we argue that part of this is due to intrinsically lower feature amplitudes in the innermost disk regions.
	
	Stronger correlation between the parameters can be seen in panel b), in the range of $2.5 < F_\mathrm{peak,\ norm} < 4.5$. These results point to fundamental differences in the dust properties in different regions of protoplanetary disks, indicating the signs of dust processing. According to earlier results in the silicate feature modeling \citep[e.g.,][]{Bouwman2001} large flux ratios and small silicate amplitudes suggest processed dust (grain coagulation and/or crystallization). Our results point out that in the inner parts ($r<1$~au) of disks the dust can be substantially more processed than in the outer parts. In addition, radiative transfer effects can also make the feature weaker, as shown by \citet{Meijer2007}, who found, by modeling Herbig Ae stars, that the bulk of the silicate emission originates beyond radii of $2-4$~au from the star, and the innermost disk region ($r < 0.55$~au) emits an almost flat spectrum. The reason for this is that the hot inner rim, which is very bright, gives nearly no emission features because the optical depth is very high and the spectrum is close to that of a blackbody. Thus, radiative transfer effects can mimic a more processed dust, leading to an overestimation of the abundance of the processed dust. 
	
	In Fig.~\ref{fig:flux_ratio_all} we plotted the distribution of parameter values extracted from all correlated spectra (panel a) and the total spectra (panel b). On panel a) the same cuts in error are applied as in Fig.~\ref{fig:flux_ratio}, for clarity. On panel b) the parameter values were averaged over all total spectra observed for each source.\footnote{Averaging was done with a robust mean algorithm (resistant\_mean by H. Freudenreich) with an outlier rejection parameter set to $3\sigma$.} The figure clearly shows the correlation between the flux ratio and feature amplitude, published by \citet{vanBoekel2003} and \citet{Przygodda2003}. For comparison, we fitted this distribution with a line ($y = ax+b$), using a weighted least-squares regression method (we fitted only points with $F_\mathrm{peak,\ norm} >0$, corresponding to silicate emission features). We used inverse-error weighting, using the root mean square of the uncertainties of the two parameters. The resulting fit parameters  are $a = -0.25 \pm 0.01$, $b = 1.24 \pm 0.02$, and the fitted line is shown on the figure (solid green line). We also plot the results of fits from \citet{vanBoekel2003} with $a = -0.28 \pm 0.04$ and $b = 1.48 \pm 0.09$ for Herbig Ae/Be disks (dot-dashed line), and from \citet{Przygodda2003} with $a = -0.16 \pm 0.04$ and $b = 1.11 \pm 0.11$ for T Tauri disks (dashed line). The three linear fits are generally consistent with each other. 
	
	On the left side of panel b), where $F_\mathrm{peak,\ norm} < 1$, data points are related to silicate absorption. Most of these have low $11.3/9.8\ \mu$m flux ratios (around and below $0.5$), indicating the presence of less-processed grains. A notable exception is SVS 13A1 ($\#1$), which has a flux ratio around $1$, showing a well-defined crystalline absorption feature. 
	
	Differences in the distribution of T Tauri and Herbig Ae spectra can be observed on Fig.~\ref{fig:flux_ratio_all} panels a) and b): there are barely any T Tauri spectra with silicate amplitudes $\mathrm{max}\left(F_{\nu\mathrm{,\ norm}}\right) \gtrsim 3$, and on panel b) points representing T Tauri objects are much more dispersed, especially in the area below the fitted lines. Stronger silicate emission in Herbig Ae stars may indicate that the disk flaring is more pronounced than in T Tauri stars, due to their higher luminosity or stellar mass ratio. Parameter values for eruptive systems are co-located with those for T Tauri objects, however, the maximum emission amplitude of eruptive spectra is smaller than that of T Tauri spectra. The eruptive subsample seems not to follow the trend between the feature amplitude and flux ratio.
	
	\subsection{Observability by MATISSE}
	
	\label{sec:matisse}
	
	\begin{figure*}
		\centering
		\includegraphics[width=0.49\textwidth]{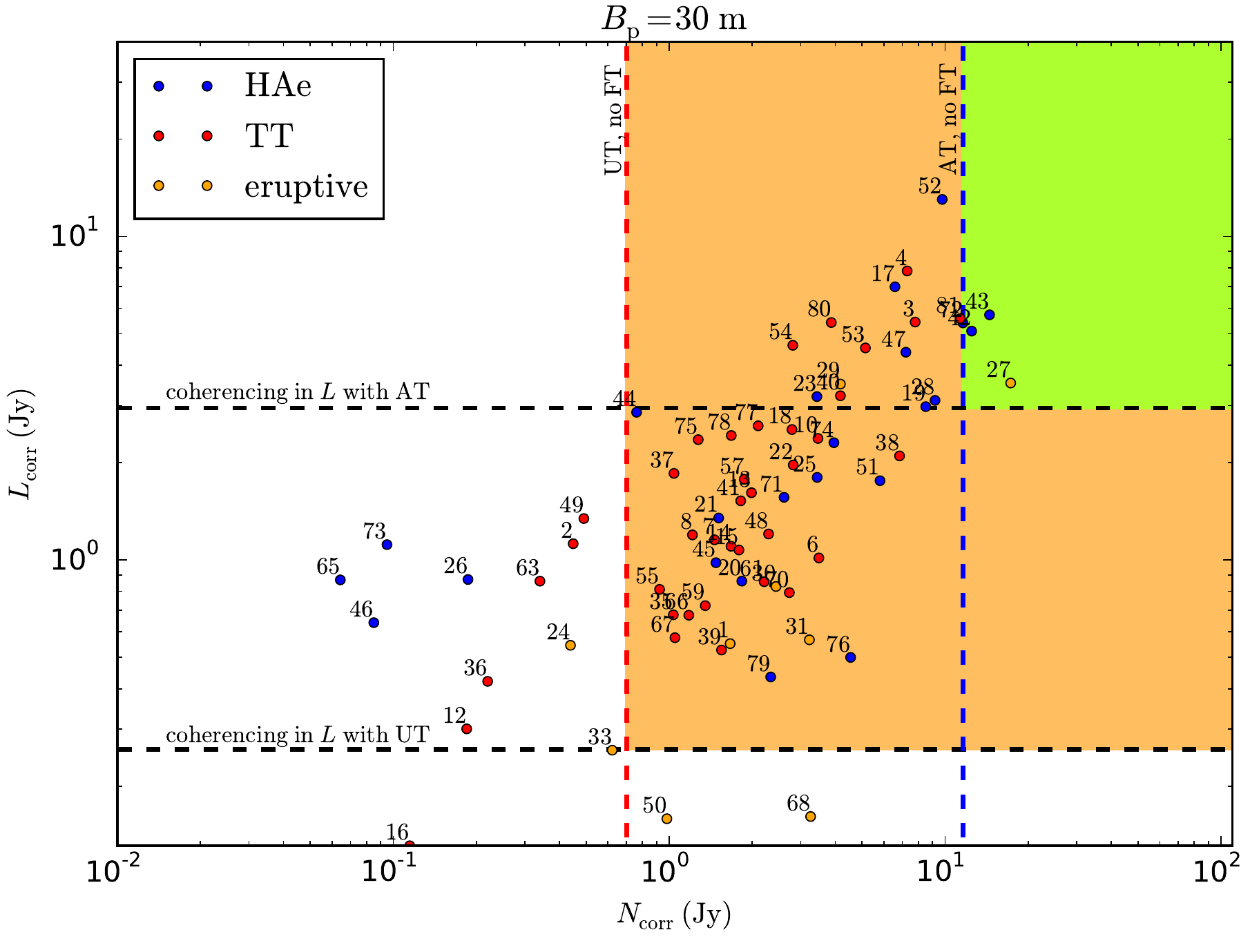}
		\includegraphics[width=0.49\textwidth]{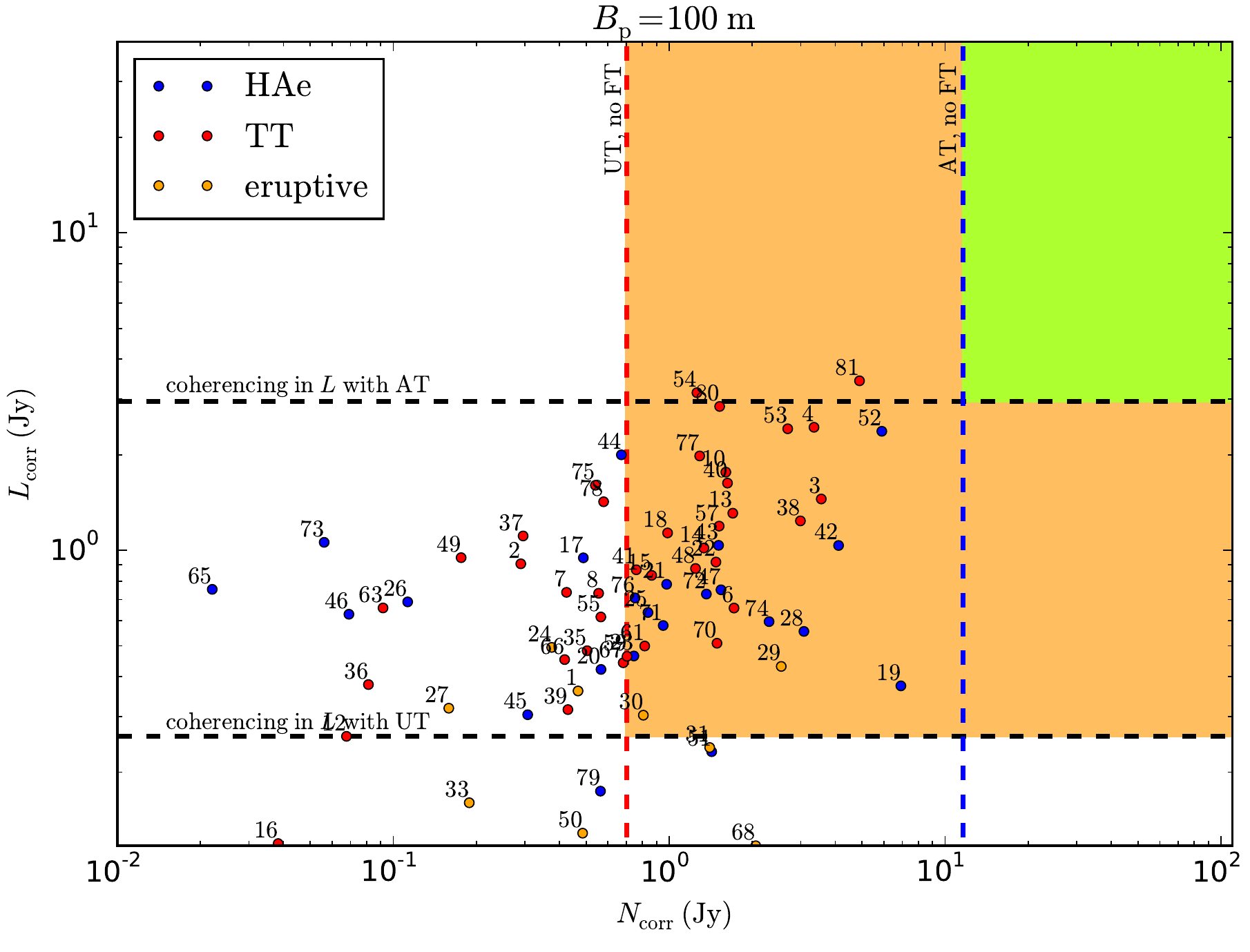}
		\includegraphics[width=0.49\textwidth]{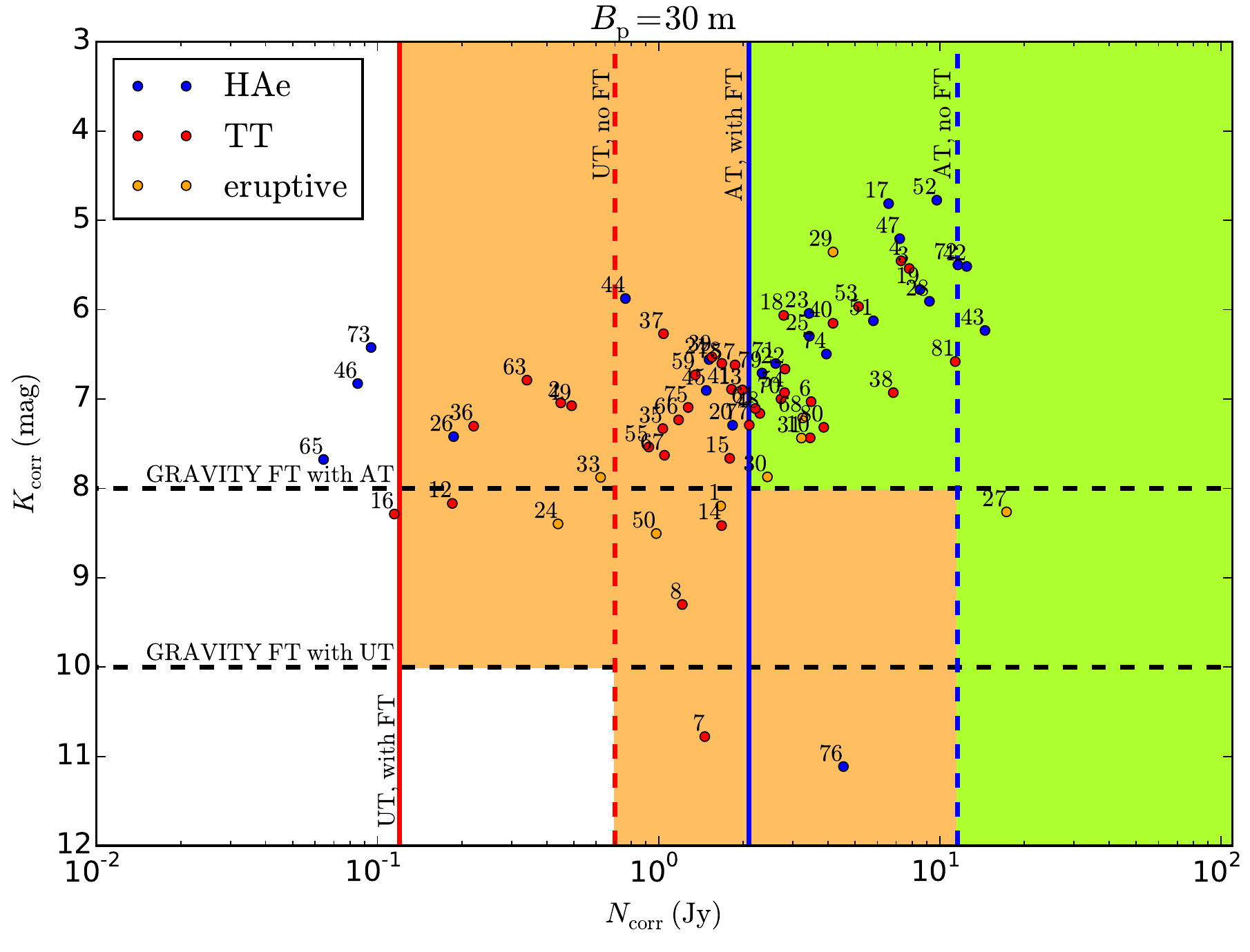}
		\includegraphics[width=0.49\textwidth]{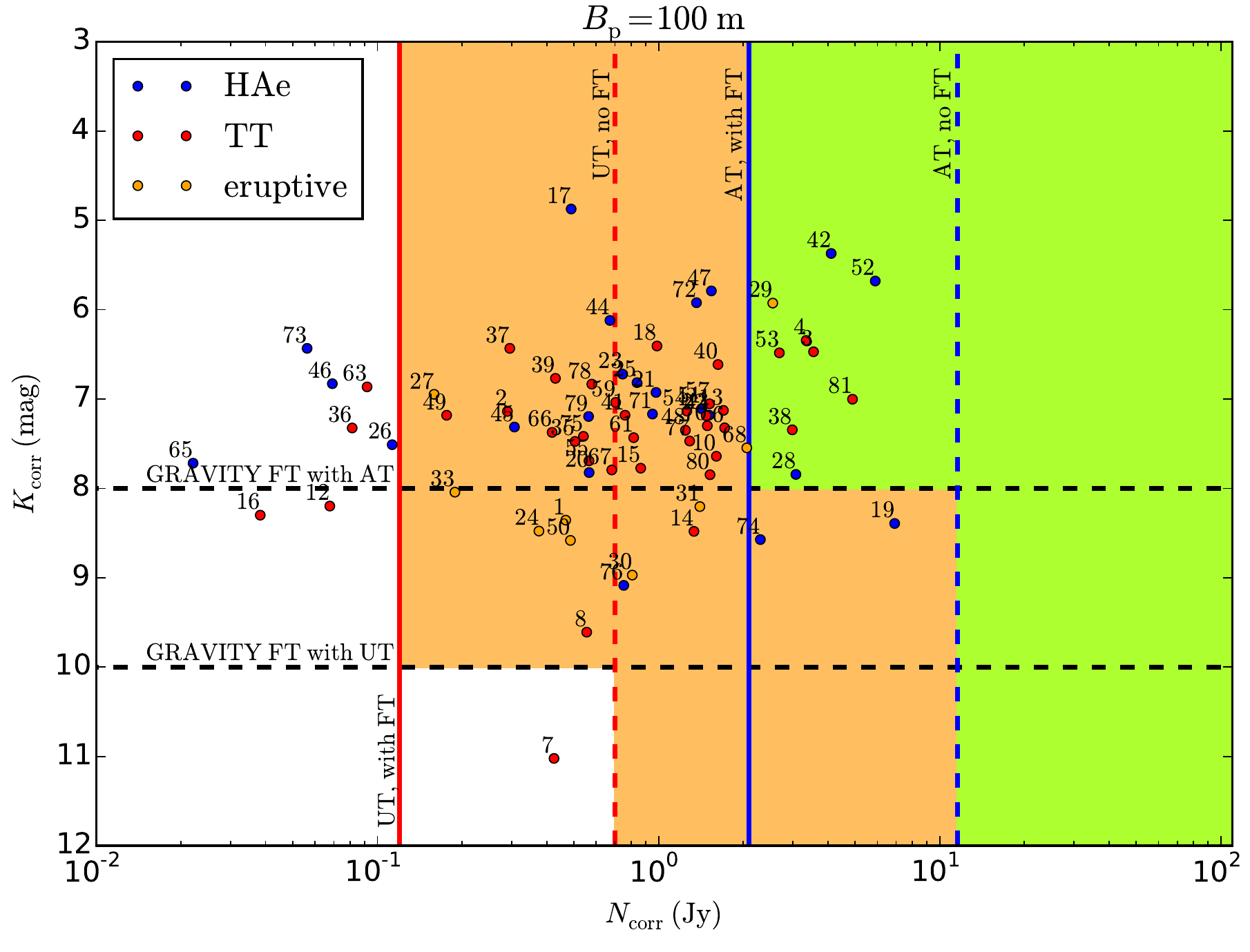}
		\includegraphics[width=0.49\textwidth]{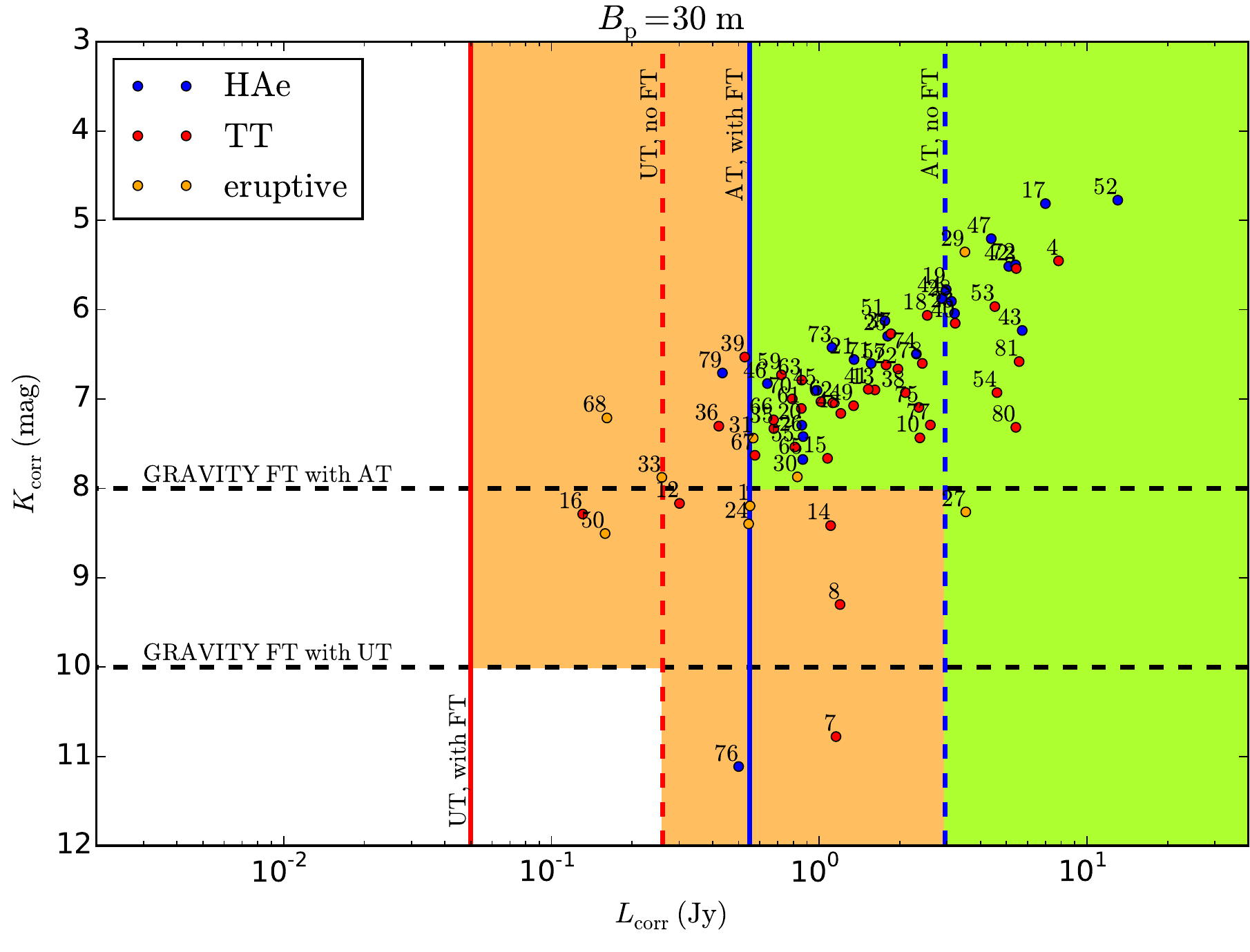}
		\includegraphics[width=0.49\textwidth]{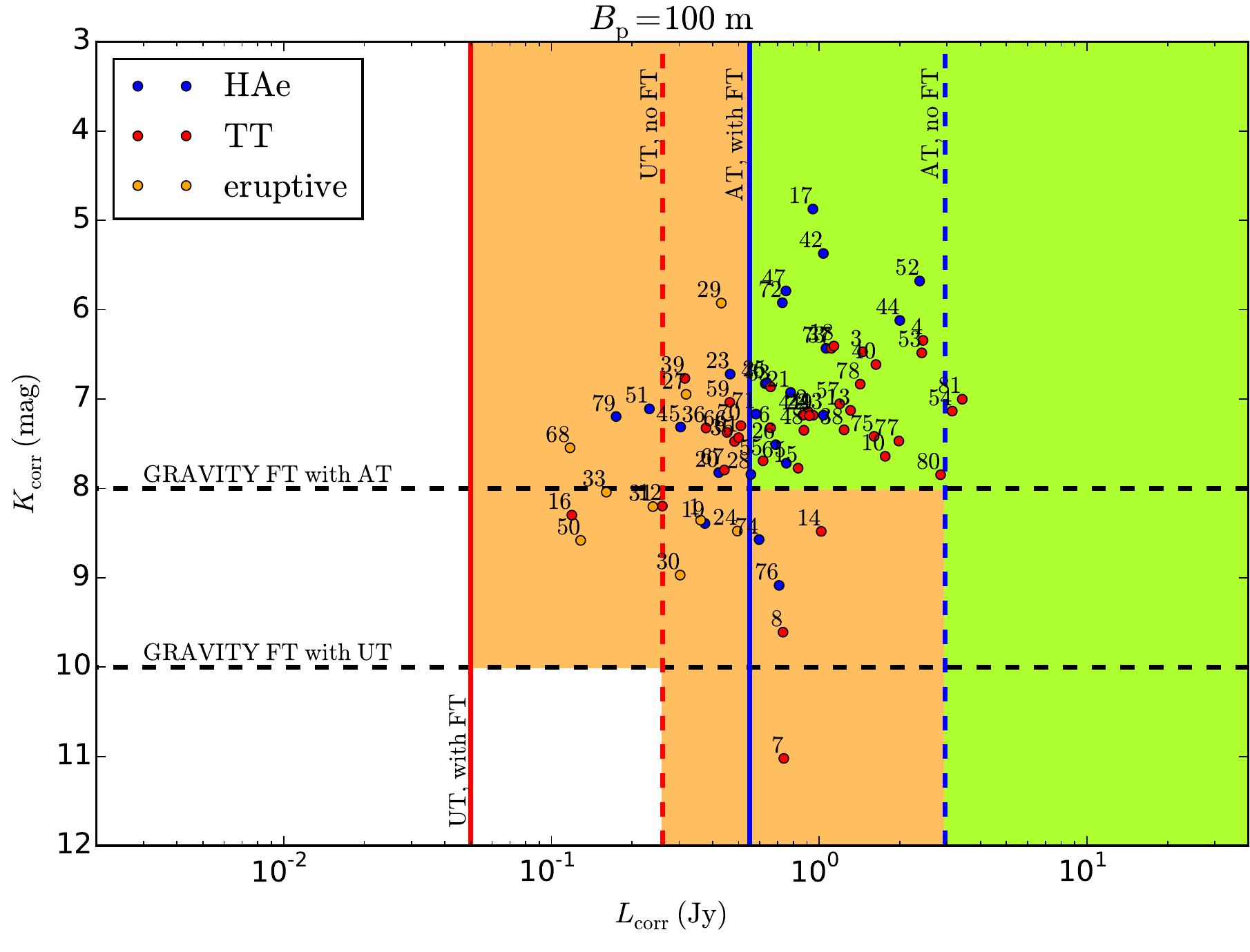}
		\caption{Observability plots for MATISSE: $L$ versus $N$ band (top row), $K$ versus $N$ band (middle row), and $K$ versus $L$ band (bottom row) correlated fluxes, estimated for our sample, from the continuous disk modeling, at 30~m (left column) and 100~m (right column) baselines. The color of the symbols indicates the object type: Herbig Ae (blue), T Tauri (red), and  eruptive systems (orange). Vertical solid and dashed lines indicate the expected MATISSE $N$ or $L$ band sensitivity limits with or without using an external fringe tracker, respectively. Blue and red lines represent the sensitivity limits with ATs and UTs, respectively. Horizontal dashed lines indicate the $K$ sensitivity limits of the external fringe tracker with ATs or UTs (in the middle and bottom rows). In the top row, horizontal lines indicate the $L$ band sensitivity limits. Sources in the green shaded area can be observed either with the UTs and ATs, objects in the orange shaded region can only be observed with UTs. Flux limits were taken from \citet{Matter2016b}.} 
		
		\label{fig:obs_plot}
	\end{figure*}
	
	One of the aims of this atlas is to provide an input database for the MATISSE instrument, from which one can derive the observability of the objects depending on the wavelength and the baseline length. The observability depends on both the correlated flux in the scientific bands ($L$, $M$, $N$), and the $K$ band correlated flux in case the GRAVITY fringe tracker were used for MATISSE. With an extrapolation of our interferometric modeling of $N$ band MIDI observations, we estimated $K$ and $L$ band correlated fluxes for all sources in the atlas, as was mentioned in Sect.~\ref{sec:int_model}. We use the continuous disk geometry, and for simplicity we assume that the optical depth of the model disk is wavelength independent and constant throughout the disk. The model takes into account the unresolved stellar emission, like in the $N$ band modeling (Sect.~\ref{sec:int_model}). The stellar fluxes were calculated in $K$, $L,$ and $N$ bands by our stellar SED fit (Sect.~\ref{sec:sedfit}). Extrapolating the $N$ band modeling, based on the MIDI observations, we predicted disk visibilities in $K$ and $L$. In order to determine correlated fluxes from the visibilities, we collected total fluxes from the literature. For $K$ band, we mainly use K$_\mathrm{s}$ magnitudes from the Two Micron All-Sky Survey (2MASS), for $L$ band we take the W1 fluxes from the catalog of the Wide-field Infrared Survey Explorer (WISE). Finally, we combined the unresolved stellar contribution and the disk visibility, and calculated the $K$ and $L$ band correlated fluxes for $15$, $30$, $60$, $100$, $130,$ and $150$~m projected baseline lengths. To validate our approach, we compared our extrapolated visibilities in the $K$ band with data from the literature (visibilities for ten objects obtained with the Keck Interferometer, Palomar Testbed Interferometer, and Astronomical Multi-Beam Combiner, AMBER). We found that our predictions are consistent with the observations, and although the correlation is somewhat low (the correlation coefficient is $0.64$), there seem to be no biases in the extrapolation. 
	
	In Fig.~\ref{fig:obs_plot} we show the observability plots for MATISSE: $L$ versus $N$ band, $K$ versus $N$ band, and $K$ versus $L$ band correlated fluxes, for two selected baseline lengths ($30$~m and $100$~m). Close binaries, where the interferometric spectra are significantly modulated, are again excluded and not shown in the figure. For other typical baselines observability plots are shown in the Appendix~\ref{app:matisse_obs}.
	
	From these figures one can see that not all sources can be observed with the longest baselines, because of low correlated fluxes. 
	%Alexis:
	The limiting $N$ and $L$ band correlated fluxes expected for MATISSE in low spectral resolution ($R=30$) mode, with and without external
	fringe tracker, are plotted as vertical lines. Those values take into account the current VLTI and MATISSE characteristics (e.g., optical transmission, adaptive optics
	performance, tip-tilt, focal laboratory) and the expected contribution of the fundamental noises. The integration and observation times with the fringe tracker are typical values, which may be increased to reach higher sensitivities. The specifications and expected performances of MATISSE, in particular with the use of an external fringe tracker, are summarized in \citet{Matter2016b}. The real on-sky sensitivity performances of MATISSE will be determined during the commissioning phase.% that will start in March 2018.
	
	As an important caveat for the interpretation of these observability plots, we recall that optical and near-IR variability among T Tauri stars is ubiquitous, which could cause uncertainties in our extrapolation. A variability of $0.5$~mag in $K$ would imply the same change in correlated flux at first order. This variability value is a rather conservative upper limit given the typical values found in several T Tauri surveys performed in the optical, near-, and mid-IR domains \citep[e.g.,][]{Eiroa2002,Rice2012,kospal_atlas,Rigon2017}. With a decrease of $0.5$~mag in the $K$ band correlated flux, only one source would become unobservable at all in the $N$ band.
	Moreover, when using these plots for planning observations, one should also be aware of the flux ratio of the unresolved stellar emission: if the disk is much fainter than the central star,  it is hardly detectable, even if the whole source is clearly observable even at very long baselines (e.g., in the case of debris disks).
	In addition, in the model fitting there were a few cases where the gapped model gave a better fit. For these objects (discussed in Sect.~\ref{sec:sizes}) the $K$ band flux may be overestimated by the continuous disk model. 
	
	\section{Summary}
	\label{sec:summ}
	
	In this study we present a mid-IR interferometric atlas focusing on low- and intermediate-mass young stellar objects, based on VLTI/MIDI data. Our sample contains $82$ sources, including $45$ T Tauri stars, $11$ young eruptive stars, and $26$ Herbig Ae stars. Our main results are as follows:
	\begin{itemize}
		
		\item We performed a homogeneous data reduction of $627$ MIDI interferometric observations from $222$ nights. The resulting interferometric atlas, containing $N$ band correlated and total spectra, visibilities, and differential phases, is published online at \url{http://www.konkoly.hu/MIDI_atlas}, and will also be made available at VizieR.
		
		\item We calculated the Gaussian sizes of the objects for each observation. In some cases (e.g., DG Tau, EX Lup, HD 144132) signs of long-term temporal change in size are observable, possibly indicating a structural rearrangement. When the data indicate an elliptical shape, it may suggest an inclined circumstellar disk (e.g., GW Ori).
		
		\item Using a simple geometry, we modeled the interferometric data in order to determine the spatial extent (half-light radius) of the disks. The majority of our data can be well fitted with a continuous disk model, except for a few objects (Elias 24, DI Cha, DK Cha, Haro 6-10N, SVS20N, TW Hya) where the gapped model gives a better match to the data.
		
		\item We compiled a list of multiple stellar systems present in the atlas. In six systems (T Tau S, GG Tau Aab, GW Ori, Z CMa, HD 142527, SR 24N) the known close binary component was detected via modulations in the interferometric observables. We also identified two systems with modulations (AB Aur, HD 72106) where there was no companion known or the known companion is either too wide or faint to cause the observed signal. These modulations could be caused by an undetected companion but could also be due to an asymmetric inner disk structure.
		
		\item We analyzed the mid-IR size--luminosity relation of our sample. The most compact ($r_\mathrm{hl} \lesssim 0.3$~au) disks around the least luminous T Tauri stars ($L_\star \lesssim 1 \mathrm{L}_\odot$) are missing from our data, possibly due to observational limitations. The distribution of sources in the size--luminosity diagram suggests that disks around T Tauri stars are generally colder and more extended with respect to the stellar luminosity than disks around Herbig Ae stars.   
		
		\item We computed a grid of radiative transfer models (with stellar luminosities between $0.5$ and $10~\mathrm{L}_\sun$ and disk inner radii between the sublimation radius and $5$~au), to interpret the distribution of sources on the mid-IR size--color diagram. Most of our data points fall into the area on the diagram defined by the models. The distribution of the T Tauri stars is very similar to that of the Herbig stars presented in \citet{Menu2015}.

		\item We compiled a list of candidate objects for mid-IR variability. Among these objects, there are known eruptive stars (EX Lup, V1647 Ori, and V346 Nor), but also normal T Tauri and Herbig Ae sources; some of them were also found to be variable at IR wavelengths in the literature. Our findings from total flux measurements suggest that the characteristic amplitude of the variations is $\sim$$0.3$~mag. 
		
		\item We identified sources with uncommon silicate spectra. In the spectra of SVS~13A1, DK~Cha, and V346 Nor, we observed a crystalline absorption feature. There are seven sources where  the feature has a different profile in the total and correlated spectra: DG Tau, LkH$\alpha$ 330, AB Aur, V1647 Ori, HD~72106, DI Cha, and HBC 639. All of these objects show absorption in their inner disk and emission feature outside. From these observations the average location of the boundary radius between the absorption and emission can be placed somewhere at $1.5-2$~au. The low amplitude or absence of silicate emission in inner disks may be related to dust grain growth, but other processes (e.g., accretion heating or an inner absorbing halo) can also play a role.
		
		\item We analyzed the relation between the $11.3/9.8\ \mu$m silicate flux ratio and the feature amplitude. We found that in the innermost part of the disks the silicate feature is generally much weaker than in the outer parts, suggesting that in the inner parts ($r \lesssim 1$~au) the dust is substantially more processed, although radiative transfer effects can also make the feature weaker. The distribution of points representing the total disk emission is in accordance with the findings of \citet{vanBoekel2003} and \citet{Przygodda2003}. Sources showing silicate absorption have low flux ratios (around and below $0.5$), indicating absorption by less-processed grains. The T Tauri stars generally have lower silicate feature amplitudes than those of Herbig stars, and the correlation between the feature amplitude and flux ratio is weaker compared to Herbig stars. Eruptive stars seem not to follow the trend between the feature amplitude and flux ratio at all.
		
		\item With the aim of determining the observability of our objects with MATISSE, we estimated $K$ and $L$ band correlated fluxes for all sources by extrapolating the interferometric modeling of $N$ band MIDI observations. Our predictions indicate that not all sources can be observed with the longest baselines because of low correlated fluxes. 
		
	\end{itemize}

	\begin{acknowledgements}
		This work was supported by the Momentum grant of the MTA CSFK Lend\"ulet Disk Research Group (LP2014-6). The authors are thankful for the support of the Fizeau Exchange Visitor Program (OPTICON/FP7) through the European Interferometry Initiative (EII) funded by WP14 OPTICON/FP7 (2013--2016, grant number 312430). This project has received funding from the European Research Council (ERC) under the European Union’s Horizon 2020 research and innovation programme under grant agreement No 716155 (SACCRED). This work is based on observations made with ESO telescopes at the Paranal Observatory. This publication makes use of data products from the Two Micron All Sky Survey, which is a joint project of the University of Massachusetts and the Infrared Processing and Analysis Center/California Institute of Technology, funded by the National Aeronautics and Space Administration and the National Science Foundation. This publication makes use of data products from the Wide-field Infrared Survey Explorer, which is a joint project of the University of California, Los Angeles, and the Jet Propulsion Laboratory/California Institute of Technology, funded by the National Aeronautics and Space Administration. This work is based in part on observations made with the Spitzer Space Telescope, which is operated by the Jet Propulsion Laboratory, California Institute of Technology under a contract with NASA. This work has made use of data from the European Space Agency (ESA)
		mission {\it Gaia} (\url{https://www.cosmos.esa.int/gaia}), processed by
		the {\it Gaia} Data Processing and Analysis Consortium (DPAC,
		\url{https://www.cosmos.esa.int/web/gaia/dpac/consortium}). Funding
		for the DPAC has been provided by national institutions, in particular
		the institutions participating in the {\it Gaia} Multilateral Agreement. We would like to thank Andres Carmona for useful discussions. Also, we thank the anonymous referee for helpful comments.
		
	\end{acknowledgements}

	%-------------------------------------------------------------------
	
	\bibliographystyle{aa}
	\bibliography{ref_MIDI_atlas}
	
	%\begin{thebibliography}{}
	
	% \bibitem[1966]{baker} Baker, N. 1966,
	%    in Stellar Evolution,
	%  ed.\ R. F. Stein,\& A. G. W. Cameron
	%(Plenum, New York) 333

	%\end{thebibliography}
	
	%-------------------------------------------------------------------

	\begin{appendix}
		
		\onecolumn
		\section{Stellar photosphere fits}
		\label{app:sed}
		\begin{figure*}[h!]
			\centering
			\caption{Stellar photosphere fits to visible and near-IR photometric data. Filled circles indicate valid data points, triangles indicate upper limits. Best fitting SEDs are plotted with gray curves.}
			\label{fig:stellar_SED}
			\includegraphics[width = 0.33\linewidth]{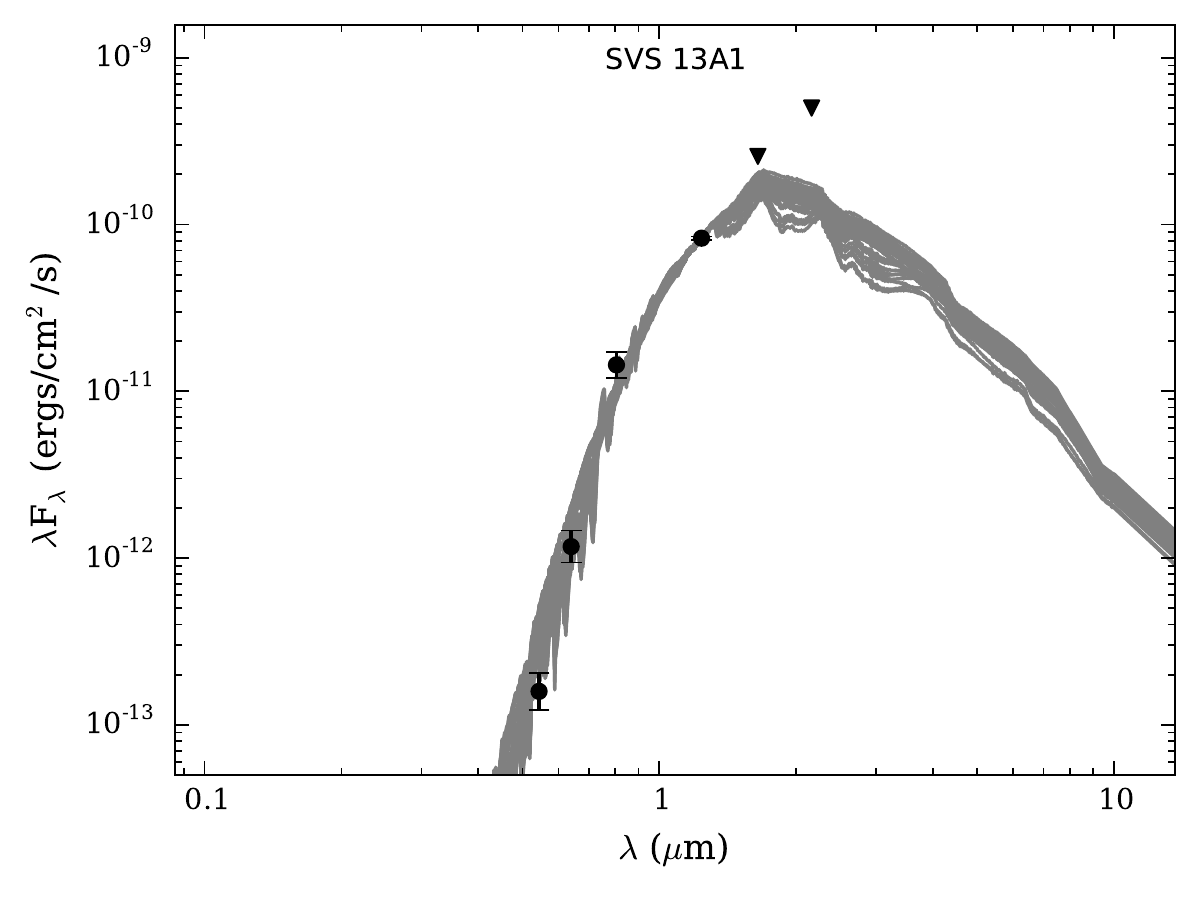}
			\includegraphics[width = 0.33\linewidth]{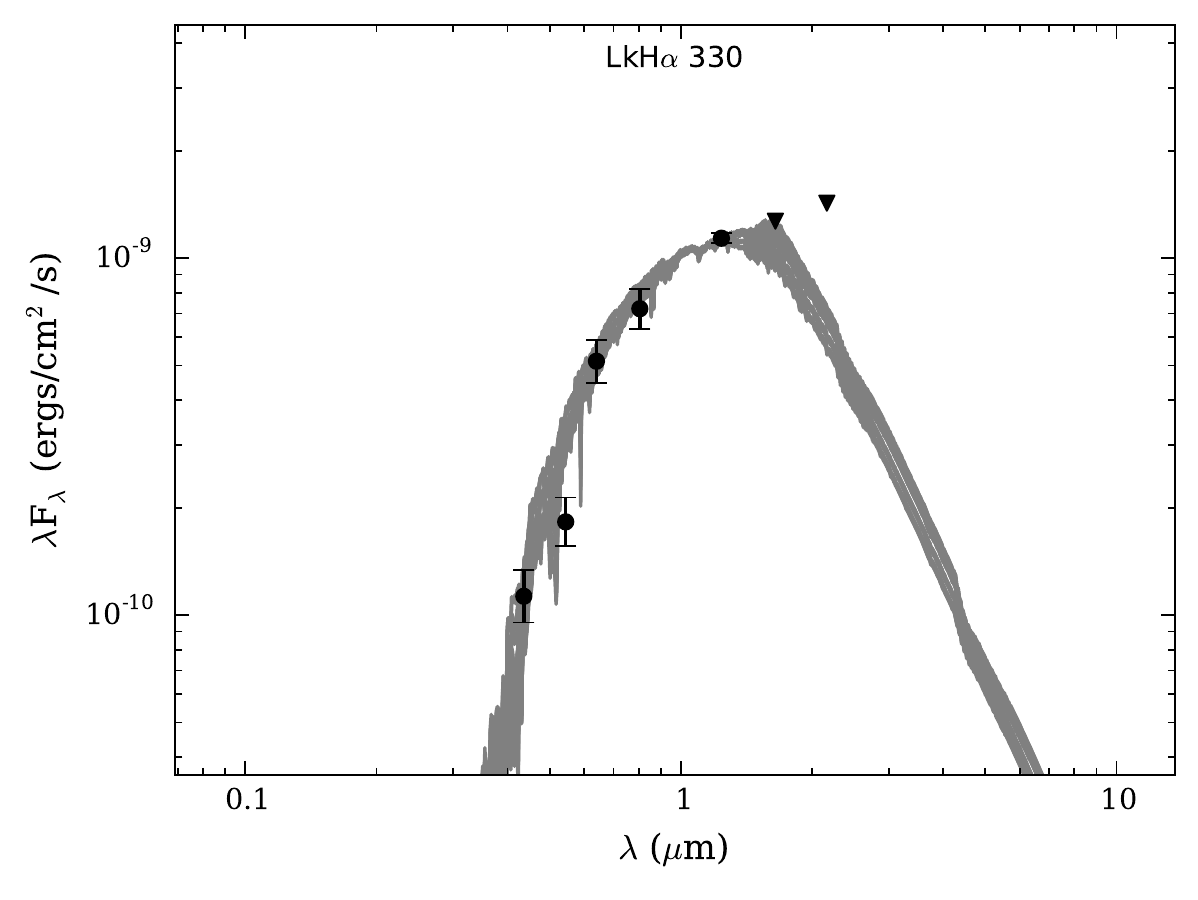}
			\includegraphics[width = 0.33\linewidth]{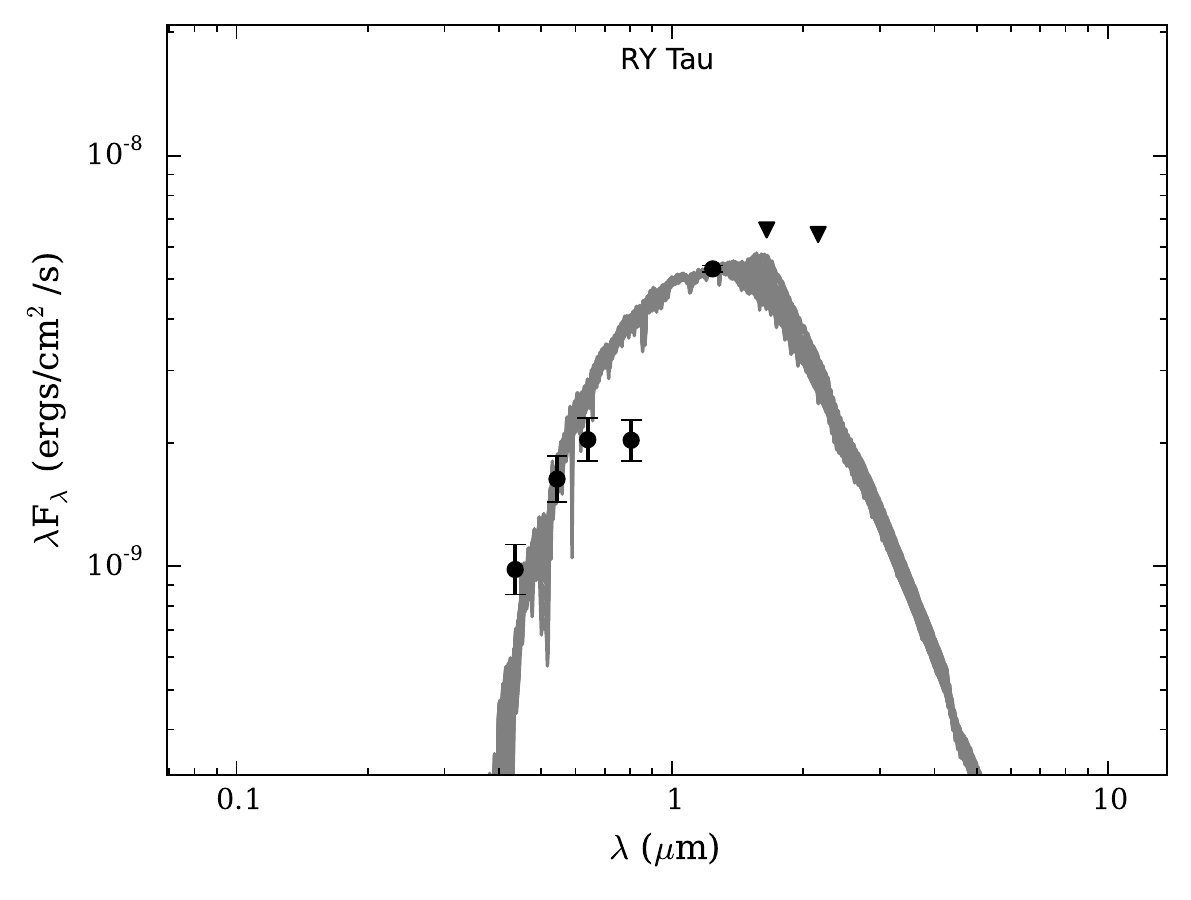}
			\includegraphics[width = 0.33\linewidth]{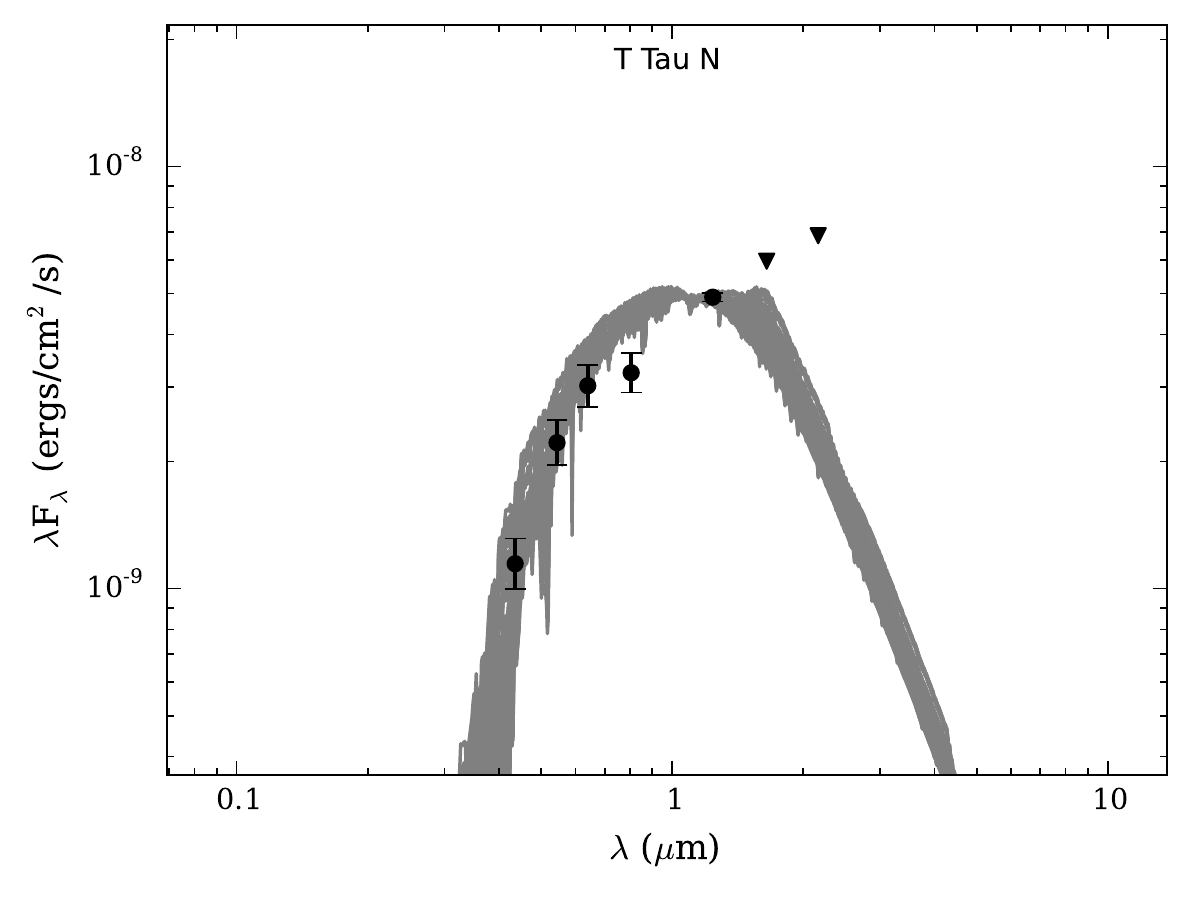}
			\includegraphics[width = 0.33\linewidth]{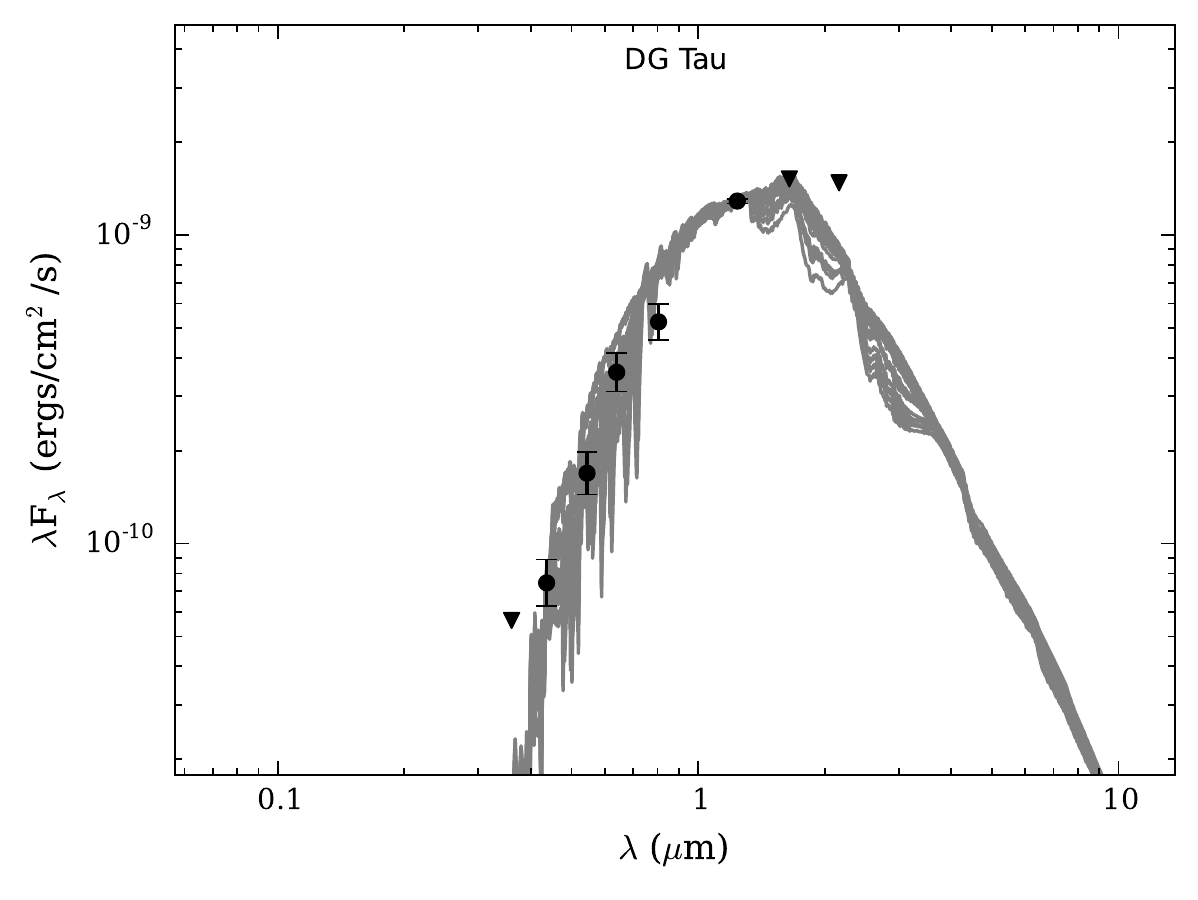}
			\includegraphics[width = 0.33\linewidth]{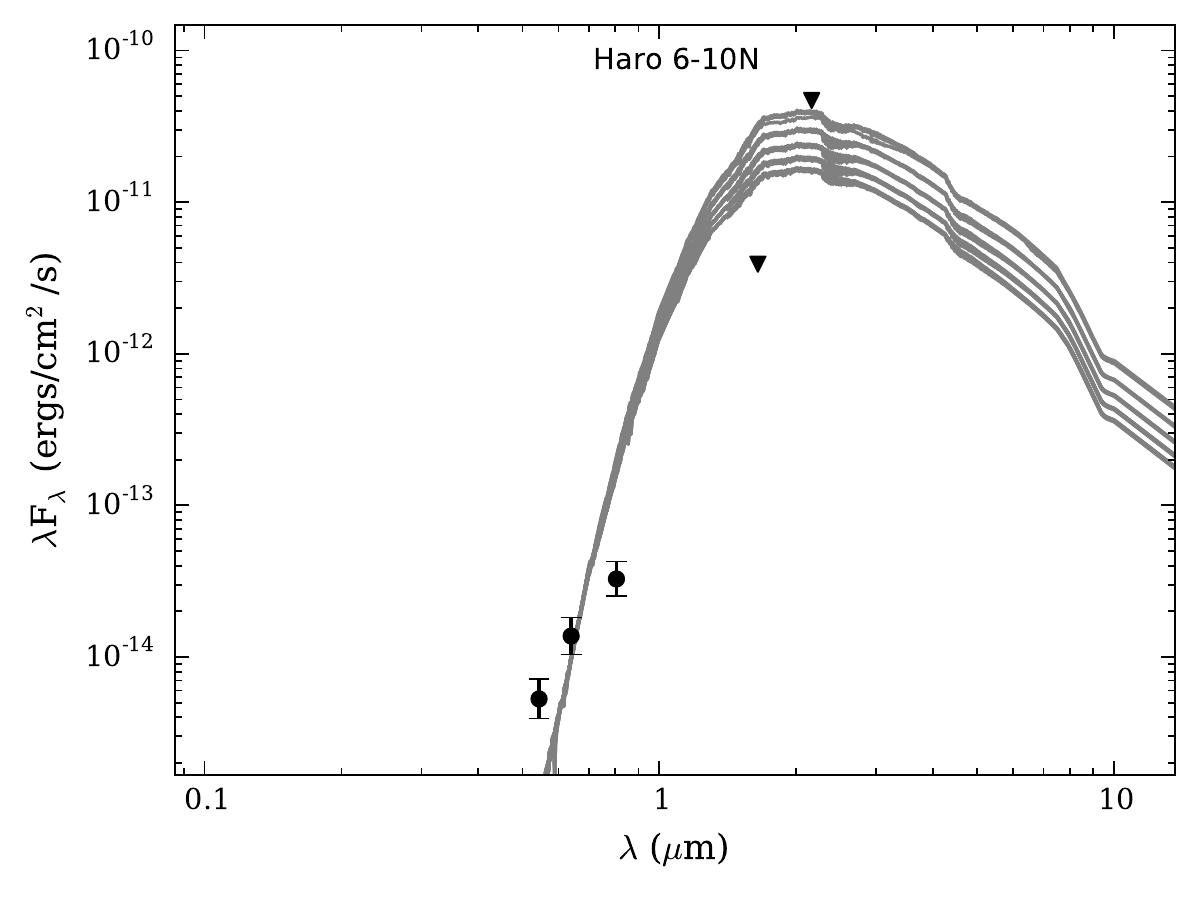}
			\includegraphics[width = 0.33\linewidth]{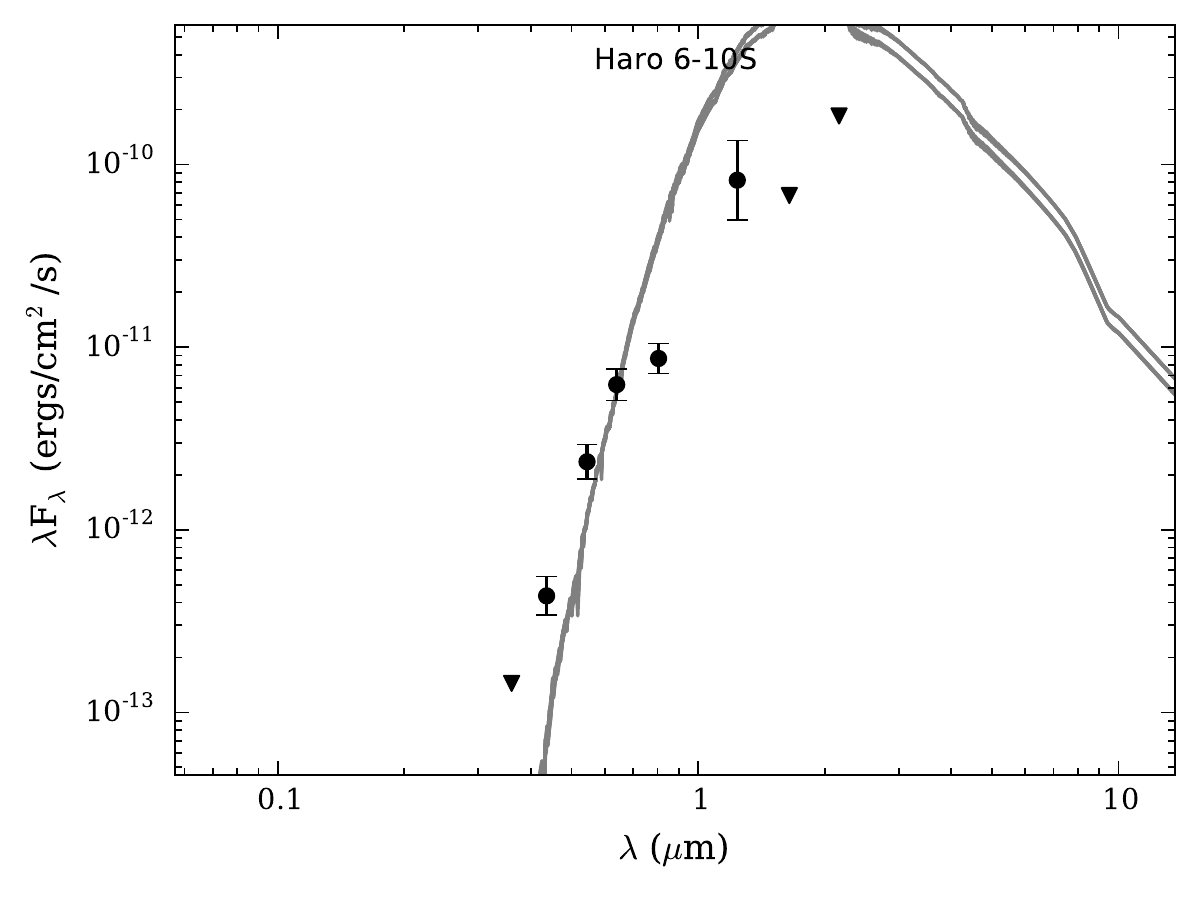}
			\includegraphics[width = 0.33\linewidth]{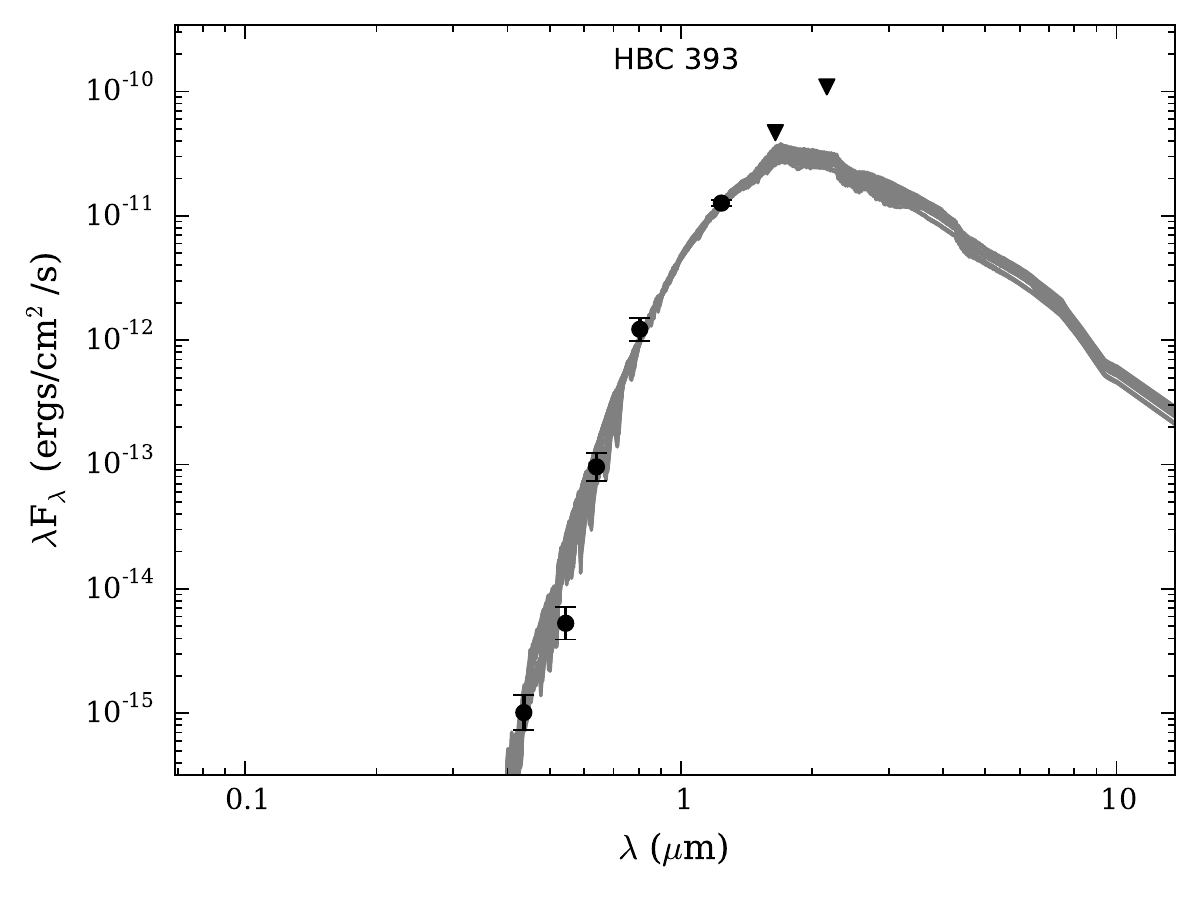}
			\includegraphics[width = 0.33\linewidth]{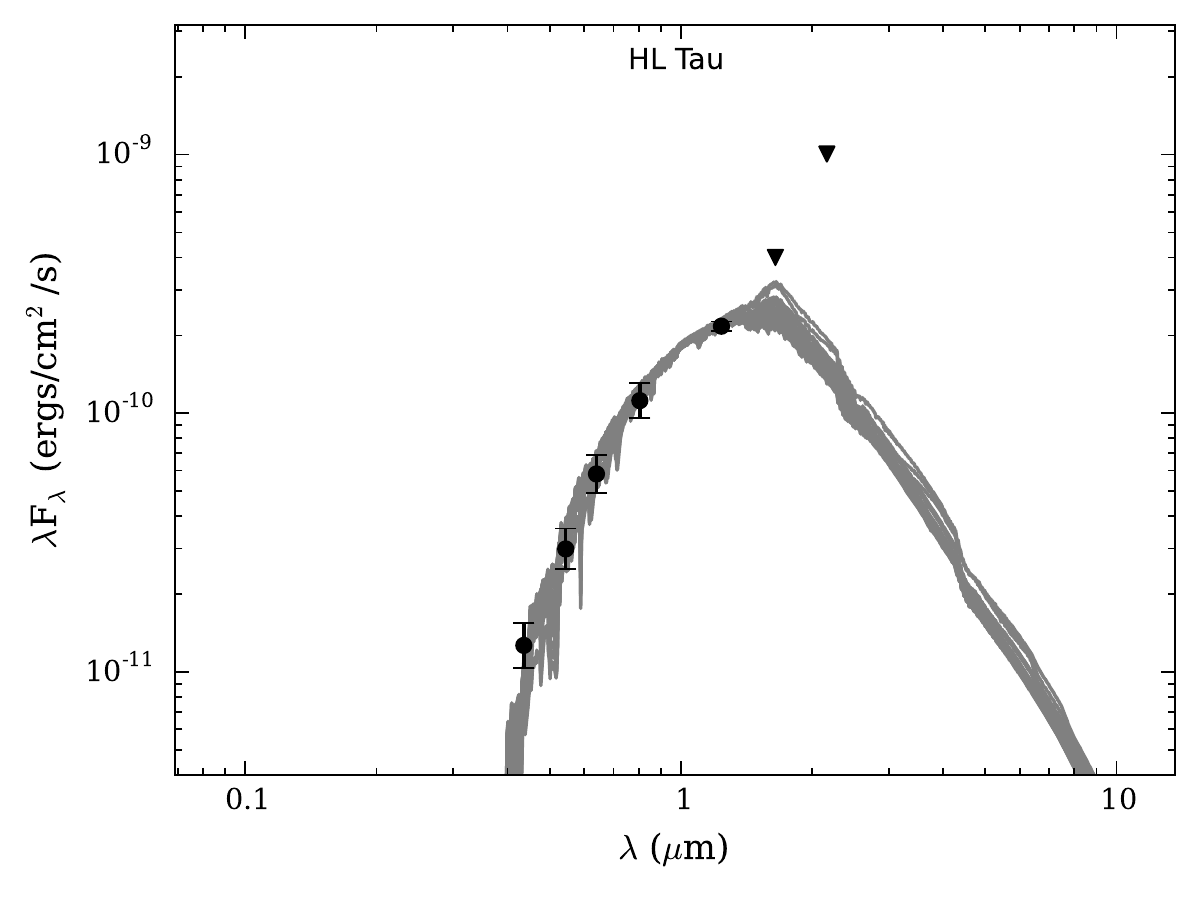}
			\includegraphics[width = 0.33\linewidth]{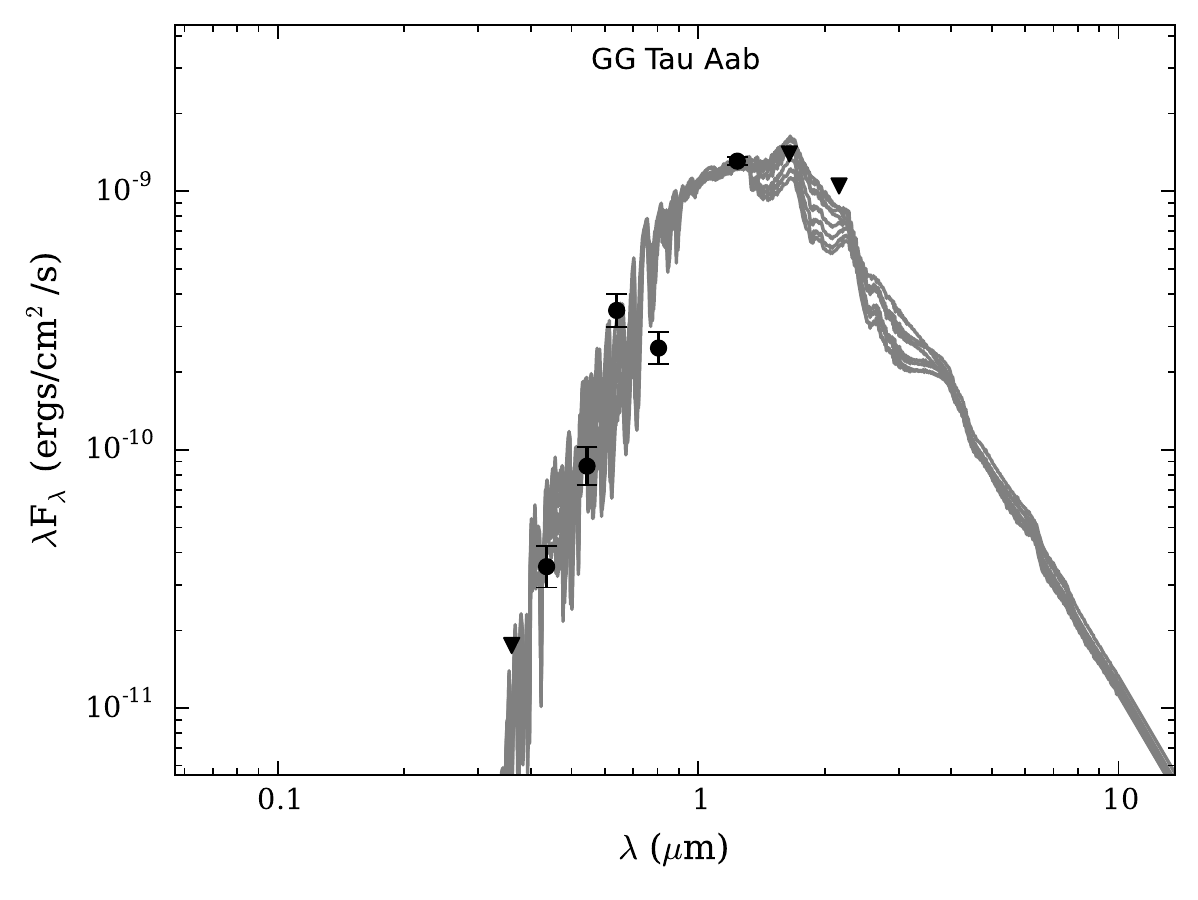}
			\includegraphics[width = 0.33\linewidth]{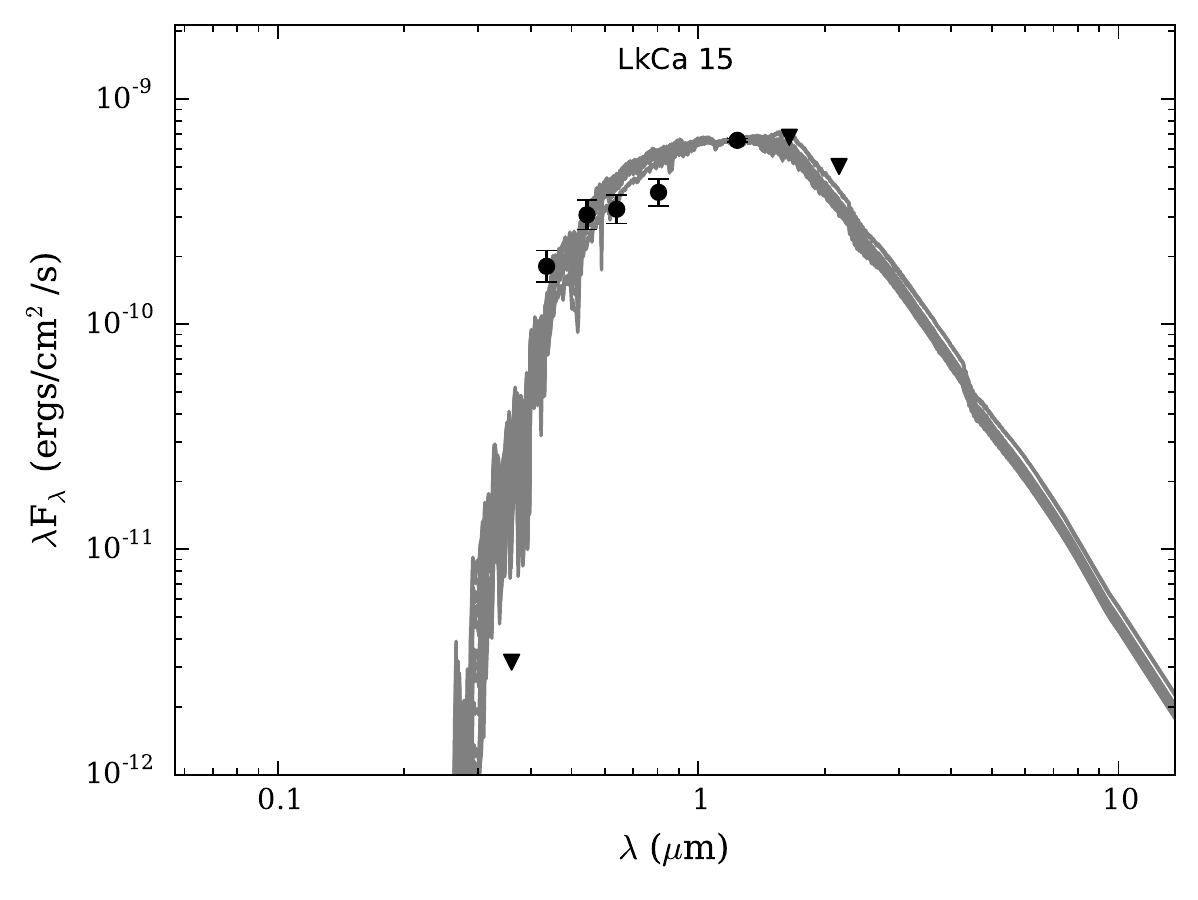}
			\includegraphics[width = 0.33\linewidth]{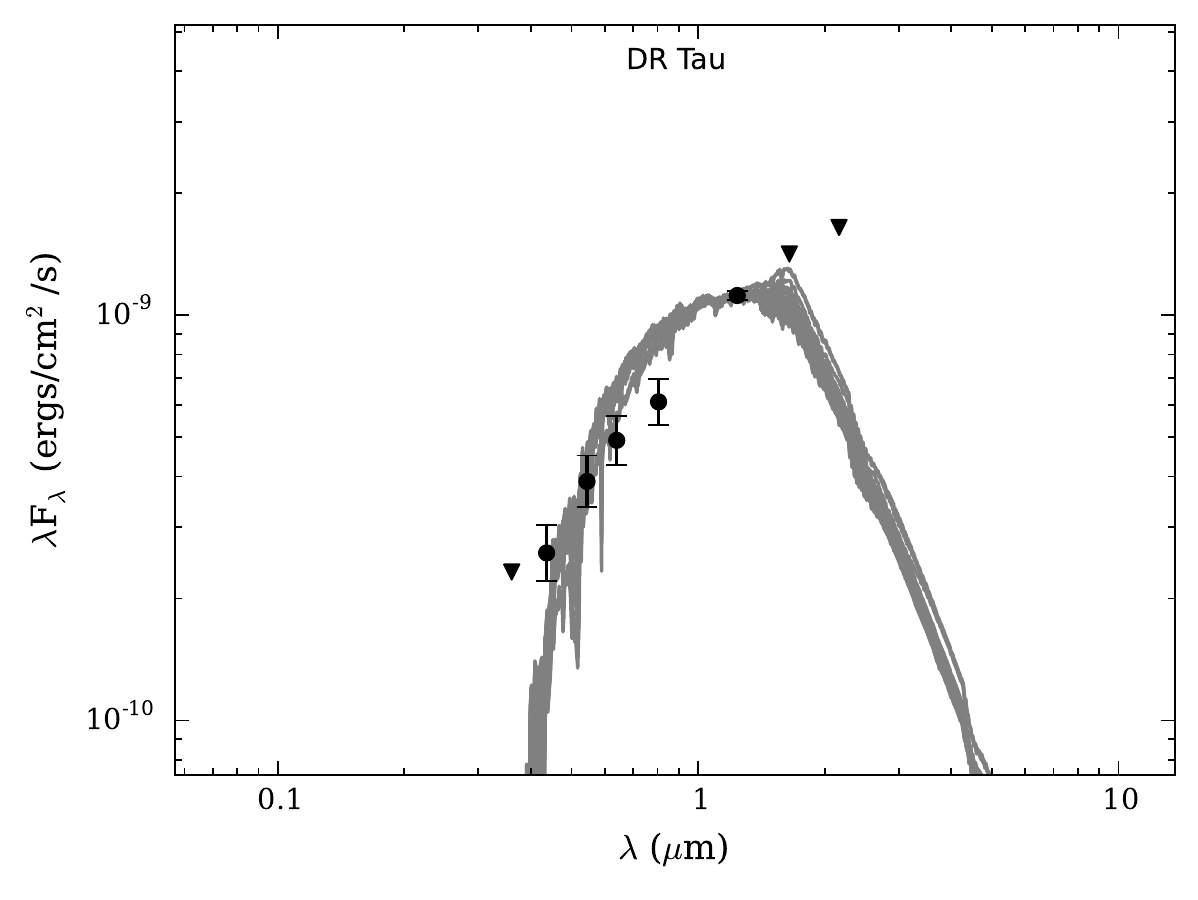}
			
		\end{figure*}
		
		\begin{figure*}[h!]
			\centering
			\includegraphics[width = 0.33\linewidth]{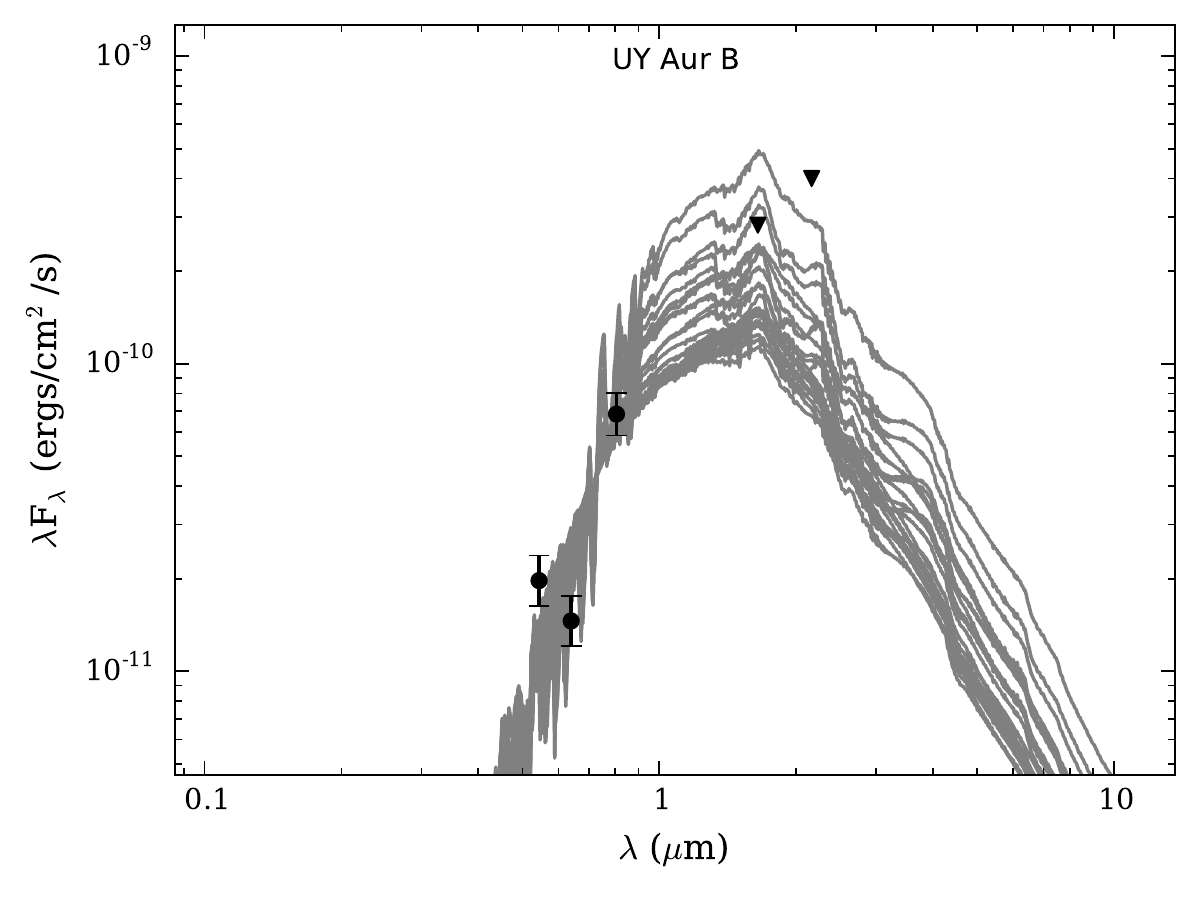}
			\includegraphics[width = 0.33\linewidth]{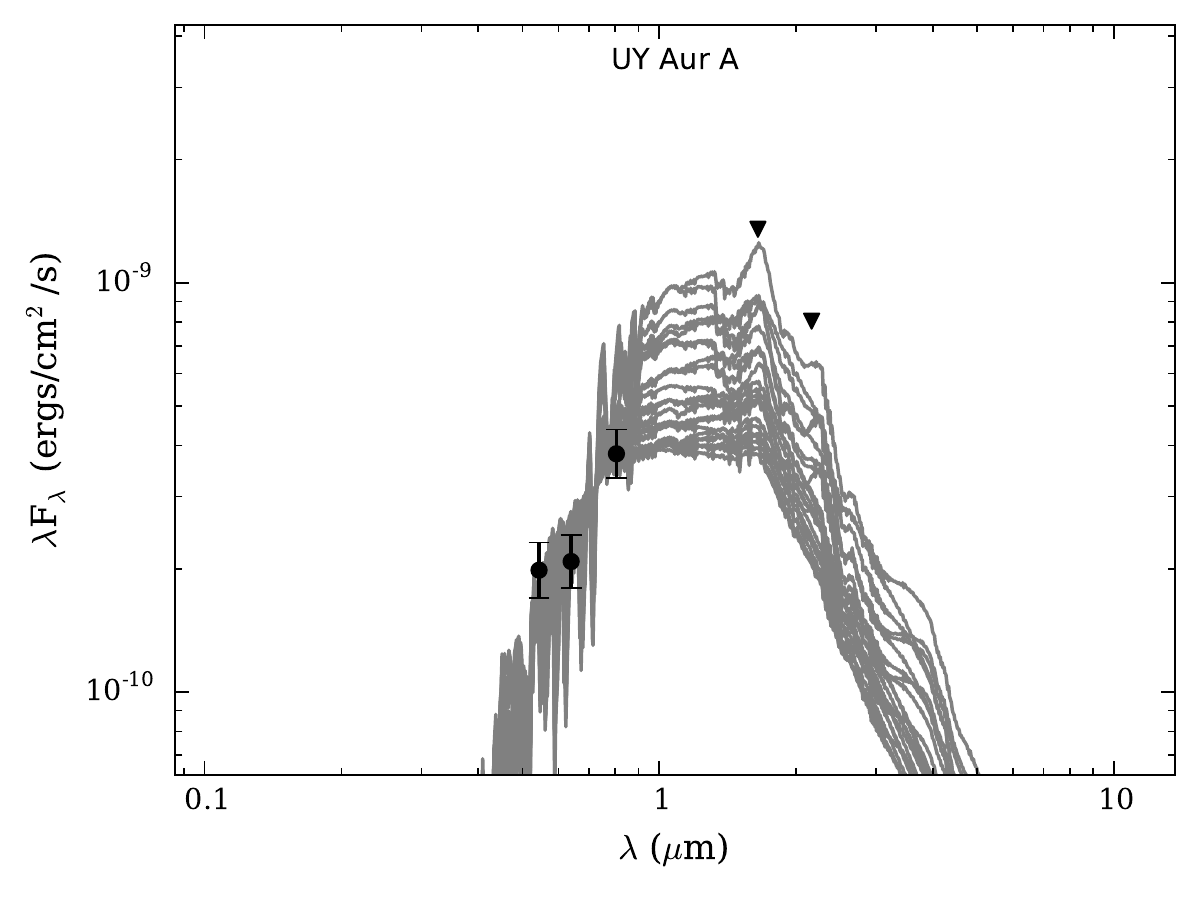}
			\includegraphics[width = 0.33\linewidth]{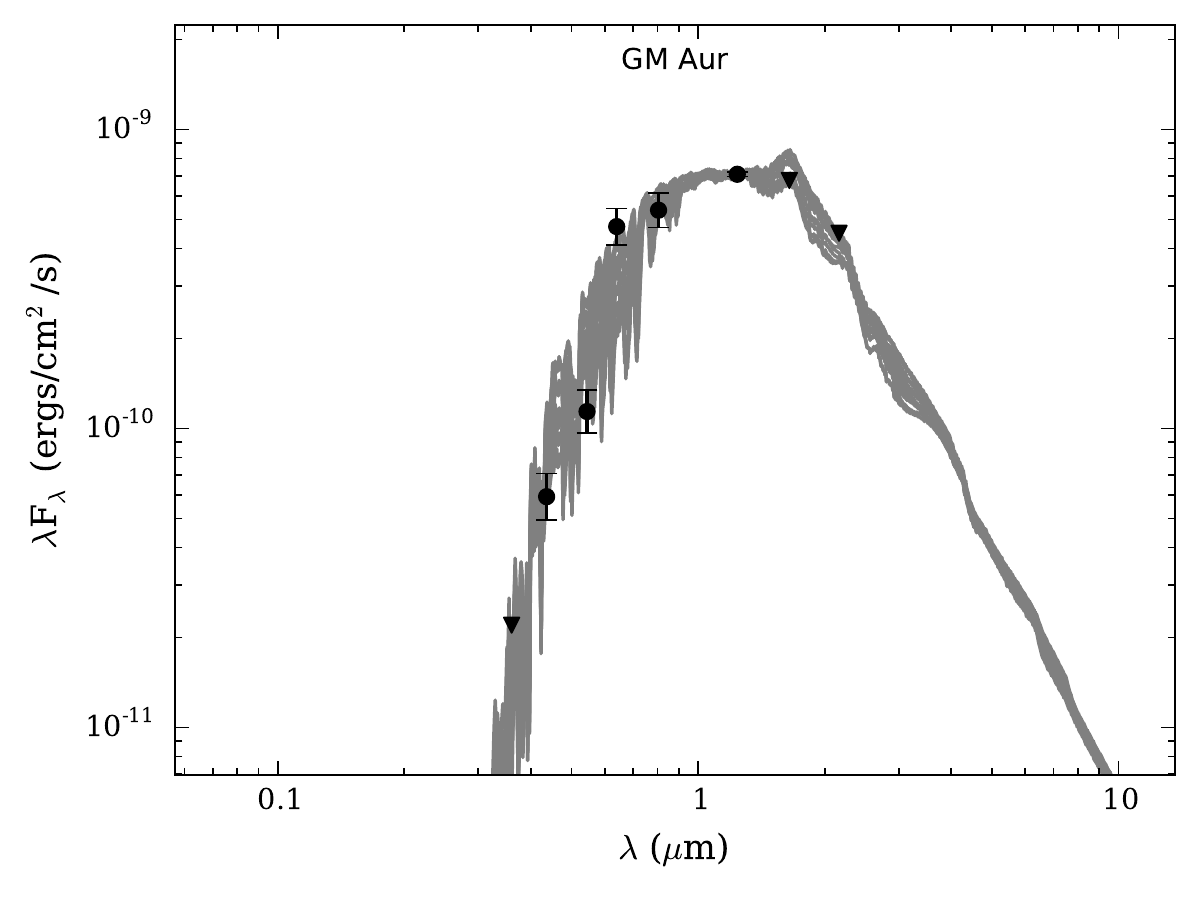}
			\includegraphics[width = 0.33\linewidth]{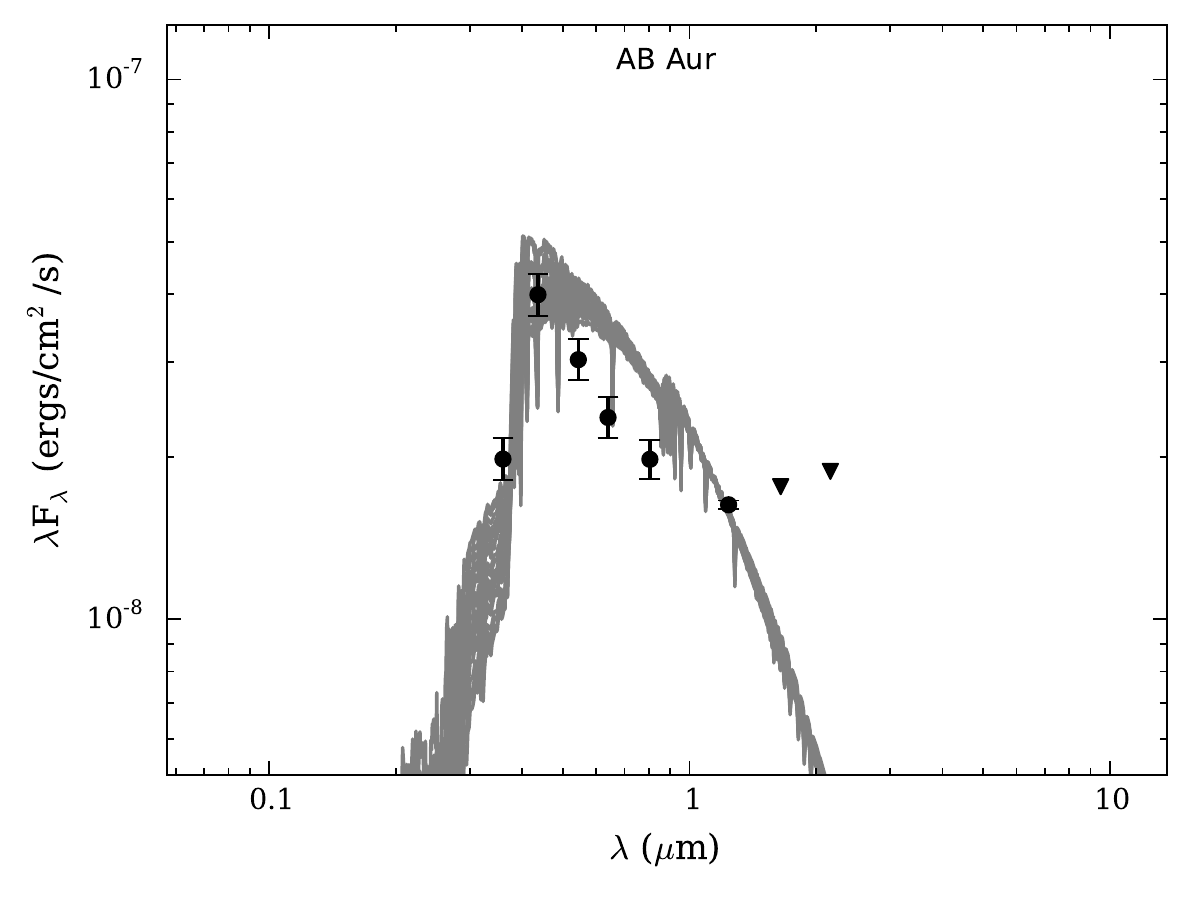}
			\includegraphics[width = 0.33\linewidth]{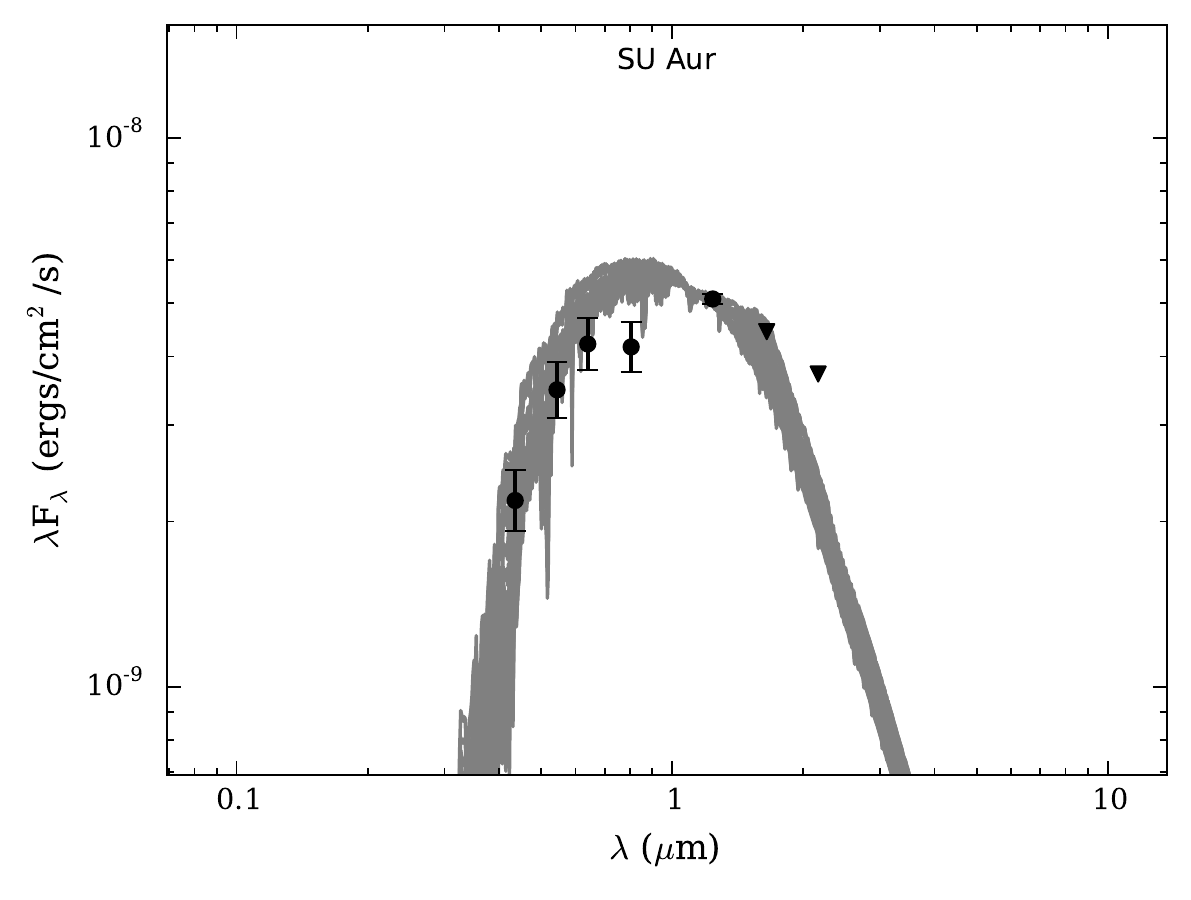}
			\includegraphics[width = 0.33\linewidth]{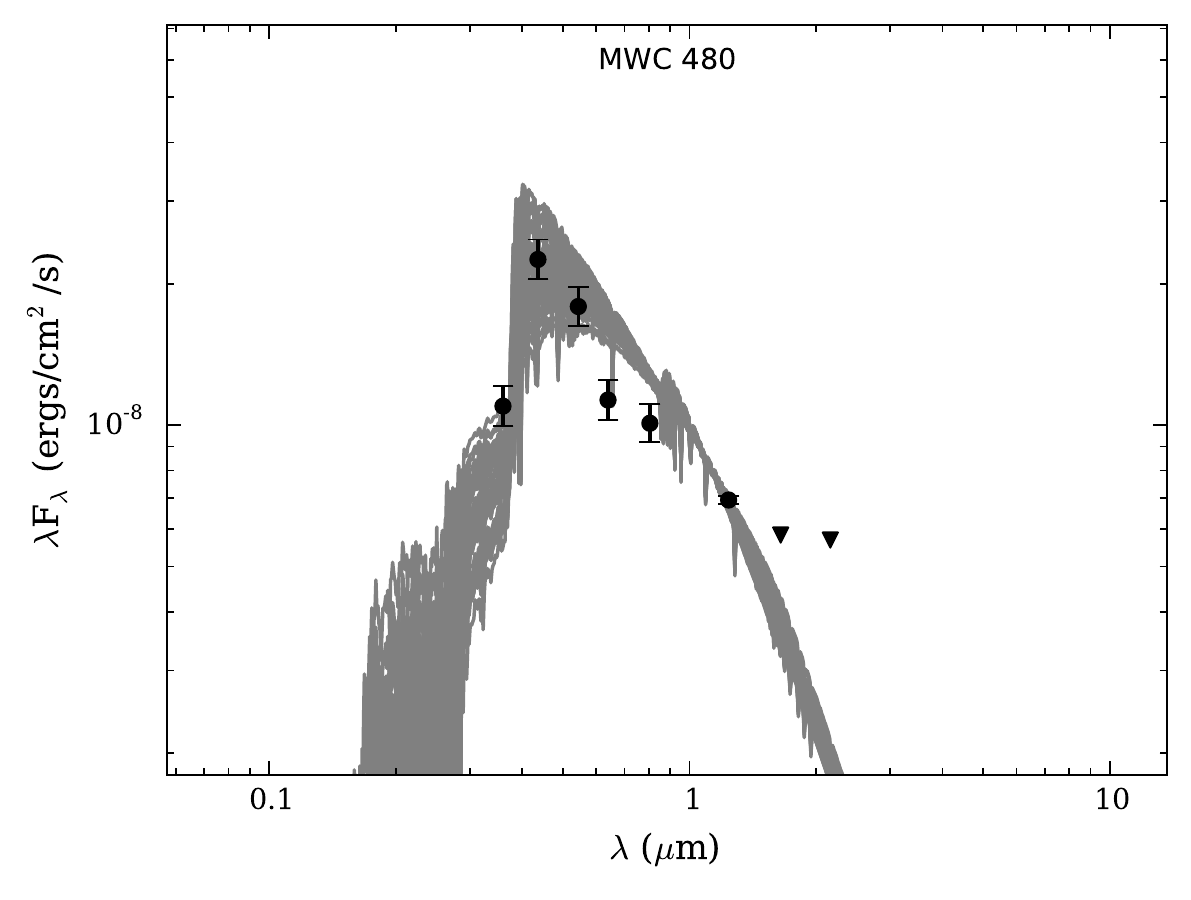}
			\includegraphics[width = 0.33\linewidth]{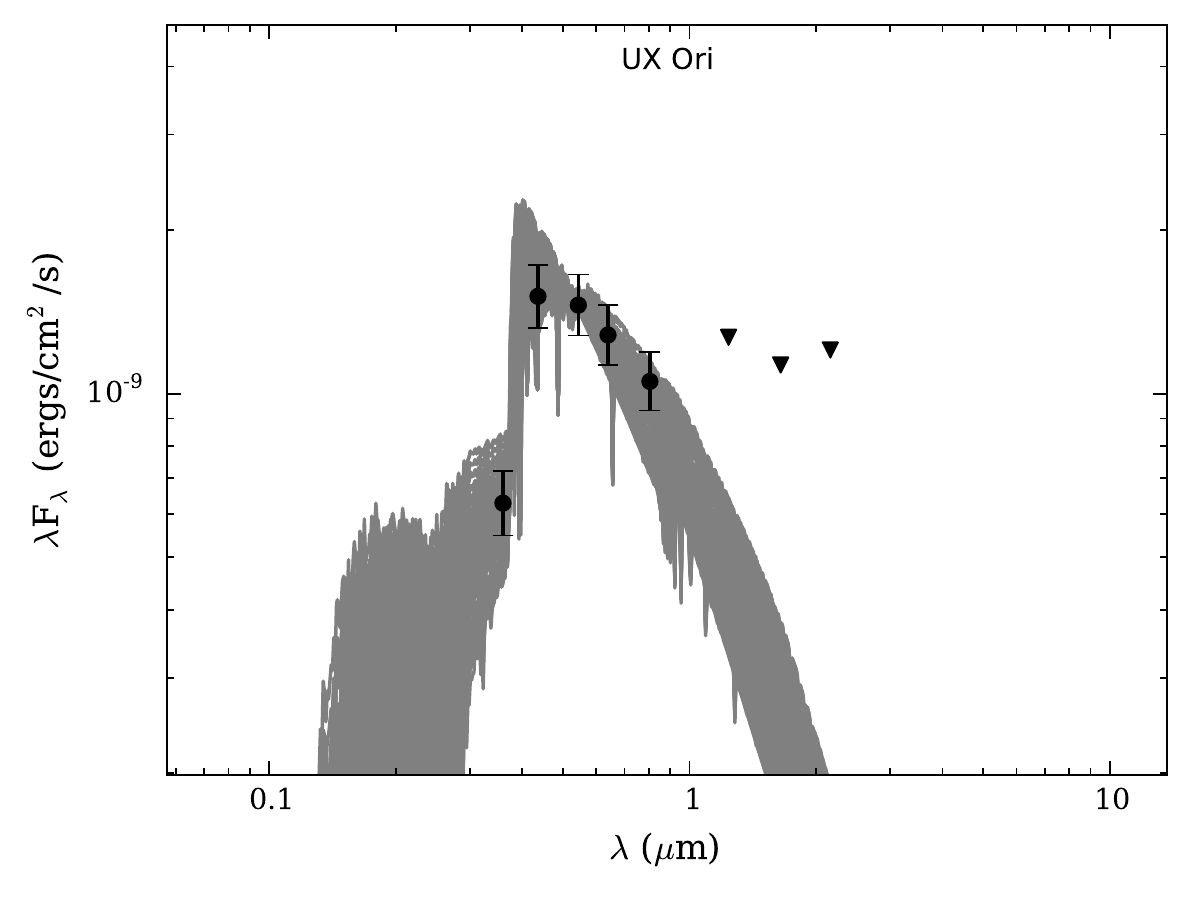}
			\includegraphics[width = 0.33\linewidth]{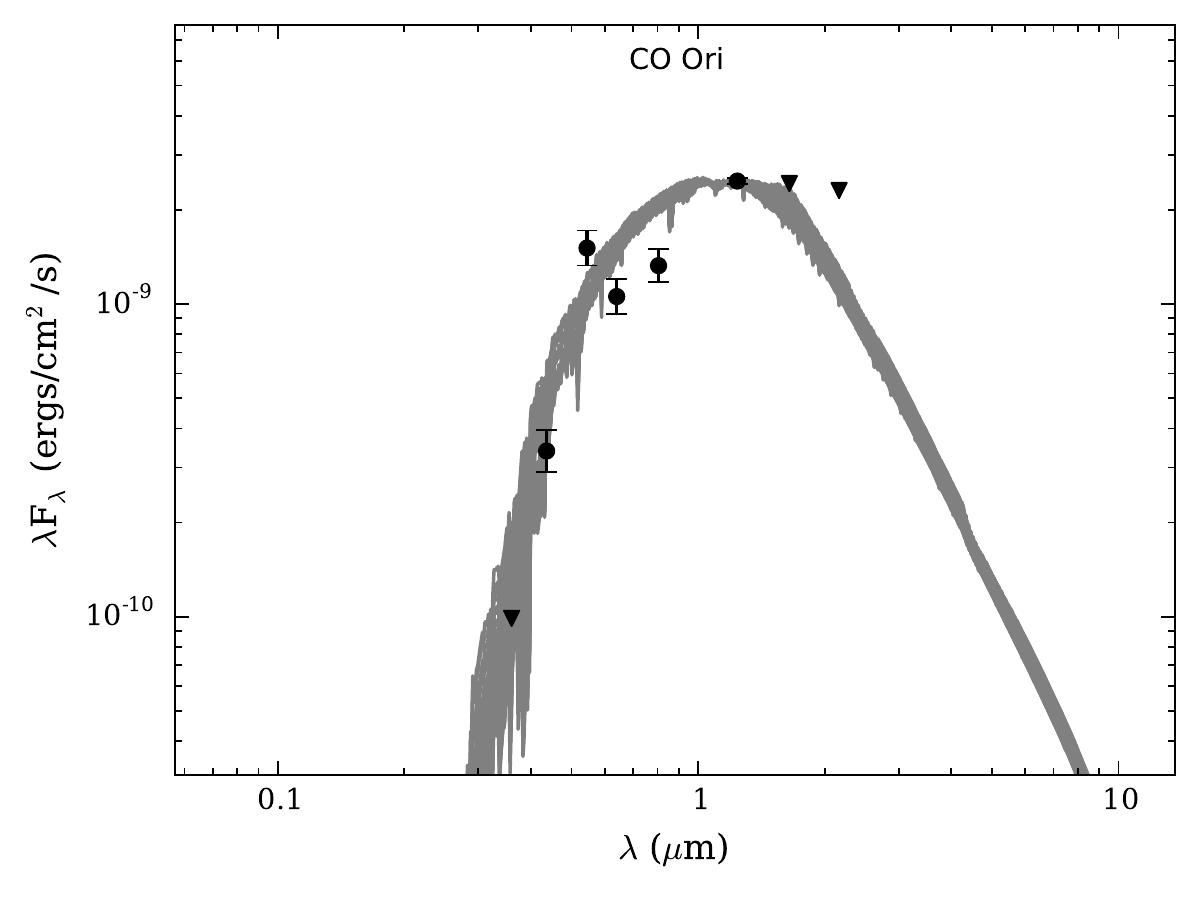}
			\includegraphics[width = 0.33\linewidth]{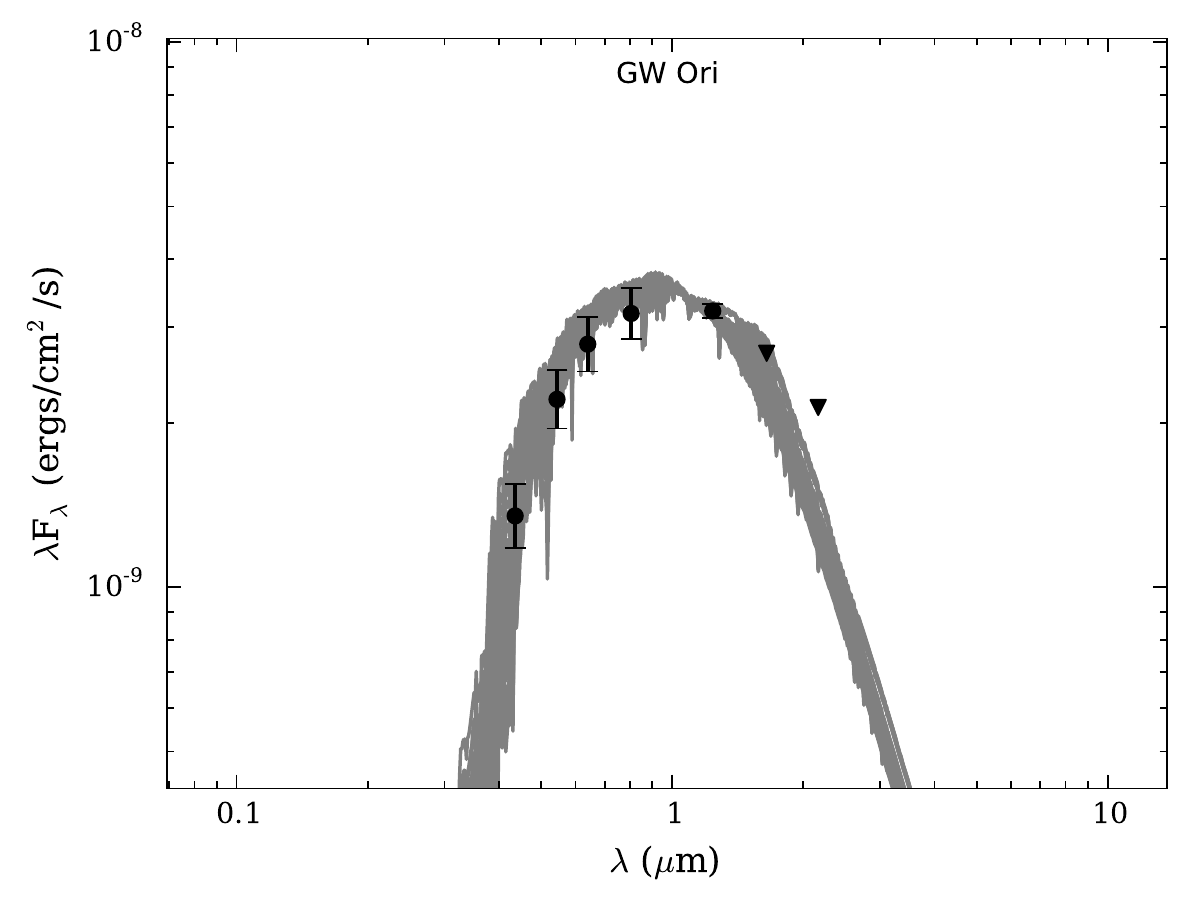}
			\includegraphics[width = 0.33\linewidth]{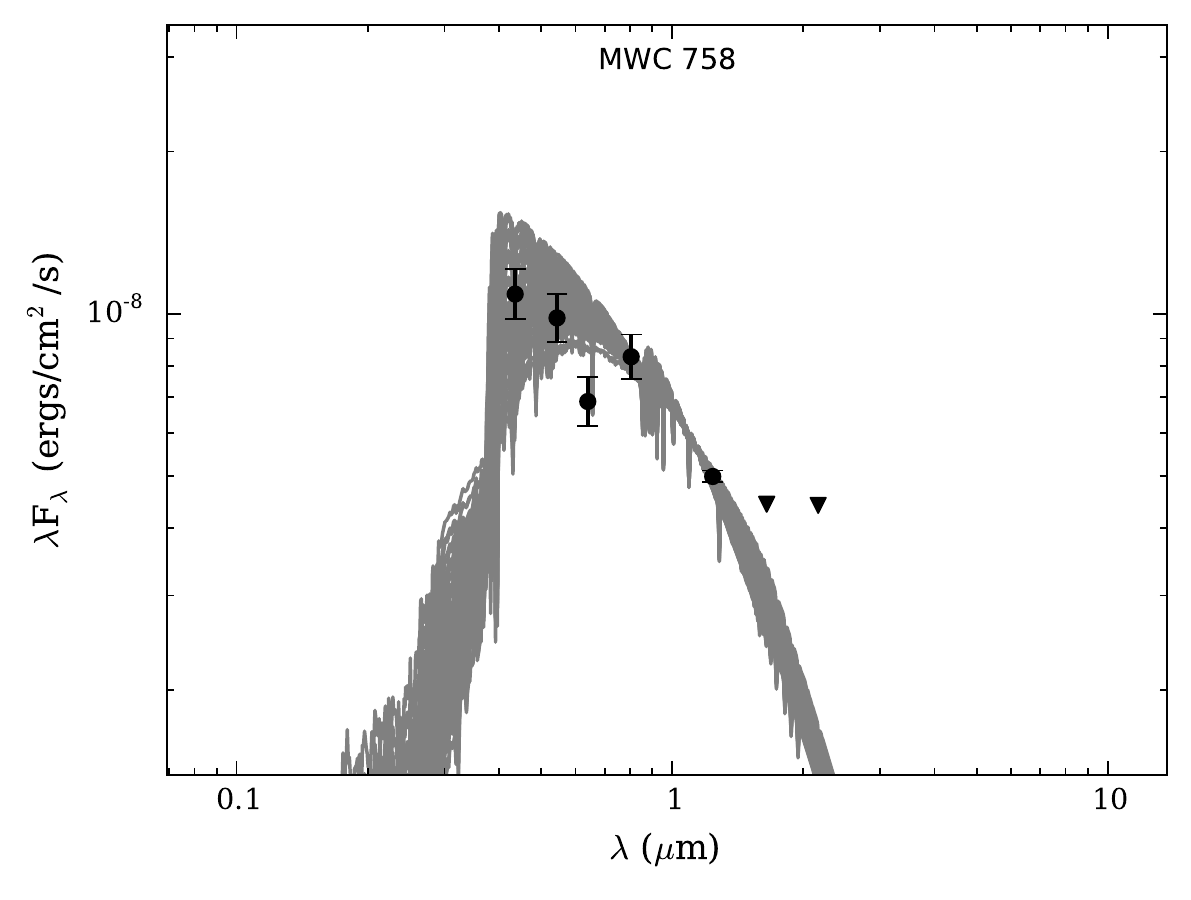}
			\includegraphics[width = 0.33\linewidth]{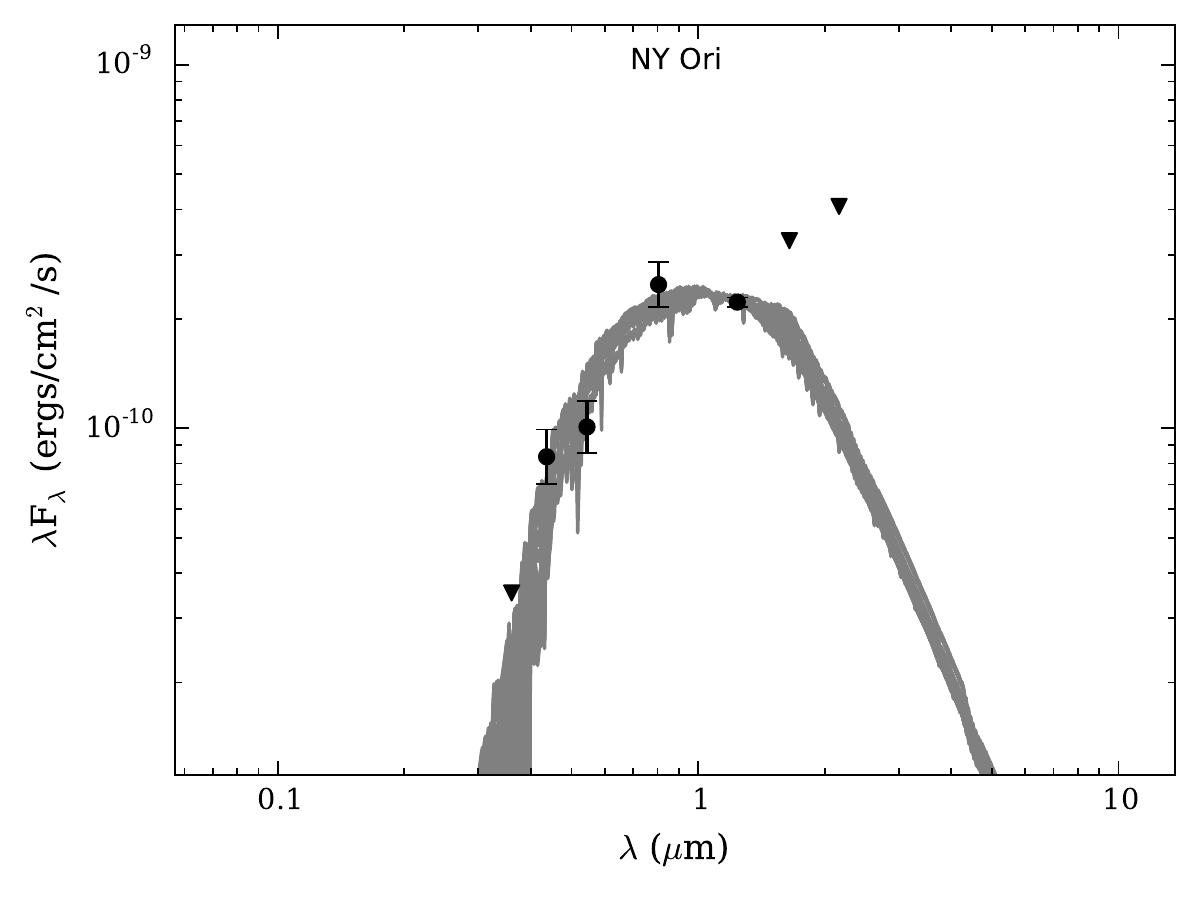}
			\includegraphics[width = 0.33\linewidth]{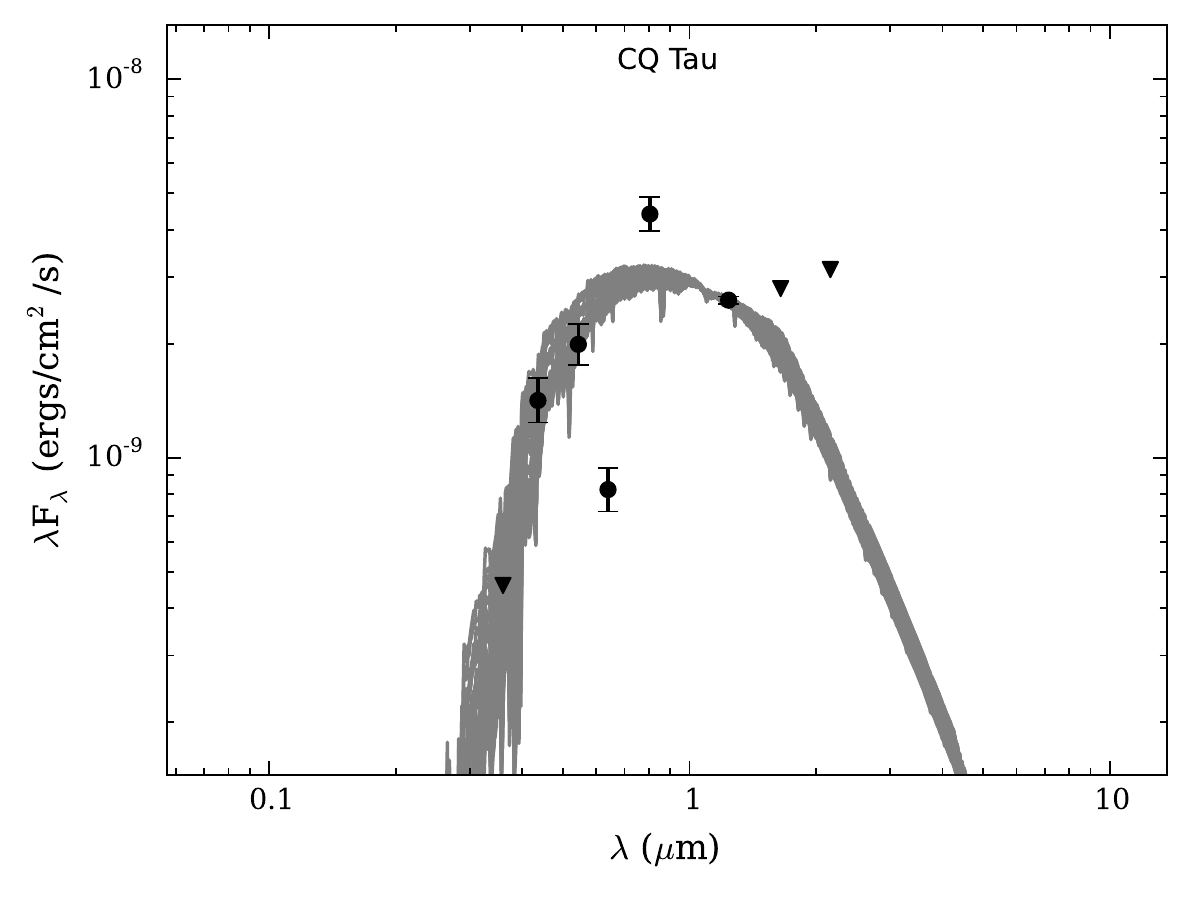}
			\includegraphics[width = 0.33\linewidth]{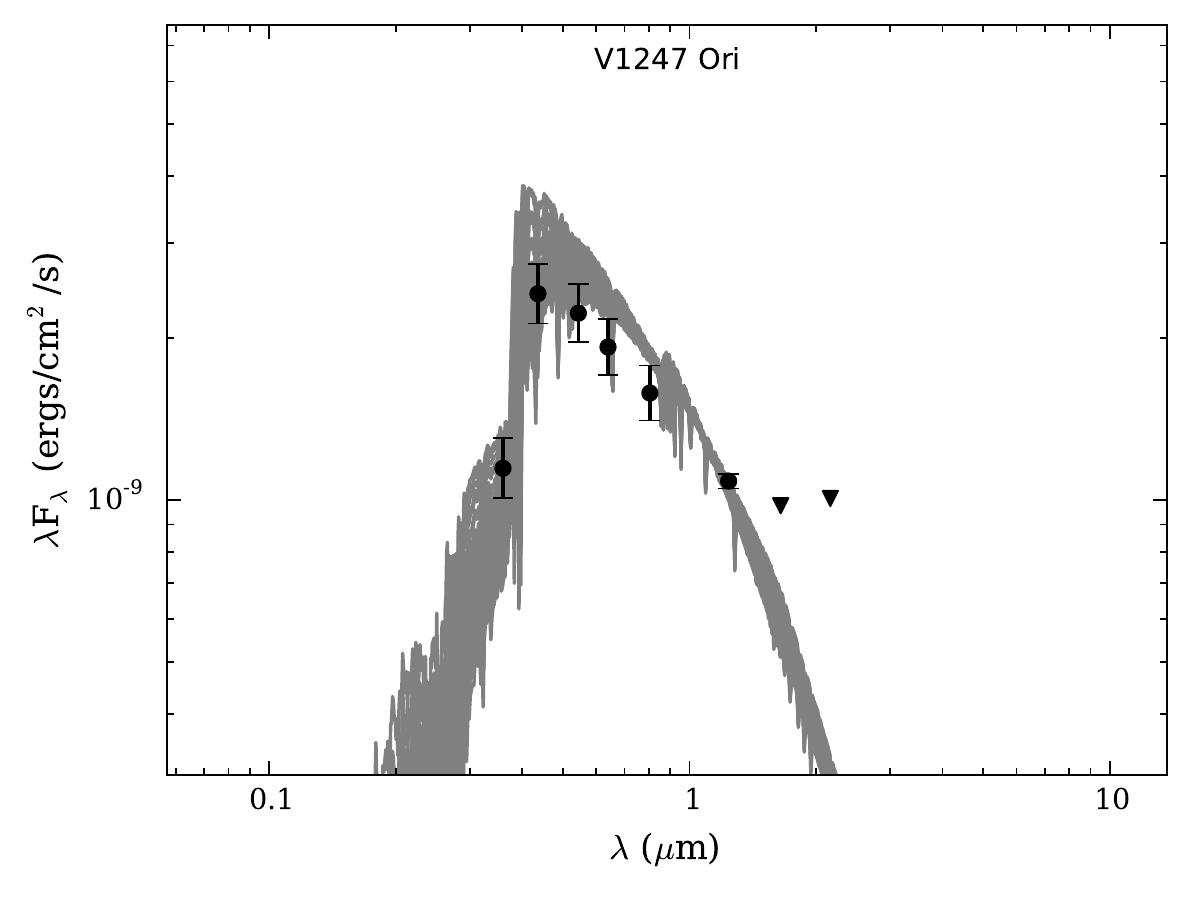}
			\includegraphics[width = 0.33\linewidth]{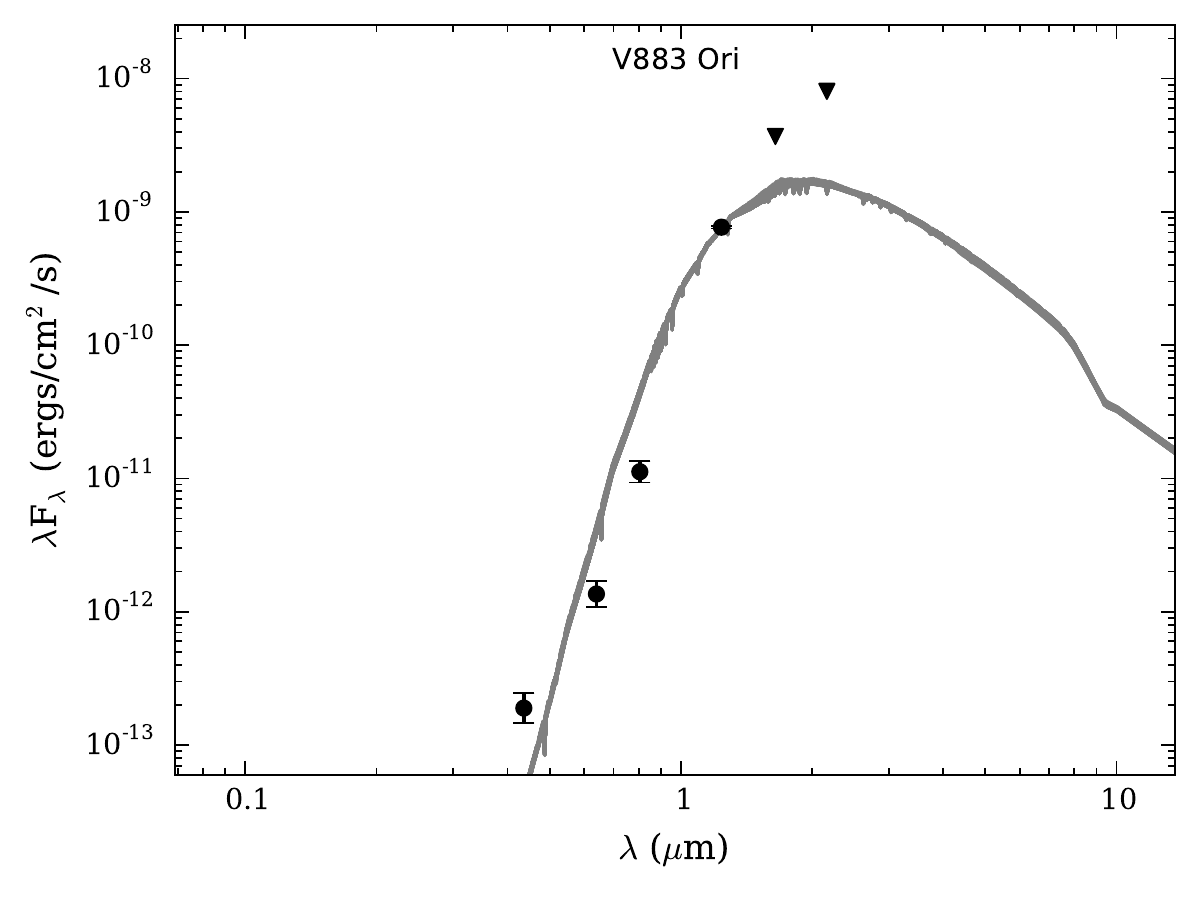}
			\includegraphics[width = 0.33\linewidth]{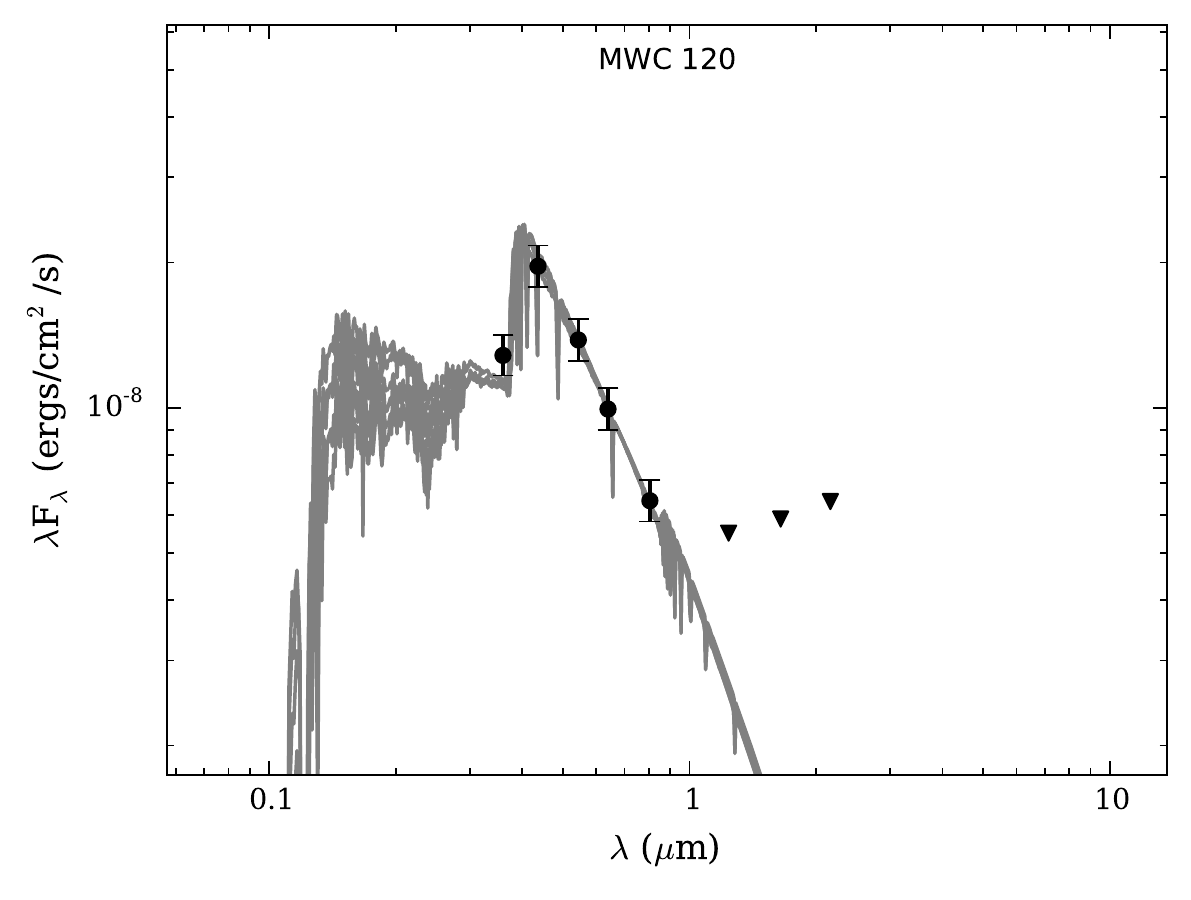}
			
		\end{figure*}
		
		\begin{figure*}[h!]
			\centering
			\includegraphics[width = 0.33\linewidth]{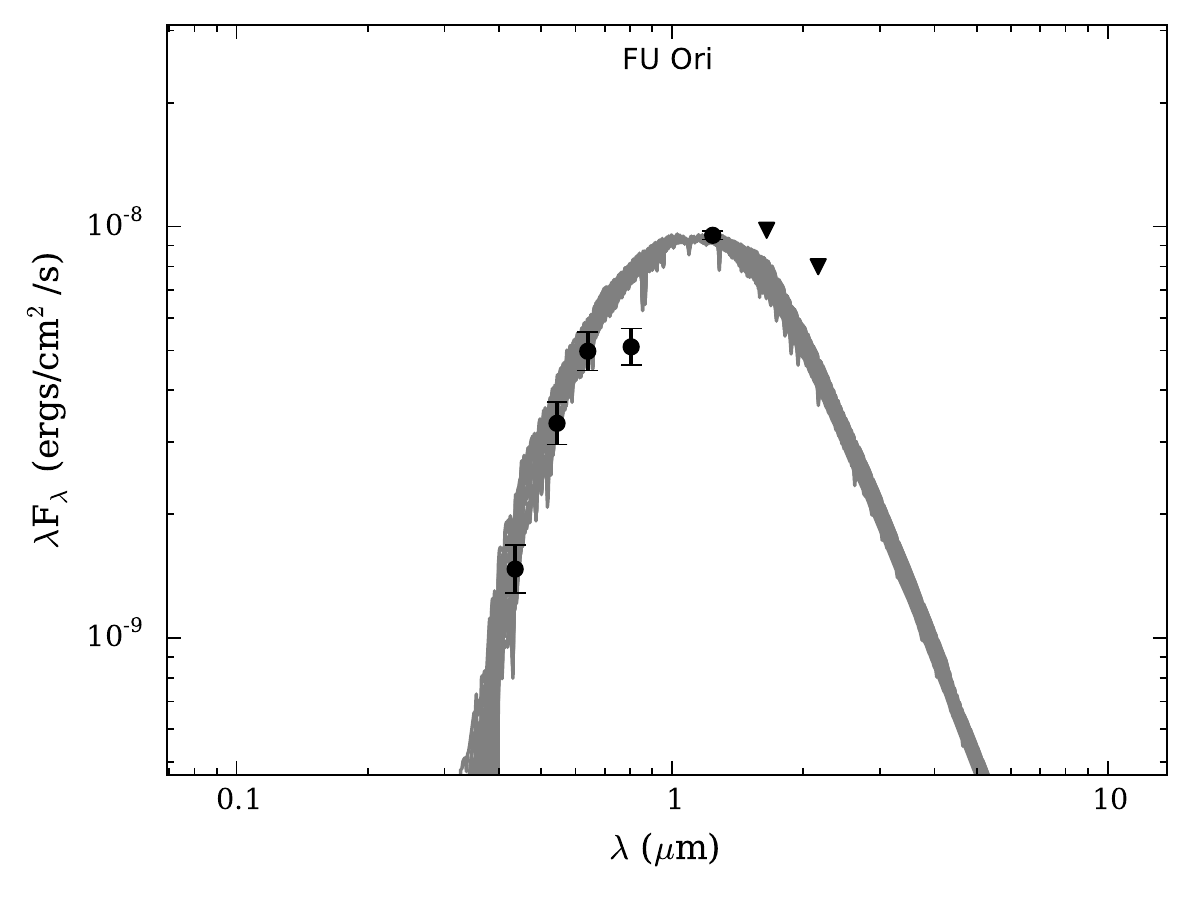}
			\includegraphics[width = 0.33\linewidth]{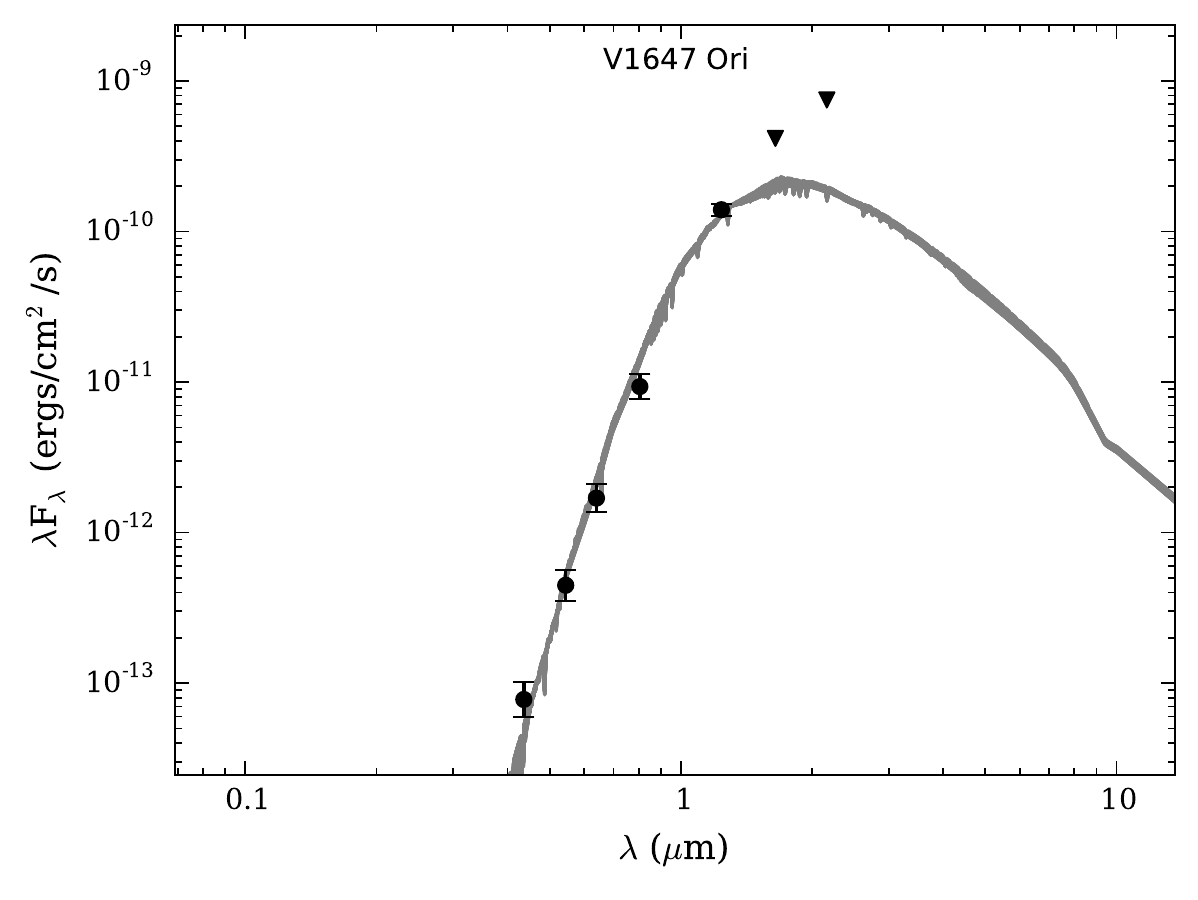}
			\includegraphics[width = 0.33\linewidth]{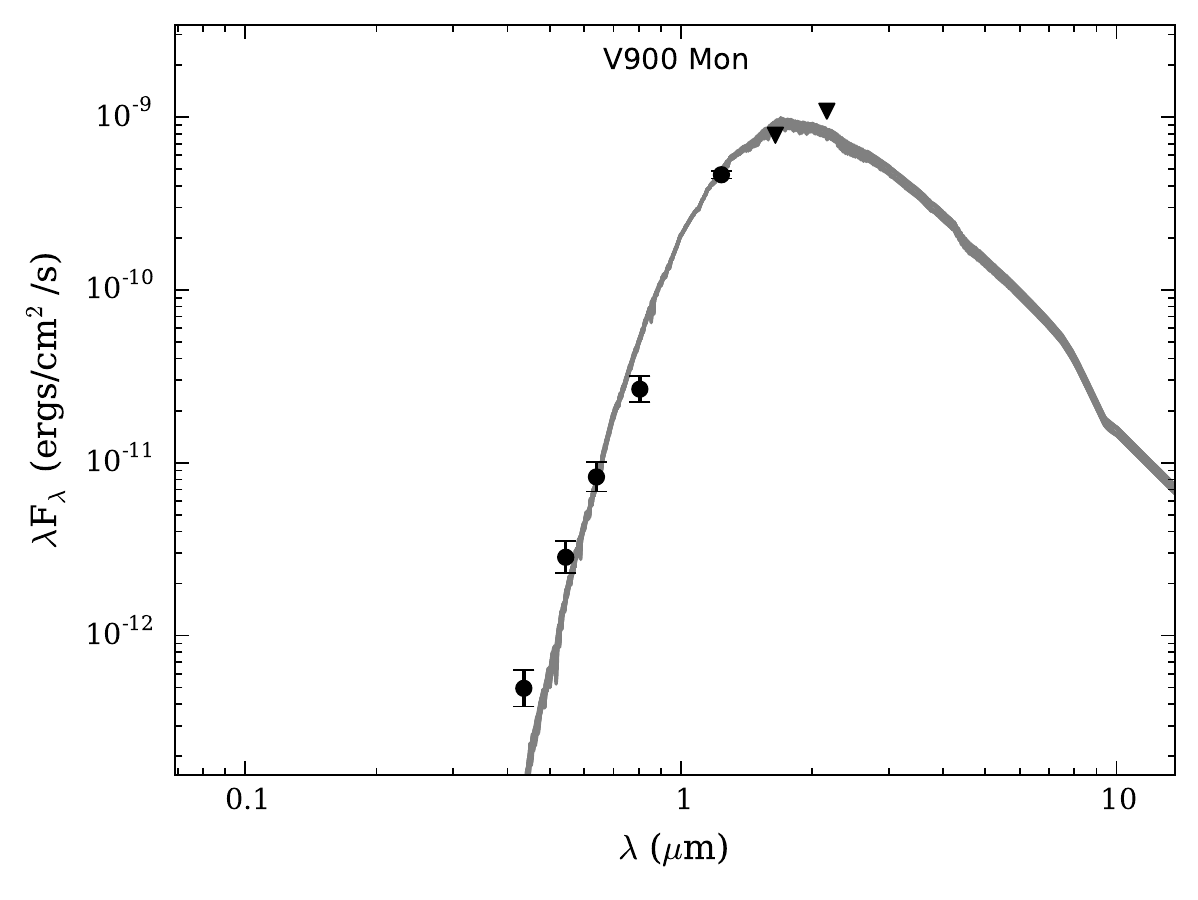}
			\includegraphics[width = 0.33\linewidth]{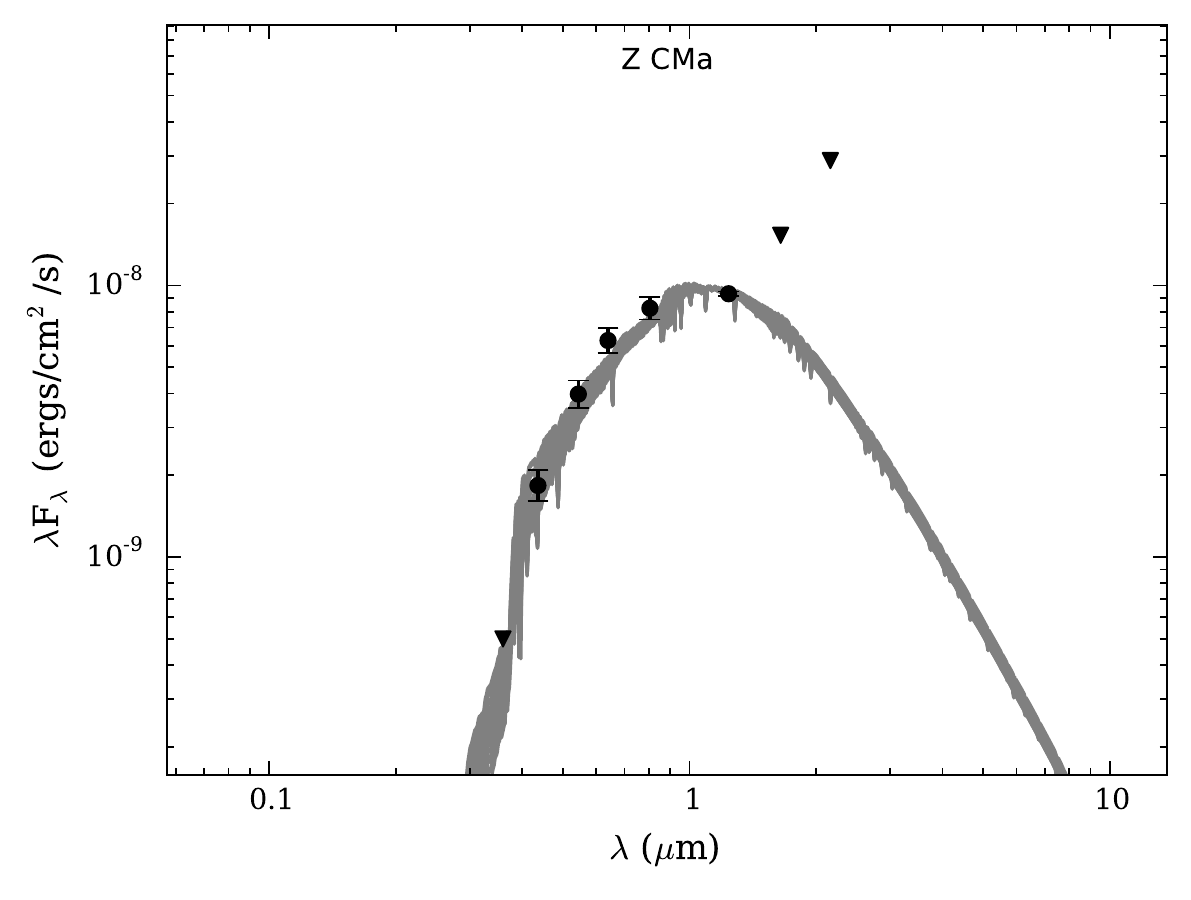}
			\includegraphics[width = 0.33\linewidth]{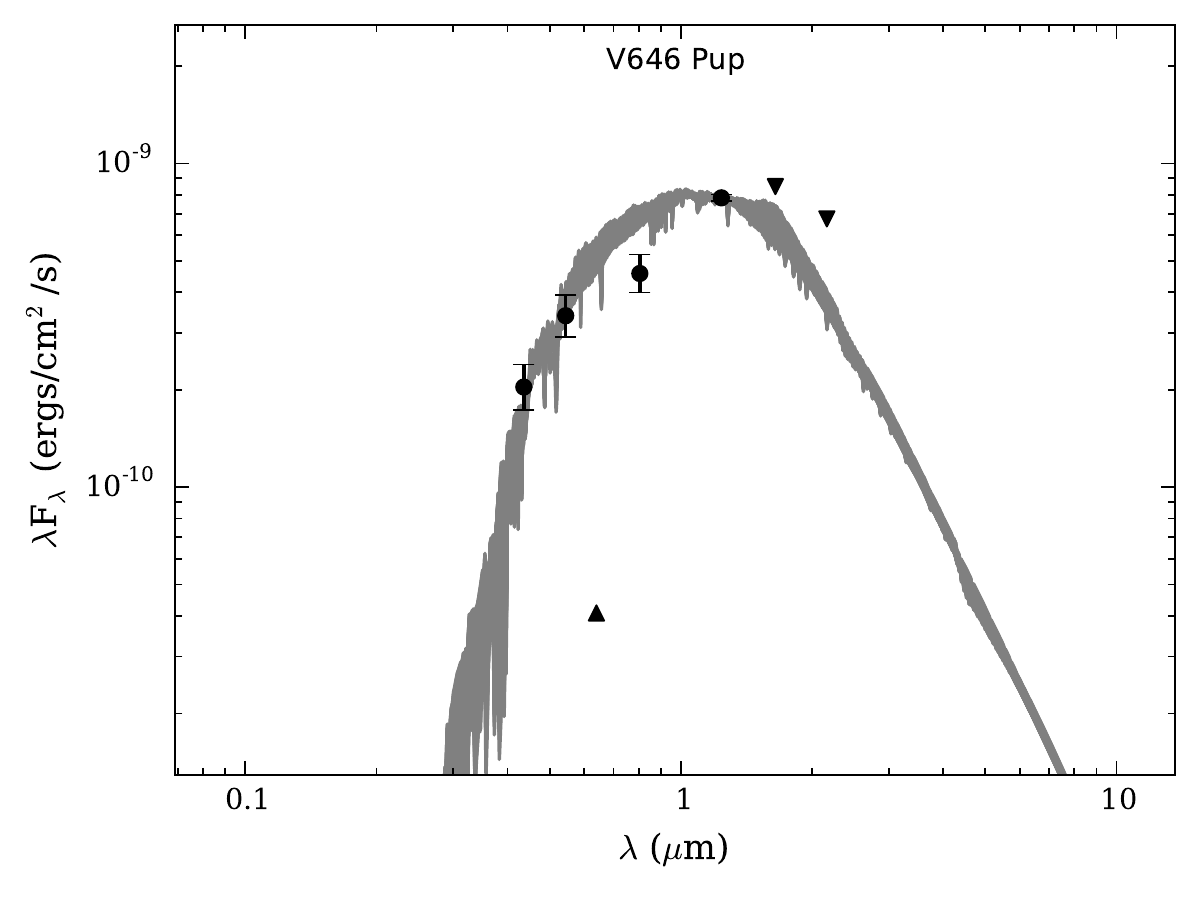}
			\includegraphics[width = 0.33\linewidth]{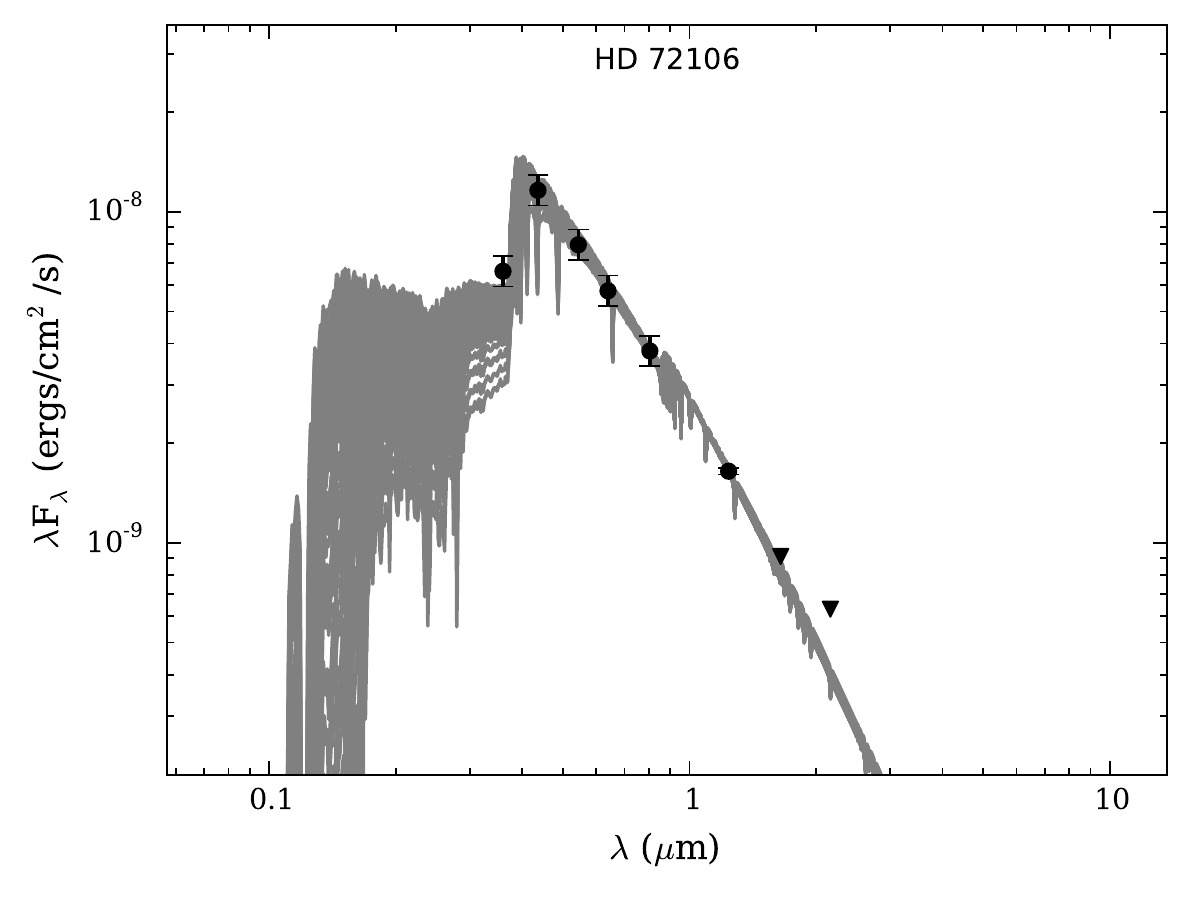}
			\includegraphics[width = 0.33\linewidth]{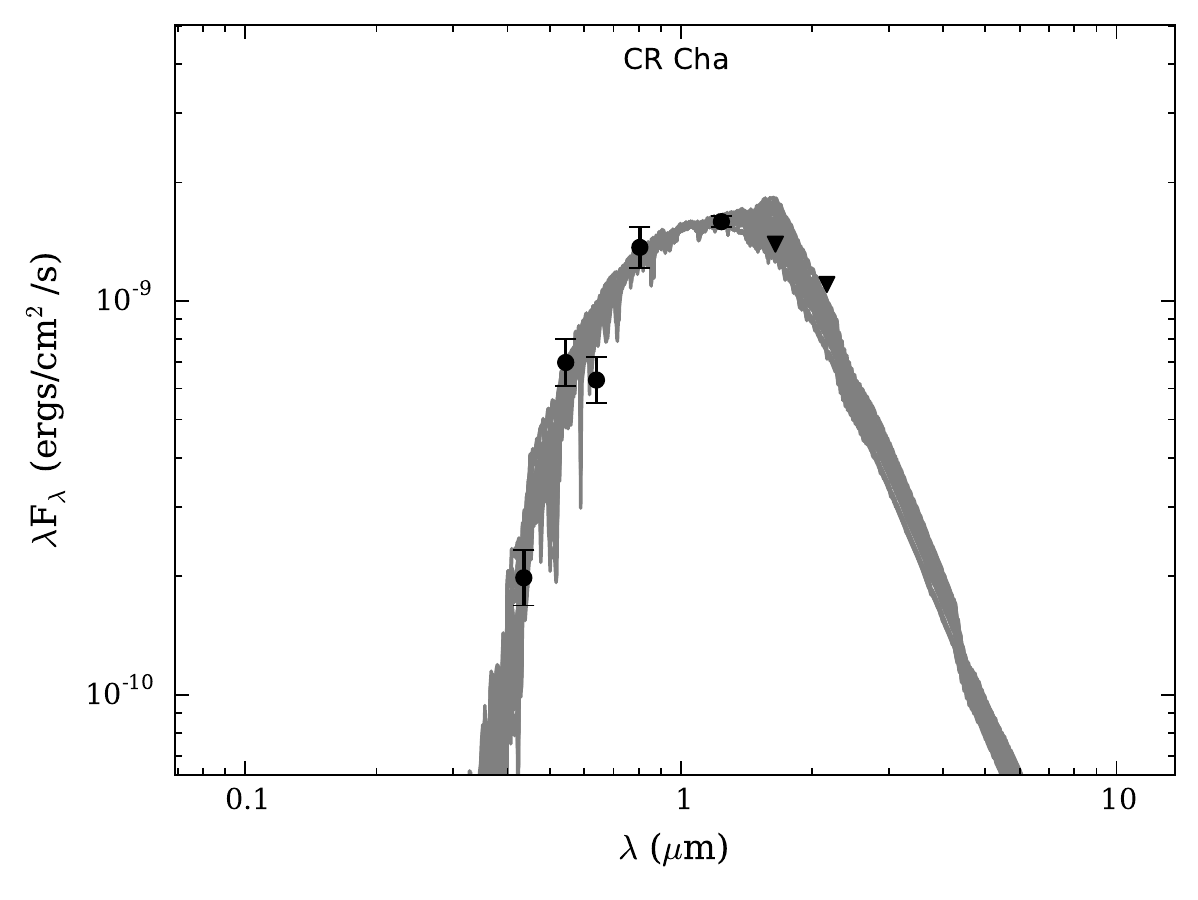}
			\includegraphics[width = 0.33\linewidth]{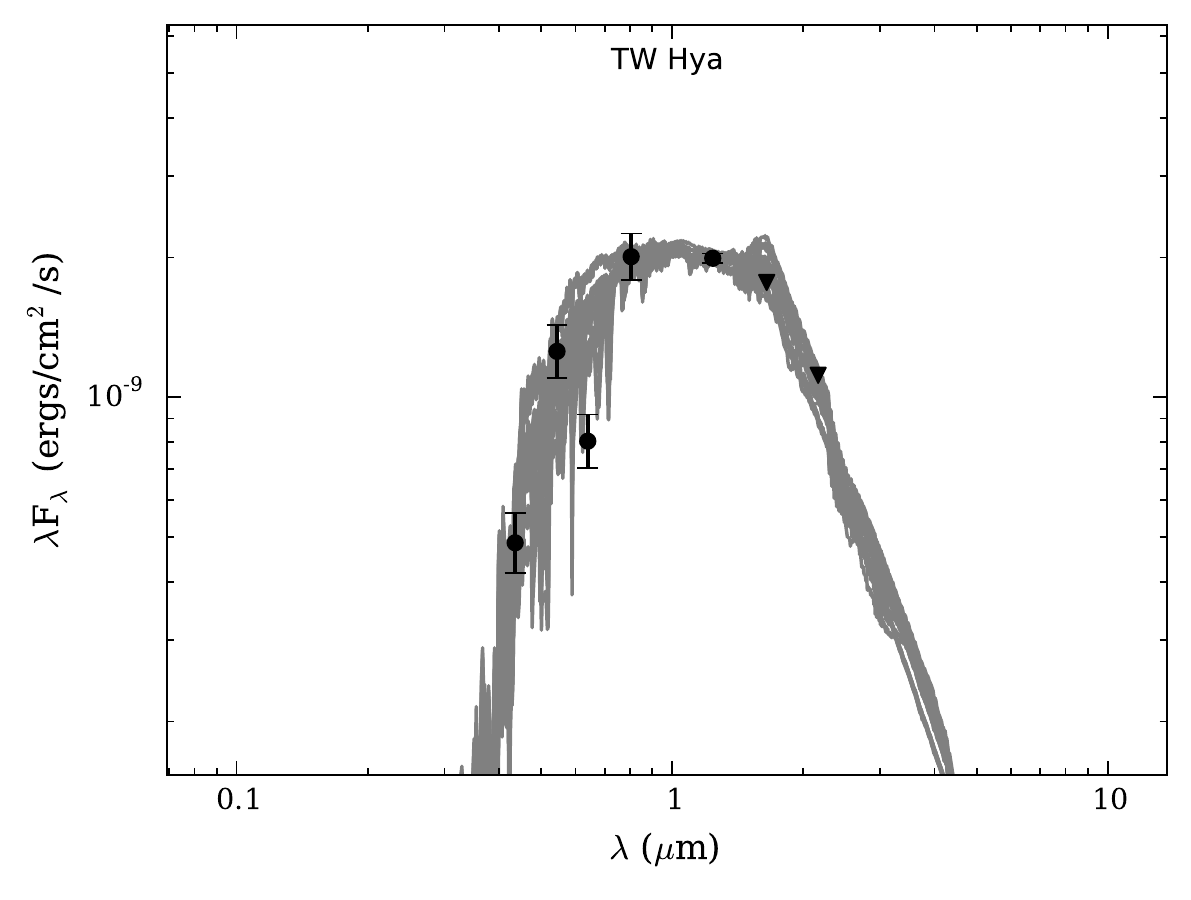}
			\includegraphics[width = 0.33\linewidth]{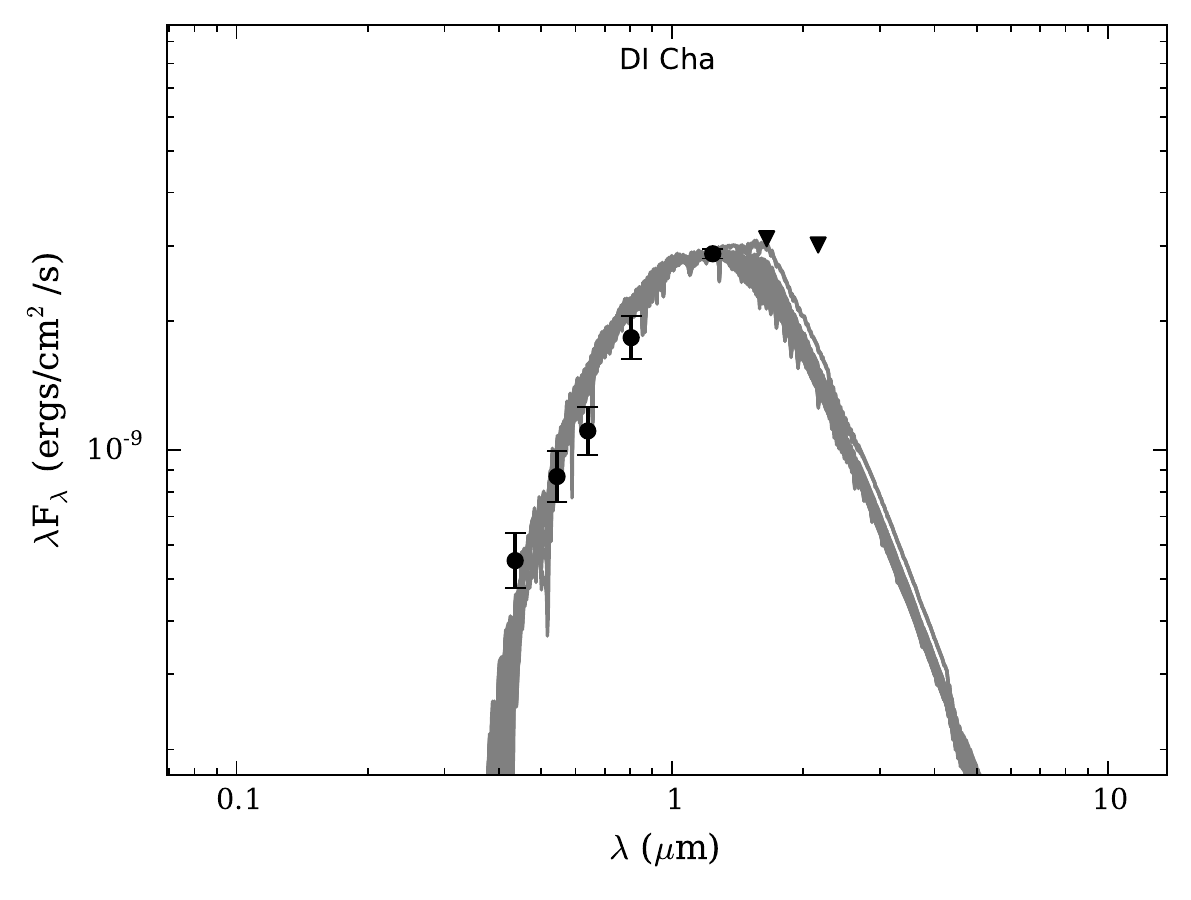}
			\includegraphics[width = 0.33\linewidth]{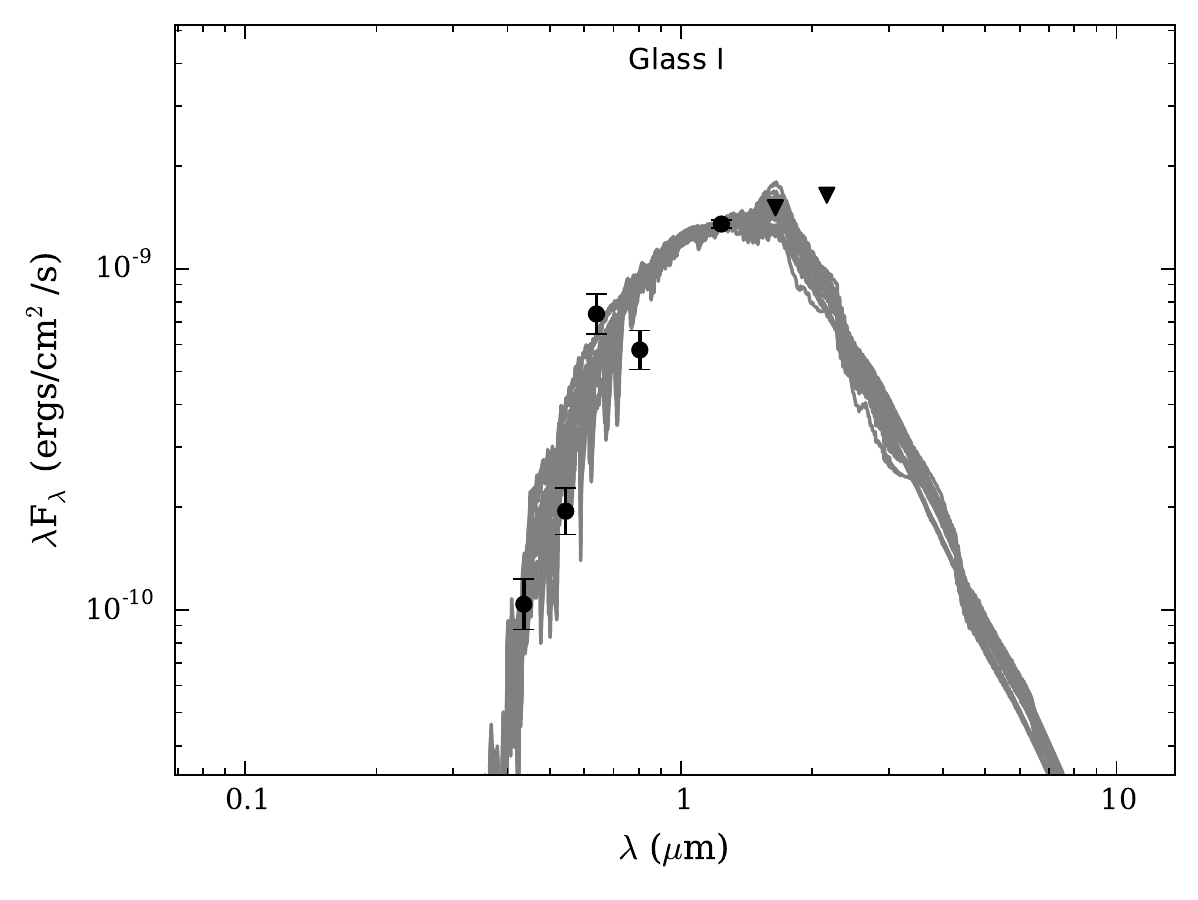}
			\includegraphics[width = 0.33\linewidth]{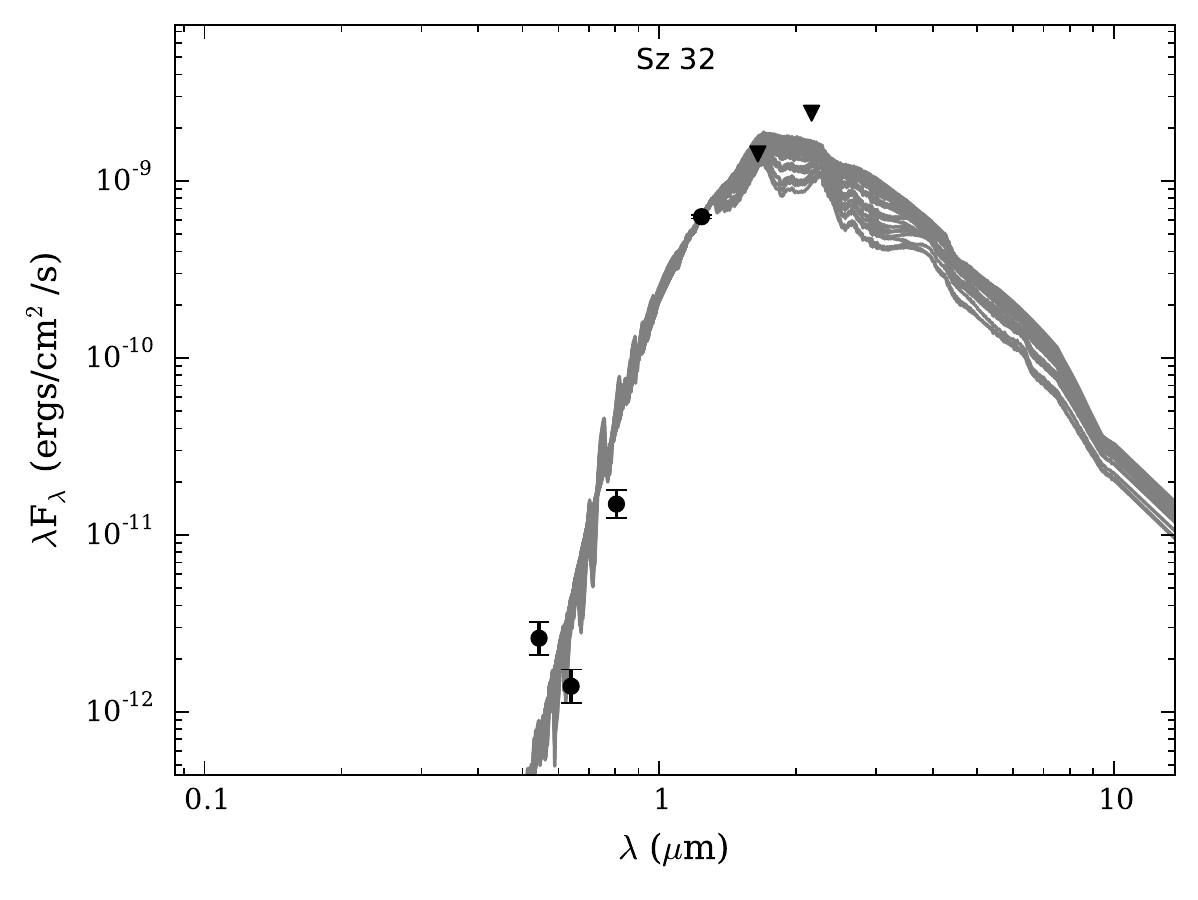}
			\includegraphics[width = 0.33\linewidth]{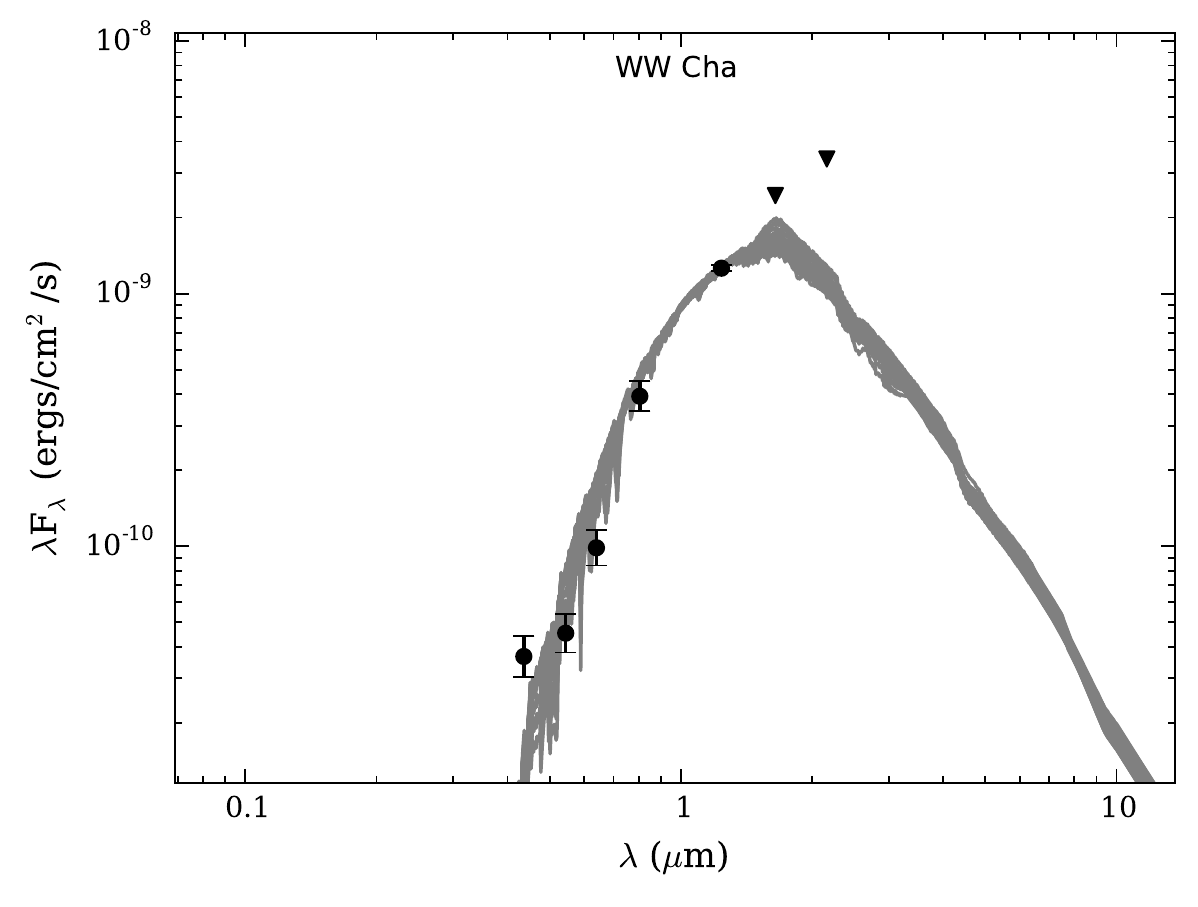}
			\includegraphics[width = 0.33\linewidth]{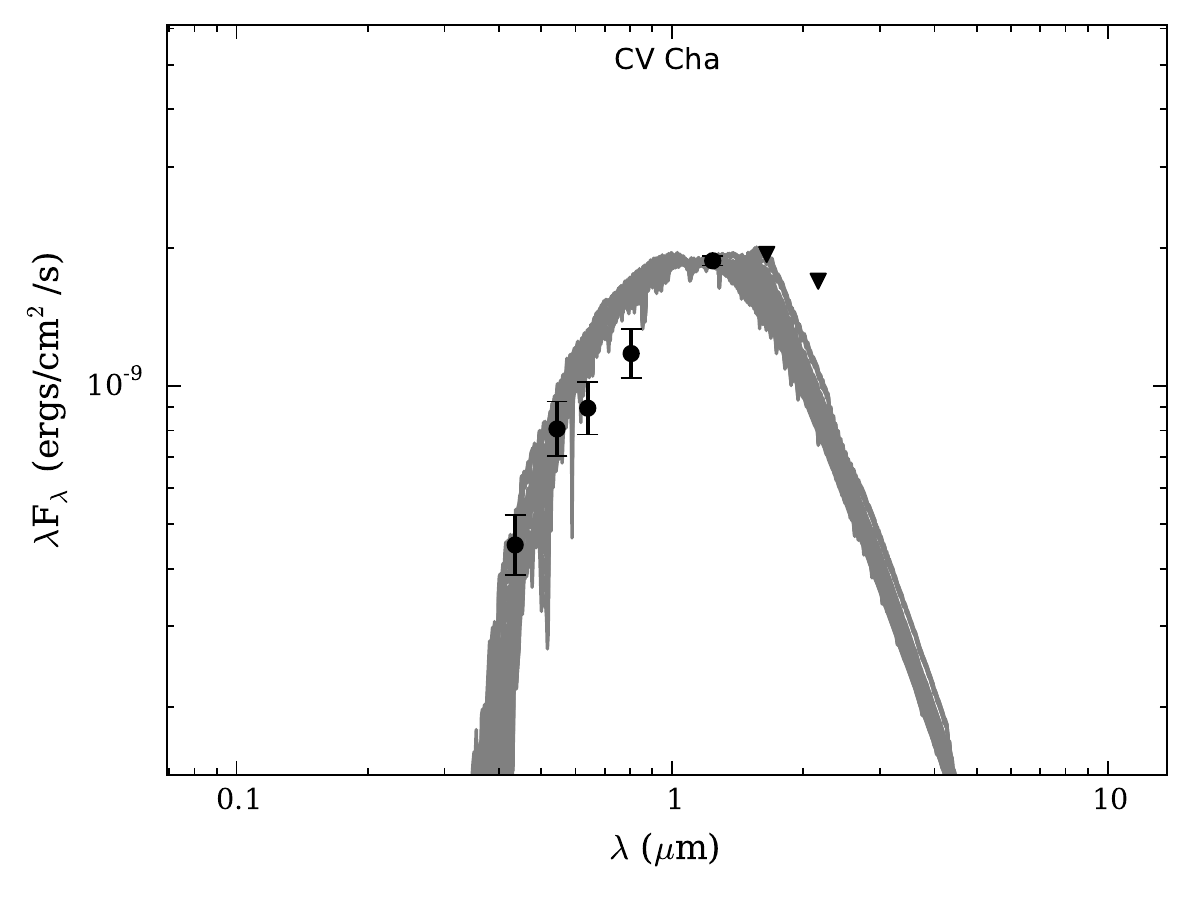}
			\includegraphics[width = 0.33\linewidth]{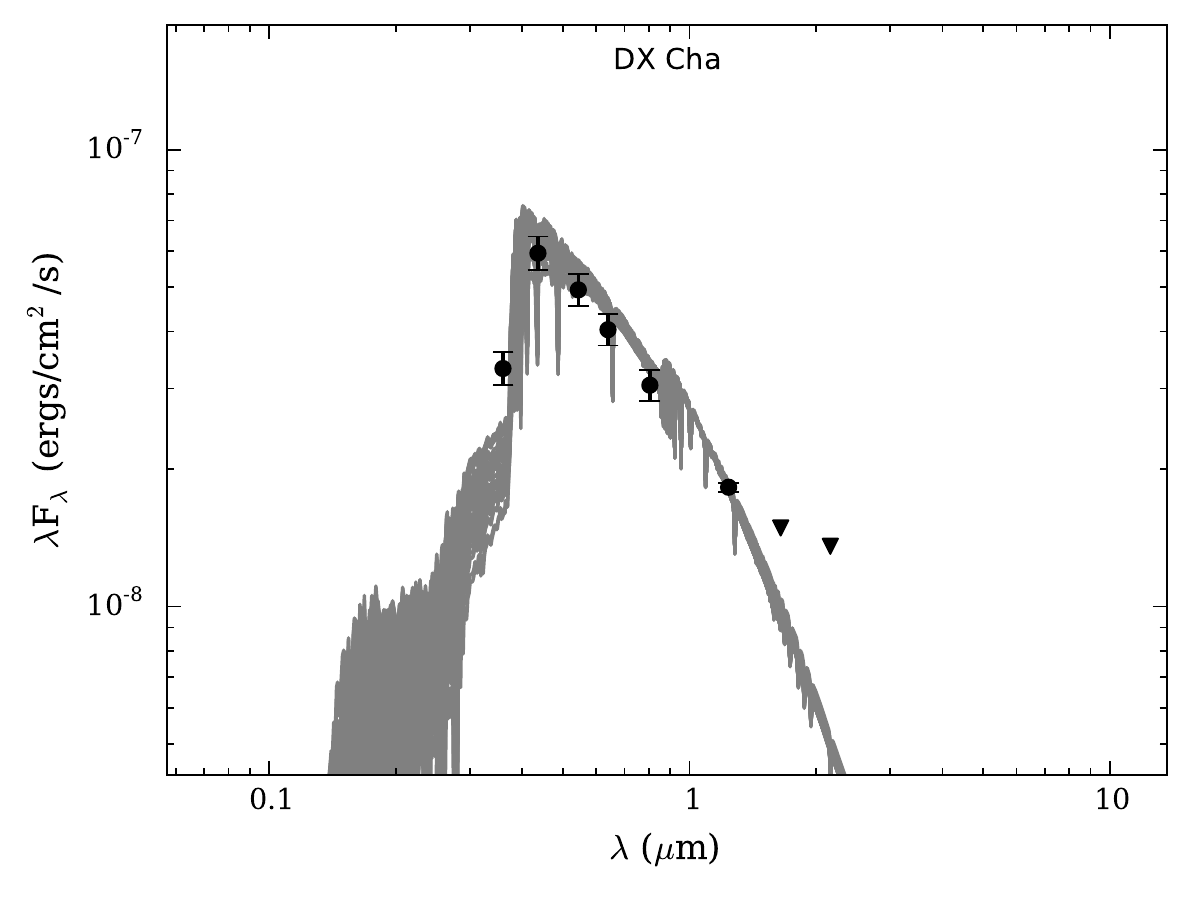}
			\includegraphics[width = 0.33\linewidth]{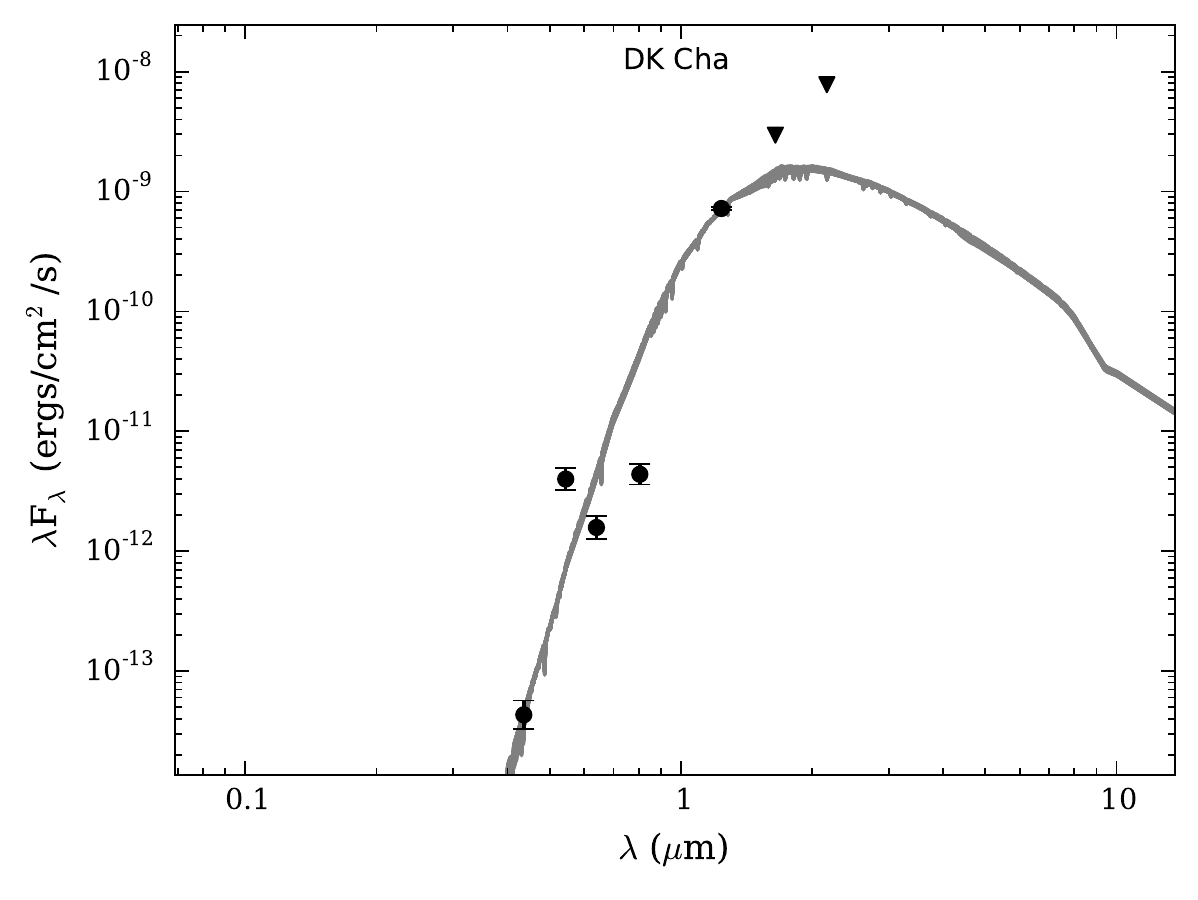}
			
		\end{figure*}
		
		\begin{figure*}[h!]
			\centering
			\includegraphics[width = 0.33\linewidth]{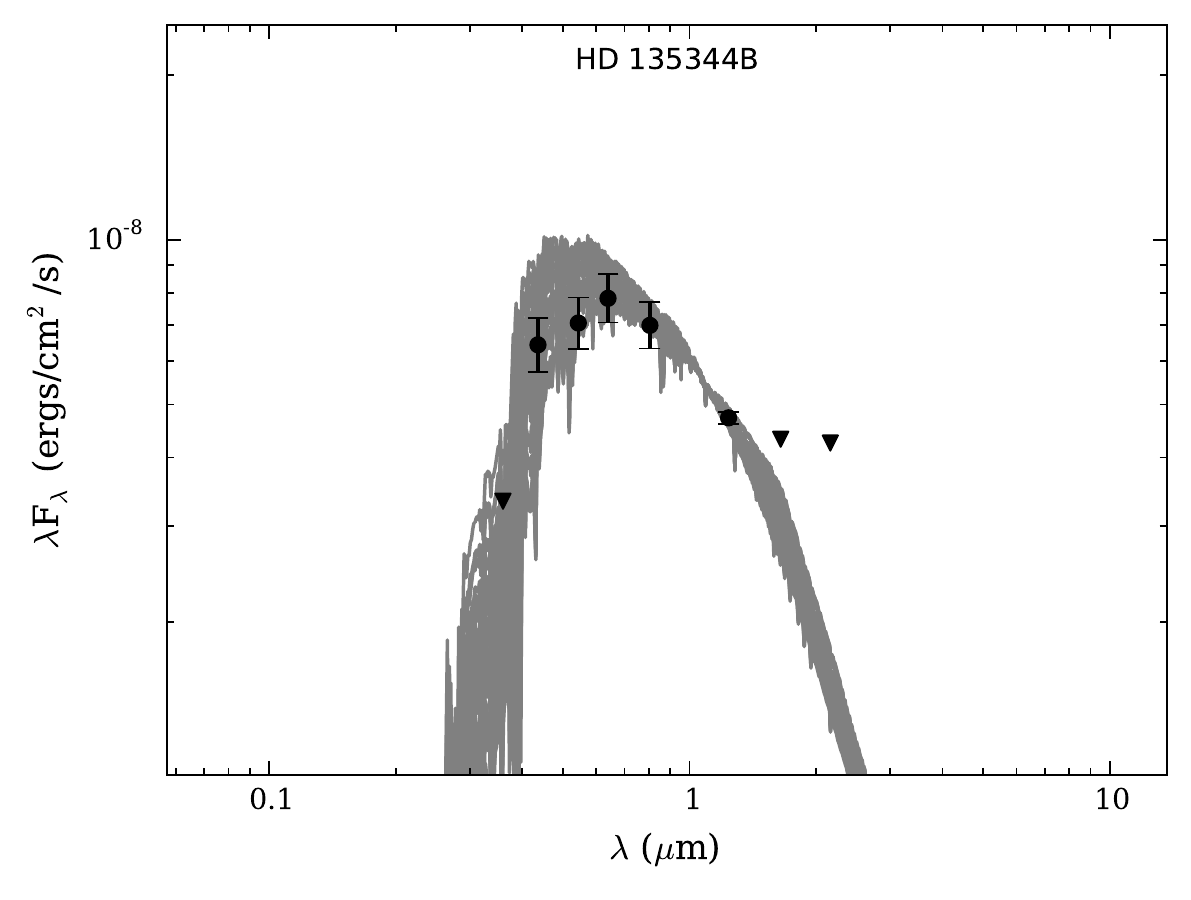}
			\includegraphics[width = 0.33\linewidth]{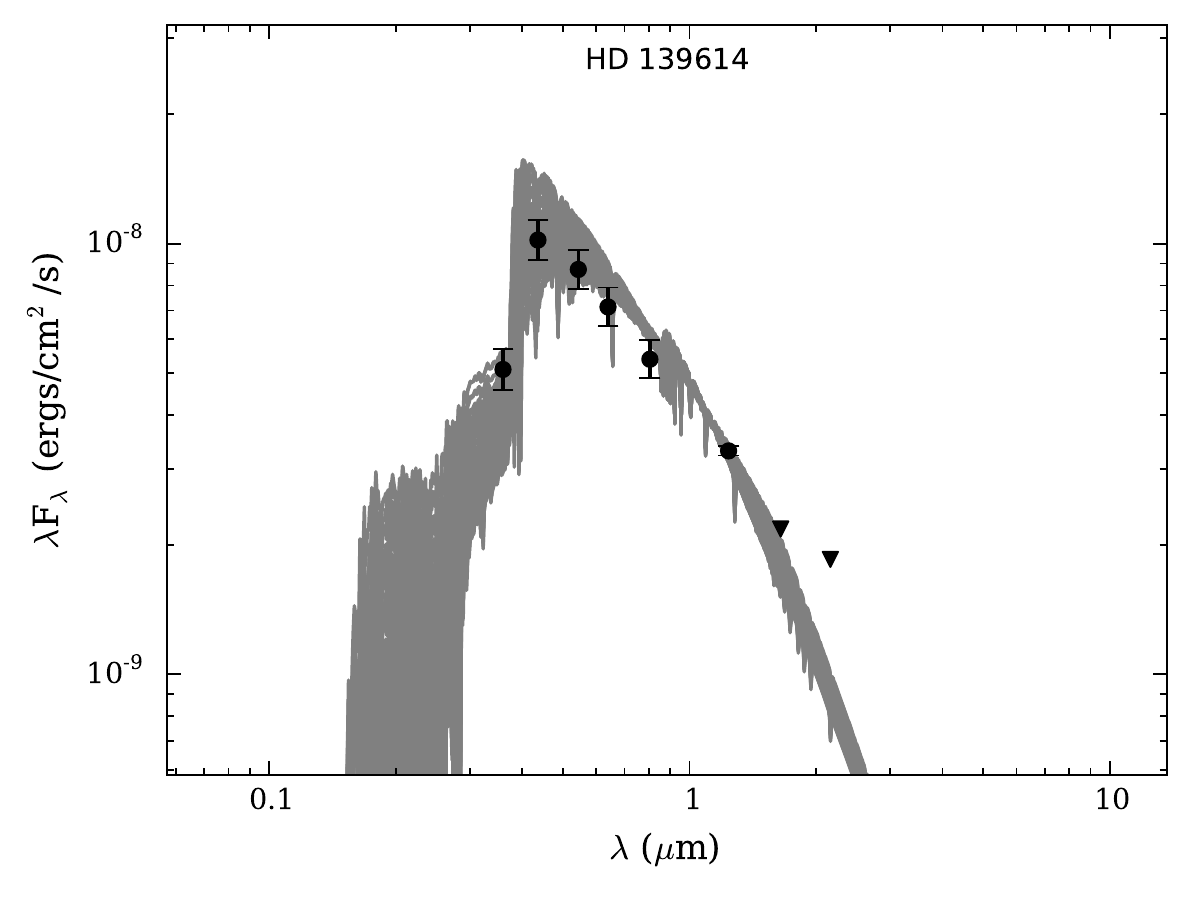}
			\includegraphics[width = 0.33\linewidth]{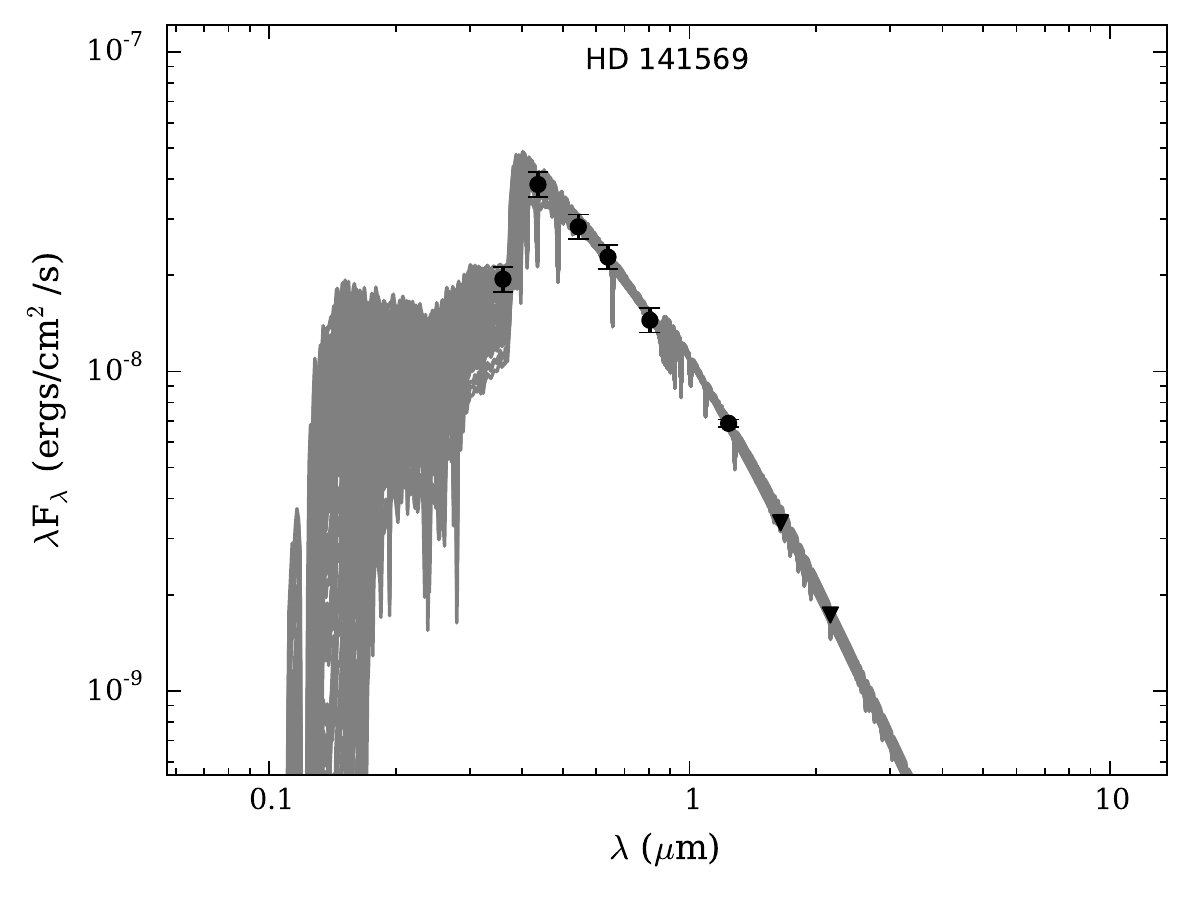}
			\includegraphics[width = 0.33\linewidth]{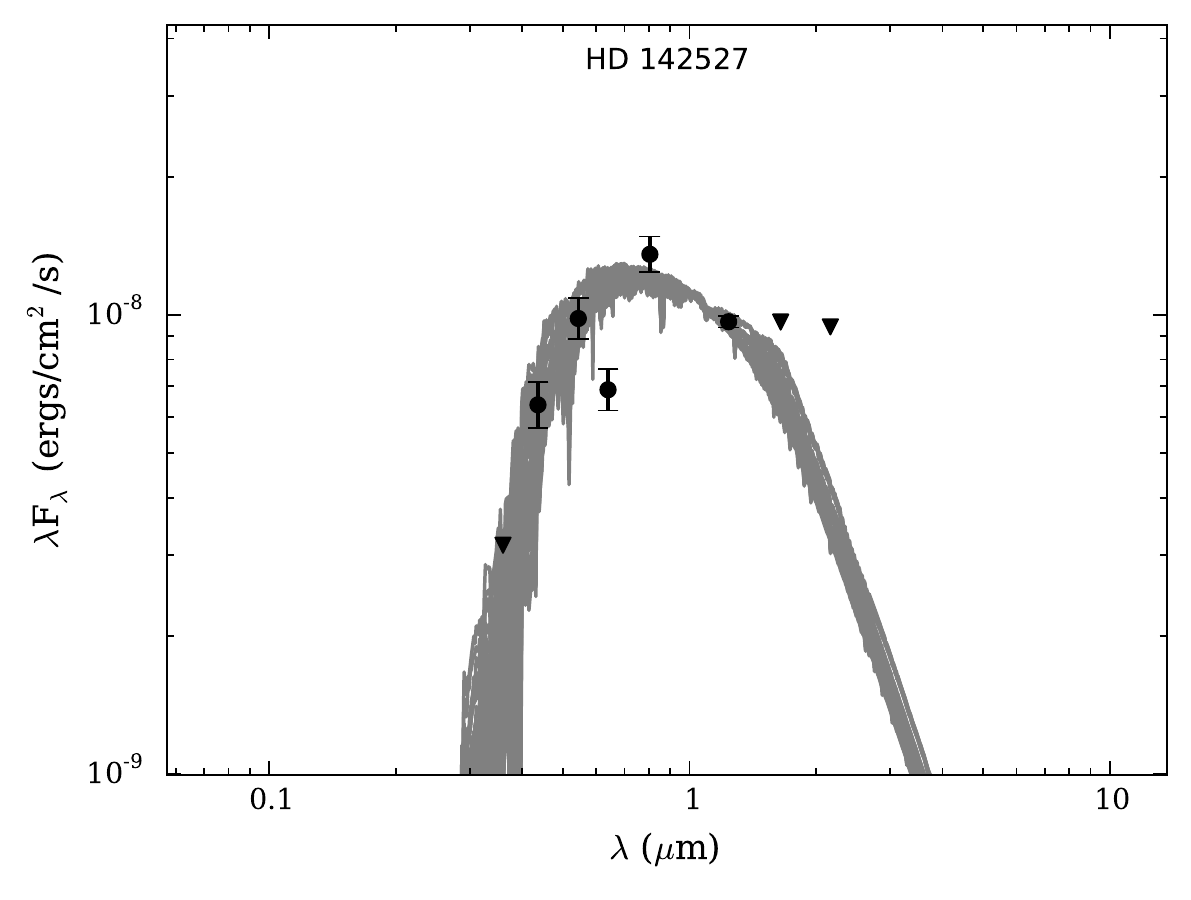}
			\includegraphics[width = 0.33\linewidth]{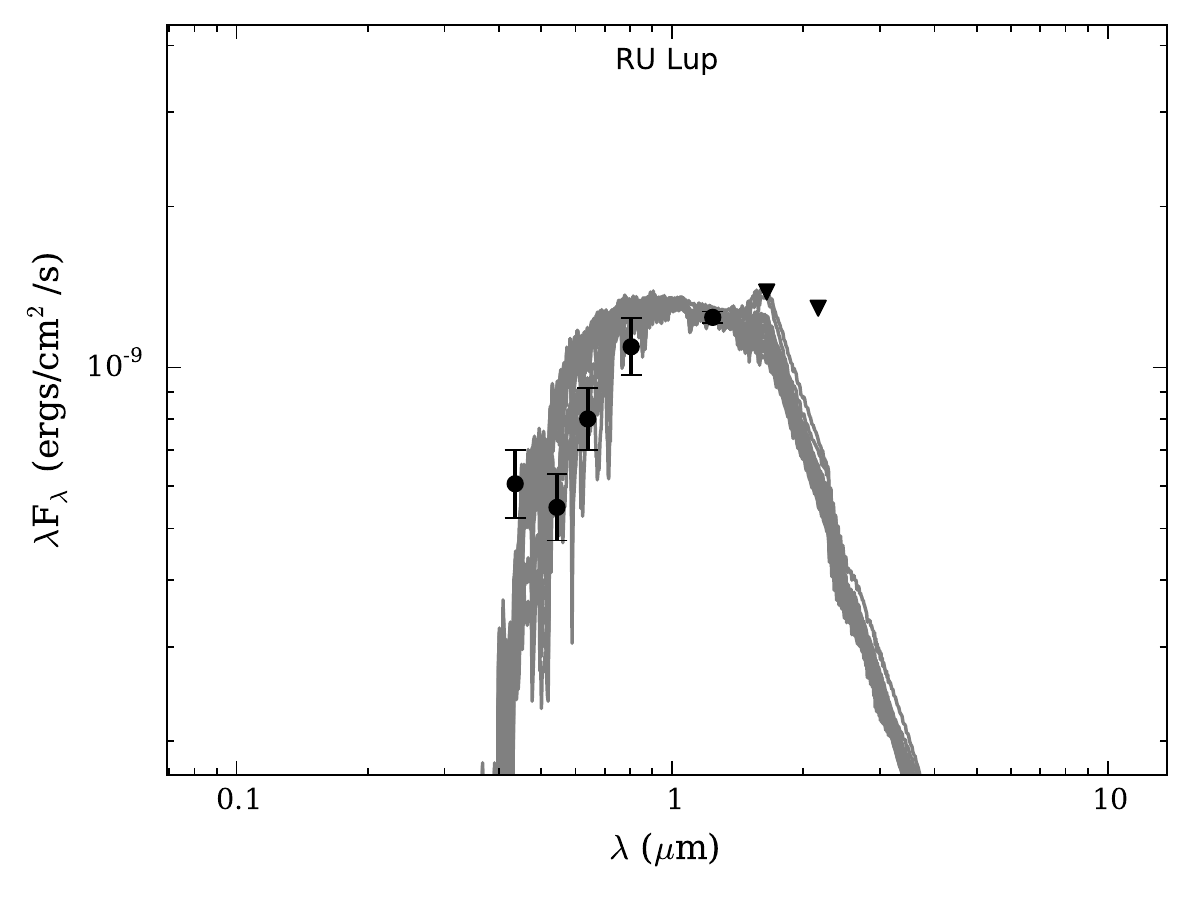}
			\includegraphics[width = 0.33\linewidth]{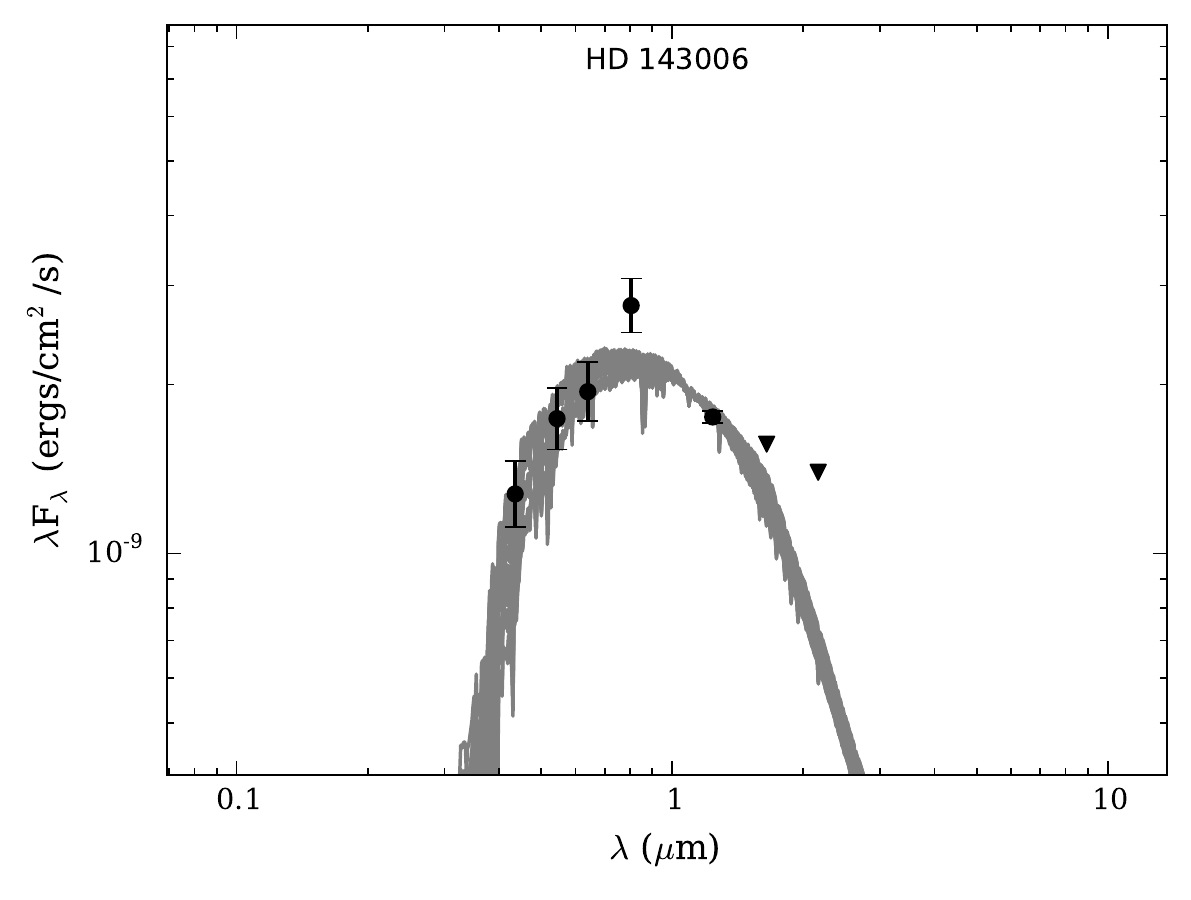}
			\includegraphics[width = 0.33\linewidth]{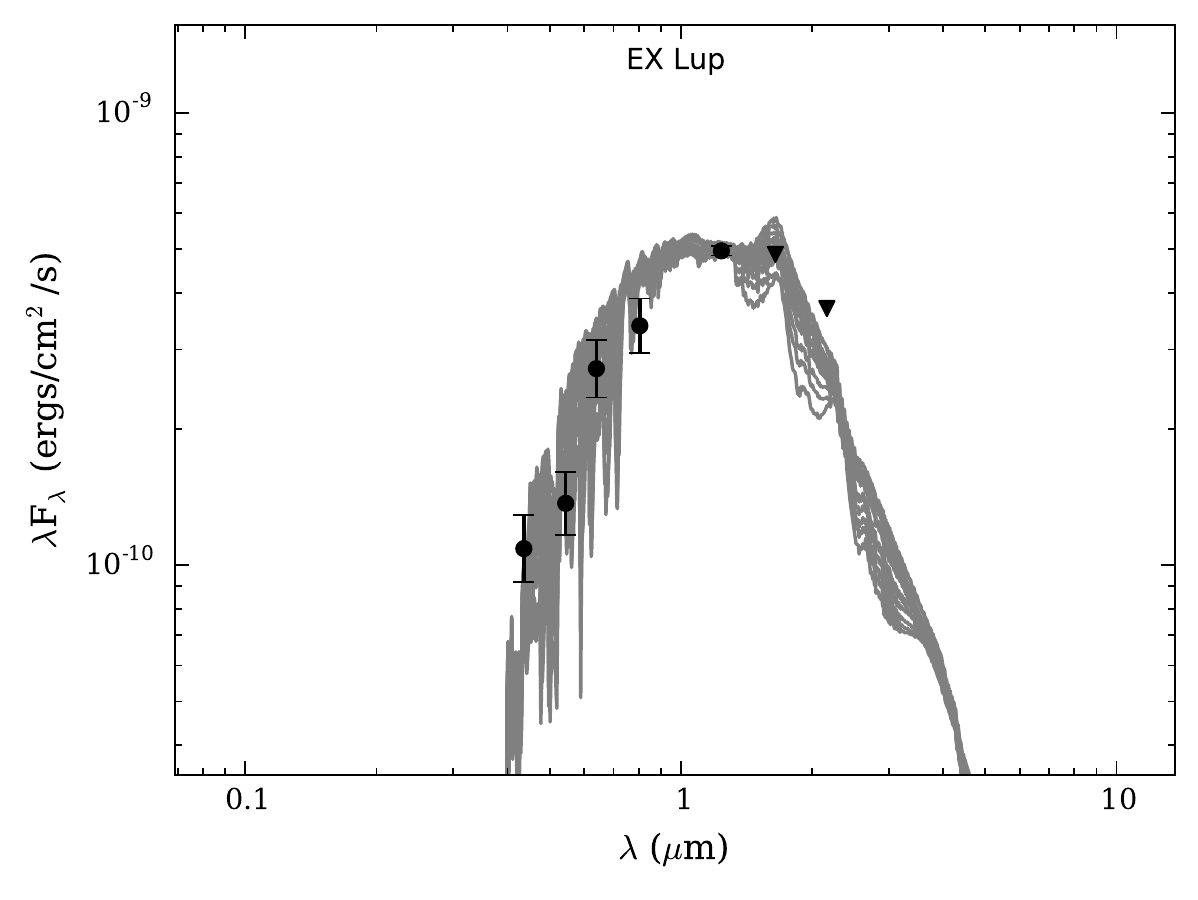}
			\includegraphics[width = 0.33\linewidth]{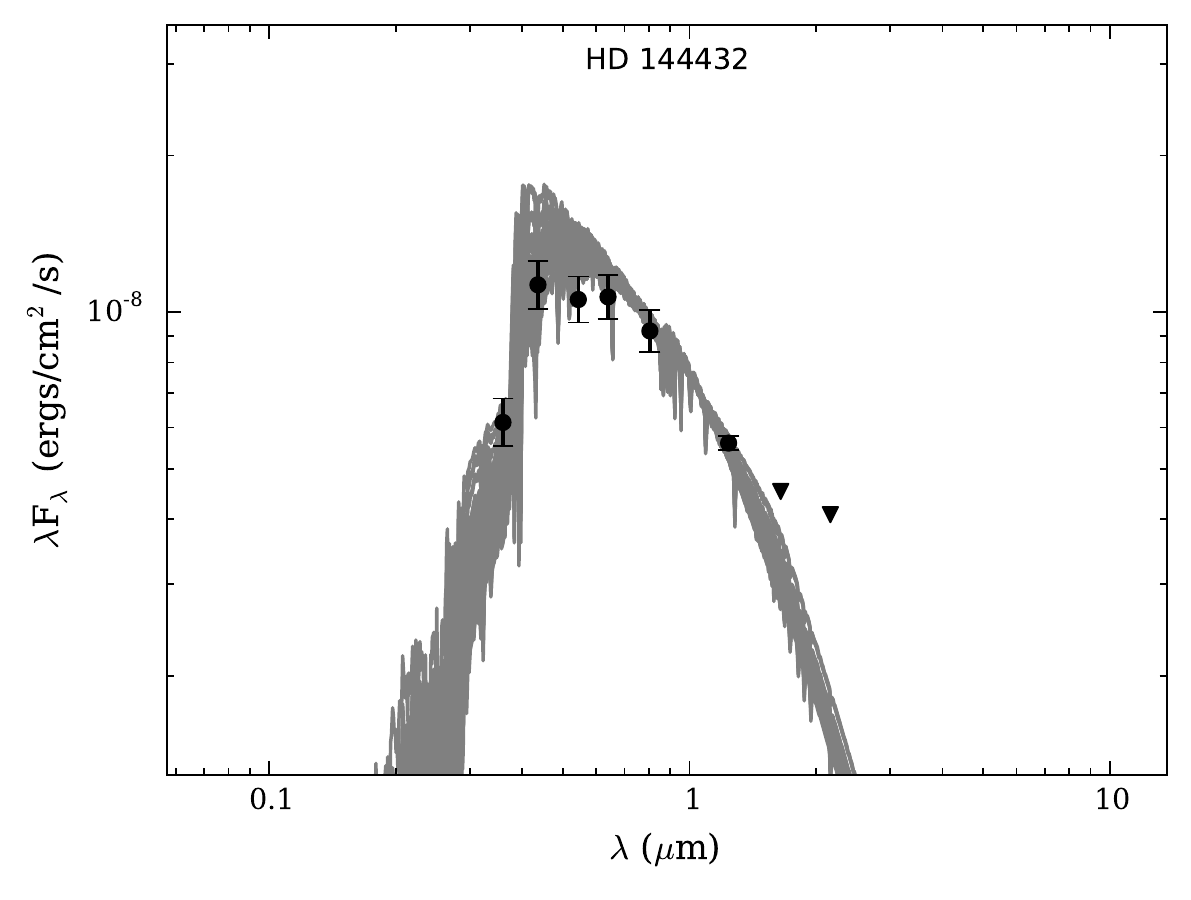}
			\includegraphics[width = 0.33\linewidth]{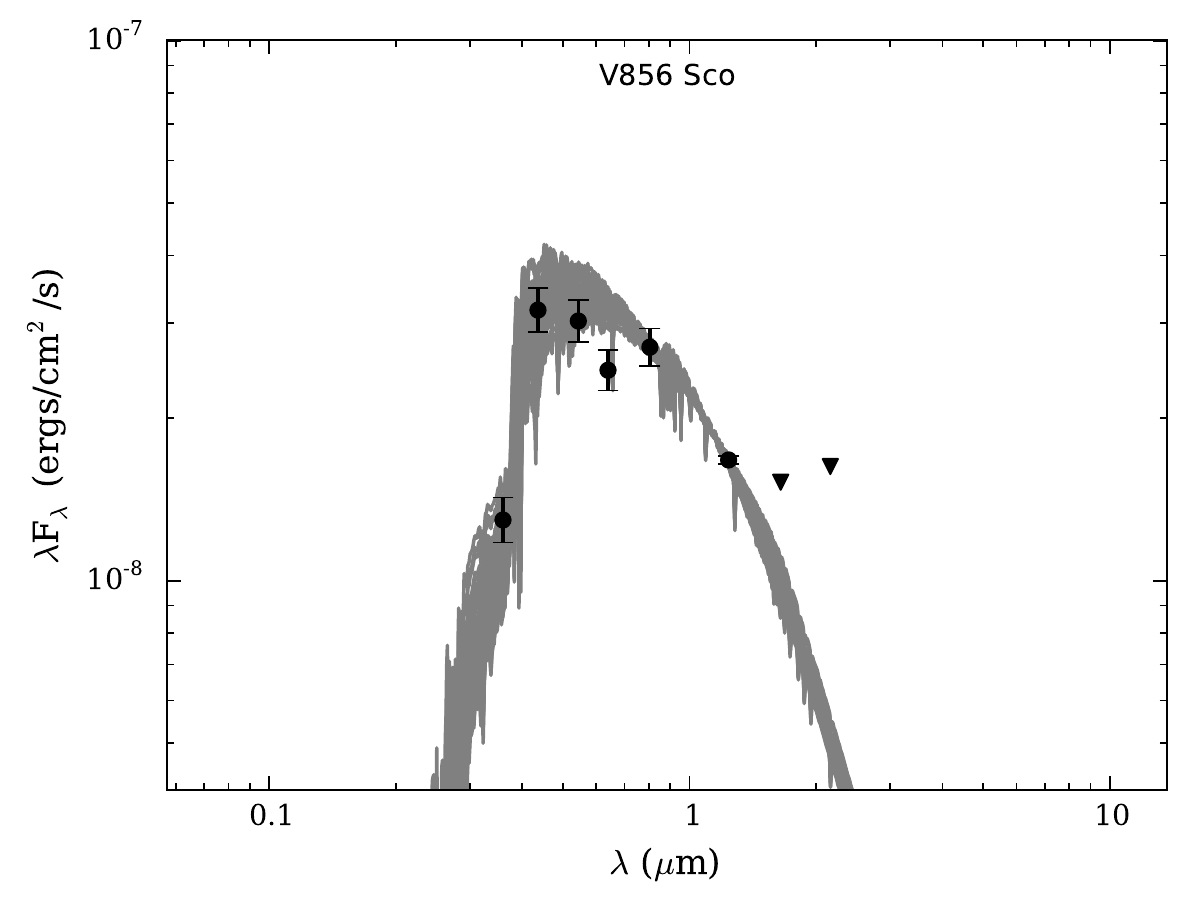}
			\includegraphics[width = 0.33\linewidth]{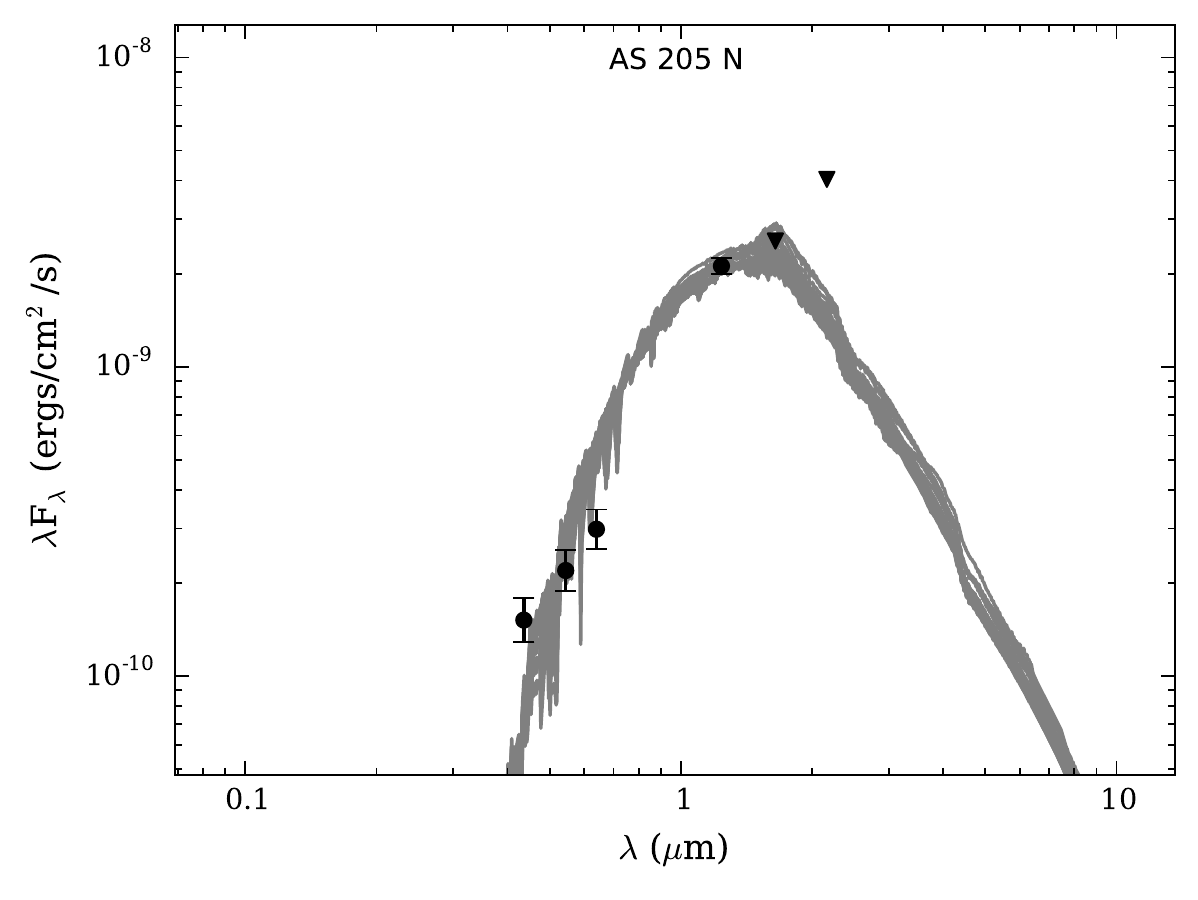}
			\includegraphics[width = 0.33\linewidth]{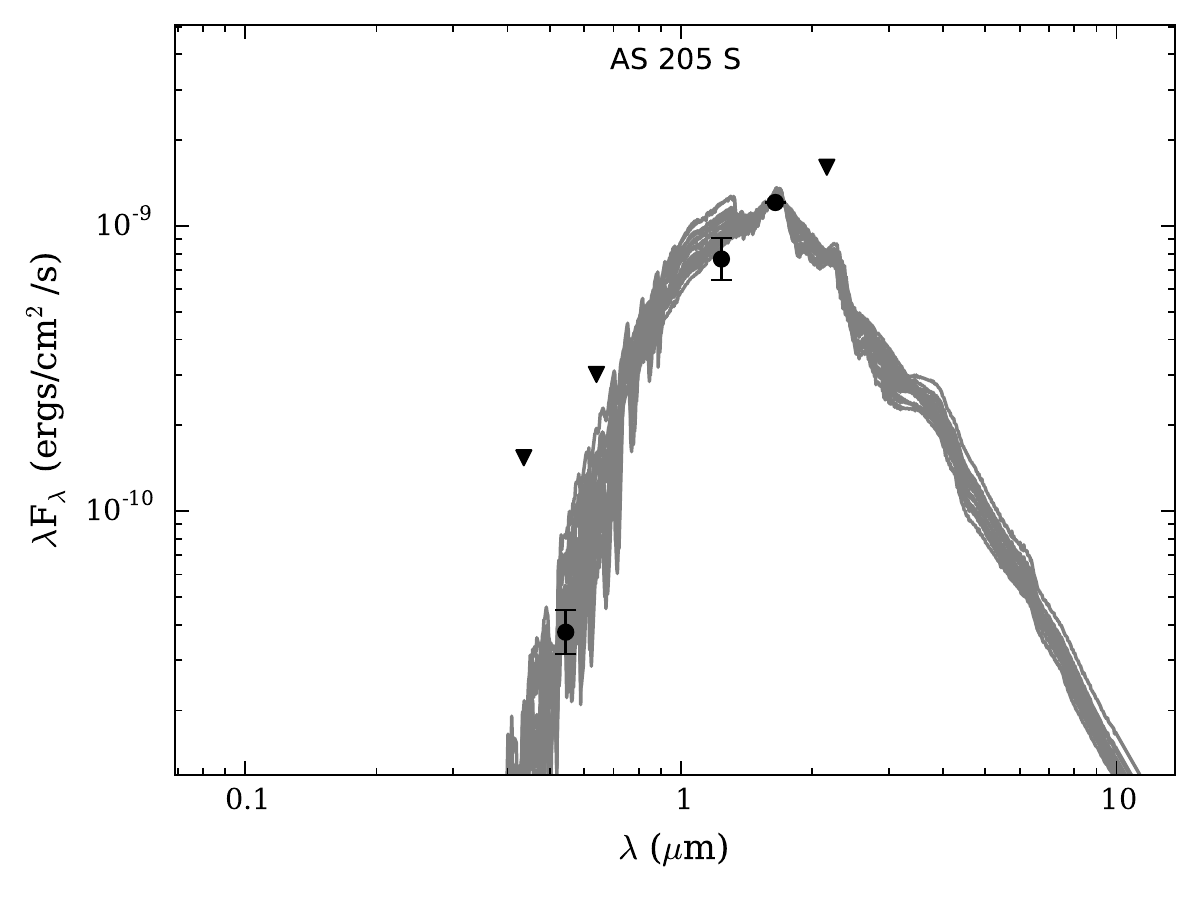}
			\includegraphics[width = 0.33\linewidth]{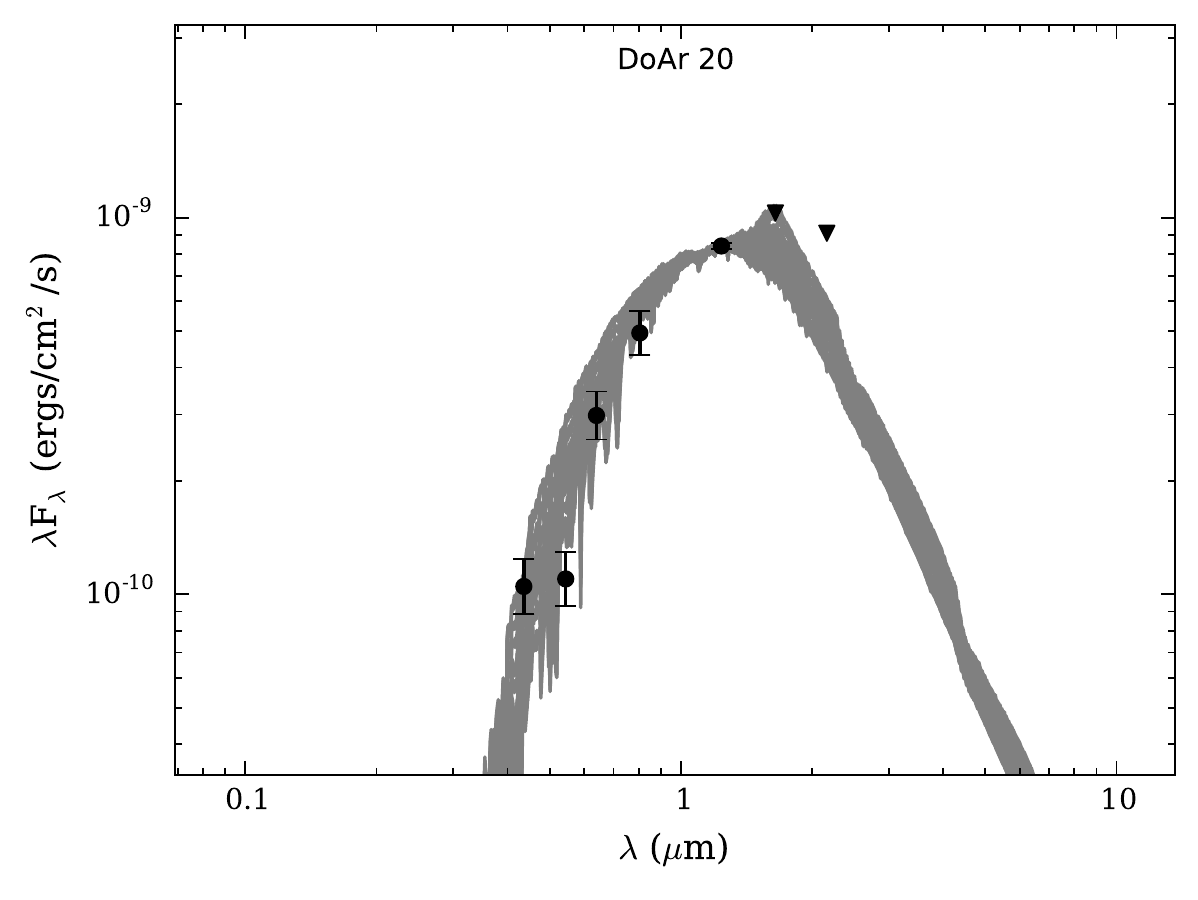}
			\includegraphics[width = 0.33\linewidth]{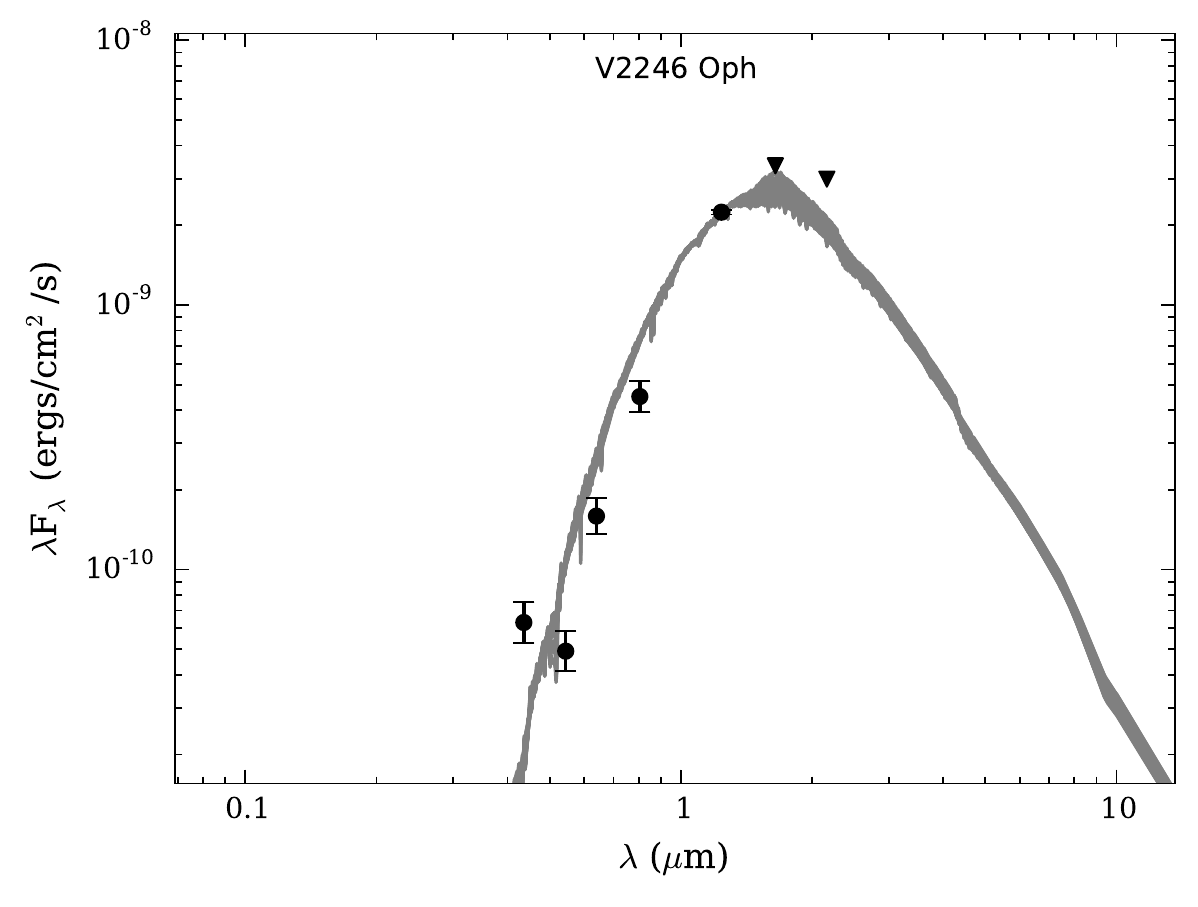}
			\includegraphics[width = 0.33\linewidth]{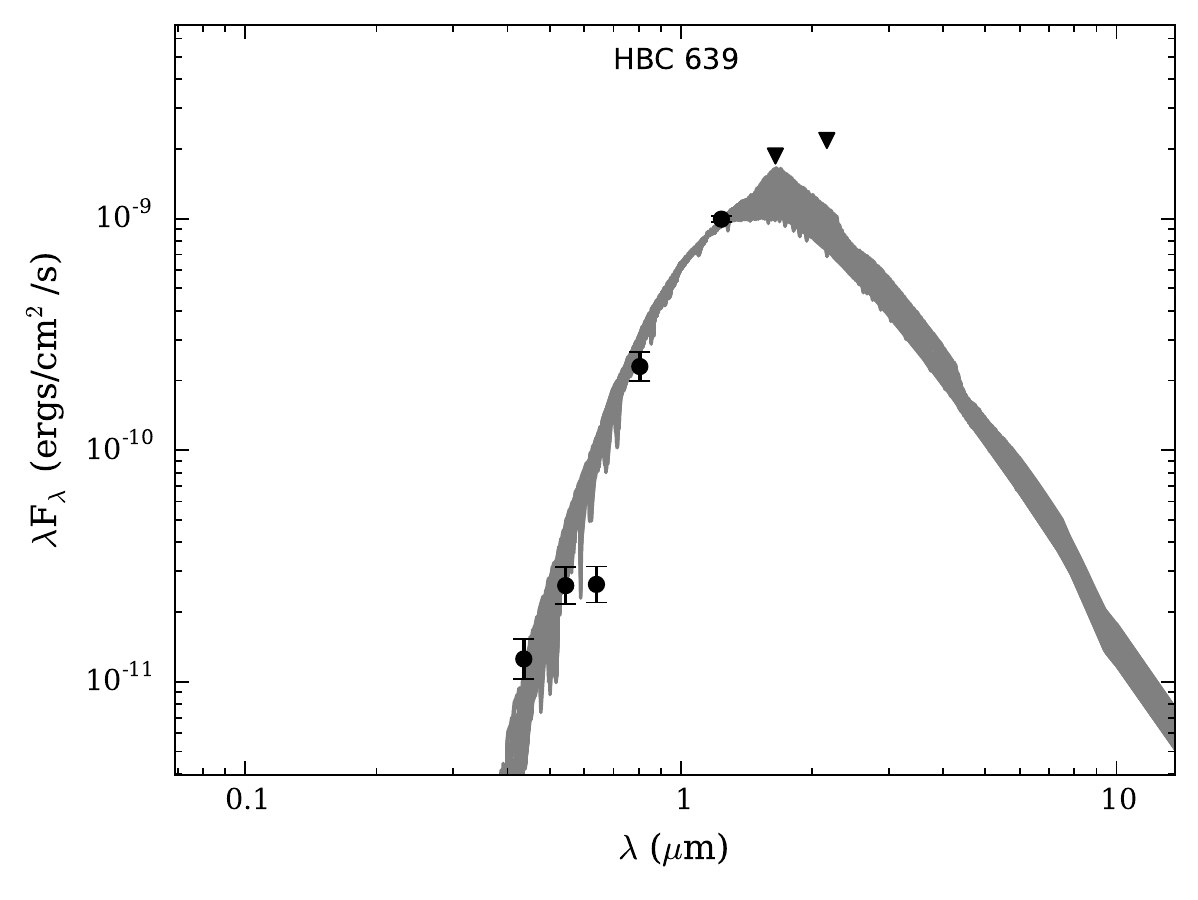}
			\includegraphics[width = 0.33\linewidth]{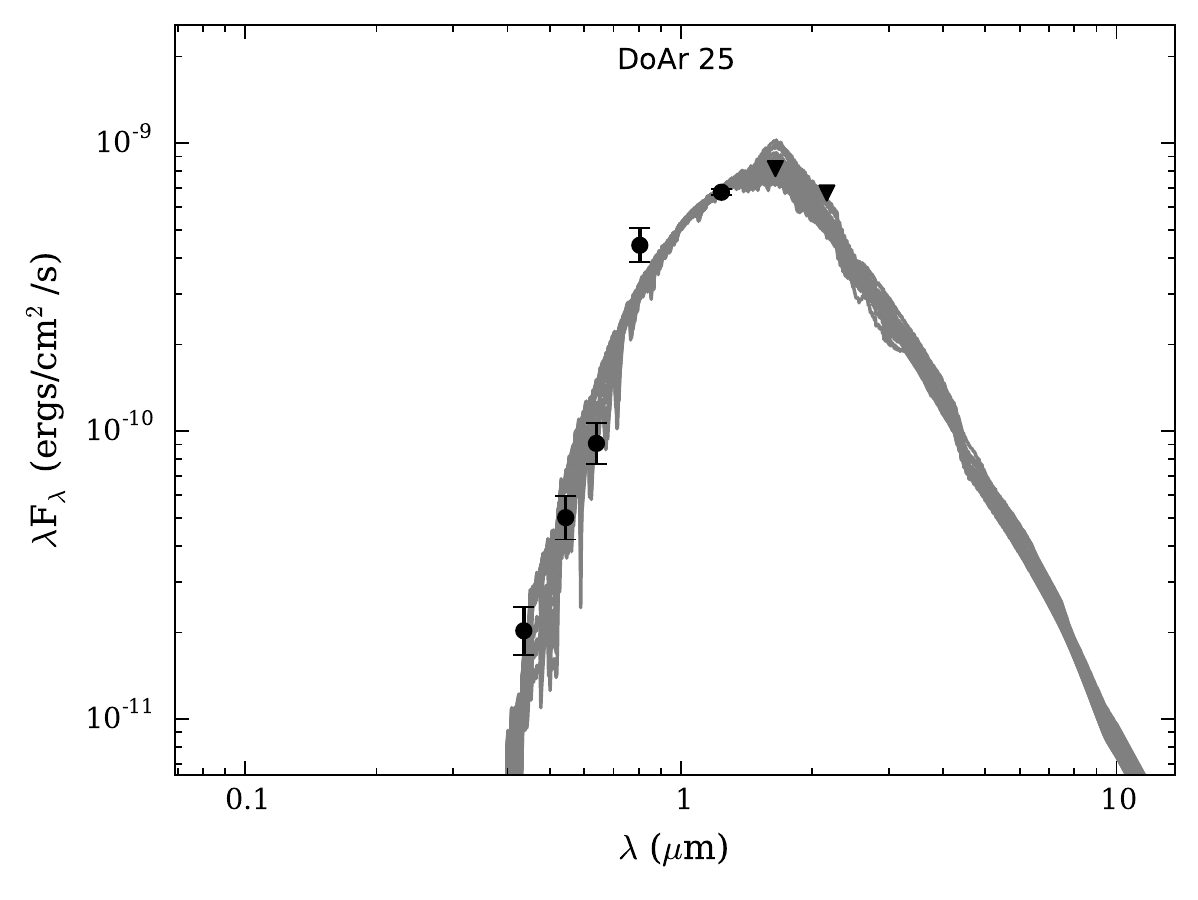}
			
		\end{figure*}
		
		\begin{figure*}[h!]
			\centering
			\includegraphics[width = 0.33\linewidth]{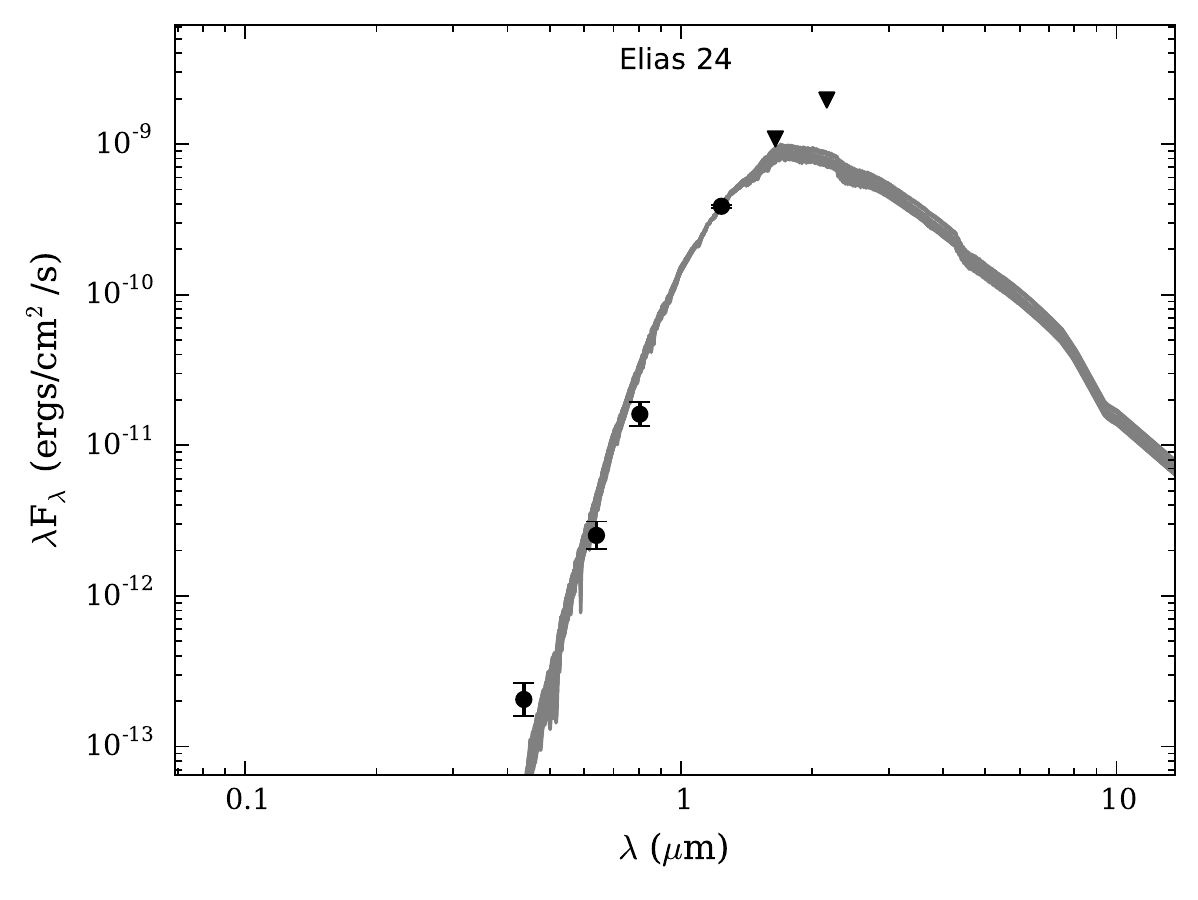}
			\includegraphics[width = 0.33\linewidth]{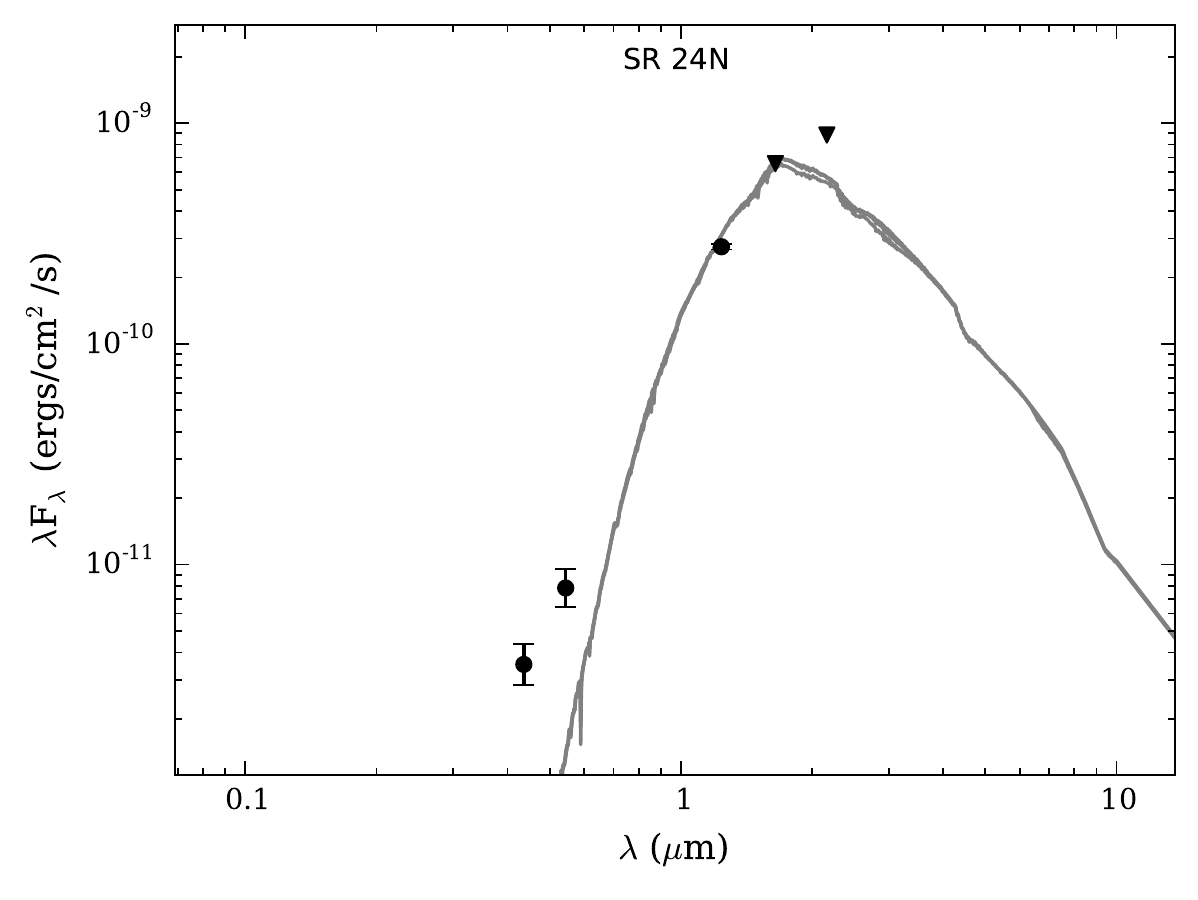}
			\includegraphics[width = 0.33\linewidth]{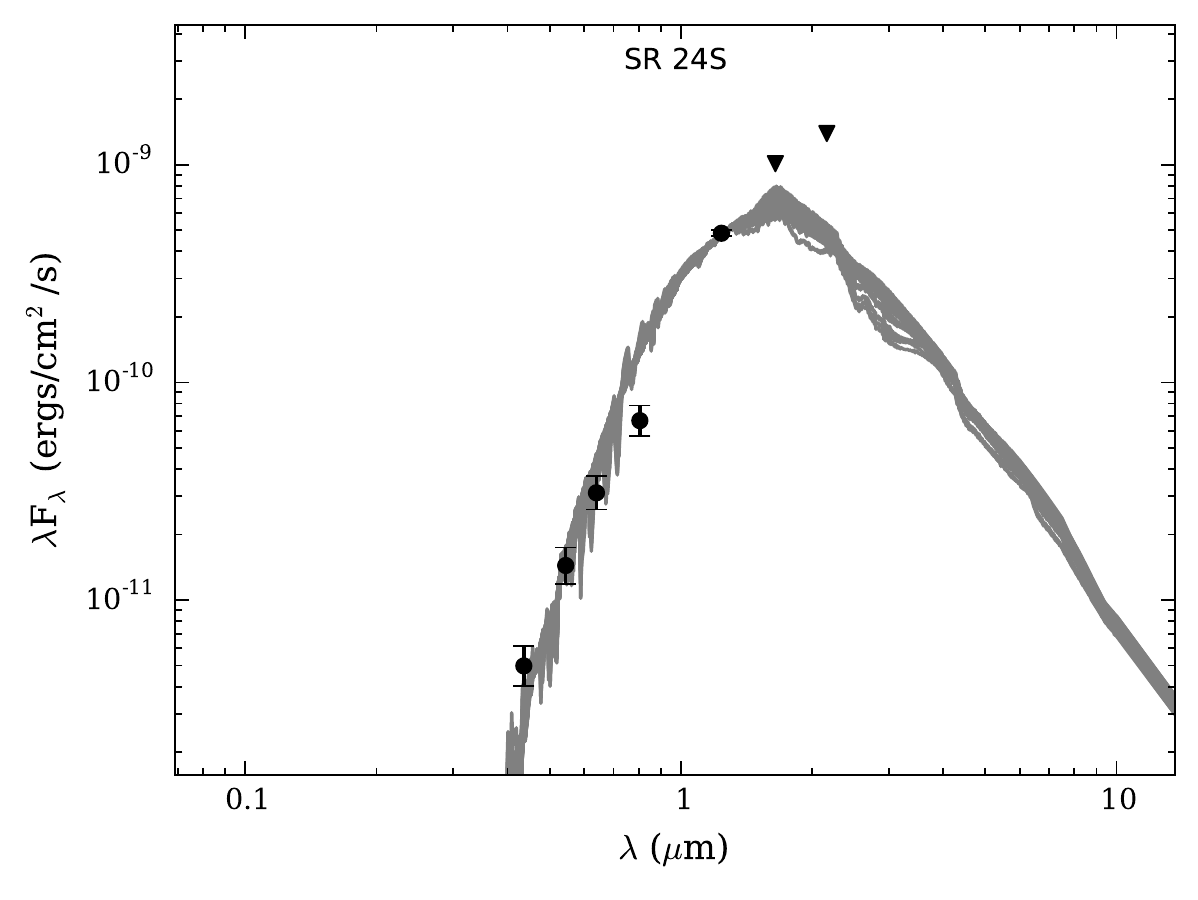}
			\includegraphics[width = 0.33\linewidth]{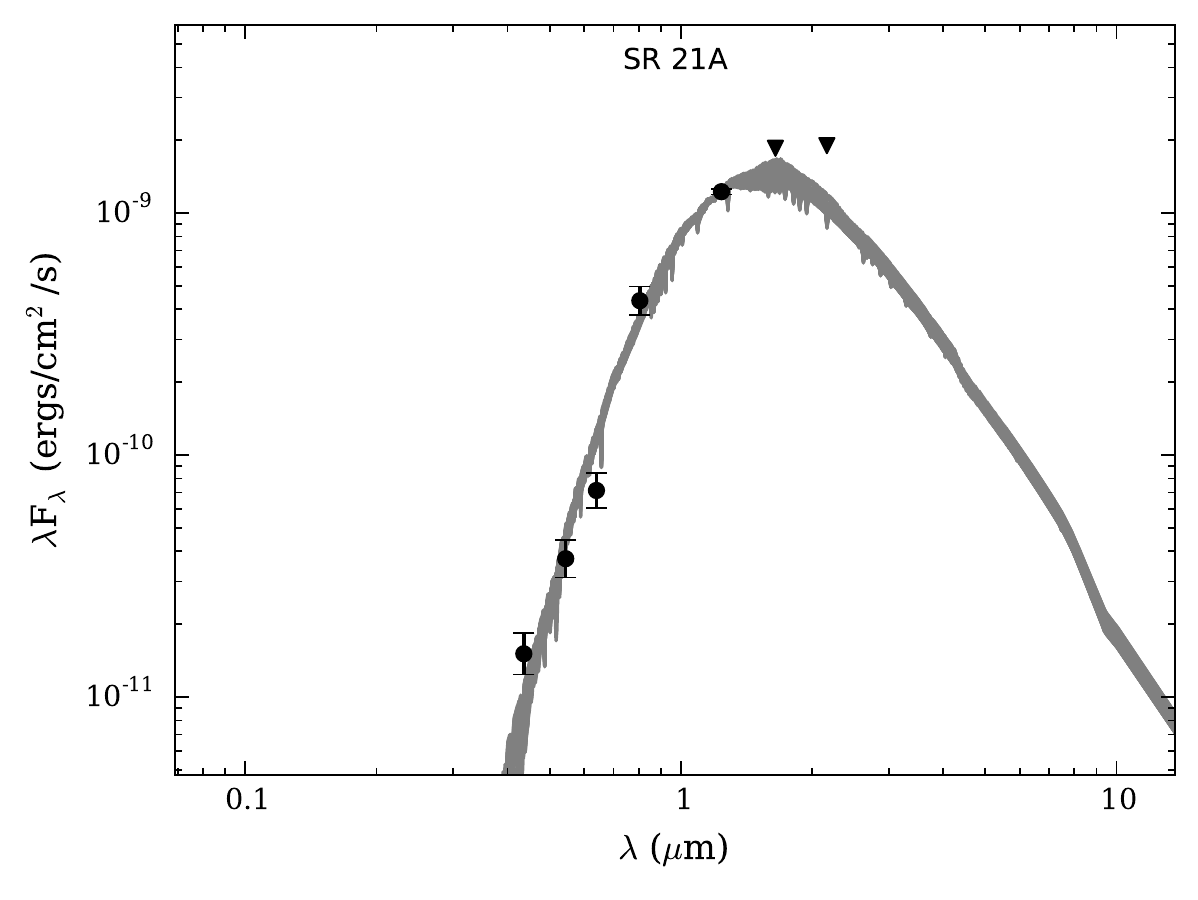}
			\includegraphics[width = 0.33\linewidth]{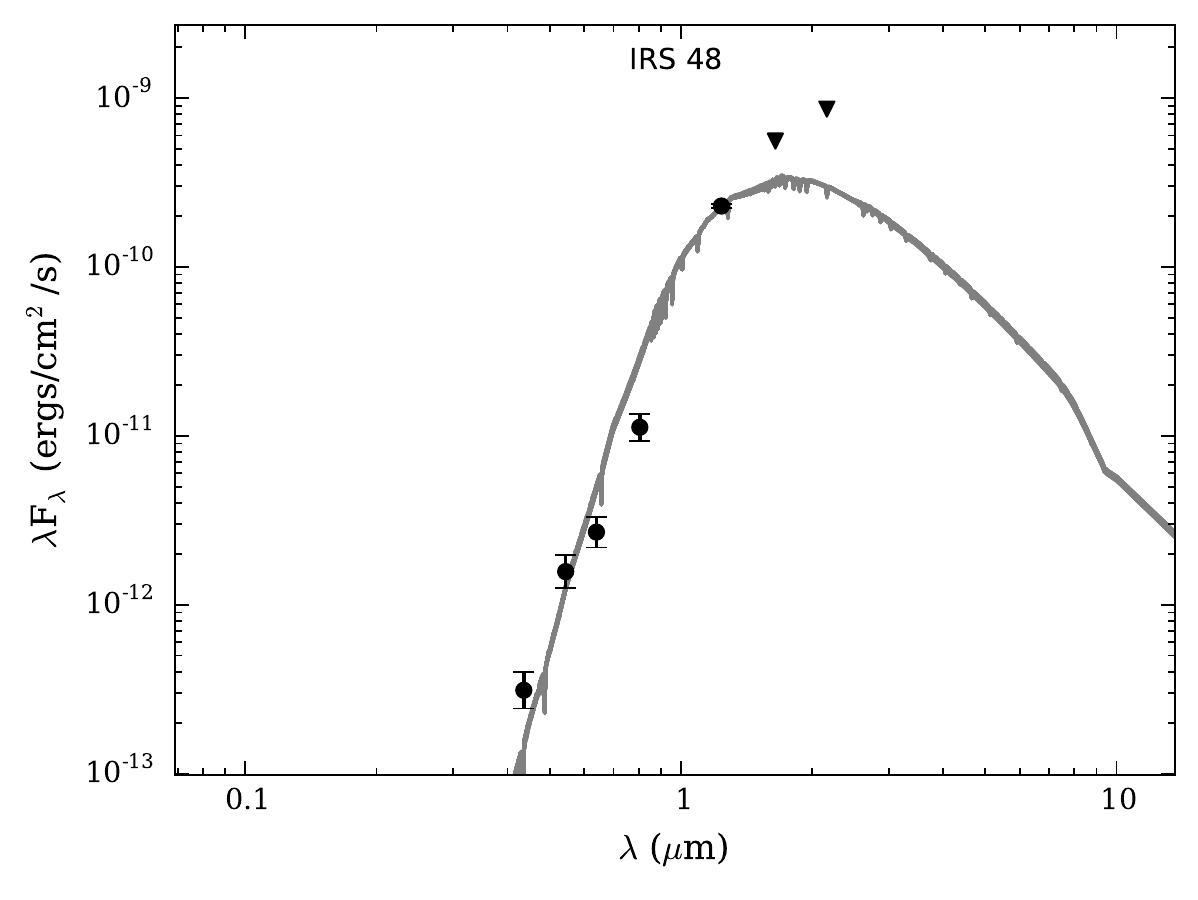}
			\includegraphics[width = 0.33\linewidth]{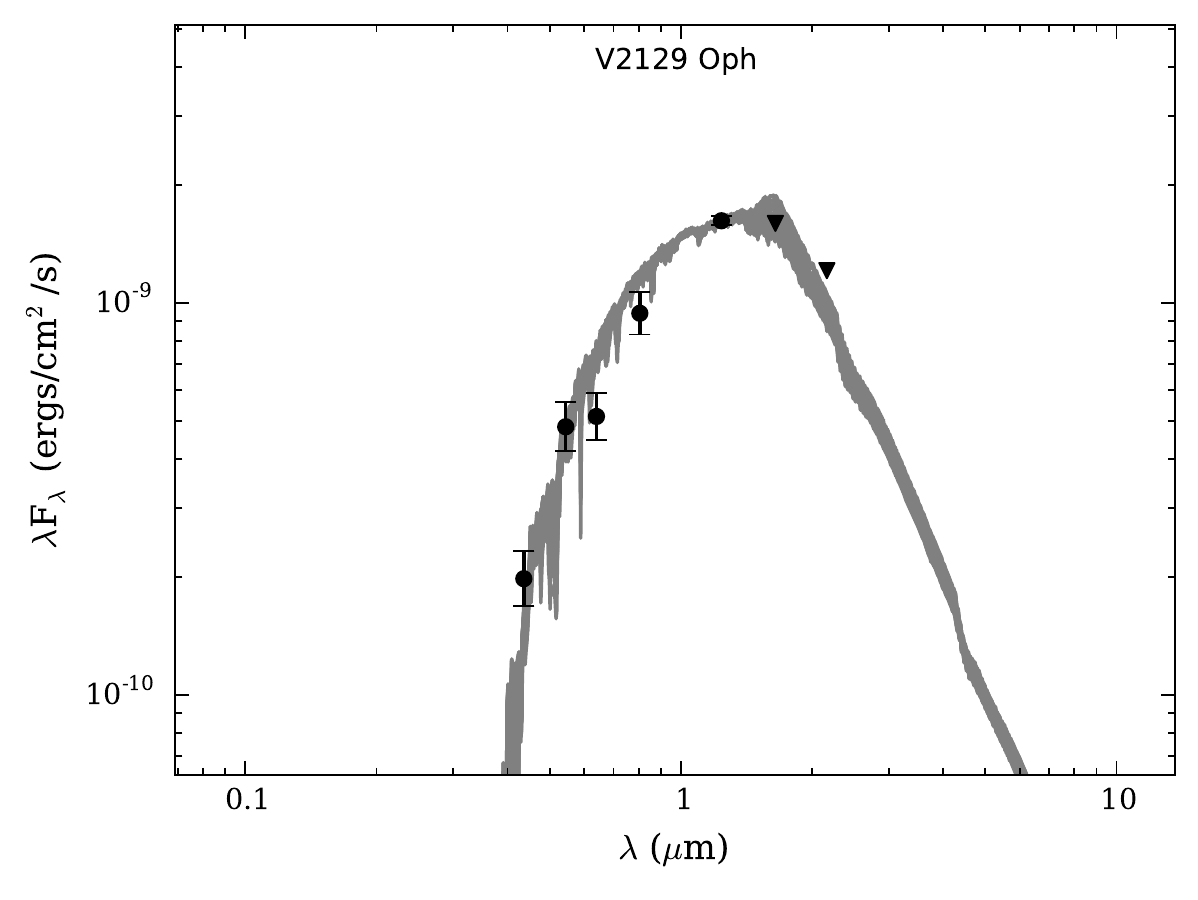}
			\includegraphics[width = 0.33\linewidth]{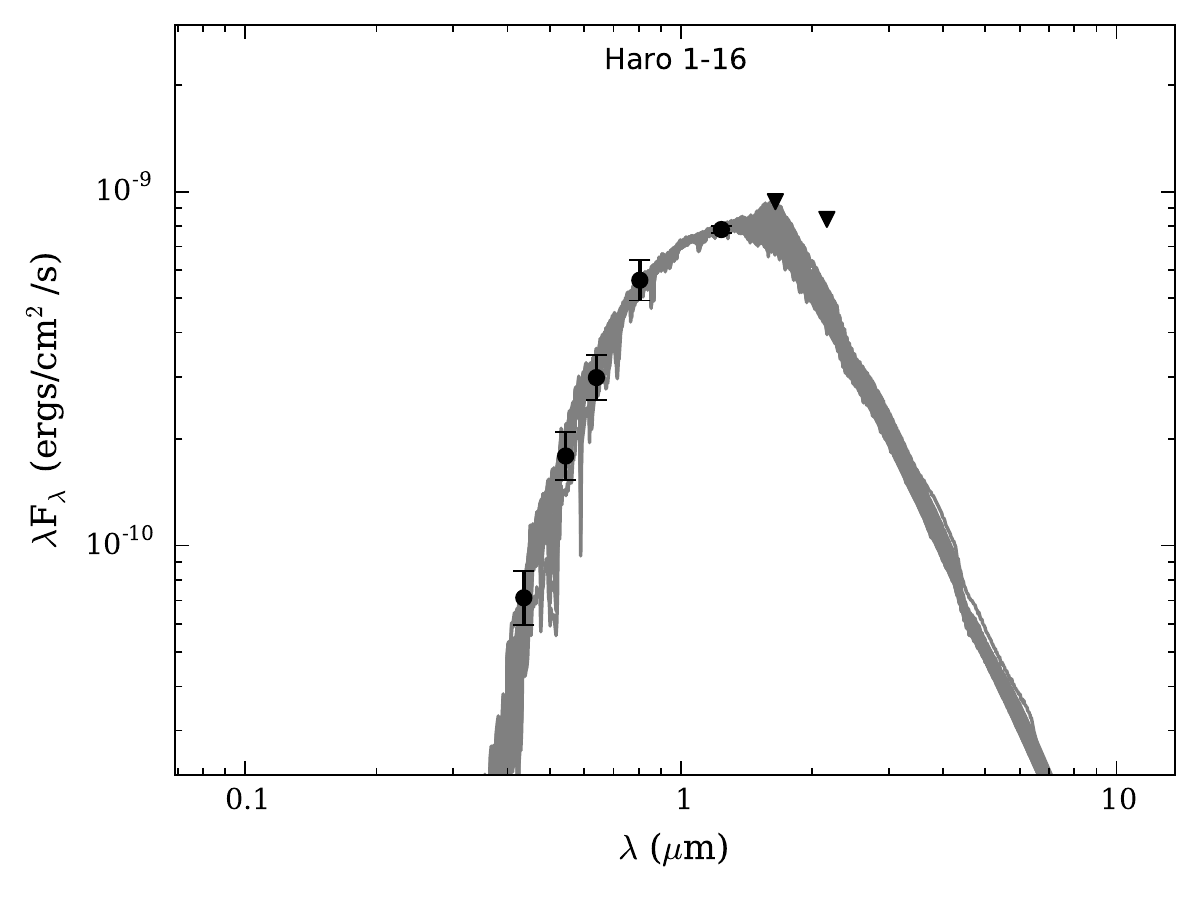}
			\includegraphics[width = 0.33\linewidth]{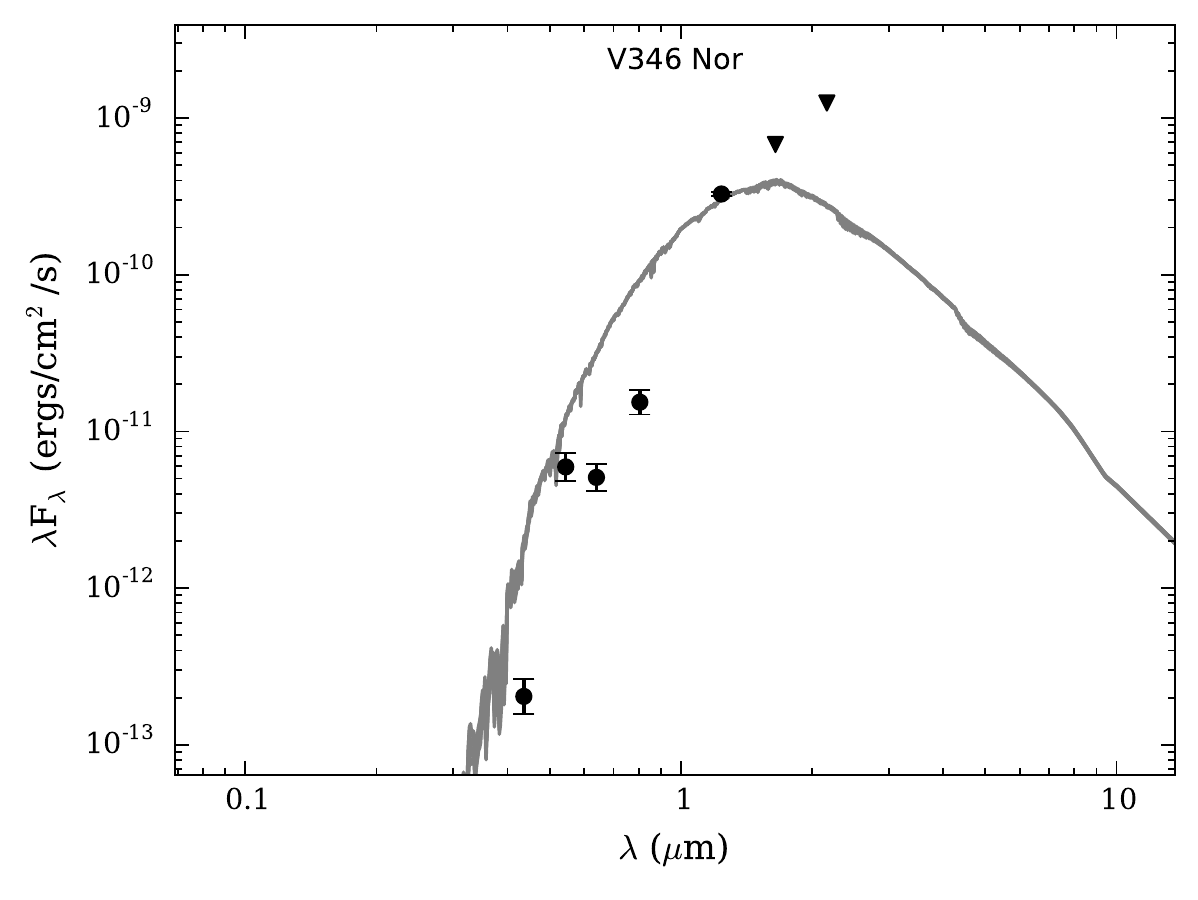}
			\includegraphics[width = 0.33\linewidth]{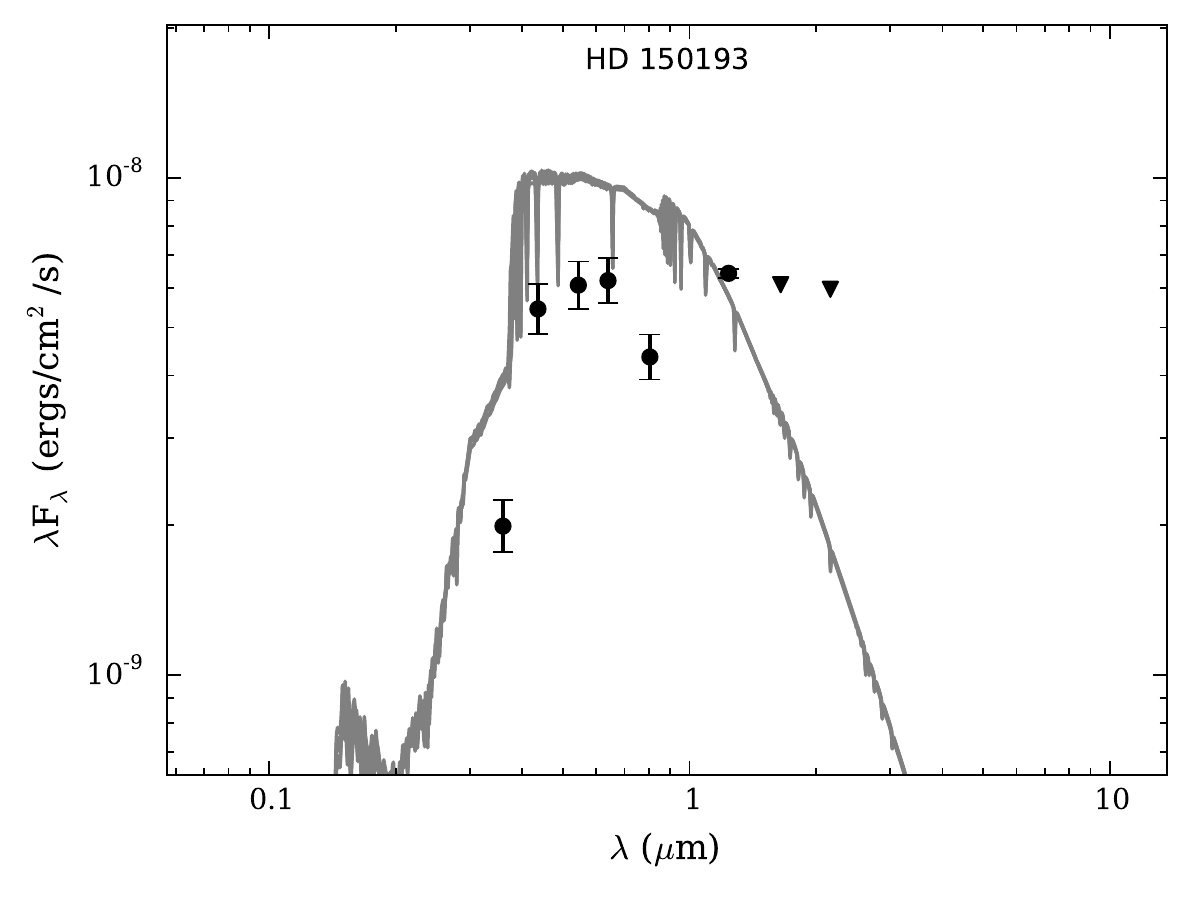}
			\includegraphics[width = 0.33\linewidth]{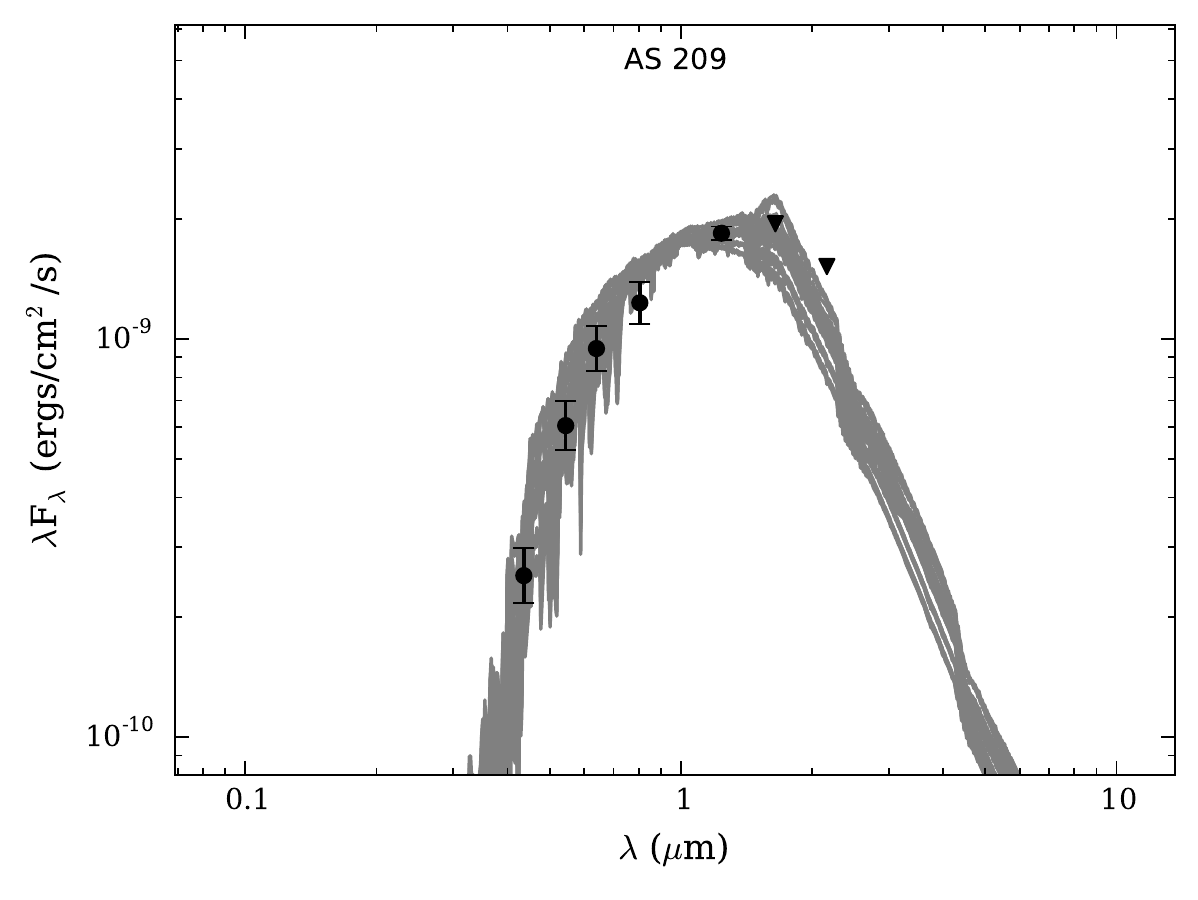}
			\includegraphics[width = 0.33\linewidth]{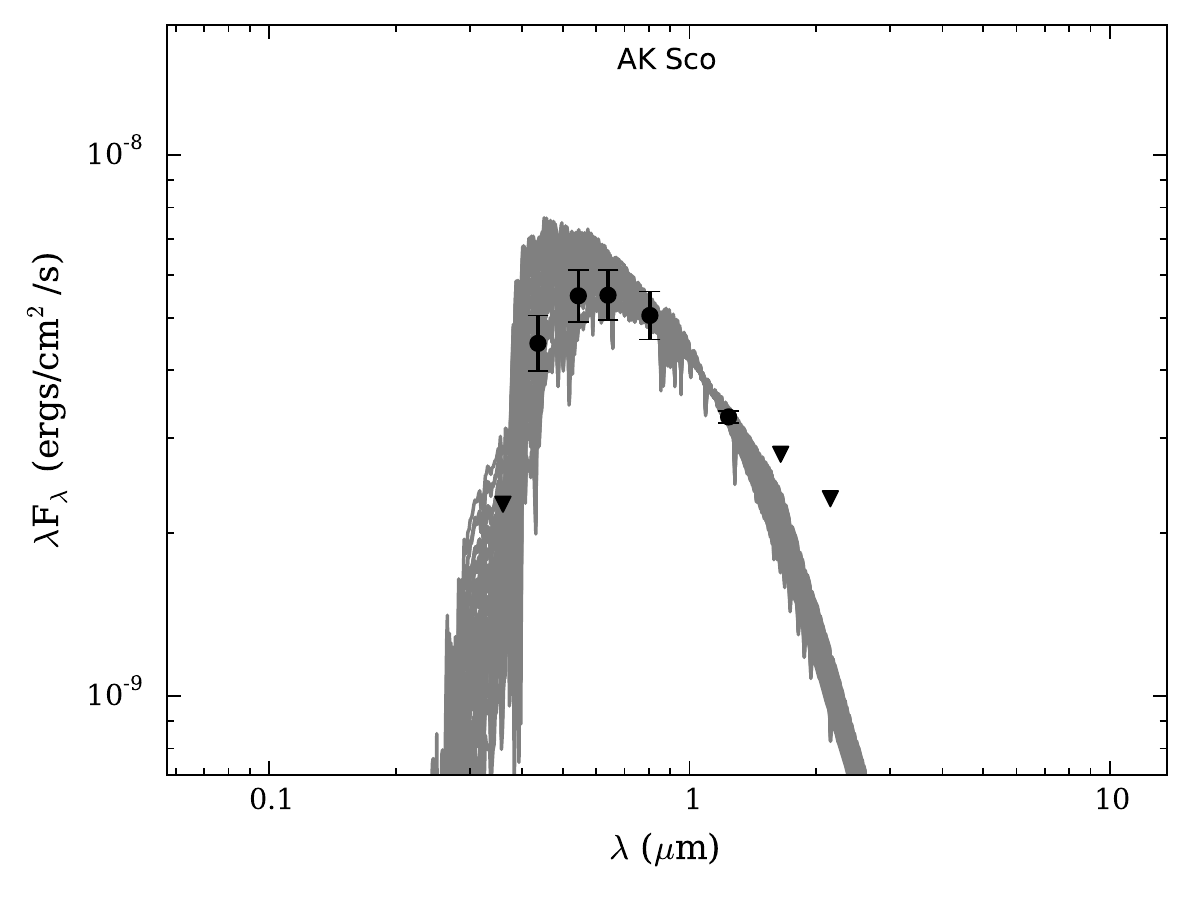}
			\includegraphics[width = 0.33\linewidth]{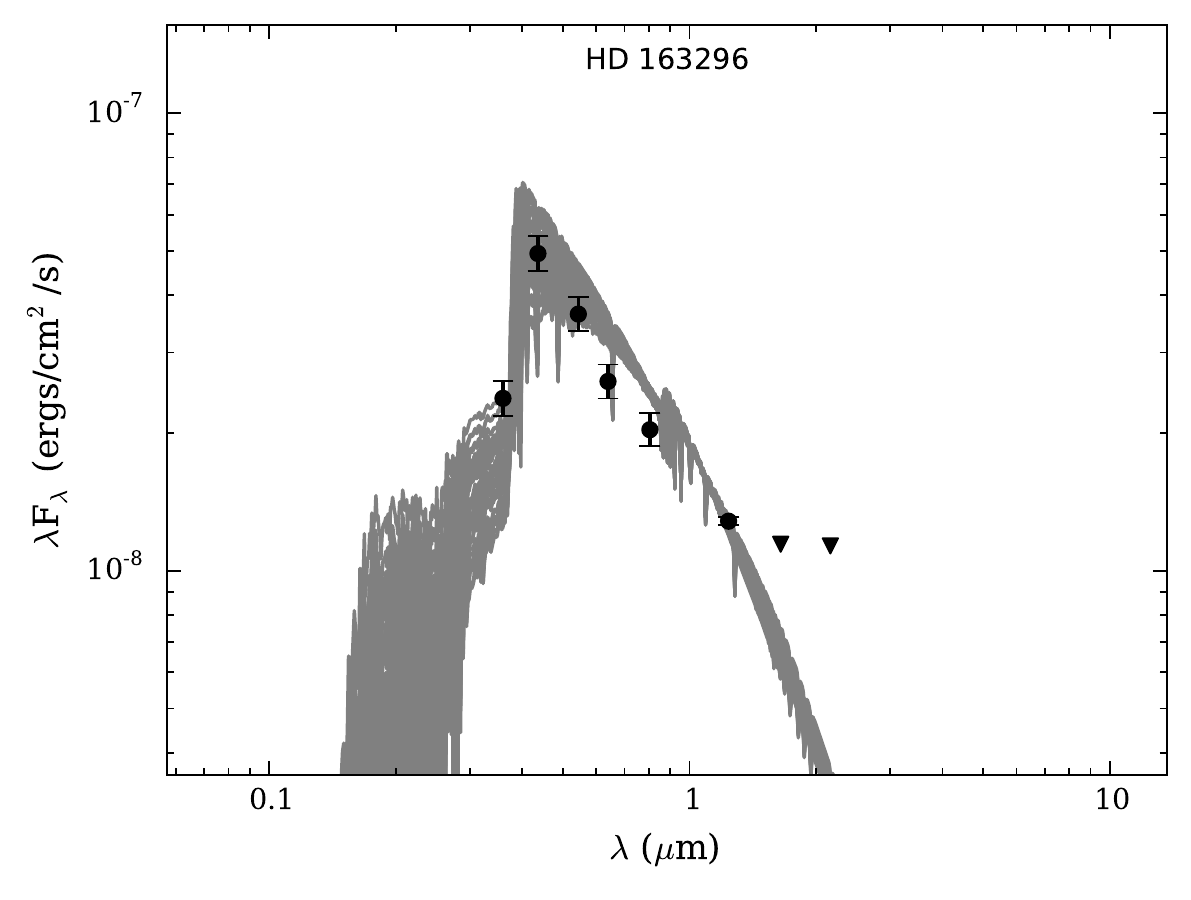}
			\includegraphics[width = 0.33\linewidth]{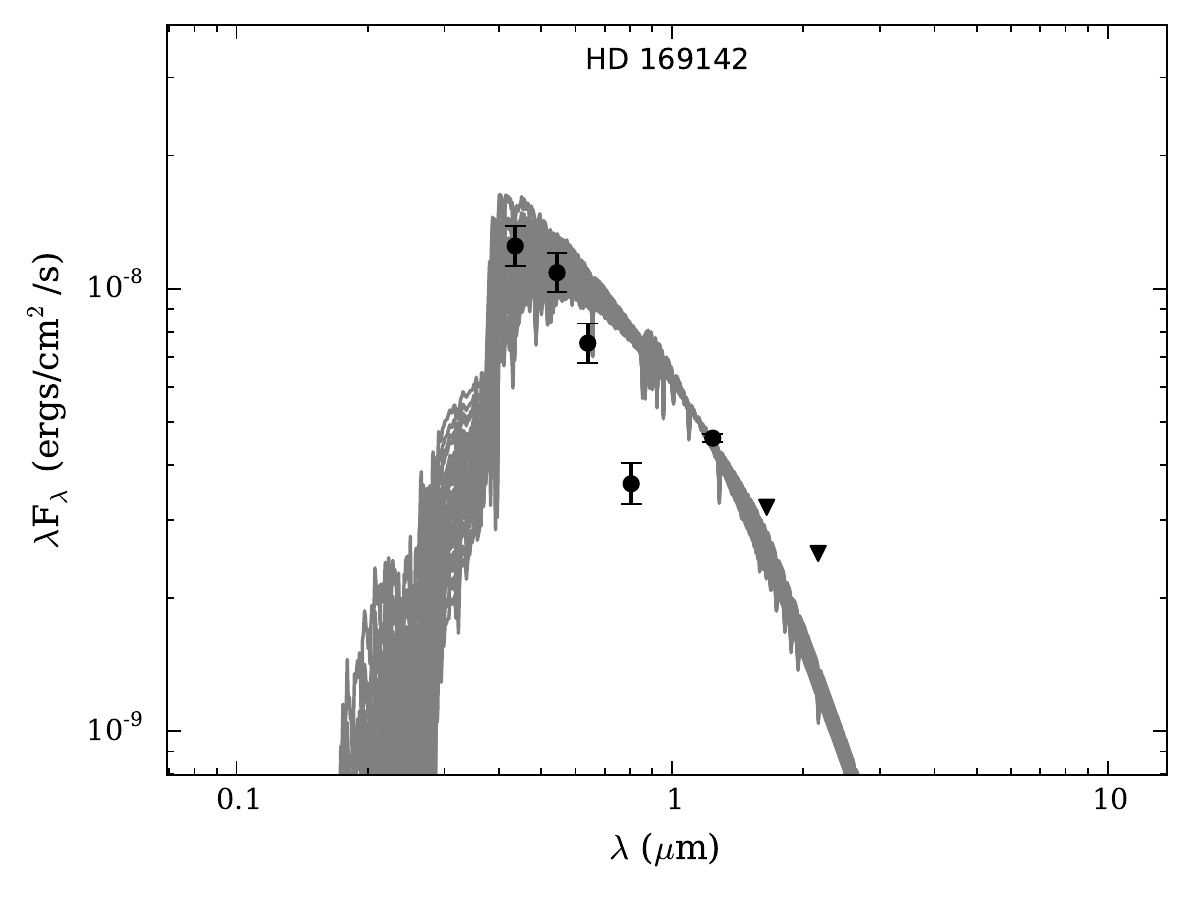}
			\includegraphics[width = 0.33\linewidth]{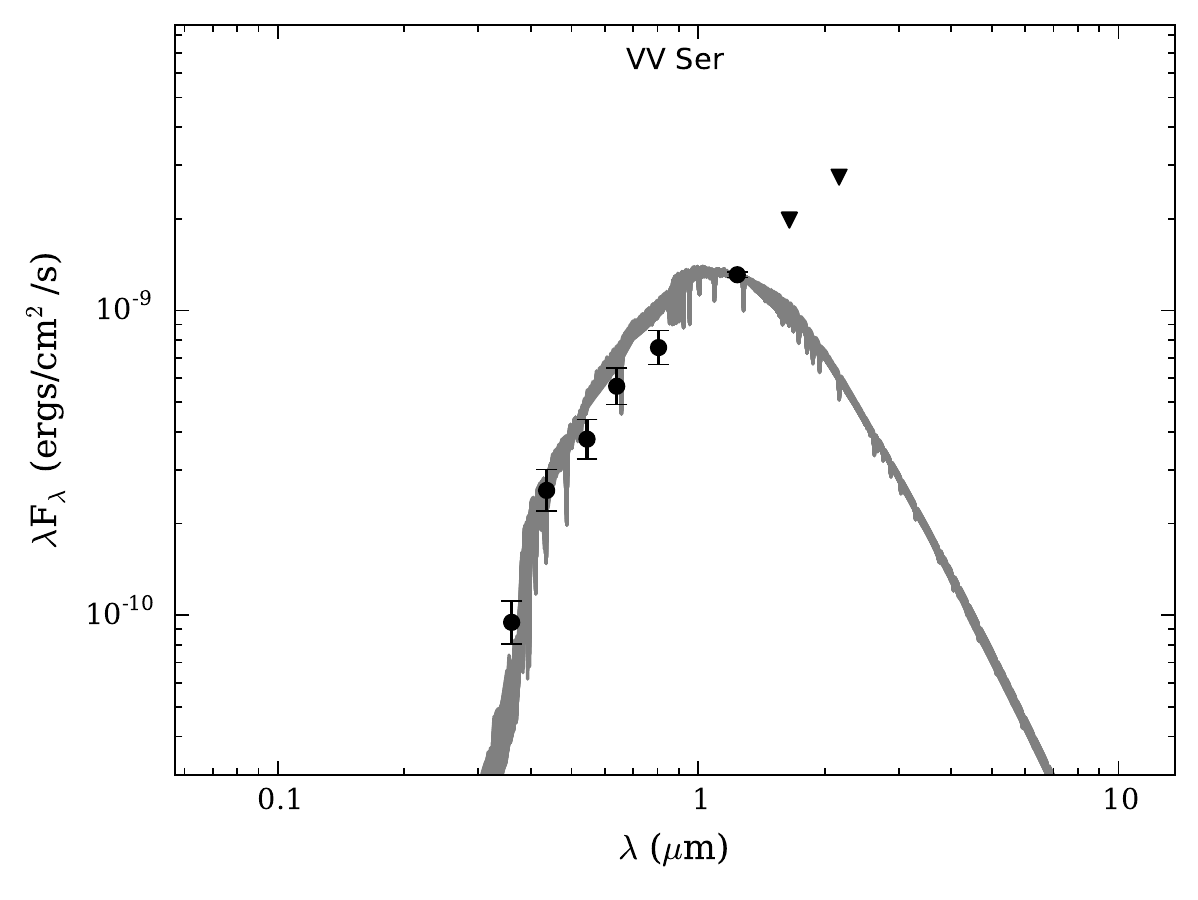}
			\includegraphics[width = 0.33\linewidth]{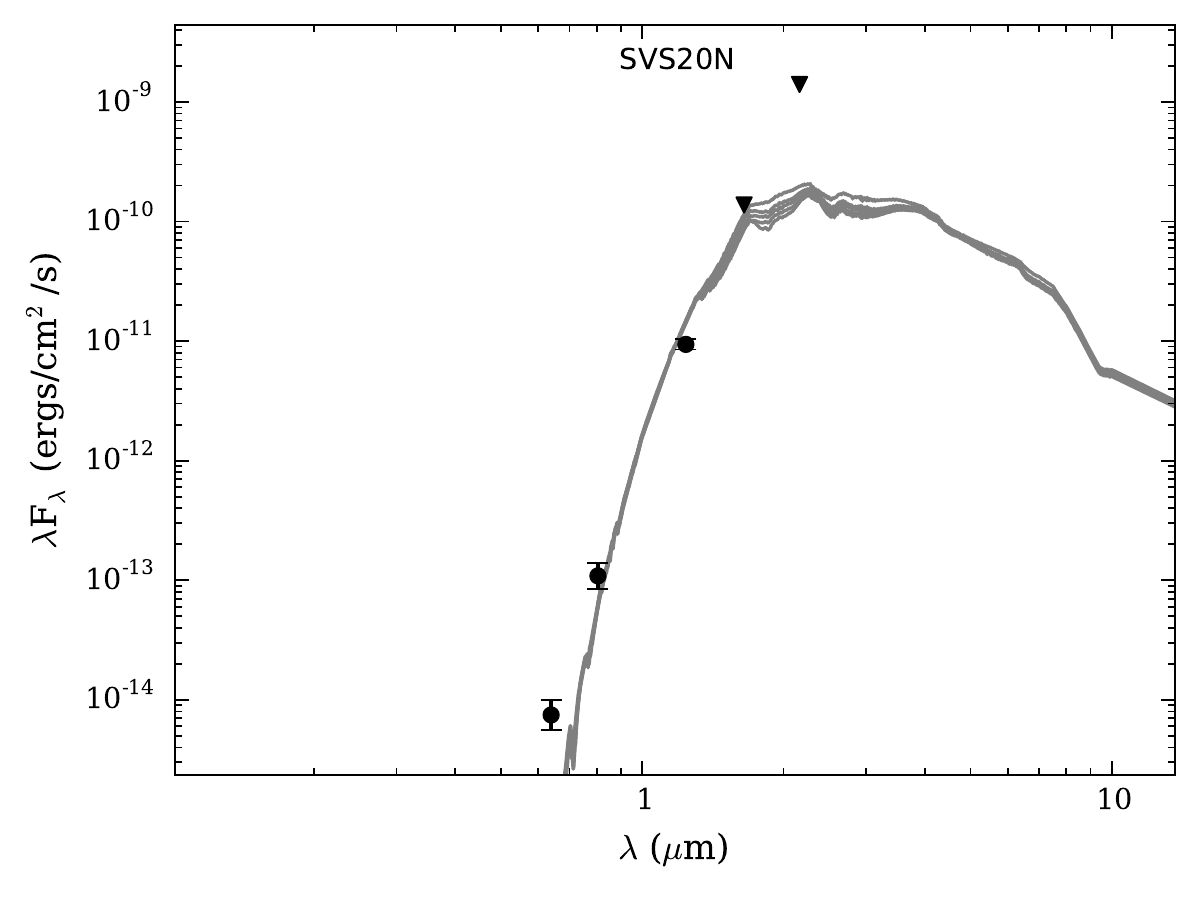}
			
		\end{figure*}
		
		\begin{figure*}[h!]
			\centering
			\includegraphics[width = 0.33\linewidth]{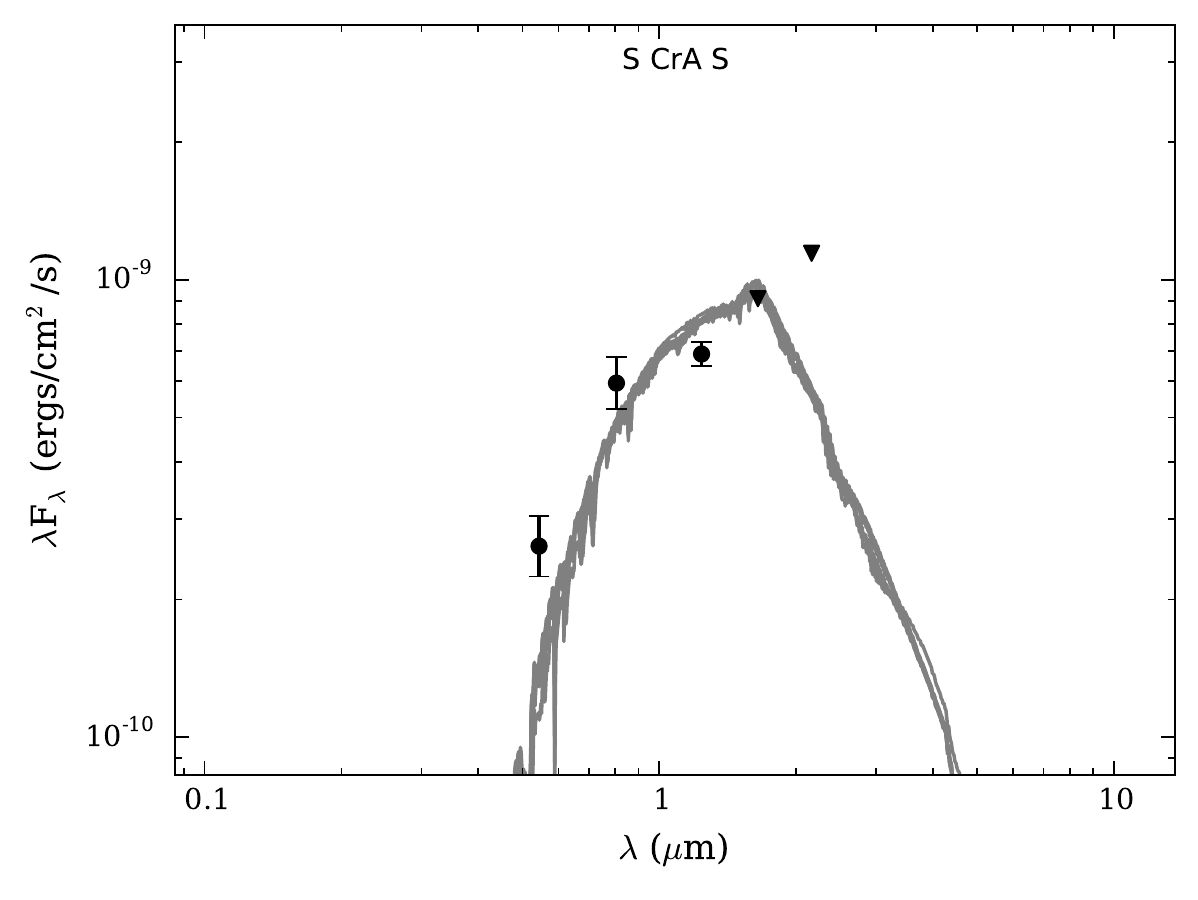}
			\includegraphics[width = 0.33\linewidth]{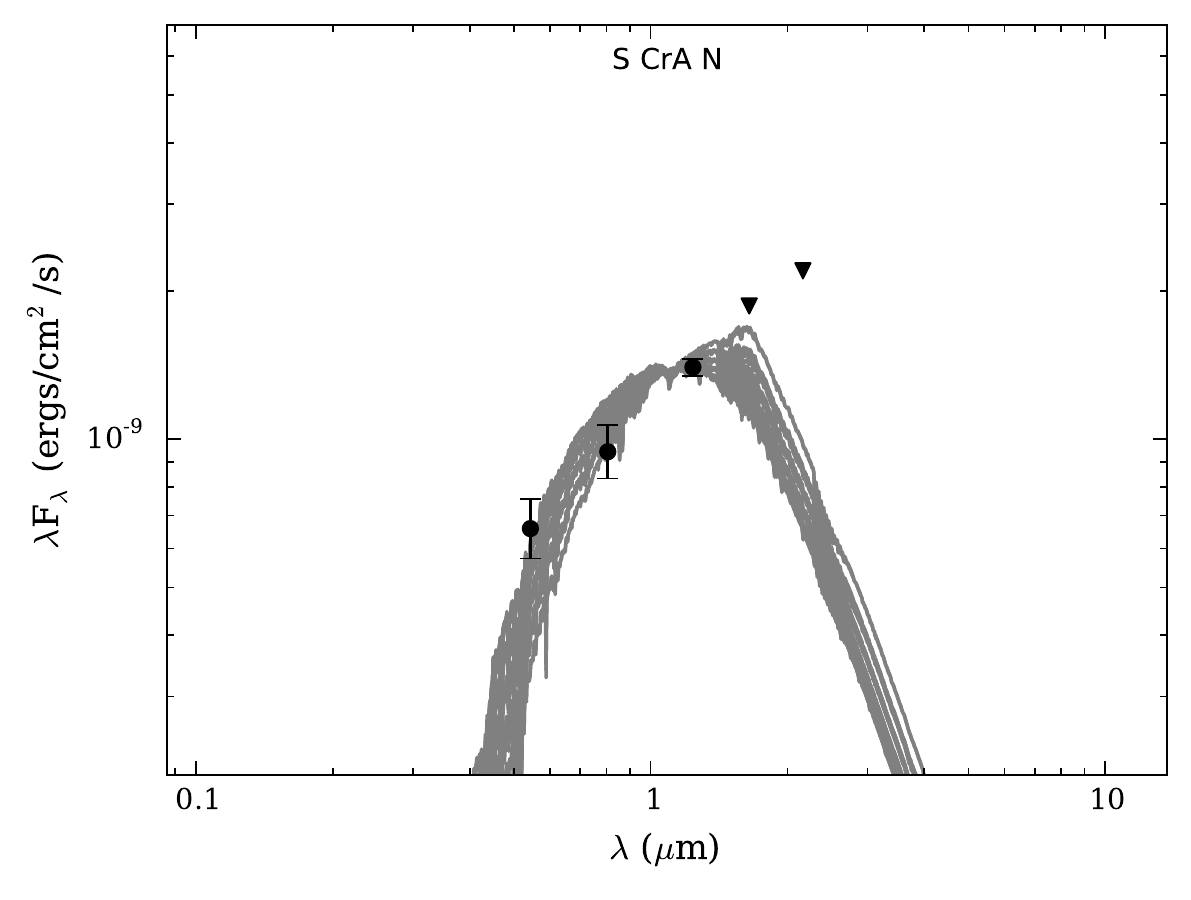}
			\includegraphics[width = 0.33\linewidth]{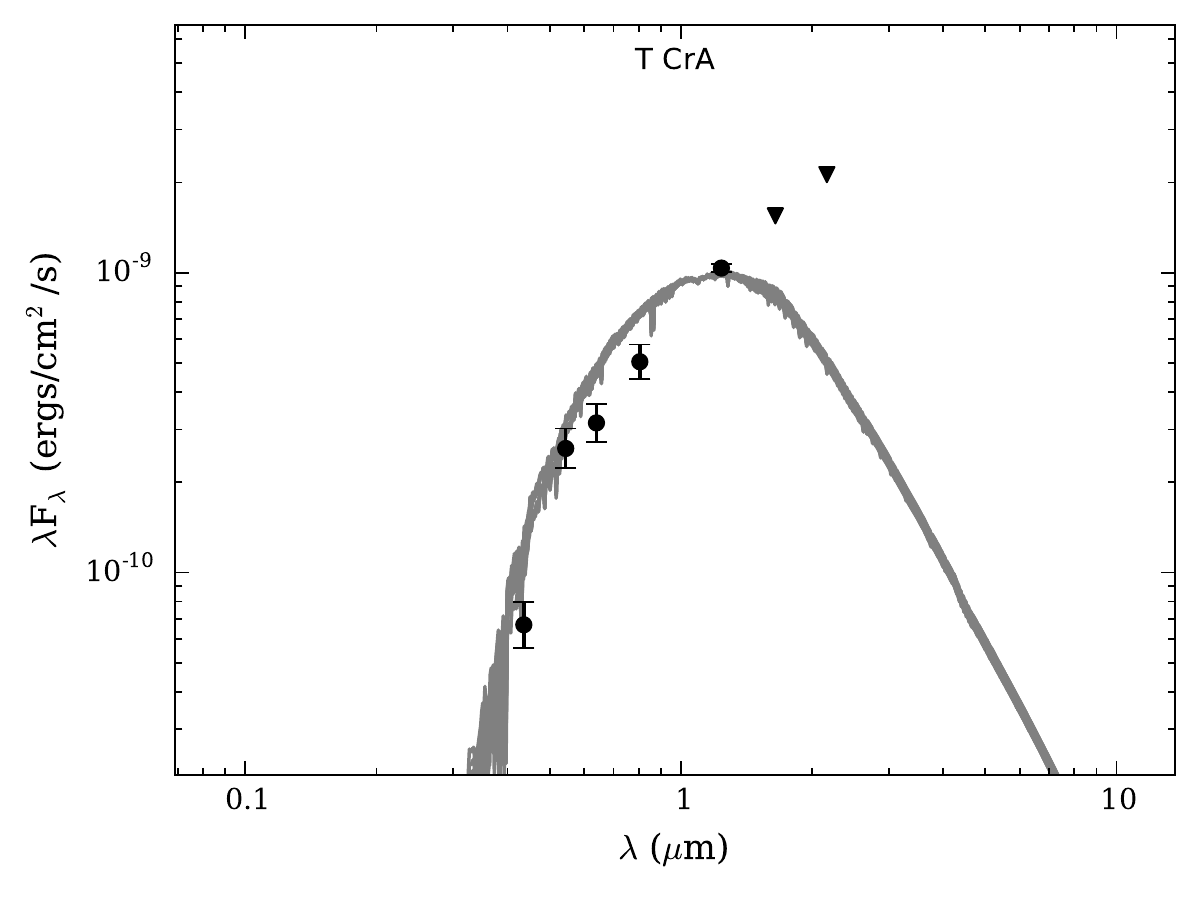}
			\includegraphics[width = 0.33\linewidth]{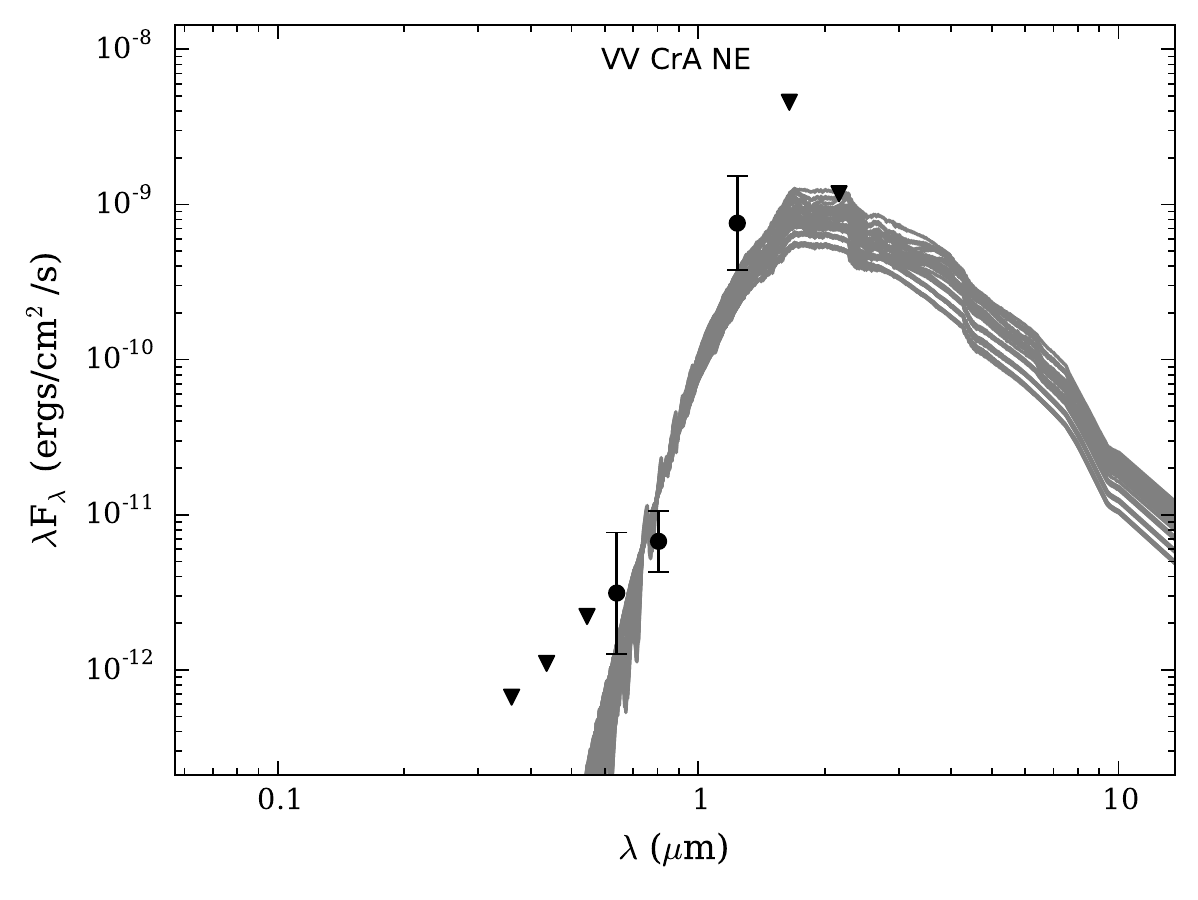}
			\includegraphics[width = 0.33\linewidth]{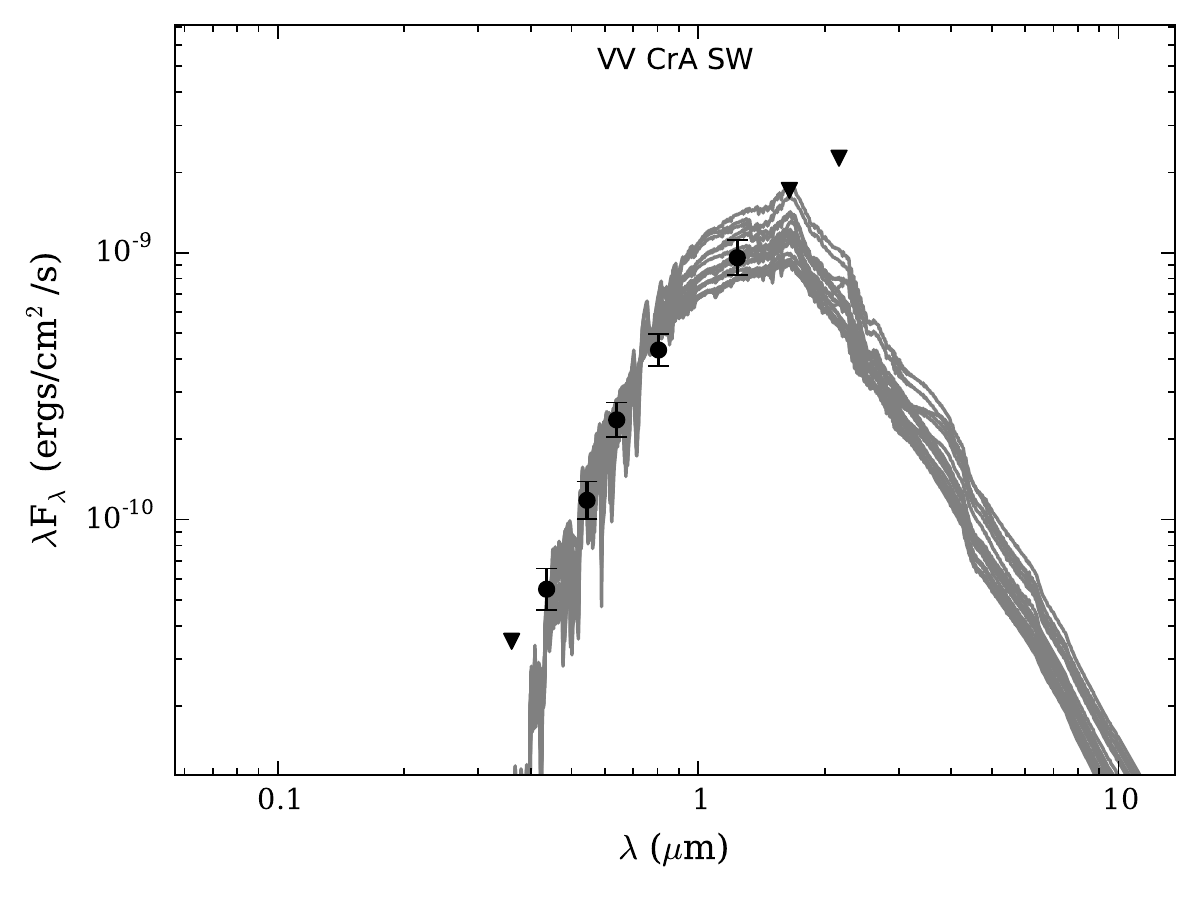}
			\includegraphics[width = 0.33\linewidth]{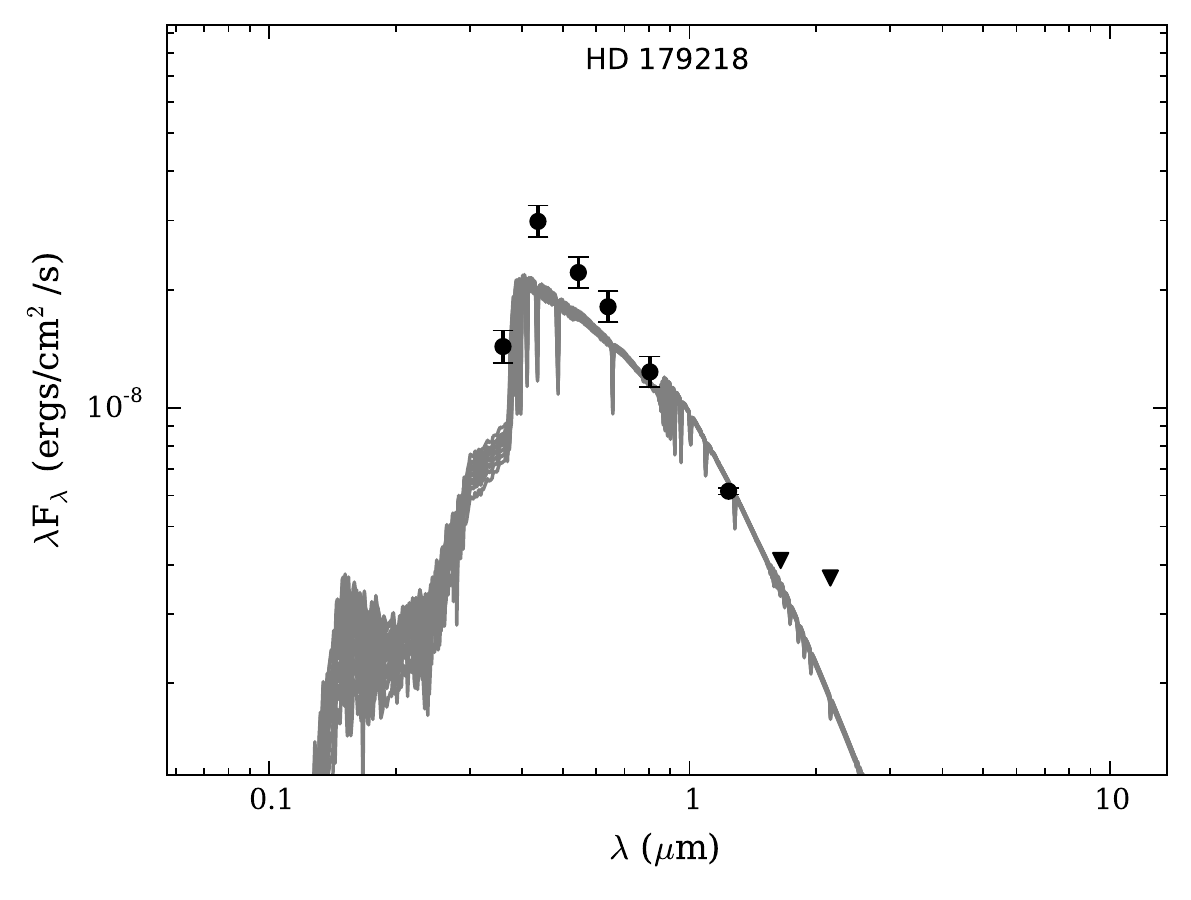}
			
		\end{figure*}

		\clearpage
		\section{Interferometric plots}
		\label{app:atlasfig} 
		
		\begin{figure*}[h!]
			\centering
			\caption{Figure outputs of the atlas. Data taken with the UTs and ATs are marked with circles and diamonds, respectively. Top: $uv$-coverage plot. Data shown with small symbols are flagged as unreliable. Middle: Gaussian size diagram. The plot shows the physical sizes of the disk (HWHM, in au, see Eq.~\ref{eq:gaussian}) individually measured for each observation corresponding to different baseline angles. Bottom: Results of the interferometric modeling to the correlated and total fluxes (shown at $0$~m baseline) at $10.7~\mu$m as a function of baseline length. Blue solid line: continuous model fit. Blue dashed line: gapped model fit. Red and green lines show the correlated fluxes extrapolated from the continuous model to the K and L bands, respectively. Dotted lines show the estimates for the unresolved stellar emission. Data points on each subplot are color-coded for observation date.}
			\label{fig:atlas_fit_app}
			\includegraphics[width = 0.21\linewidth]{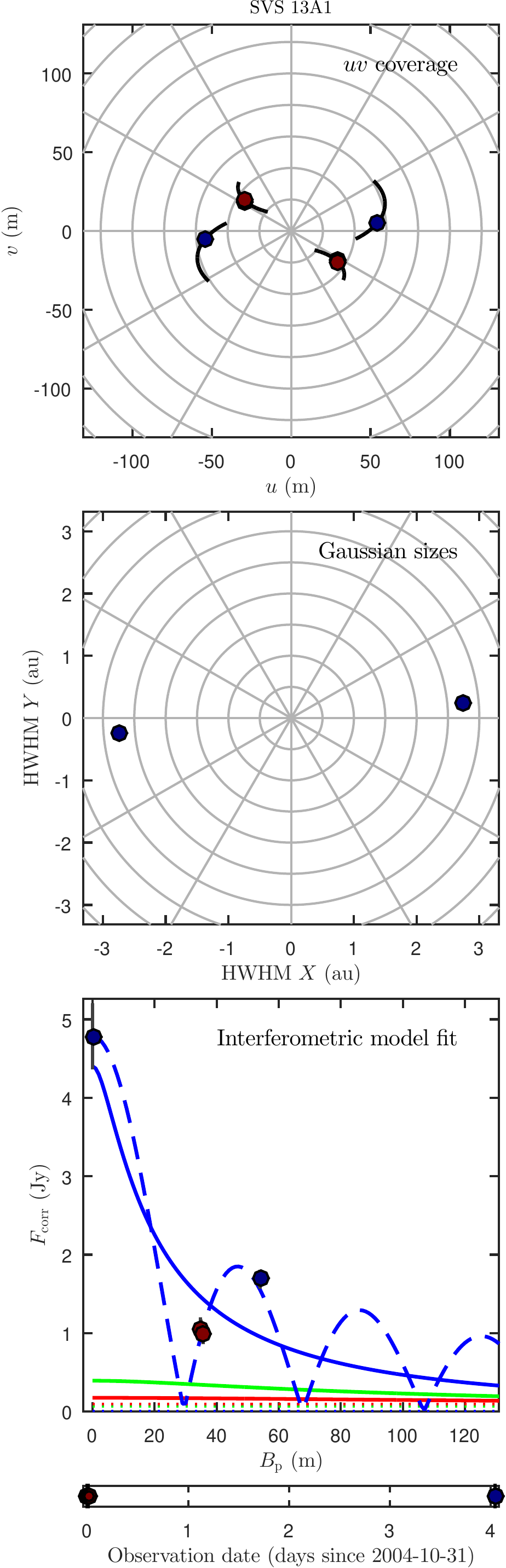}
			\includegraphics[width = 0.21\linewidth]{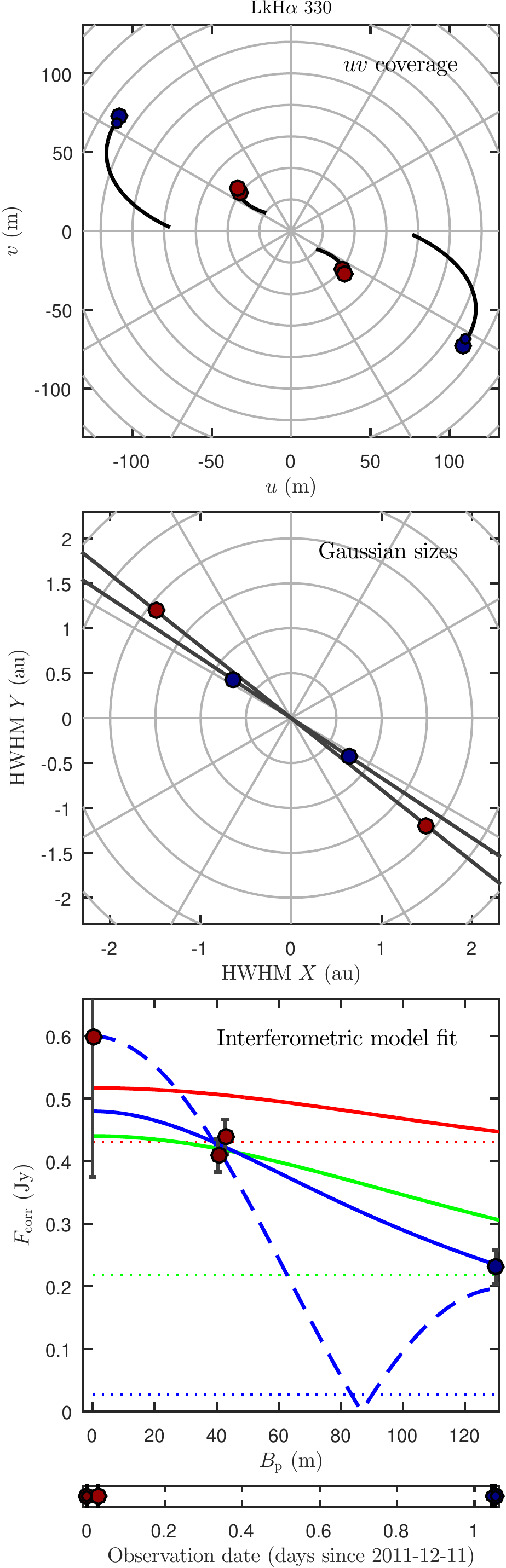}
			\includegraphics[width = 0.21\linewidth]{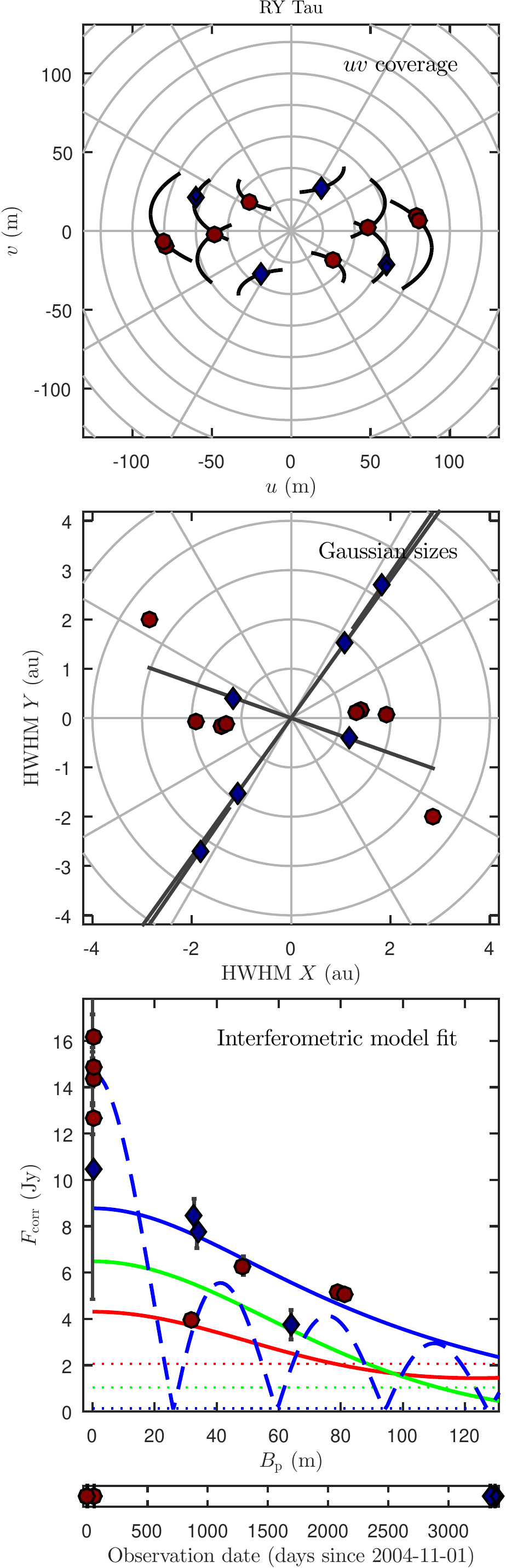}
			\includegraphics[width = 0.21\linewidth]{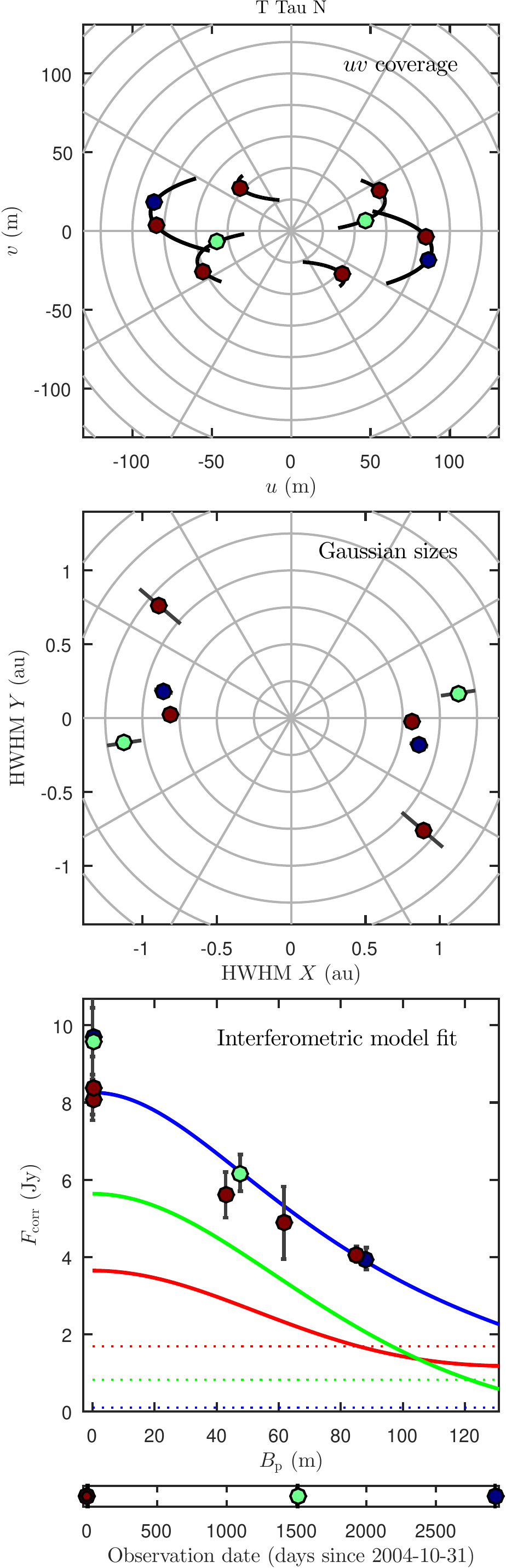}
			
		\end{figure*}
		
		\begin{figure*}[h!]
			\centering
			\includegraphics[width = 0.21\linewidth]{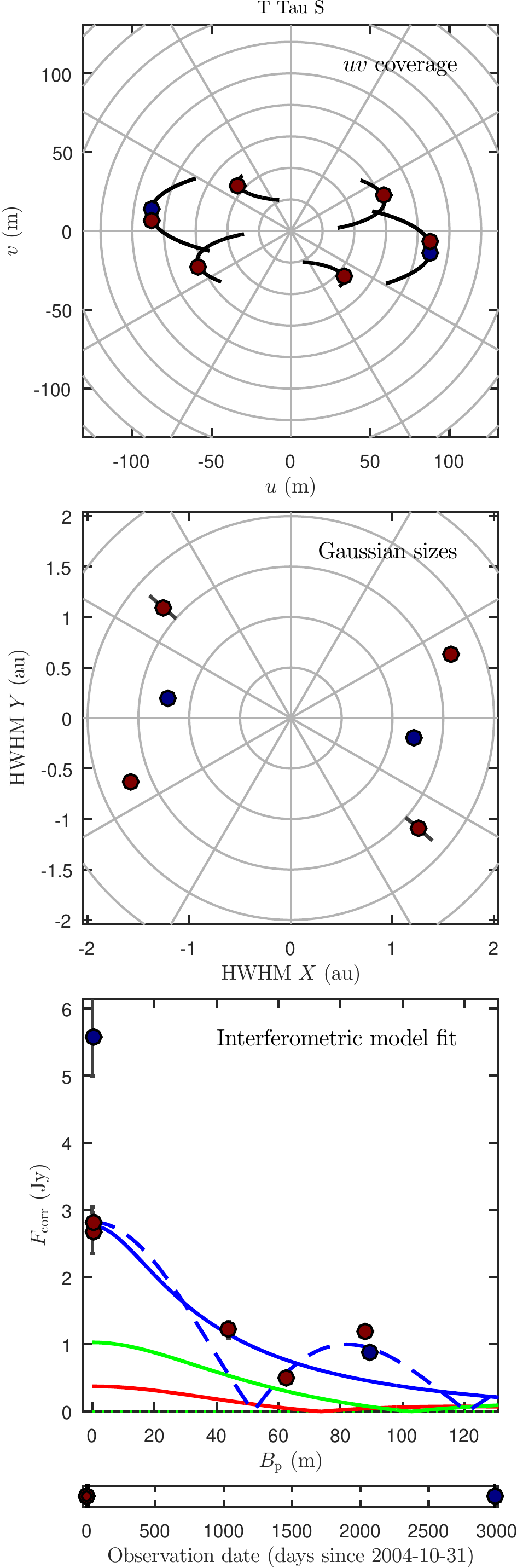}
			\includegraphics[width = 0.21\linewidth]{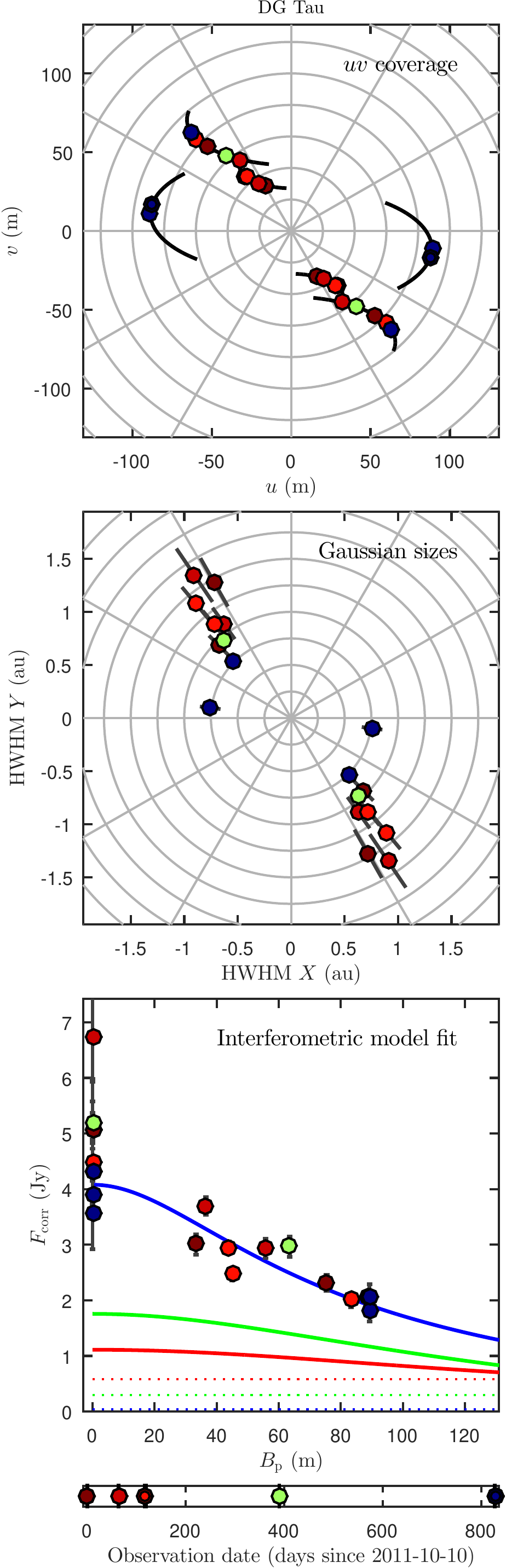}
			\includegraphics[width = 0.21\linewidth]{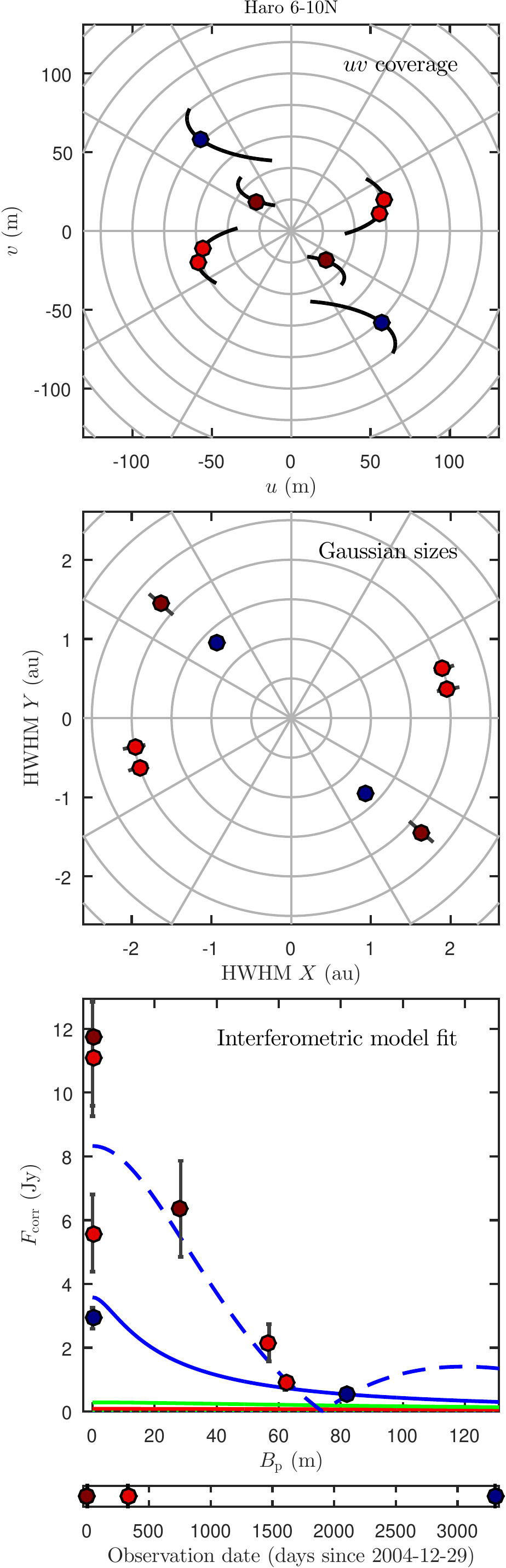}
			\includegraphics[width = 0.21\linewidth]{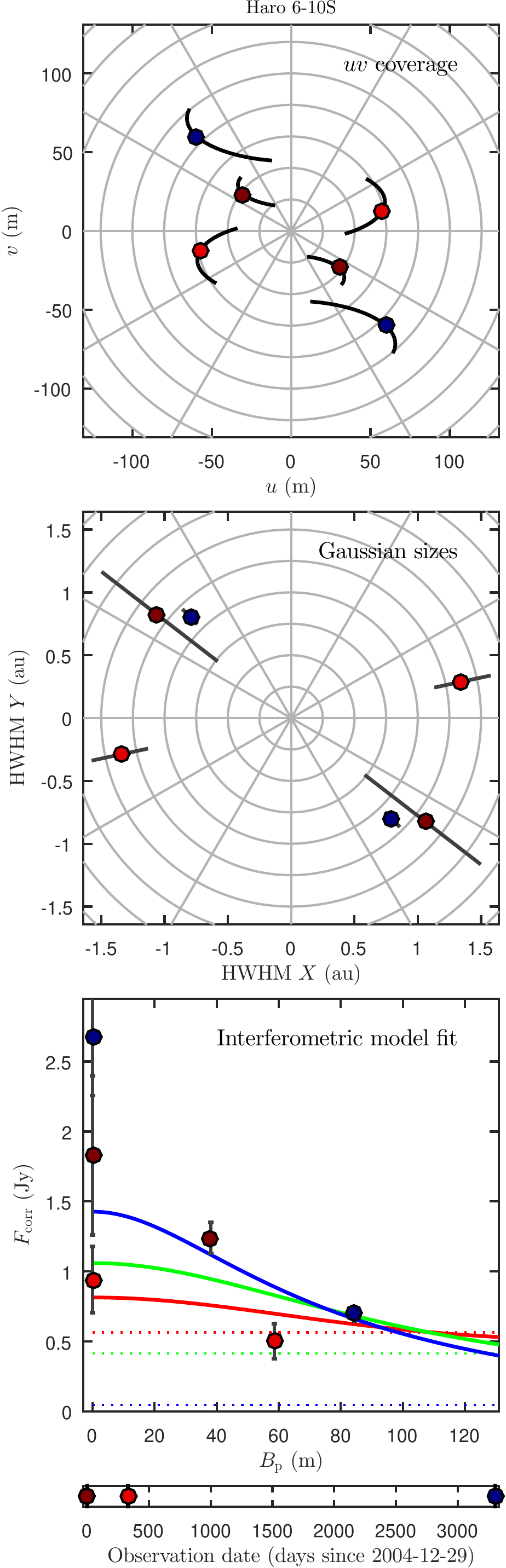}
			\includegraphics[width = 0.21\linewidth]{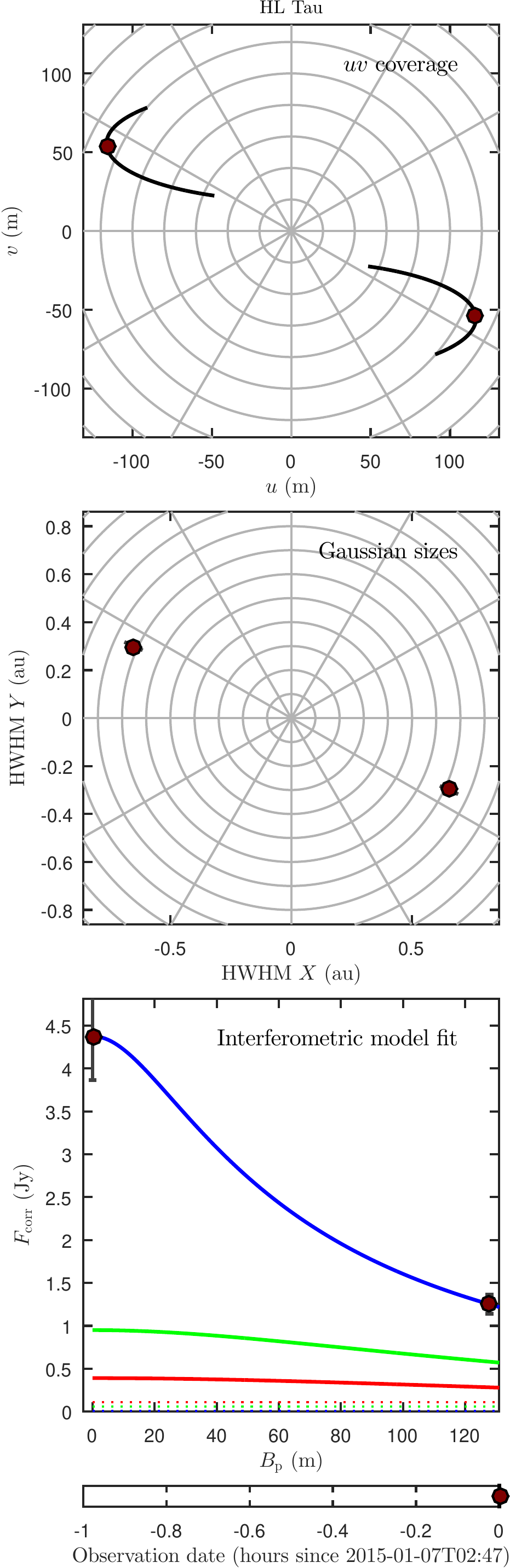}
			\includegraphics[width = 0.21\linewidth]{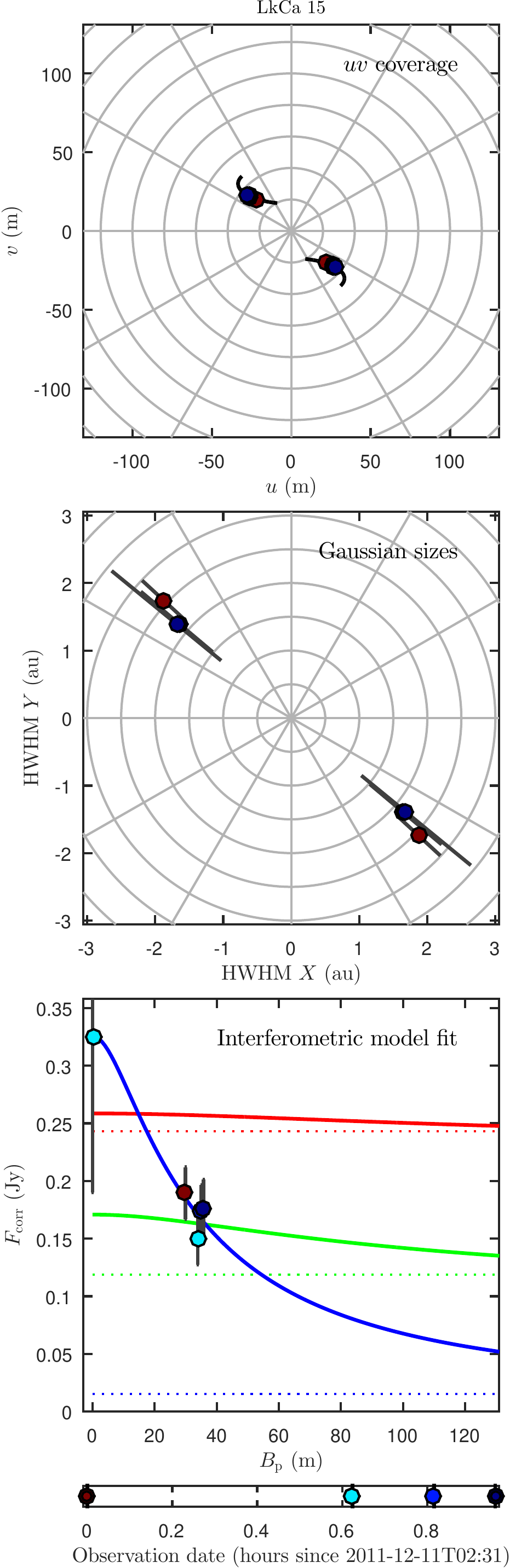}
			\includegraphics[width = 0.21\linewidth]{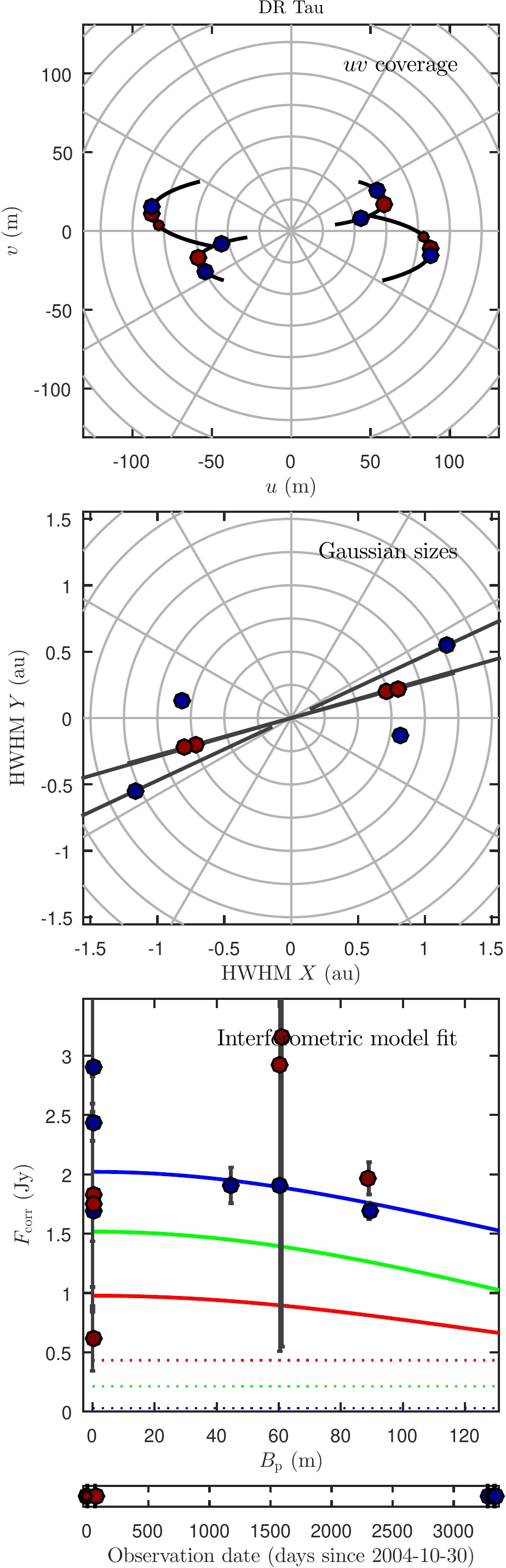}
			\includegraphics[width = 0.21\linewidth]{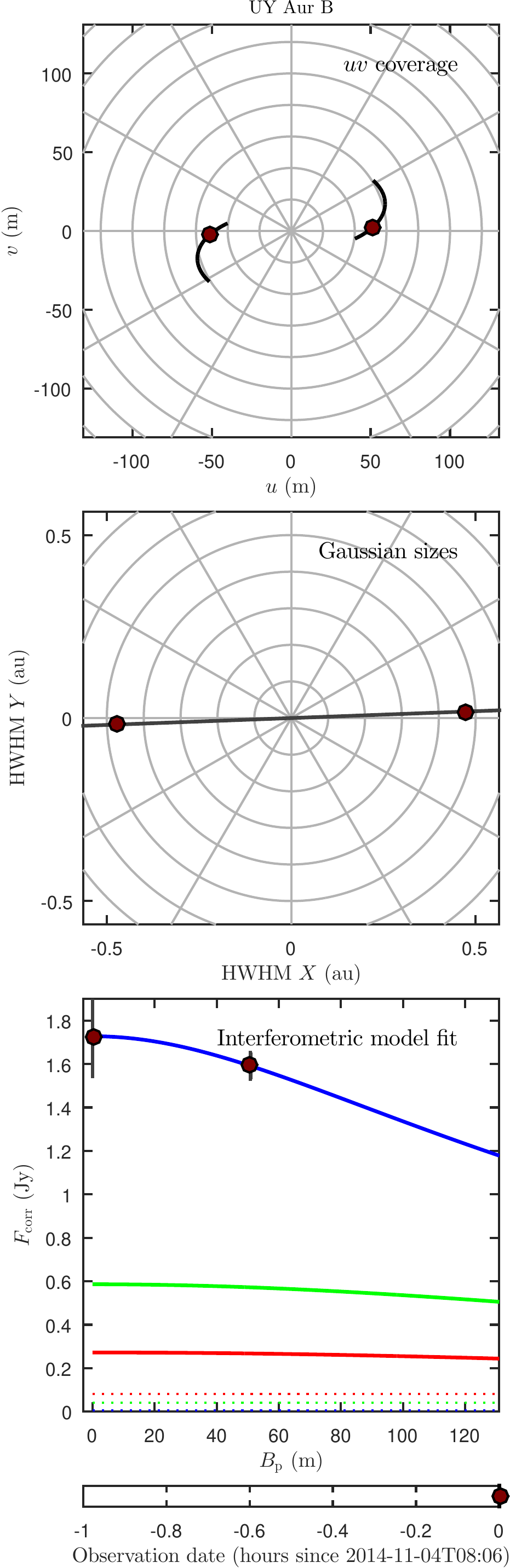}
		\end{figure*}
		
		\begin{figure*}[h!]
			\centering
			\includegraphics[width = 0.21\linewidth]{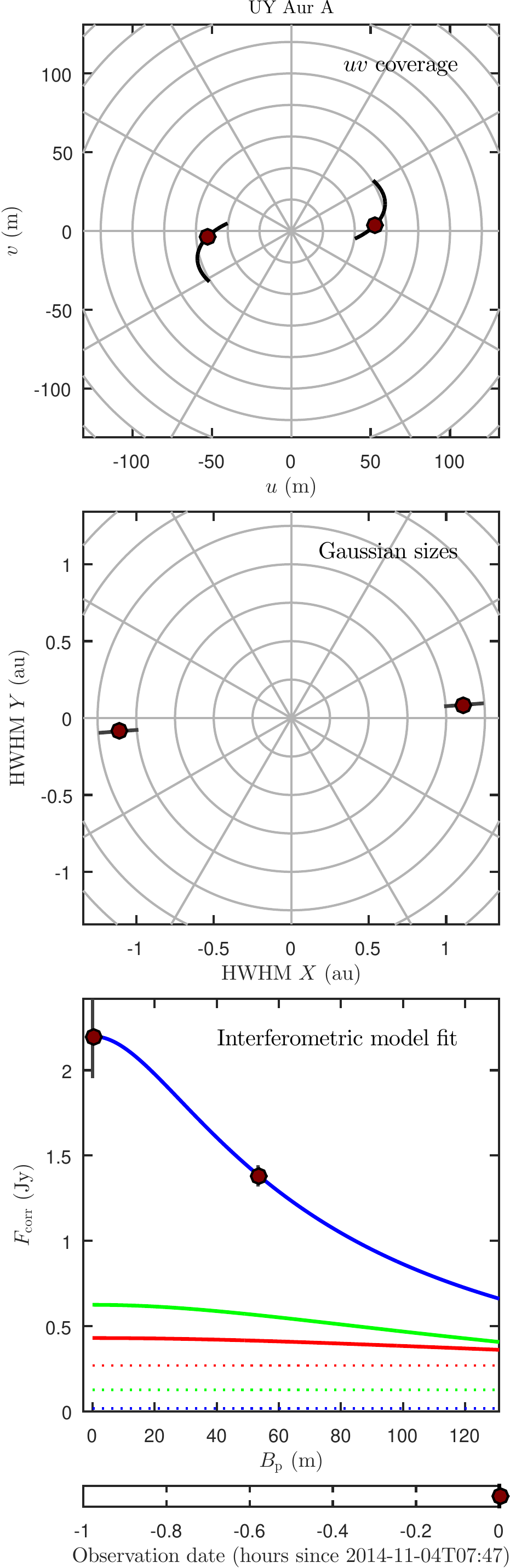}
			\includegraphics[width = 0.21\linewidth]{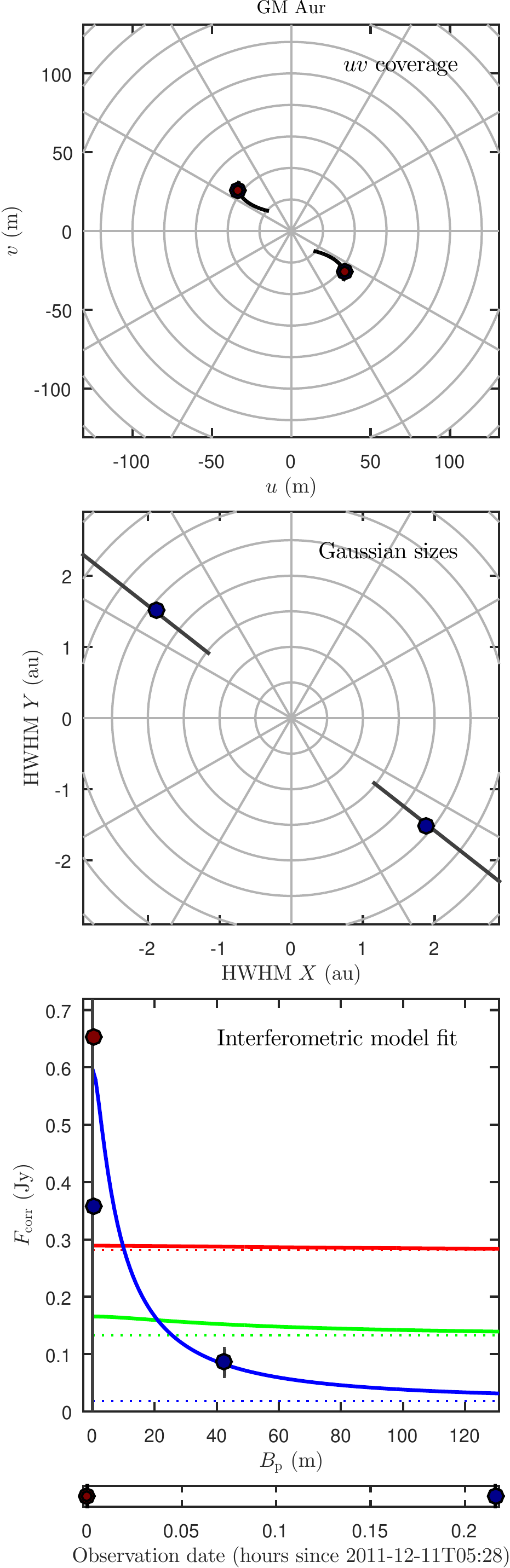}
			\includegraphics[width = 0.21\linewidth]{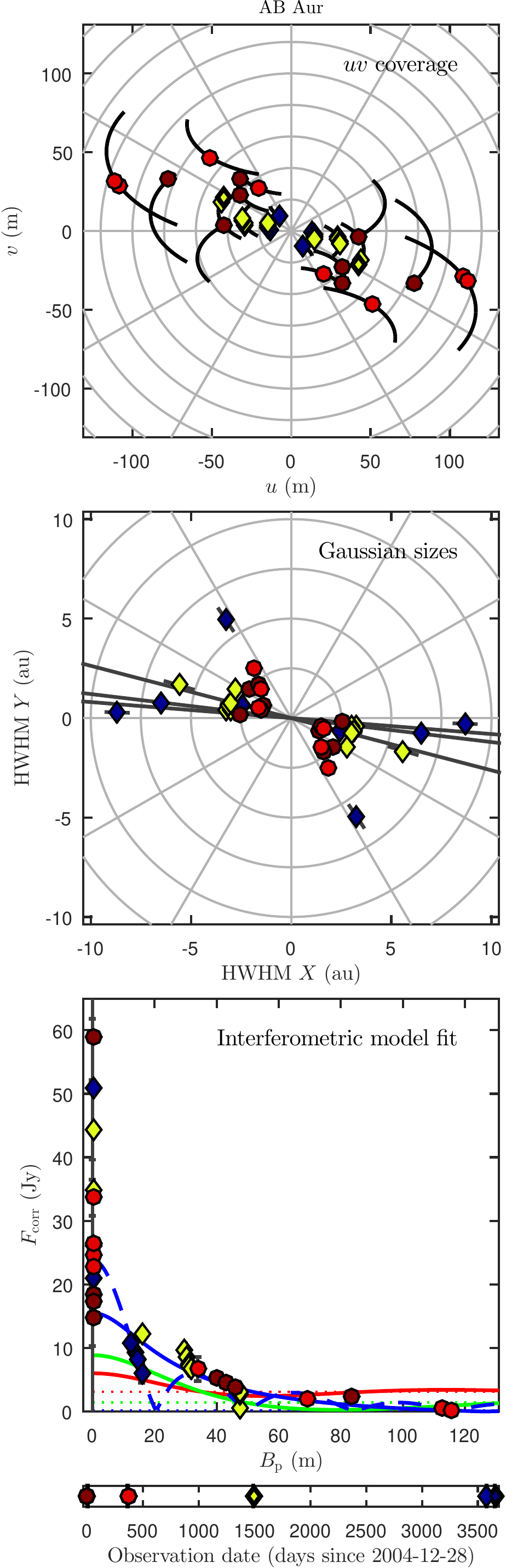}
			\includegraphics[width = 0.21\linewidth]{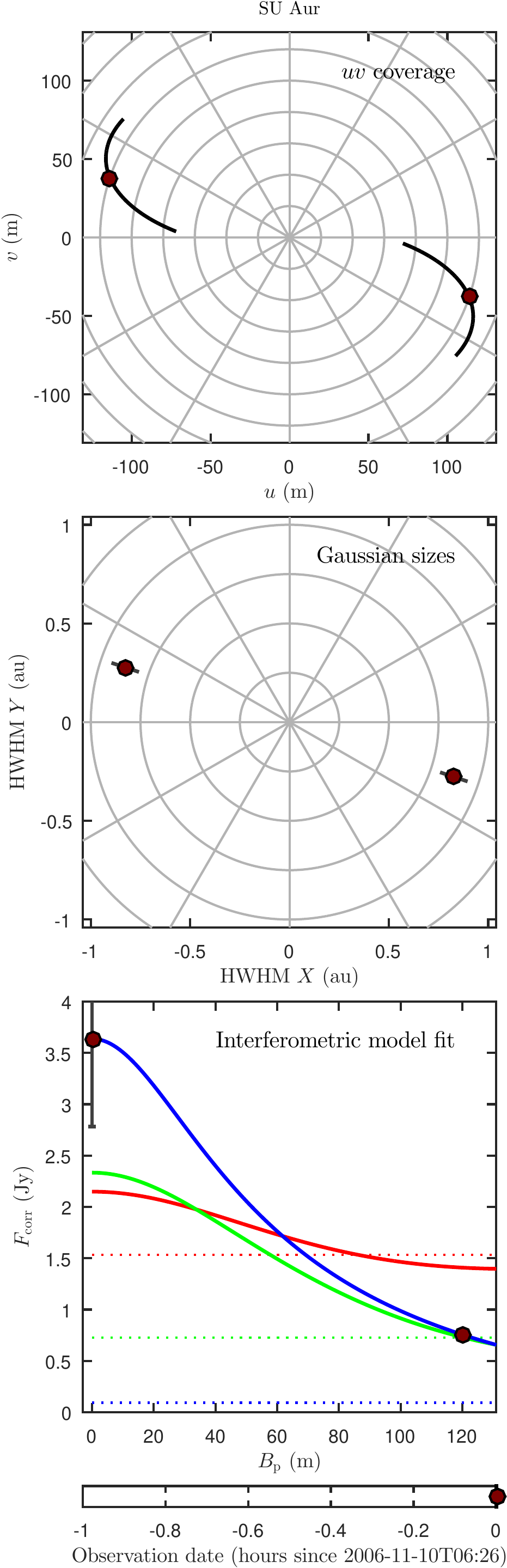}
			\includegraphics[width = 0.21\linewidth]{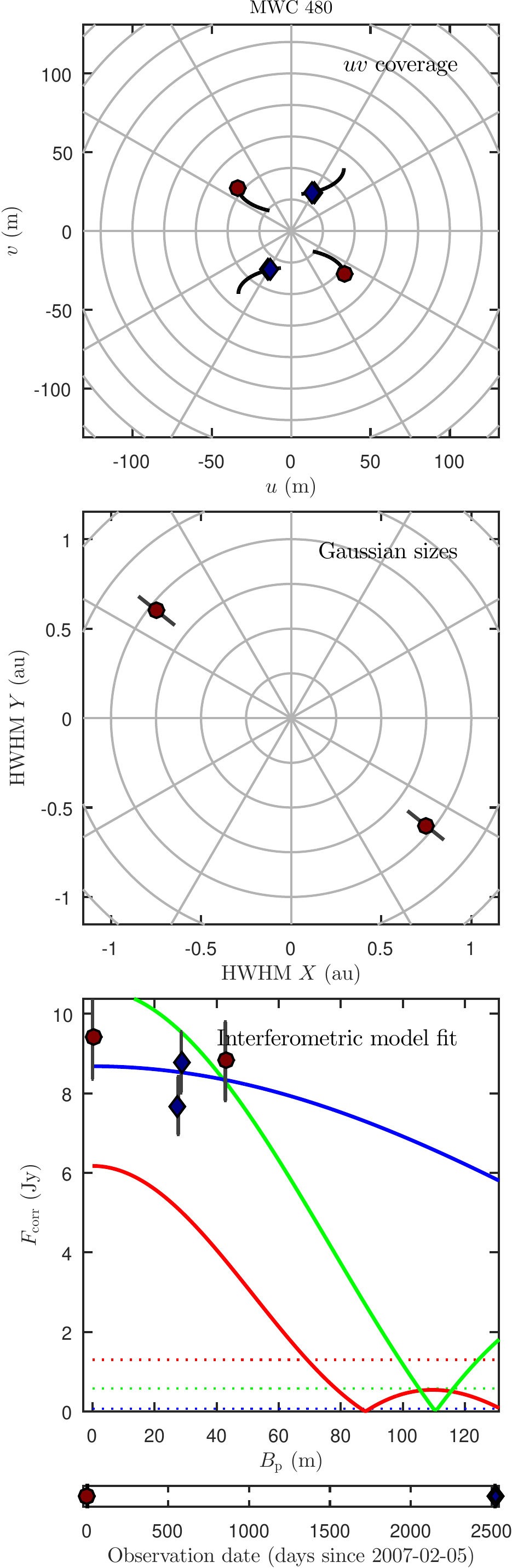}
			\includegraphics[width = 0.21\linewidth]{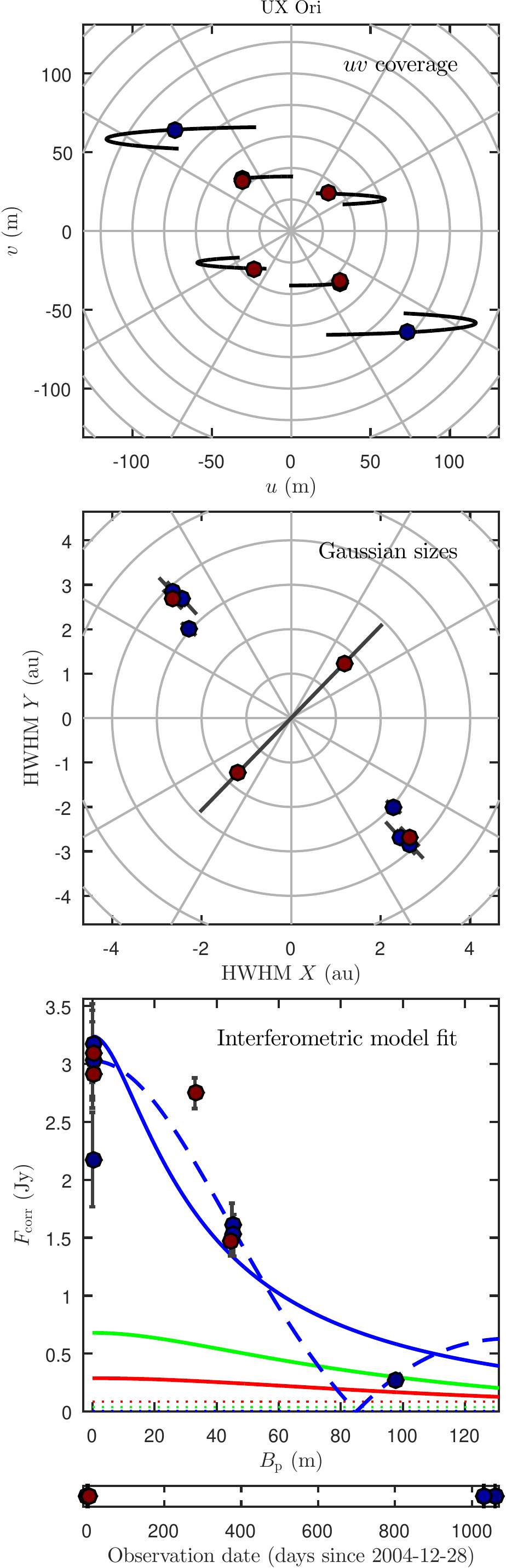}
			\includegraphics[width = 0.21\linewidth]{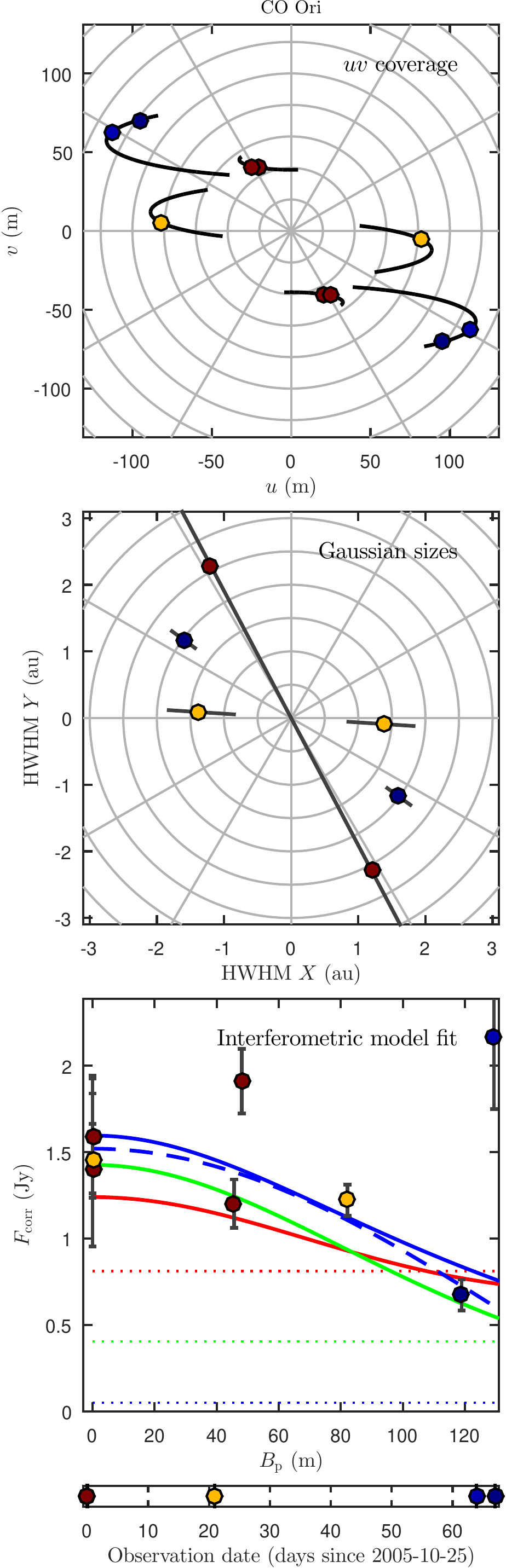}
			\includegraphics[width = 0.21\linewidth]{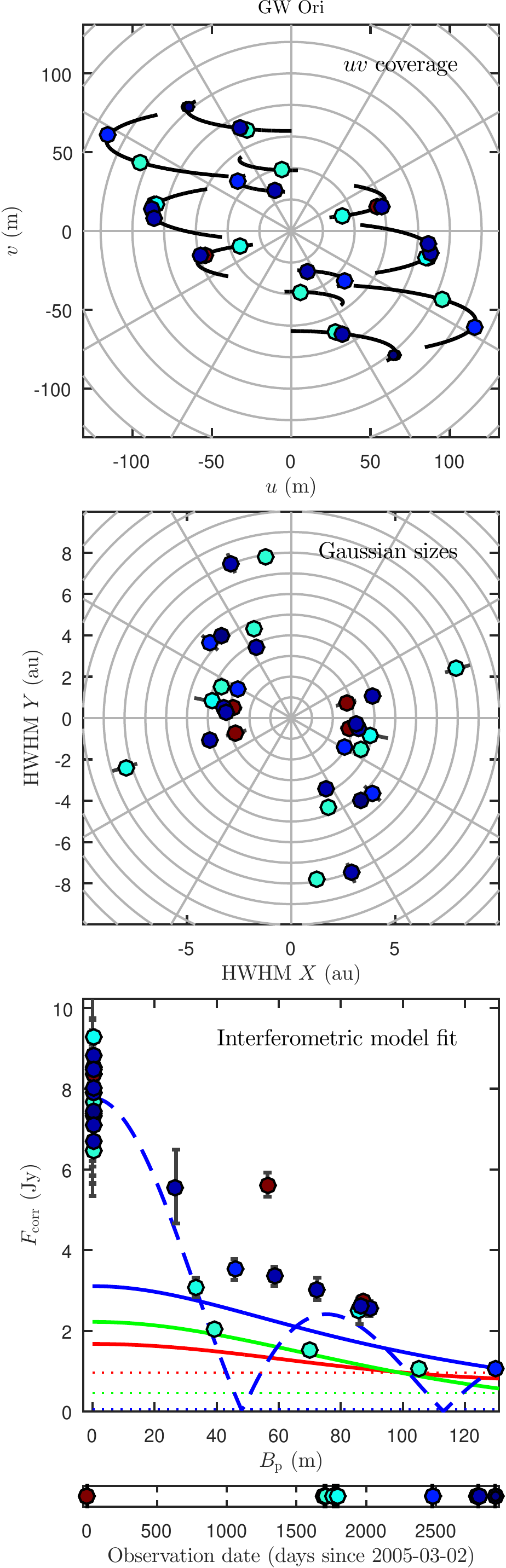}
			
		\end{figure*}
		
		\begin{figure*}[h!]
			\centering
			\includegraphics[width = 0.21\linewidth]{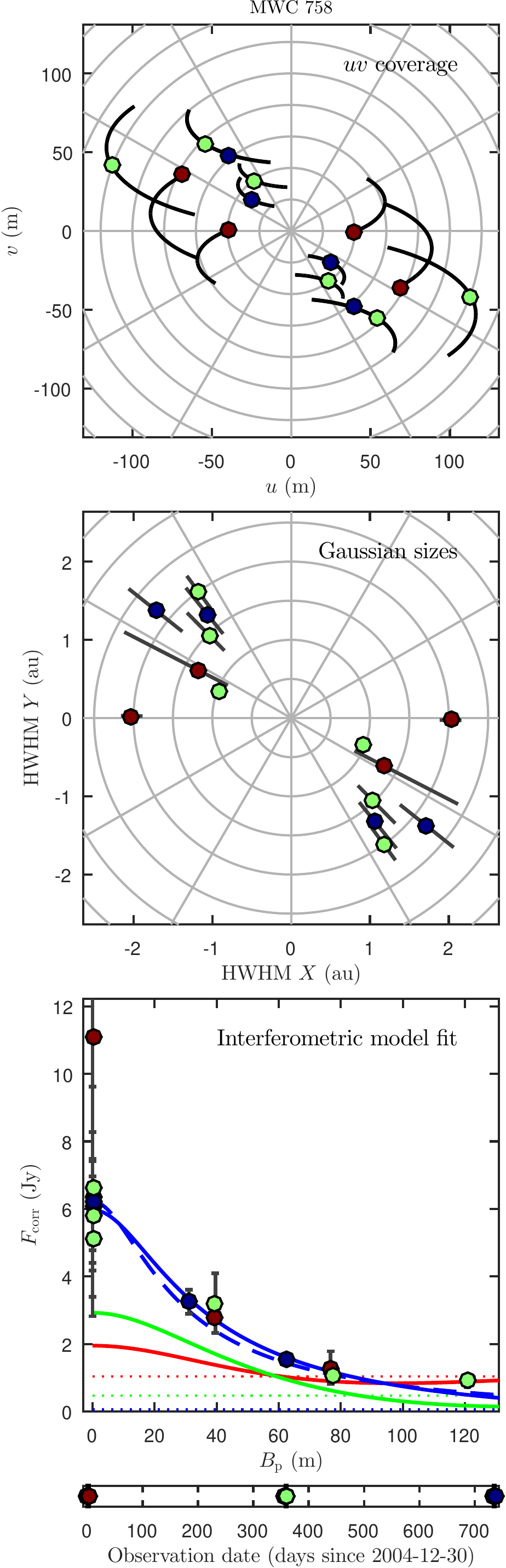}
			\includegraphics[width = 0.21\linewidth]{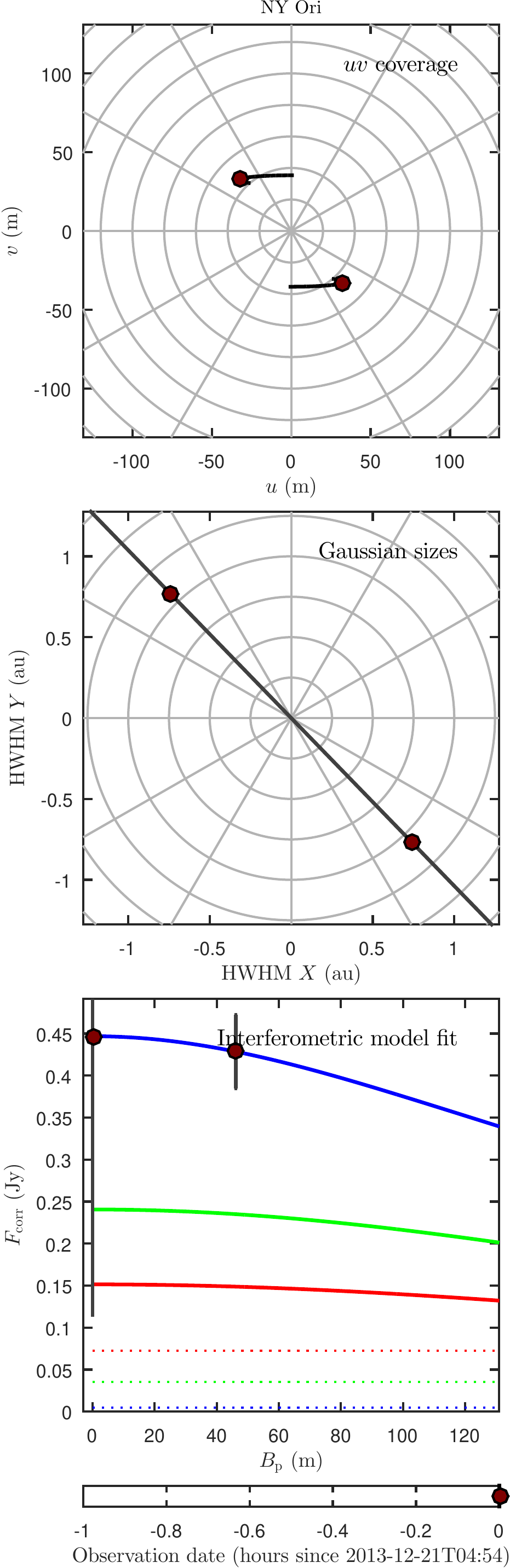}
			\includegraphics[width = 0.21\linewidth]{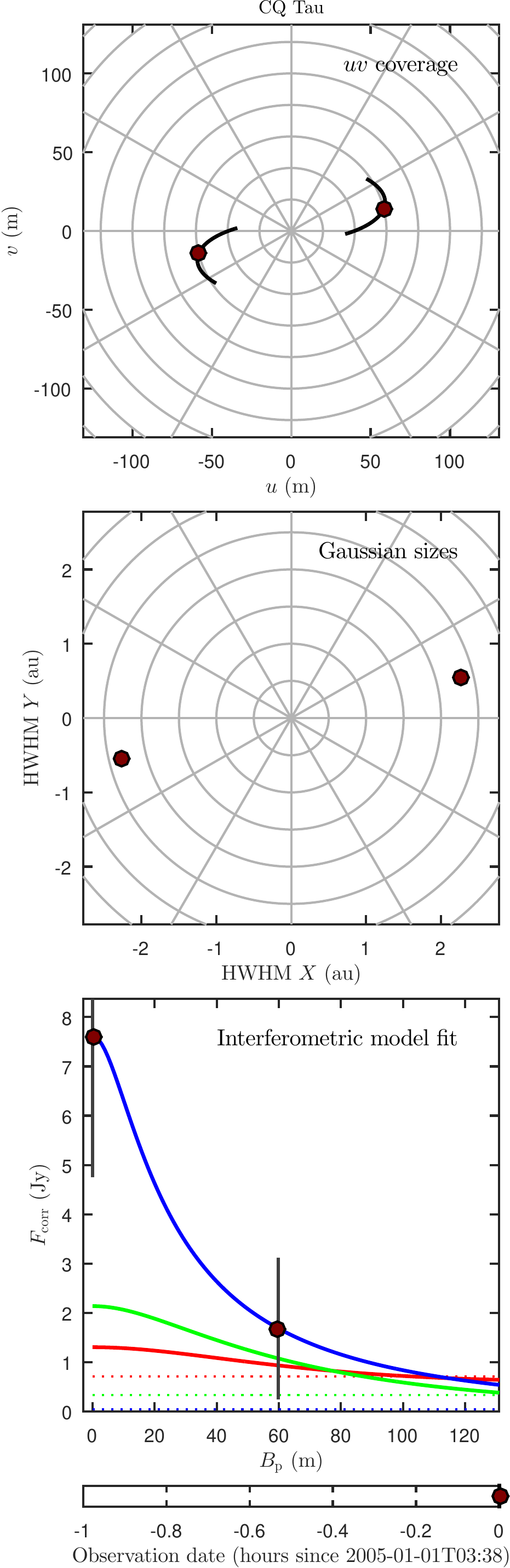}
			\includegraphics[width = 0.21\linewidth]{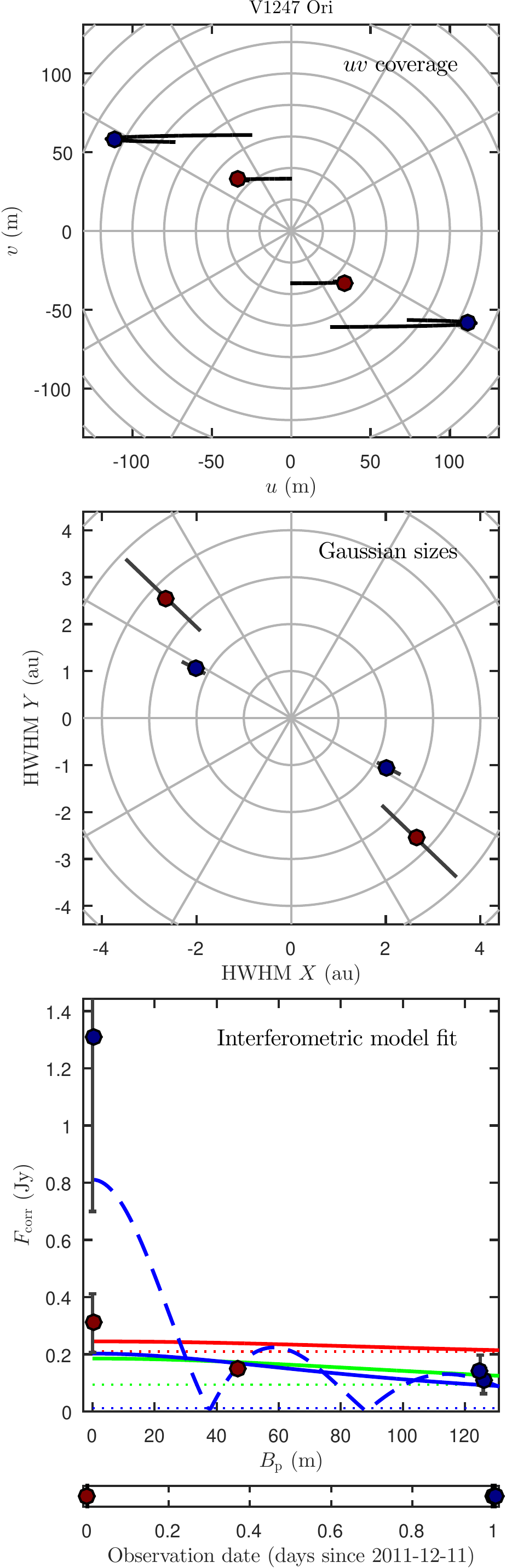}
			\includegraphics[width = 0.21\linewidth]{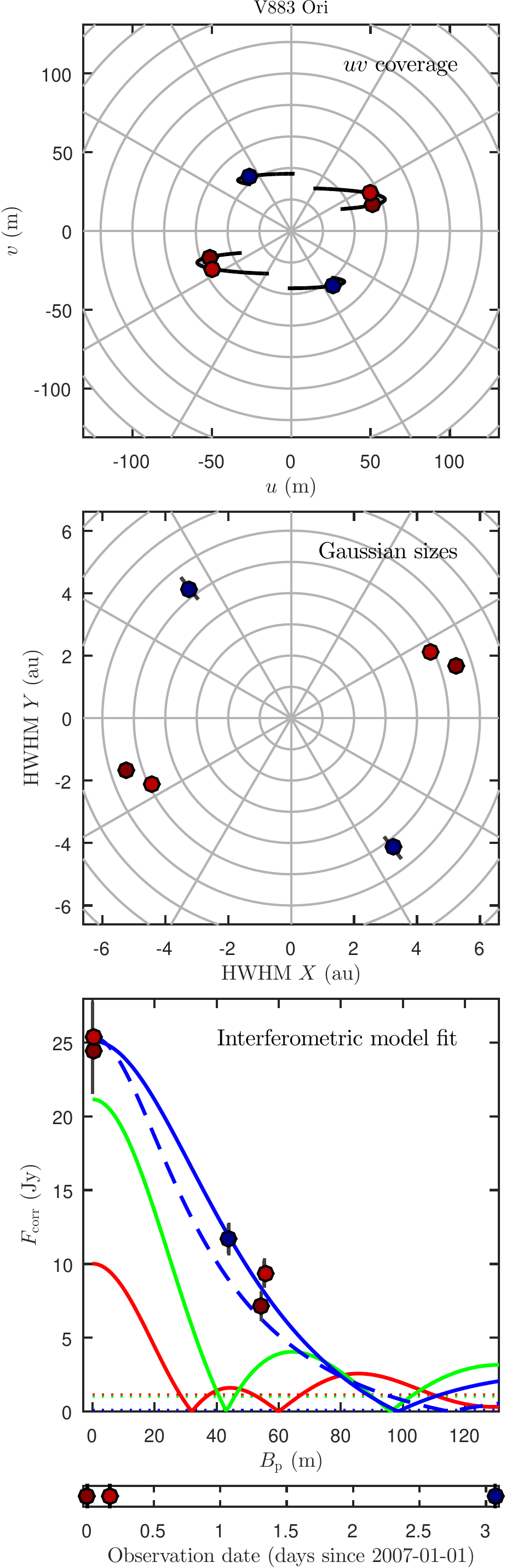}
			\includegraphics[width = 0.21\linewidth]{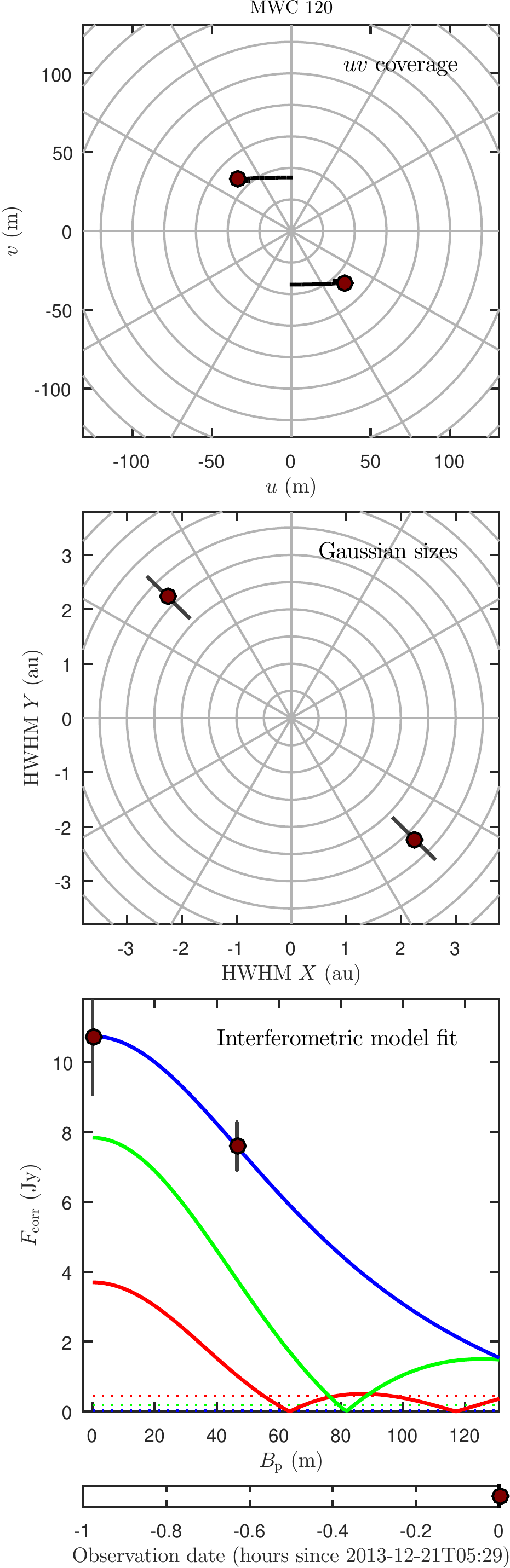}
			\includegraphics[width = 0.21\linewidth]{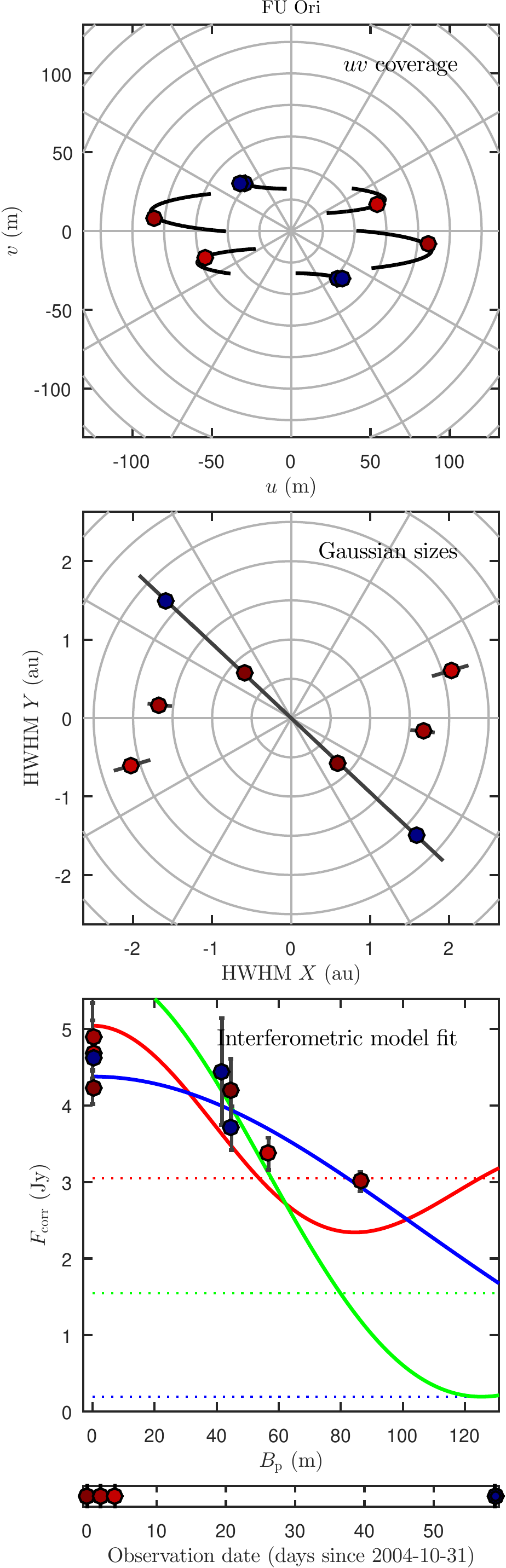}
			\includegraphics[width = 0.21\linewidth]{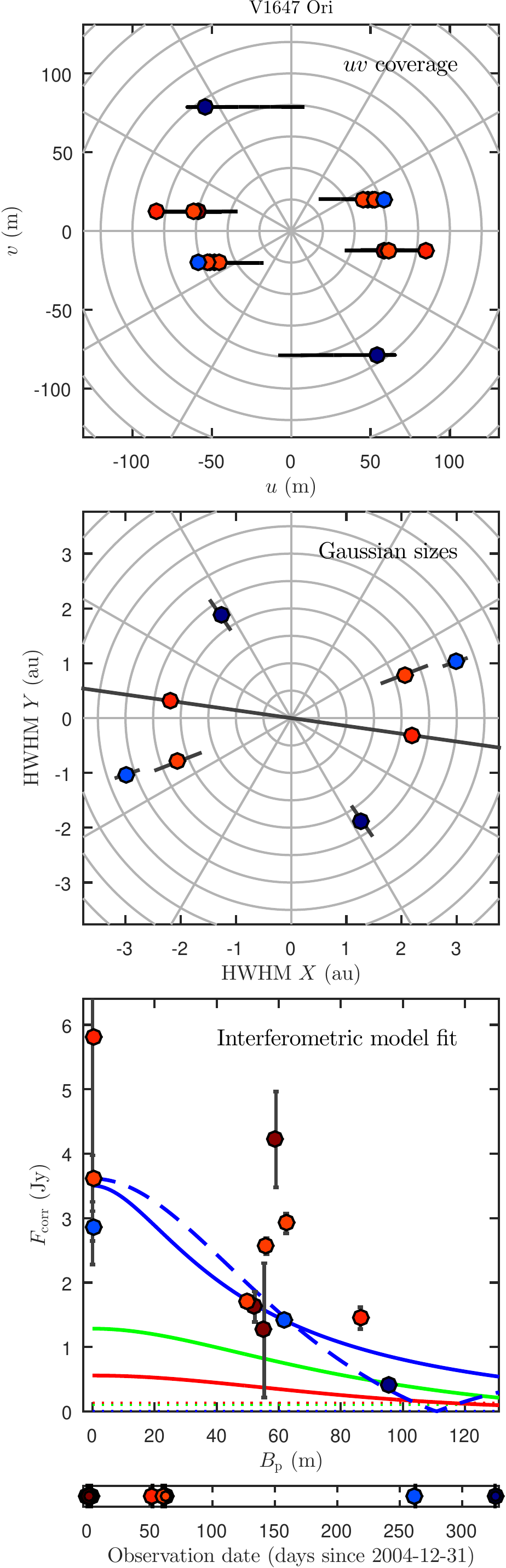}
			
		\end{figure*}
		
		\begin{figure*}[h!]
			\centering
			\includegraphics[width = 0.21\linewidth]{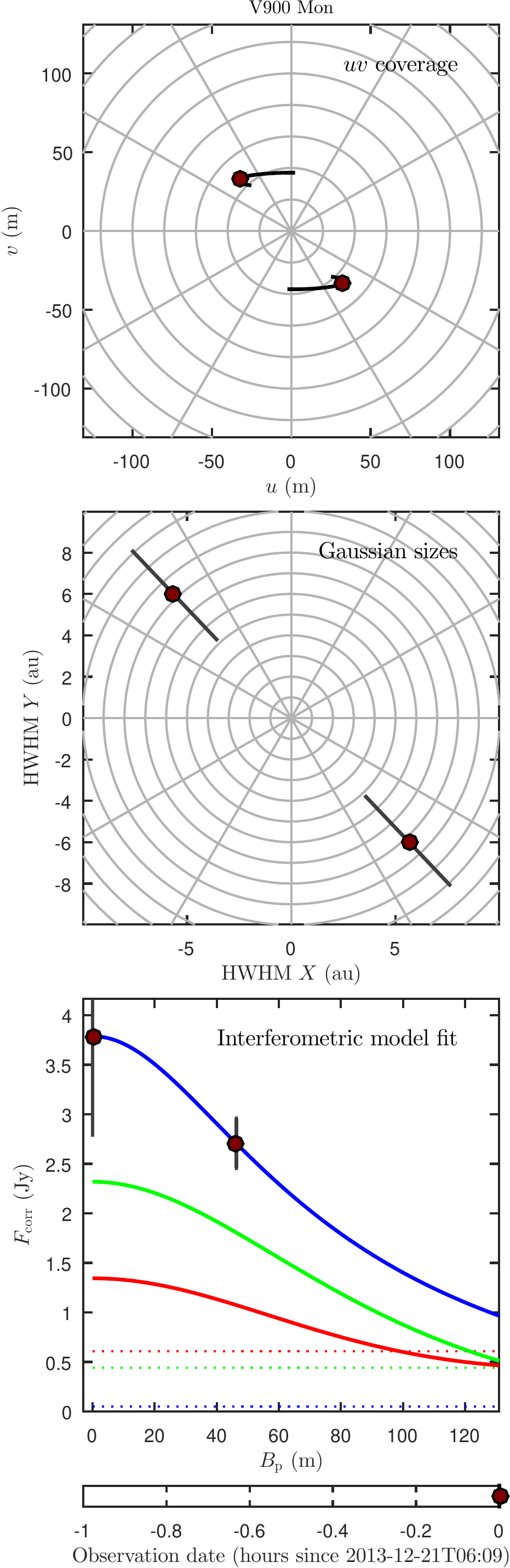}
			\includegraphics[width = 0.21\linewidth]{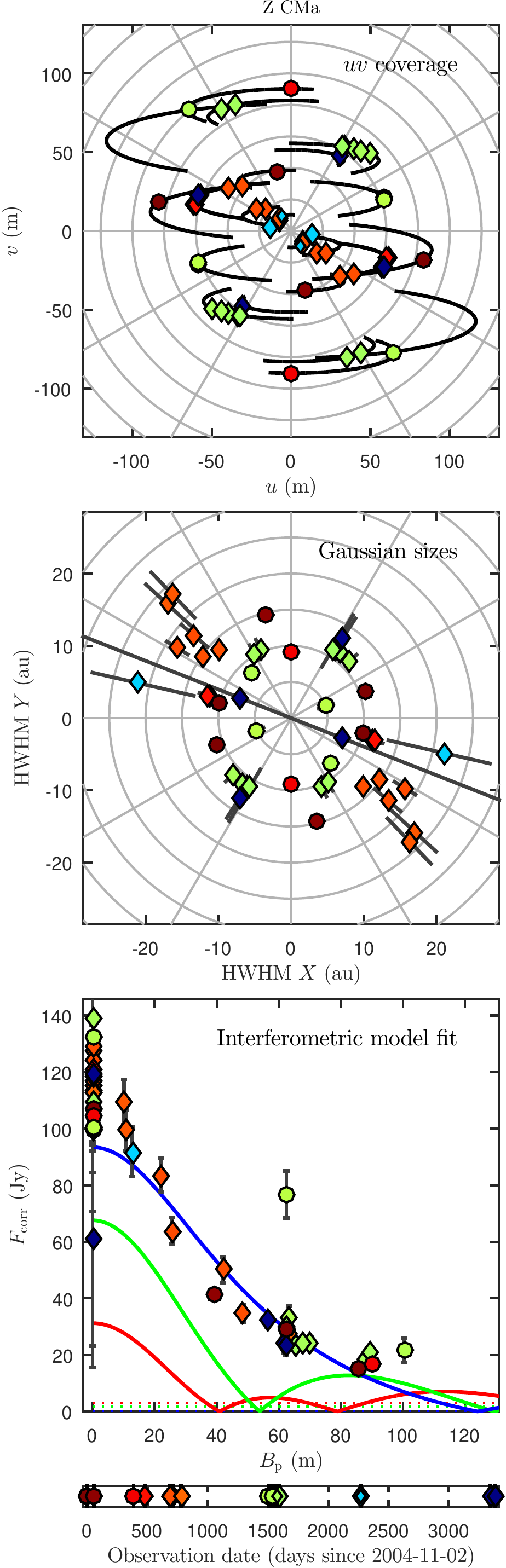}
			\includegraphics[width = 0.21\linewidth]{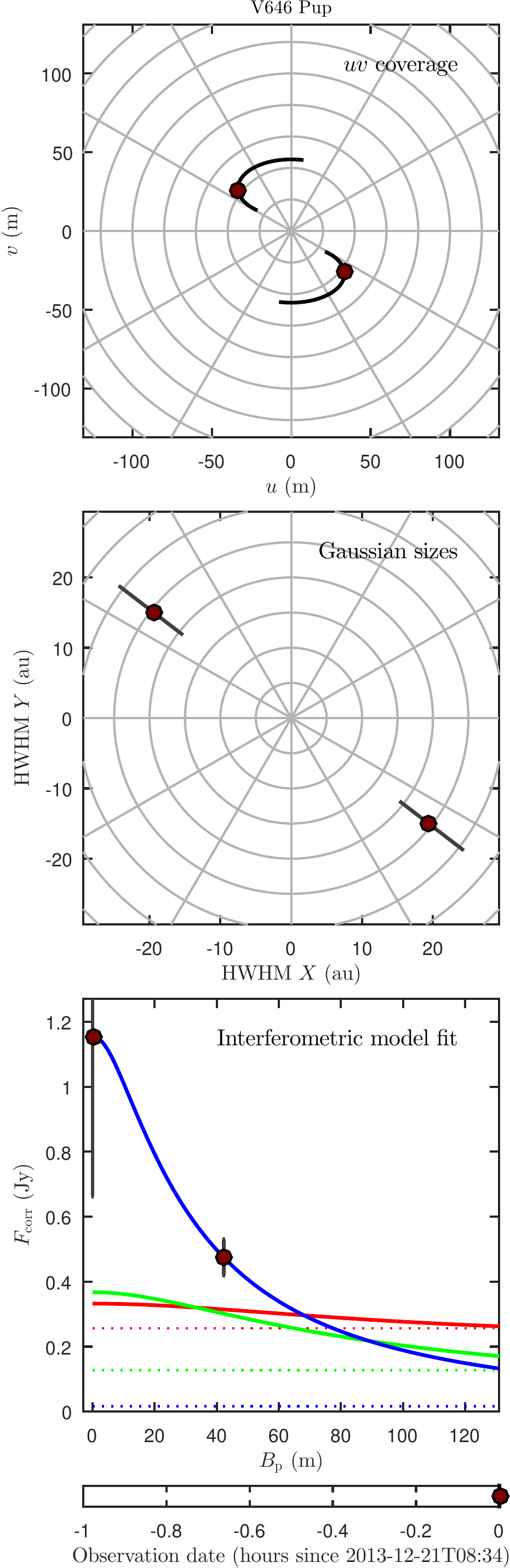}
			\includegraphics[width = 0.21\linewidth]{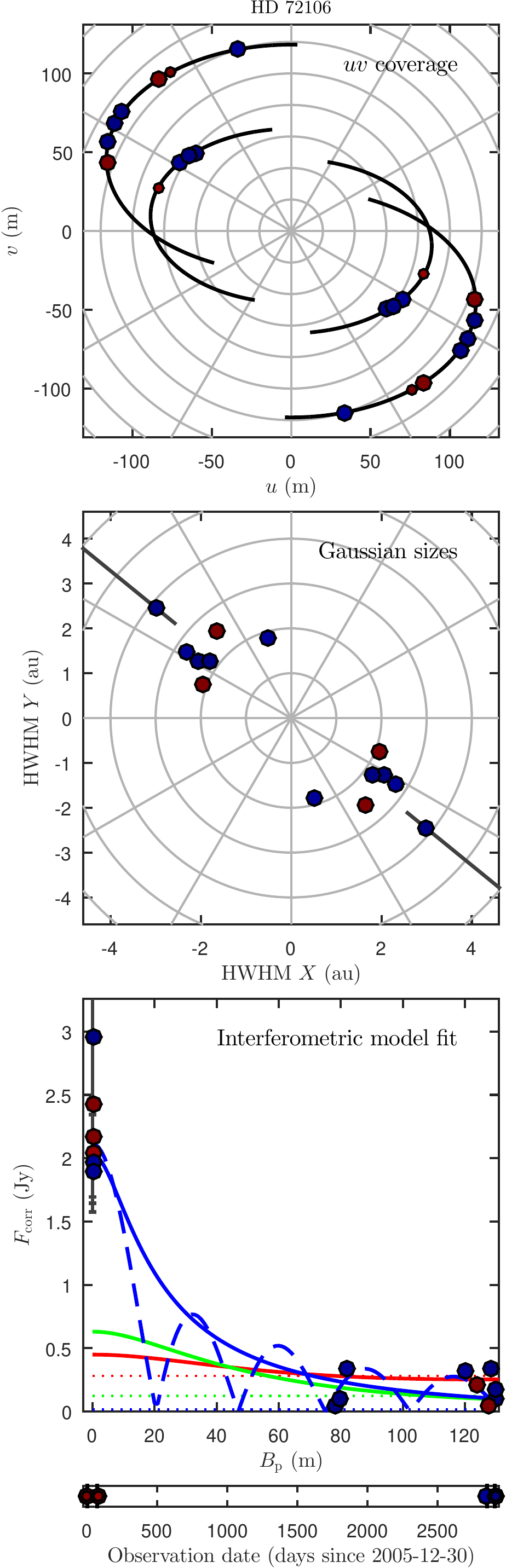}
			\includegraphics[width = 0.21\linewidth]{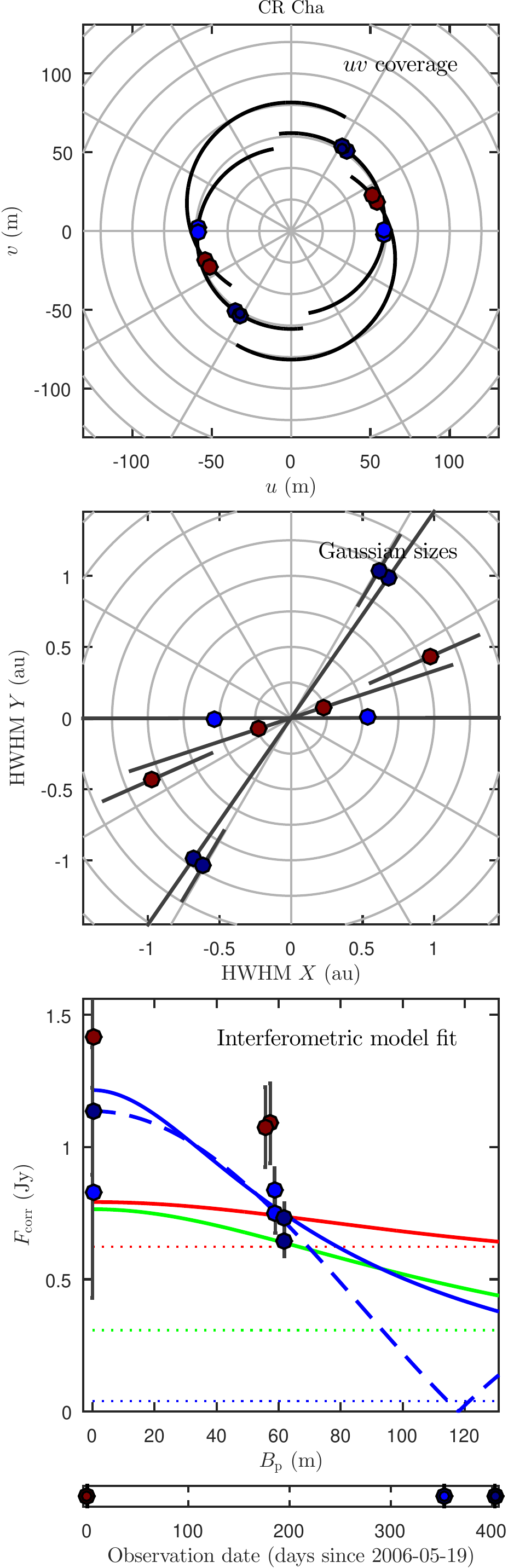}
			\includegraphics[width = 0.21\linewidth]{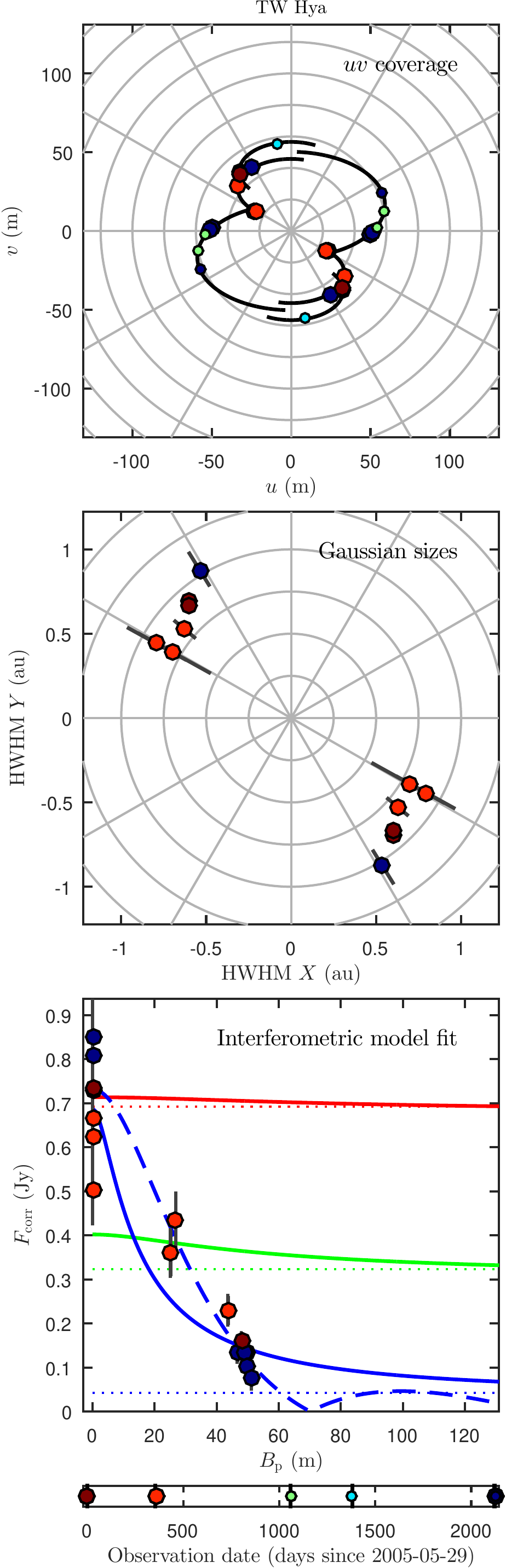}
			\includegraphics[width = 0.21\linewidth]{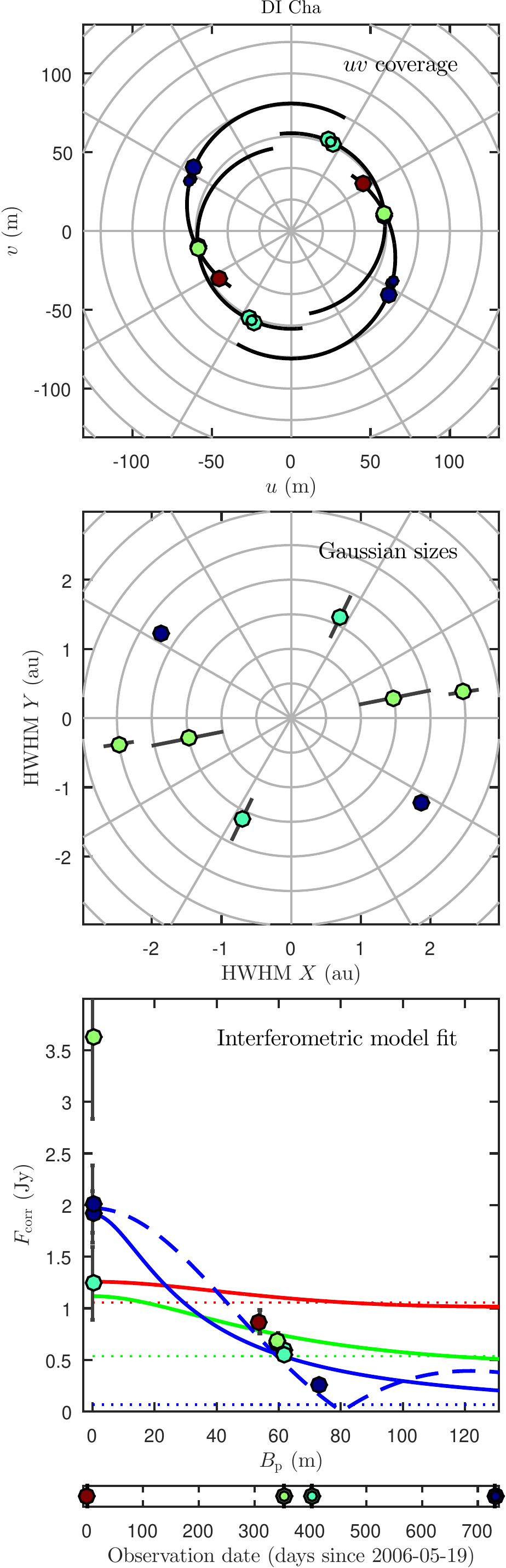}
			\includegraphics[width = 0.21\linewidth]{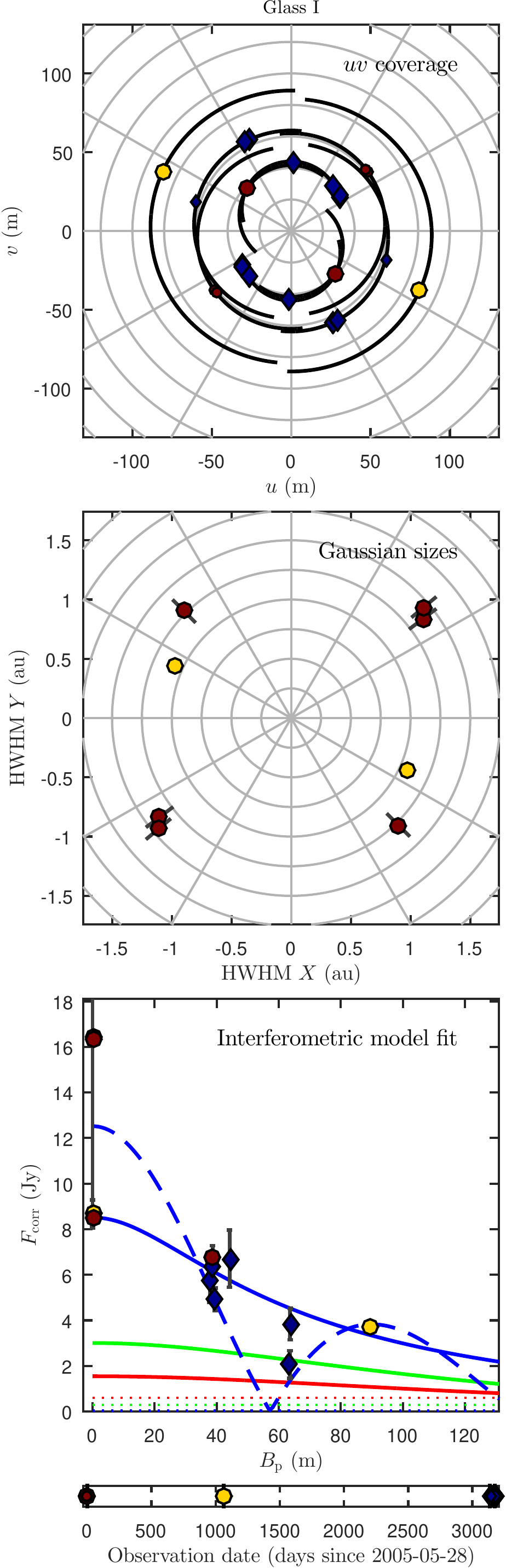}
			
		\end{figure*}
		
		\begin{figure*}[h!]
			\centering
			\includegraphics[width = 0.21\linewidth]{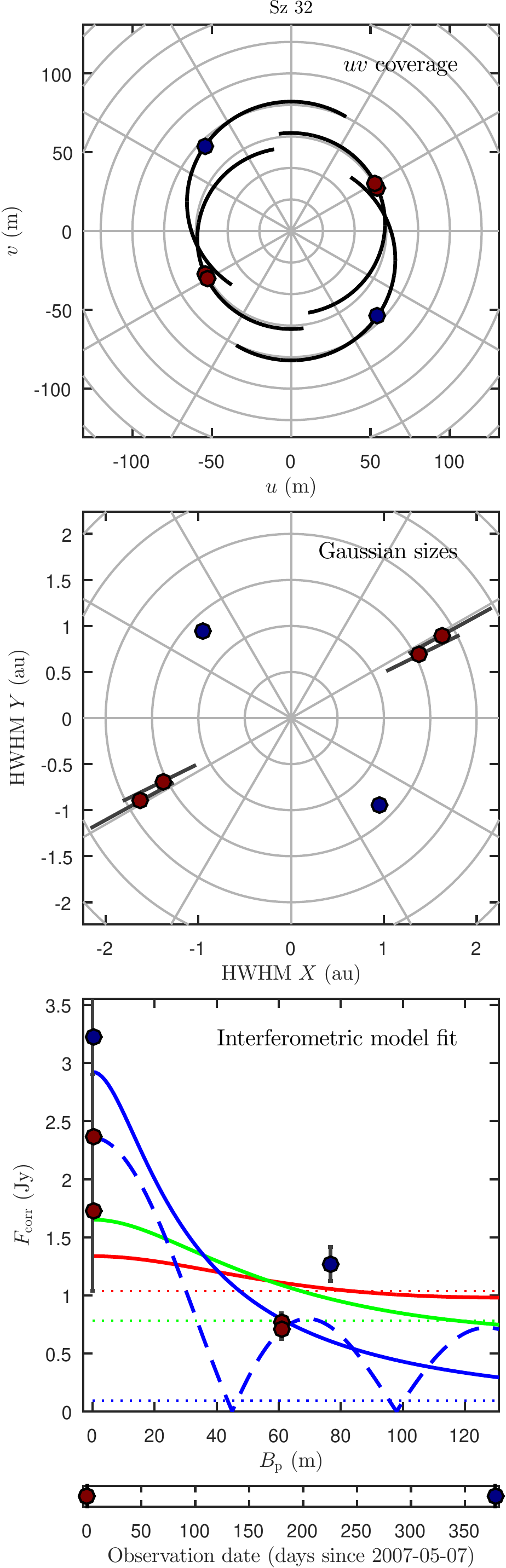}
			\includegraphics[width = 0.21\linewidth]{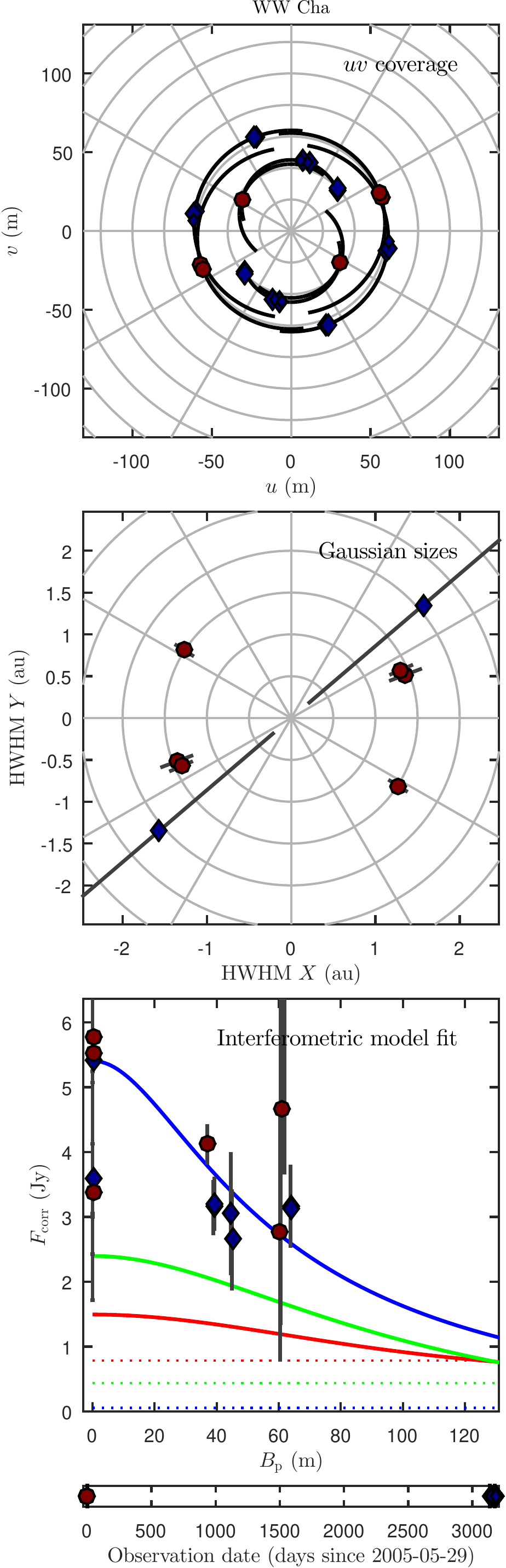}
			\includegraphics[width = 0.21\linewidth]{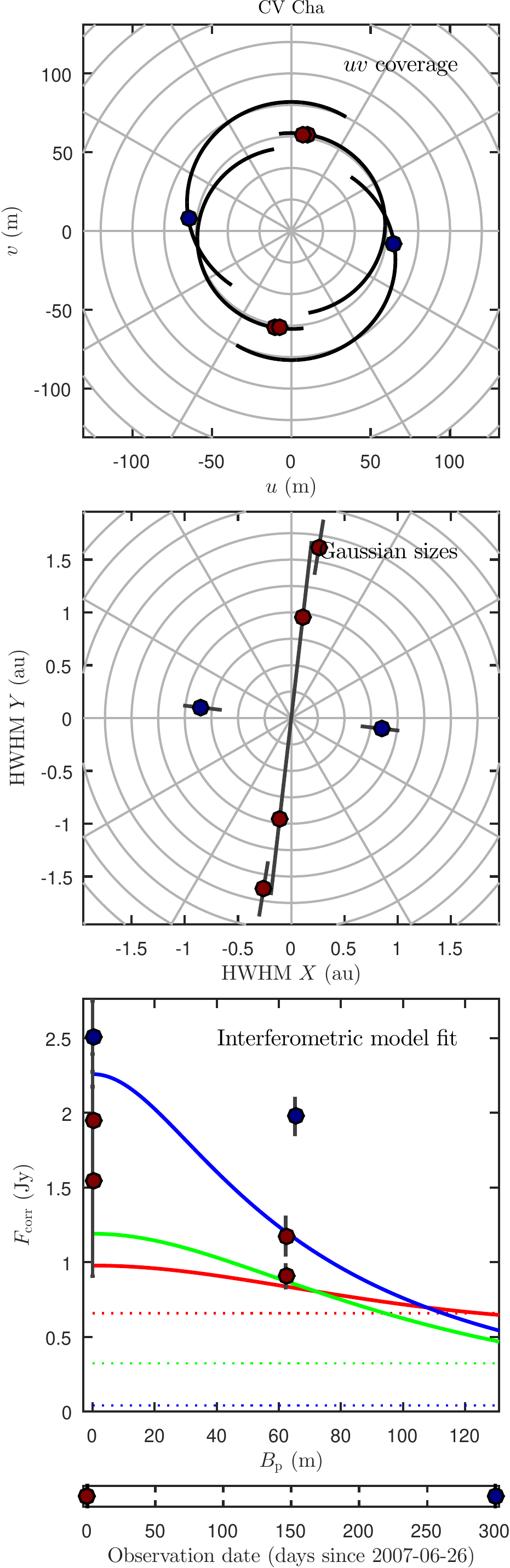}
			\includegraphics[width = 0.21\linewidth]{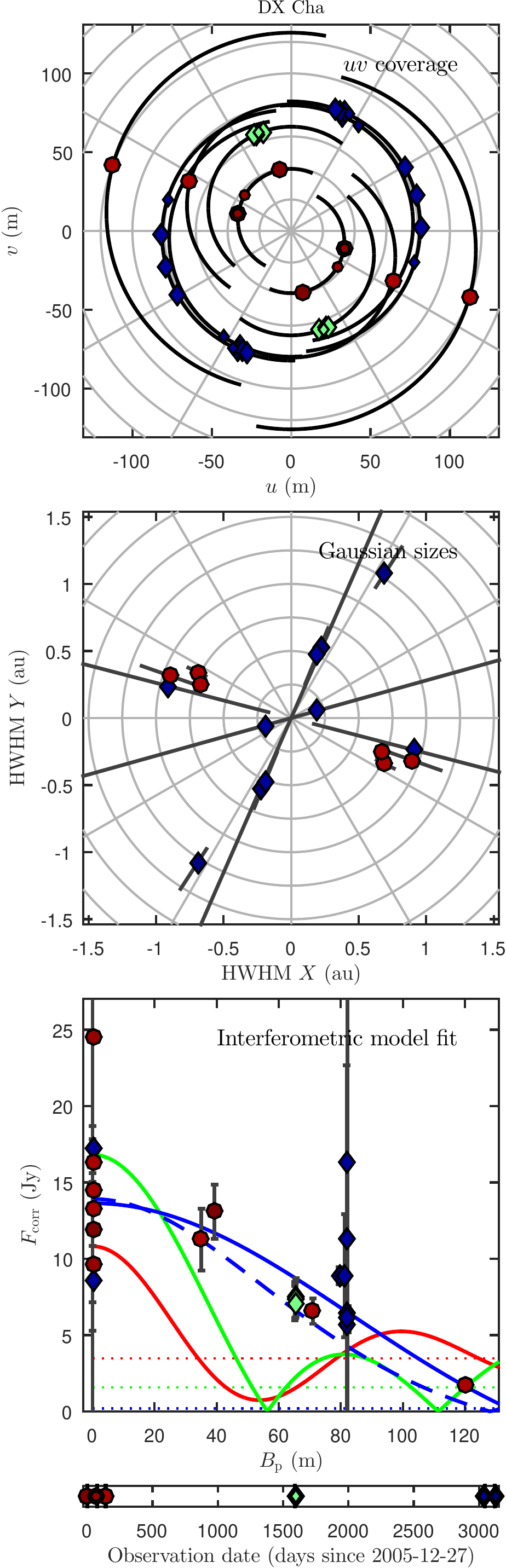}
			\includegraphics[width = 0.21\linewidth]{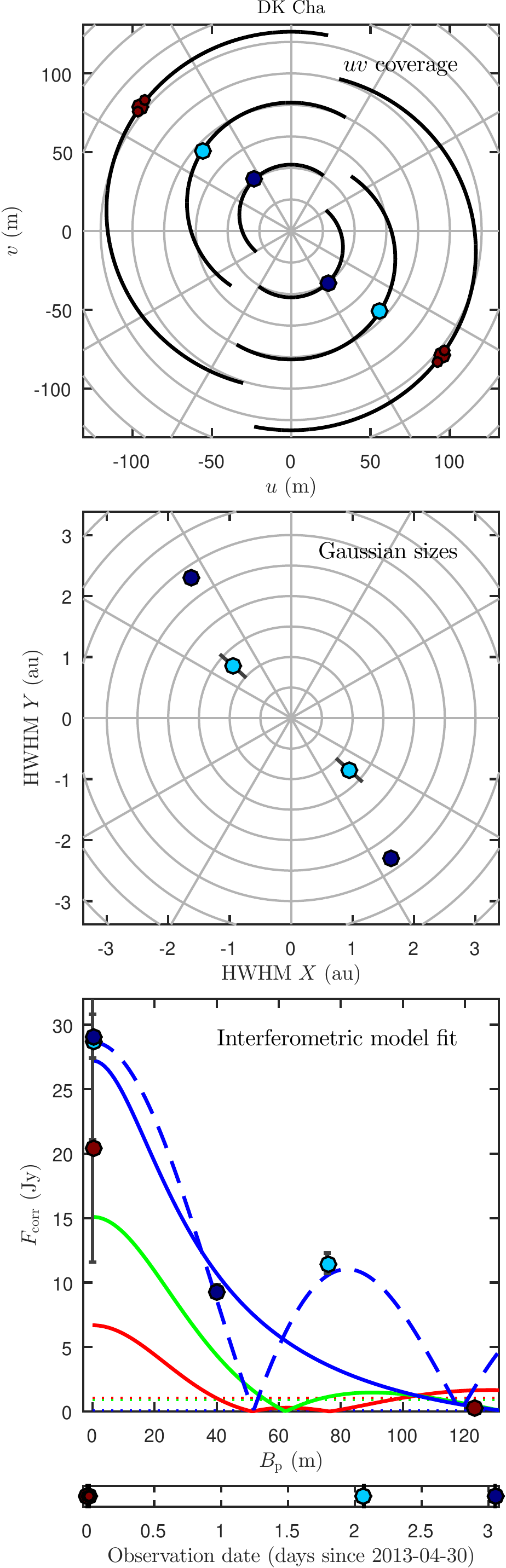}
			\includegraphics[width = 0.21\linewidth]{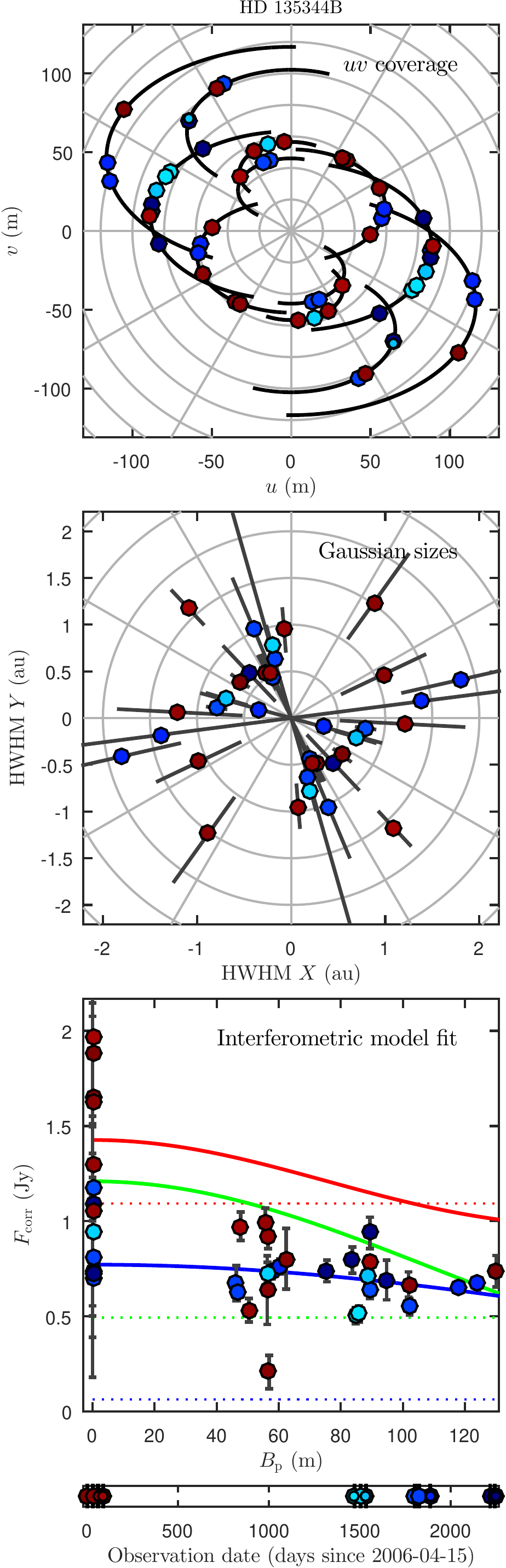}
			\includegraphics[width = 0.21\linewidth]{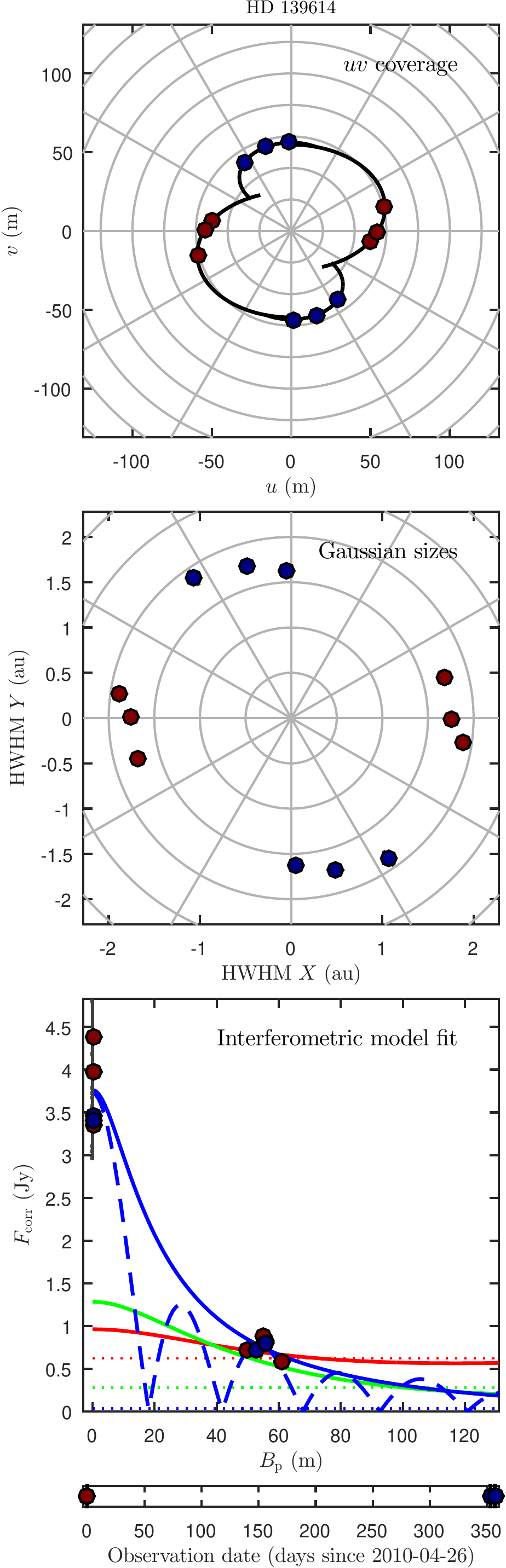}
			\includegraphics[width = 0.21\linewidth]{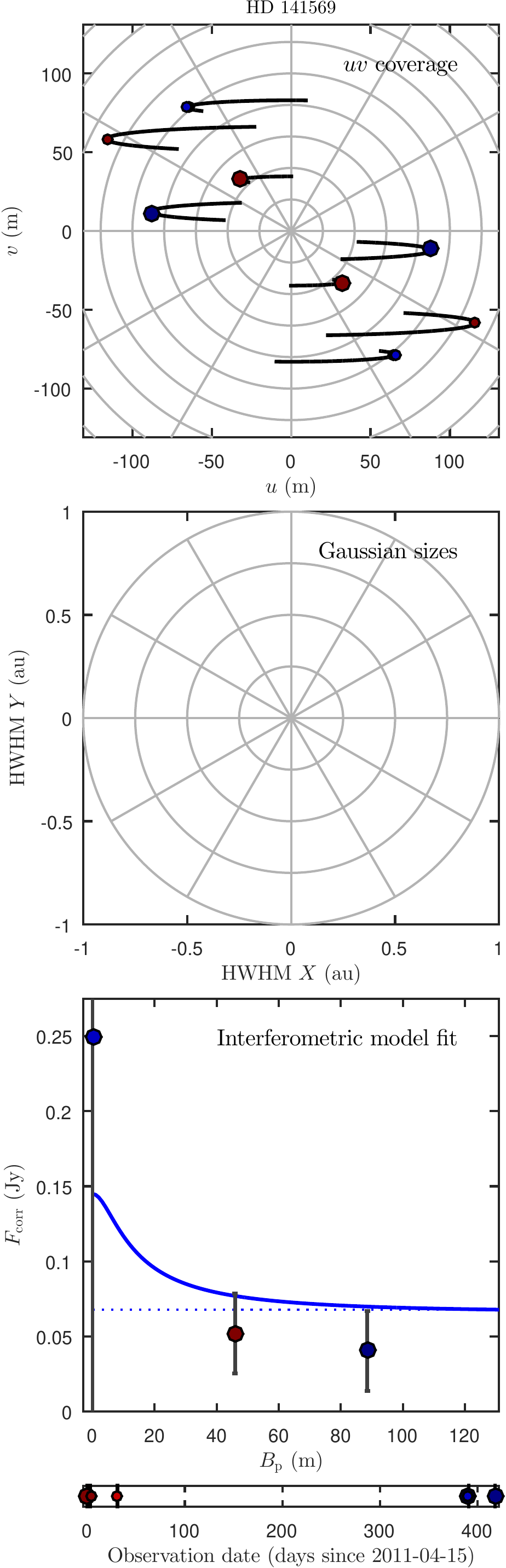}
			
		\end{figure*}
		
		\begin{figure*}[h!]
			\centering
			\includegraphics[width = 0.21\linewidth]{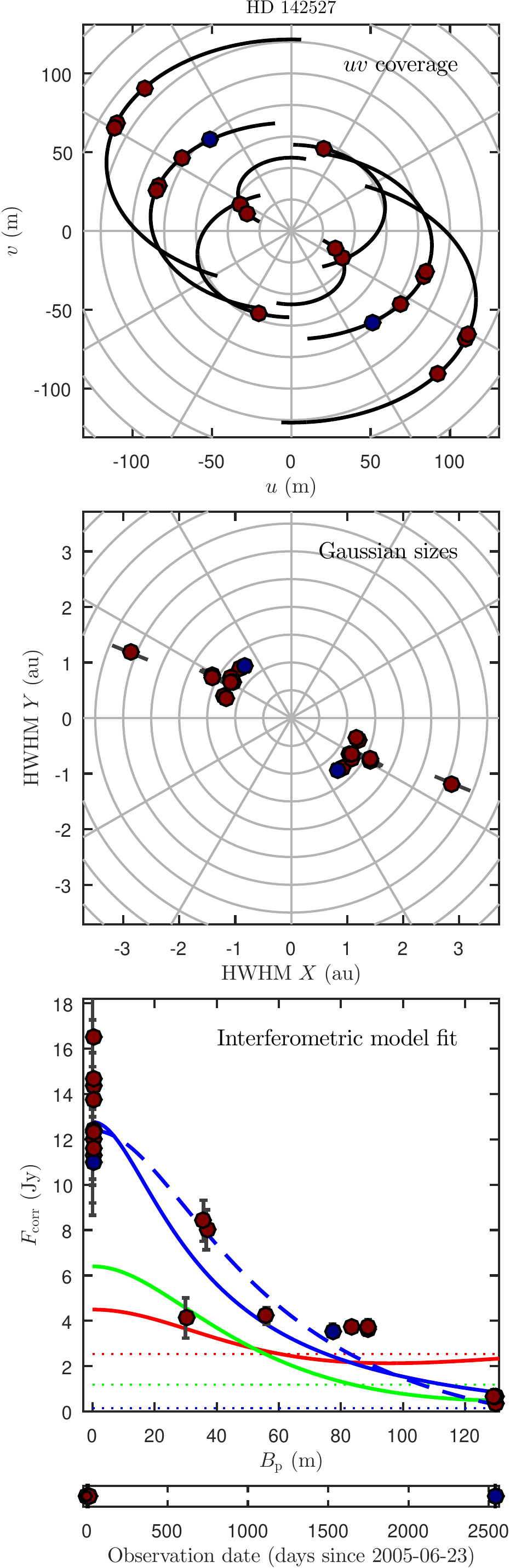}
			\includegraphics[width = 0.21\linewidth]{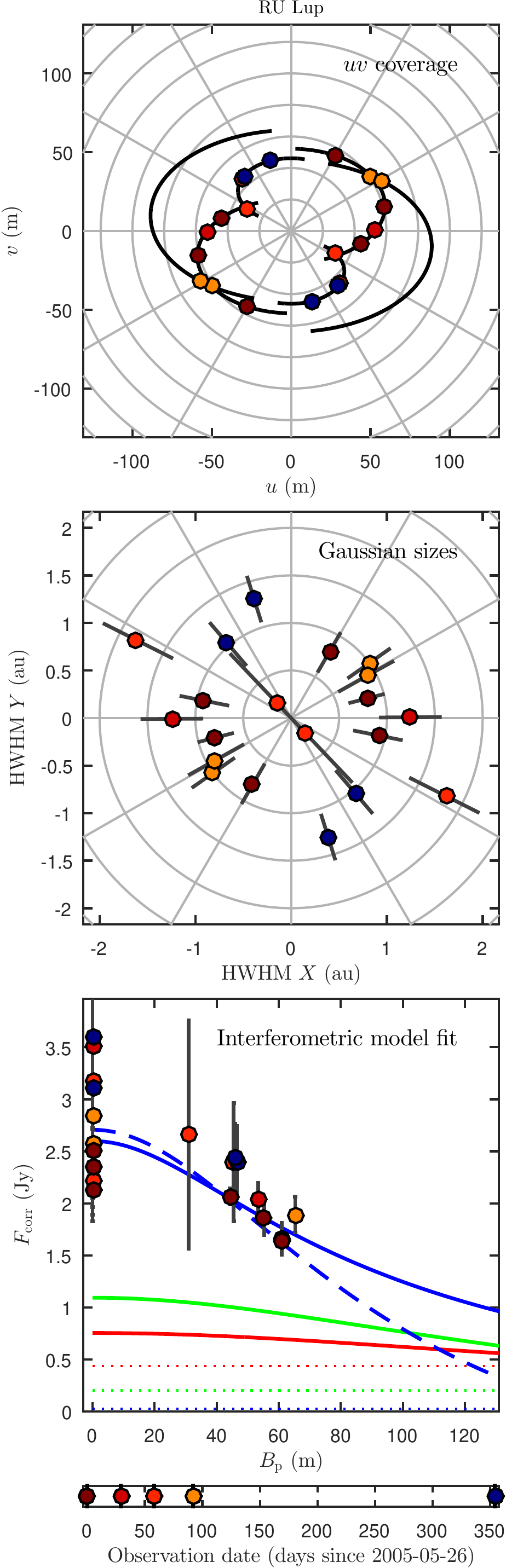}
			\includegraphics[width = 0.21\linewidth]{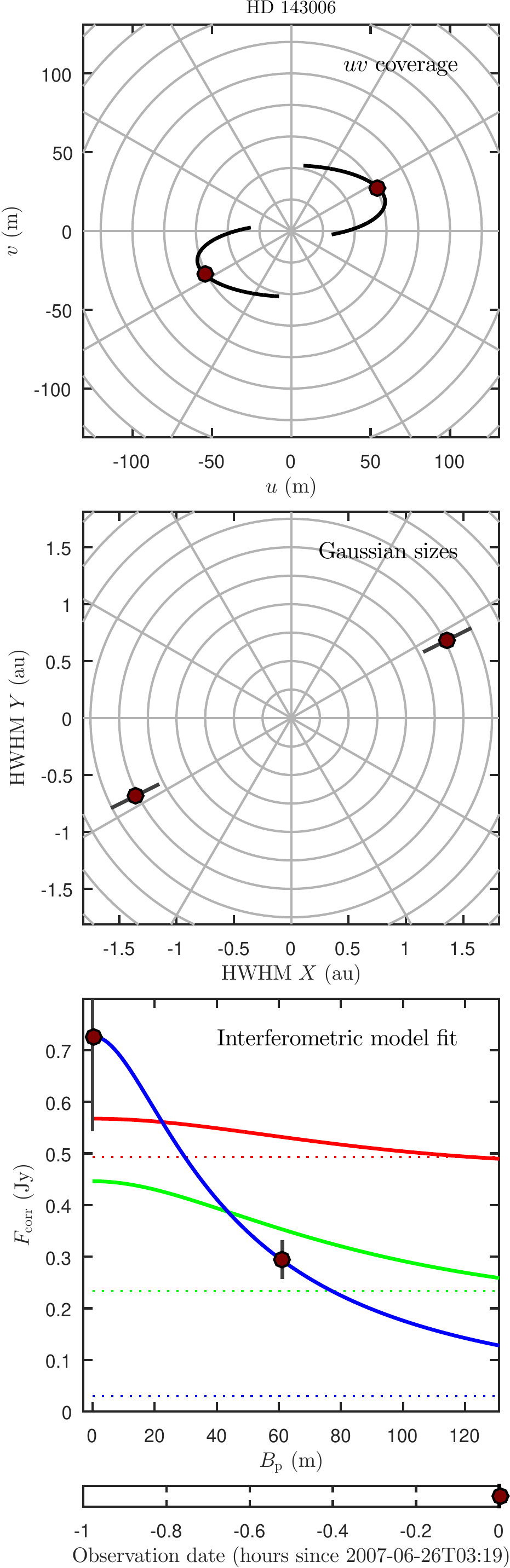}
			\includegraphics[width = 0.21\linewidth]{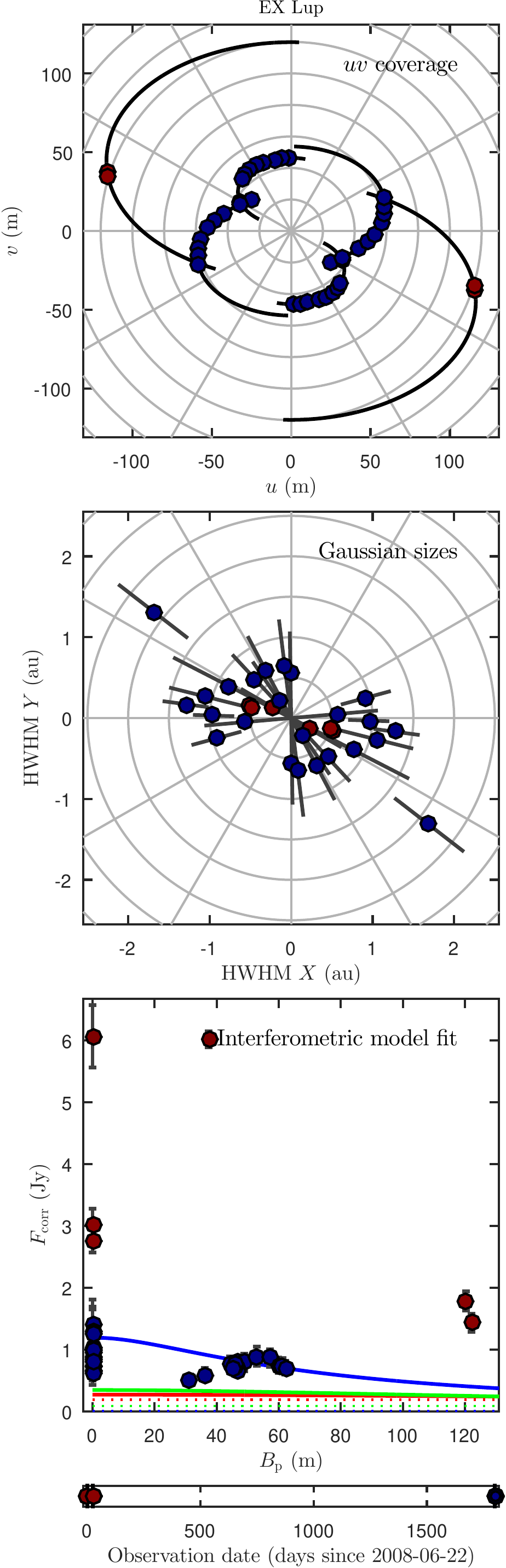}
			\includegraphics[width = 0.21\linewidth]{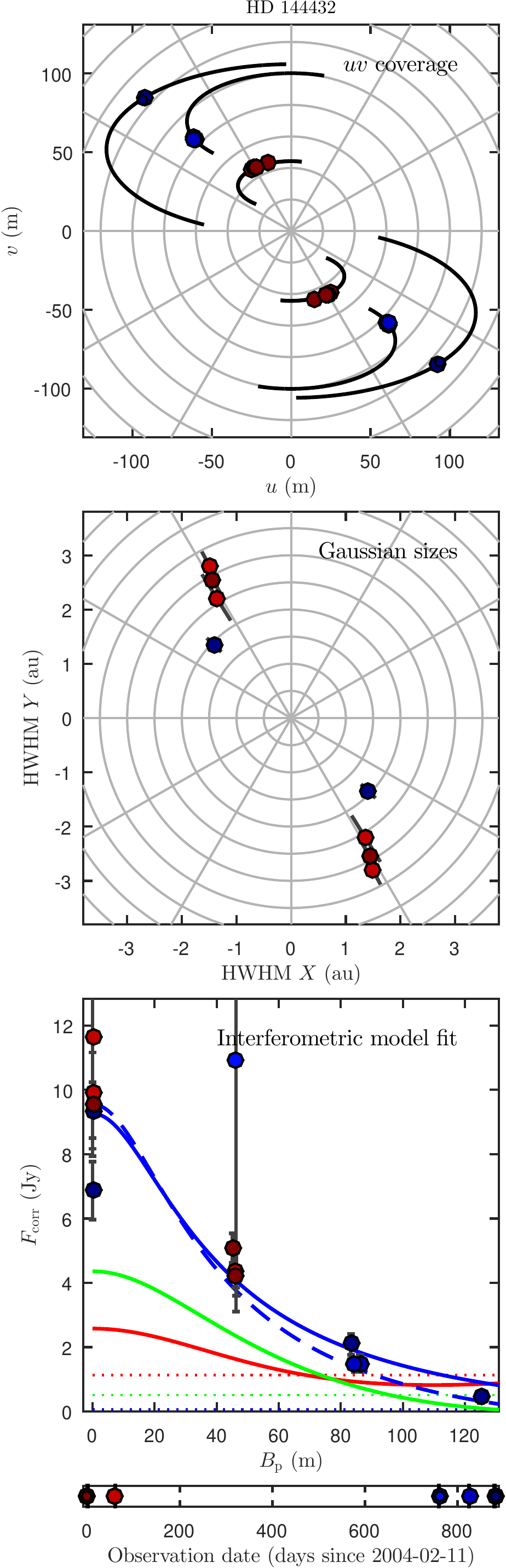}
			\includegraphics[width = 0.21\linewidth]{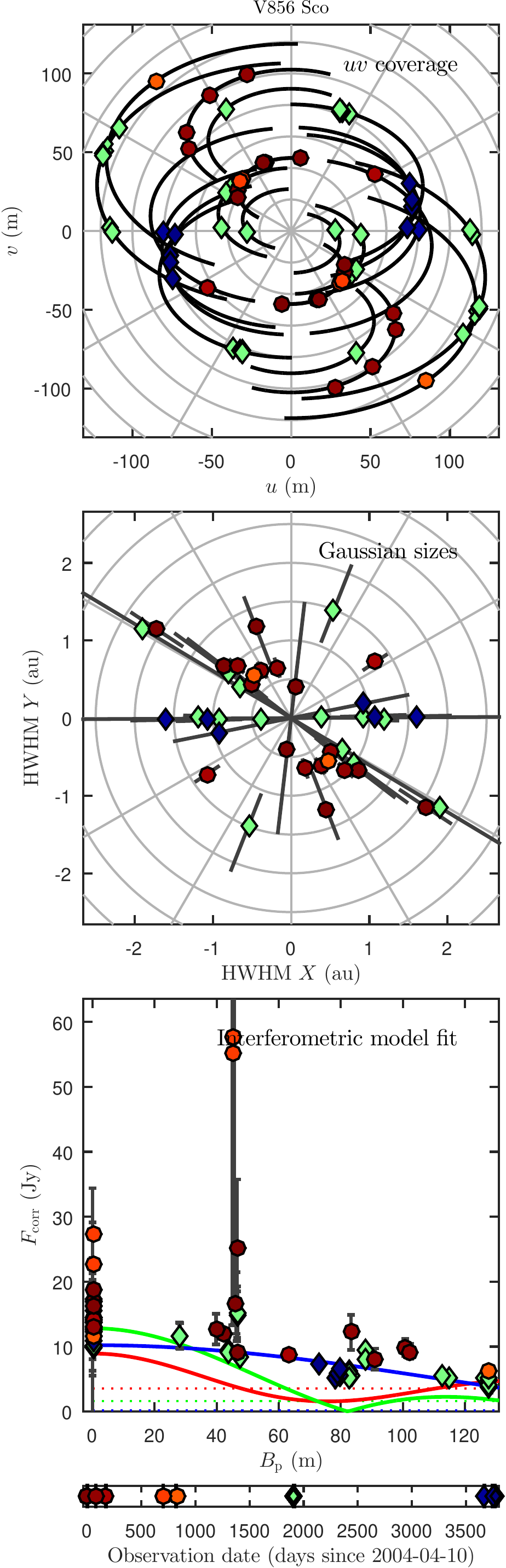}
			\includegraphics[width = 0.21\linewidth]{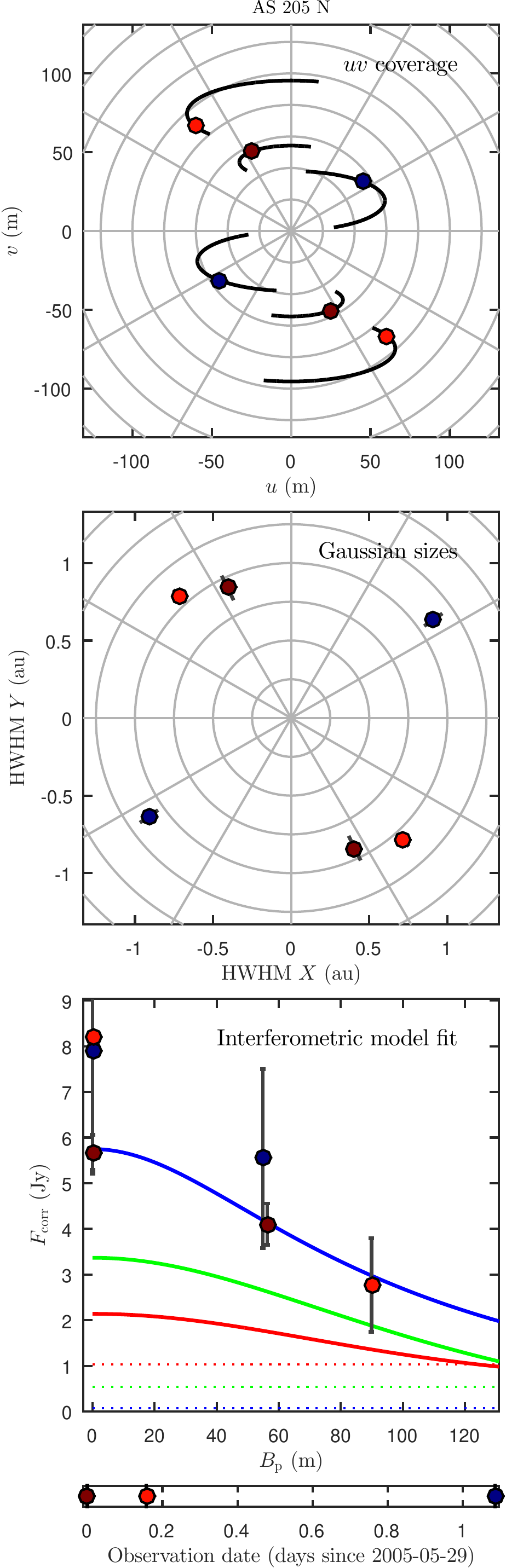}
			\includegraphics[width = 0.21\linewidth]{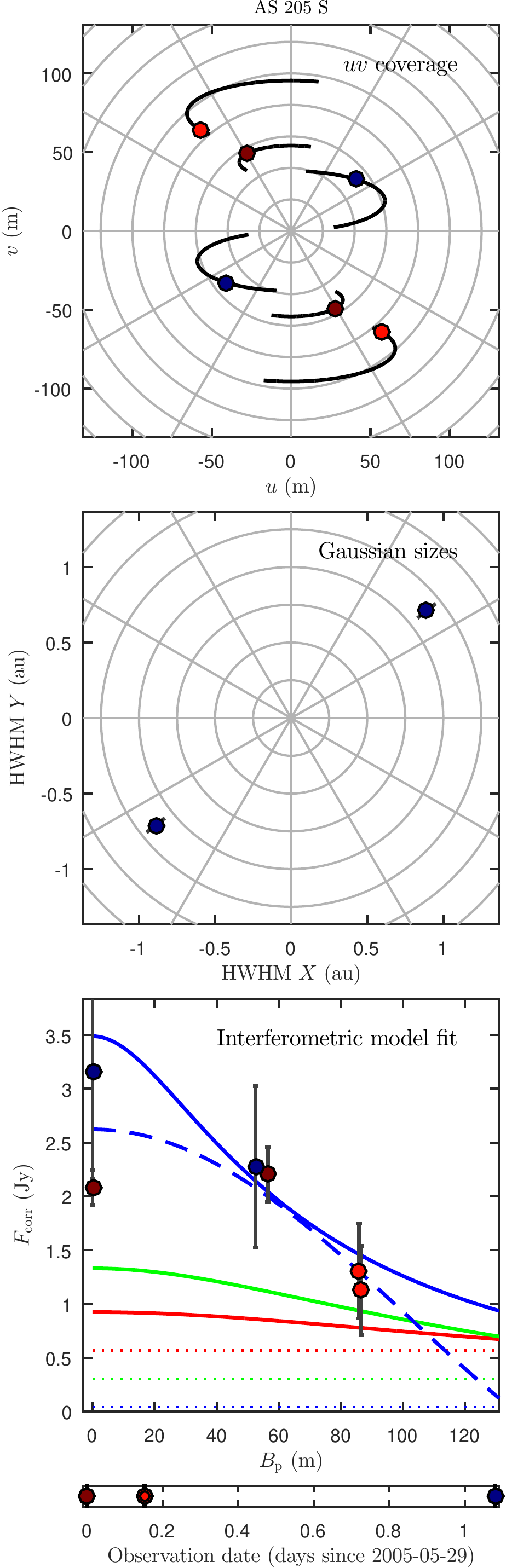}
			
		\end{figure*}
		
		\begin{figure*}[h!]
			\centering
			\includegraphics[width = 0.21\linewidth]{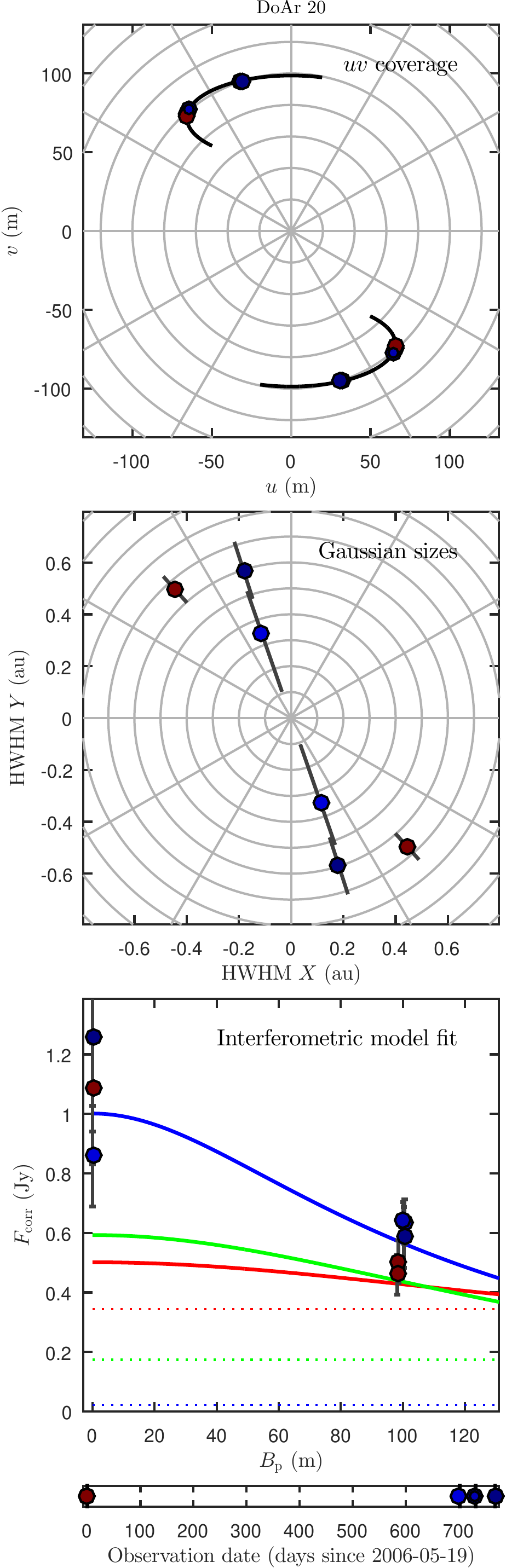}
			\includegraphics[width = 0.21\linewidth]{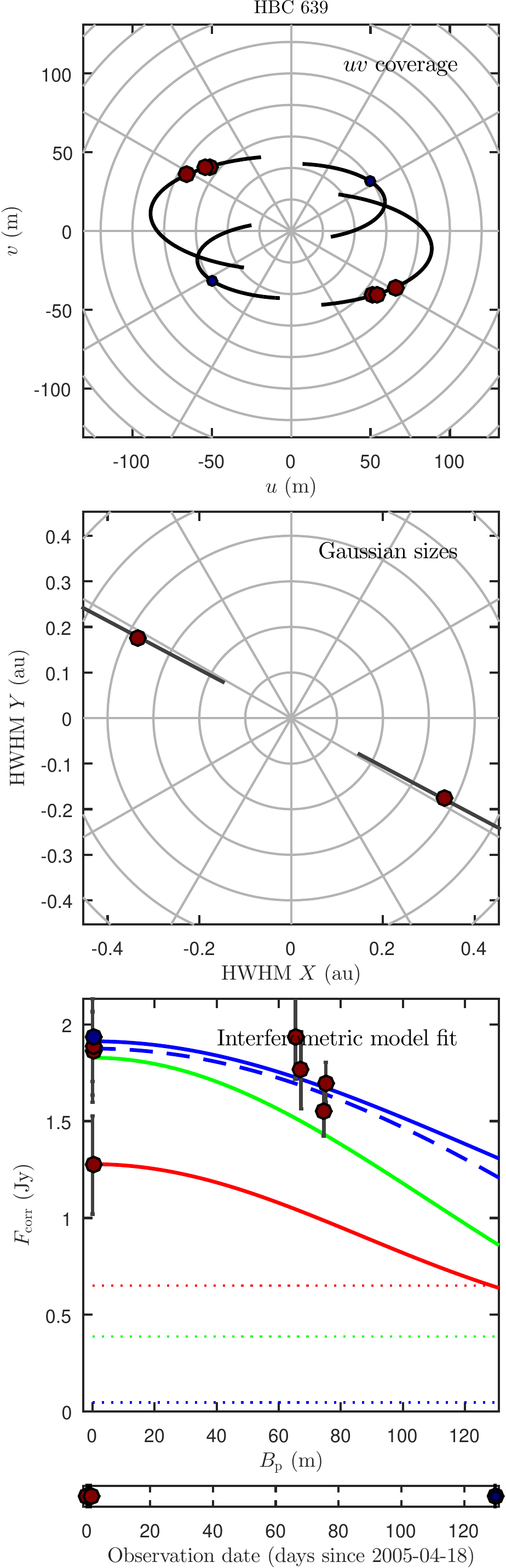}
			\includegraphics[width = 0.21\linewidth]{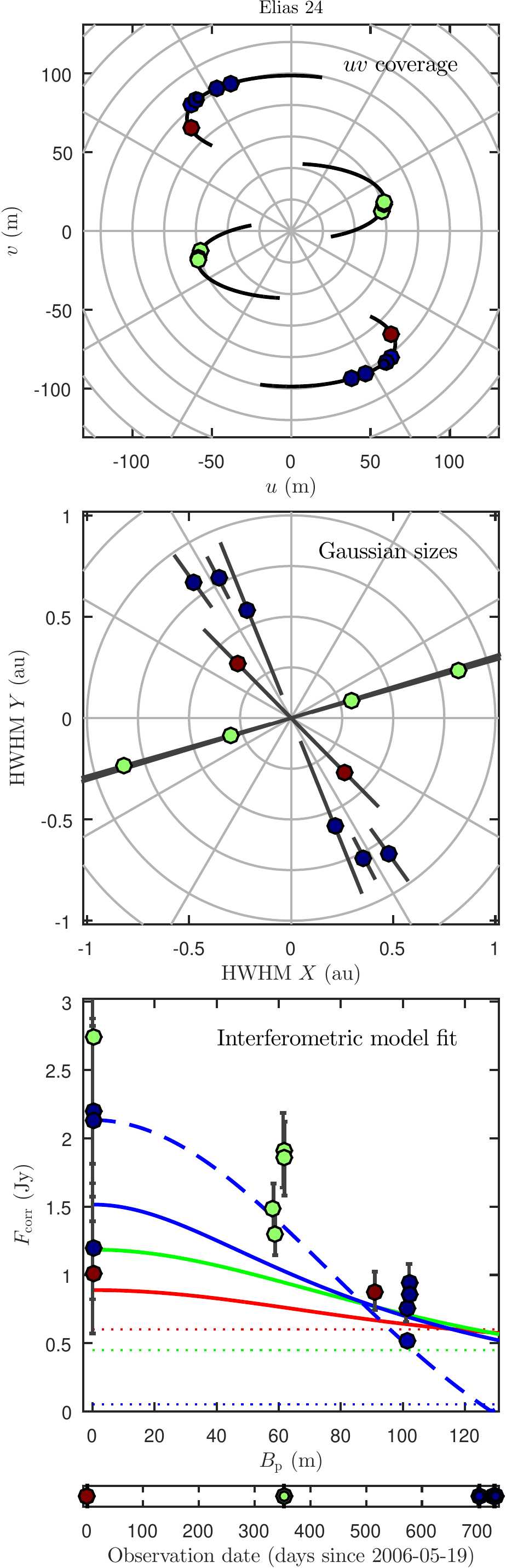}
			\includegraphics[width = 0.21\linewidth]{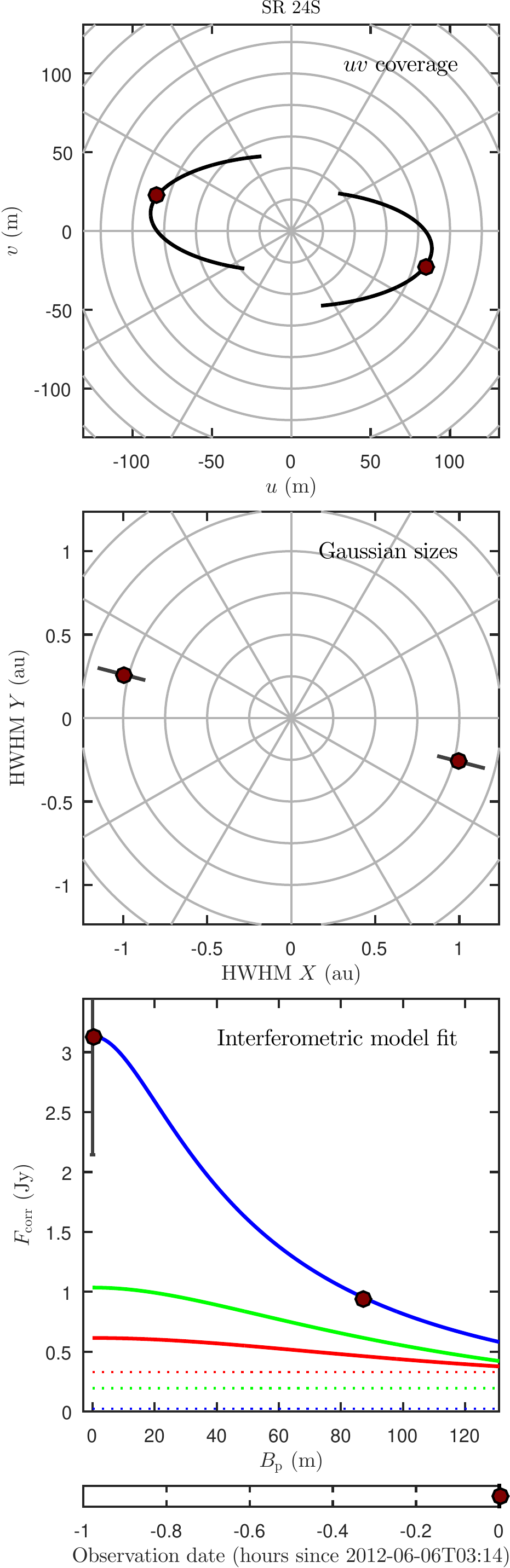}
			\includegraphics[width = 0.21\linewidth]{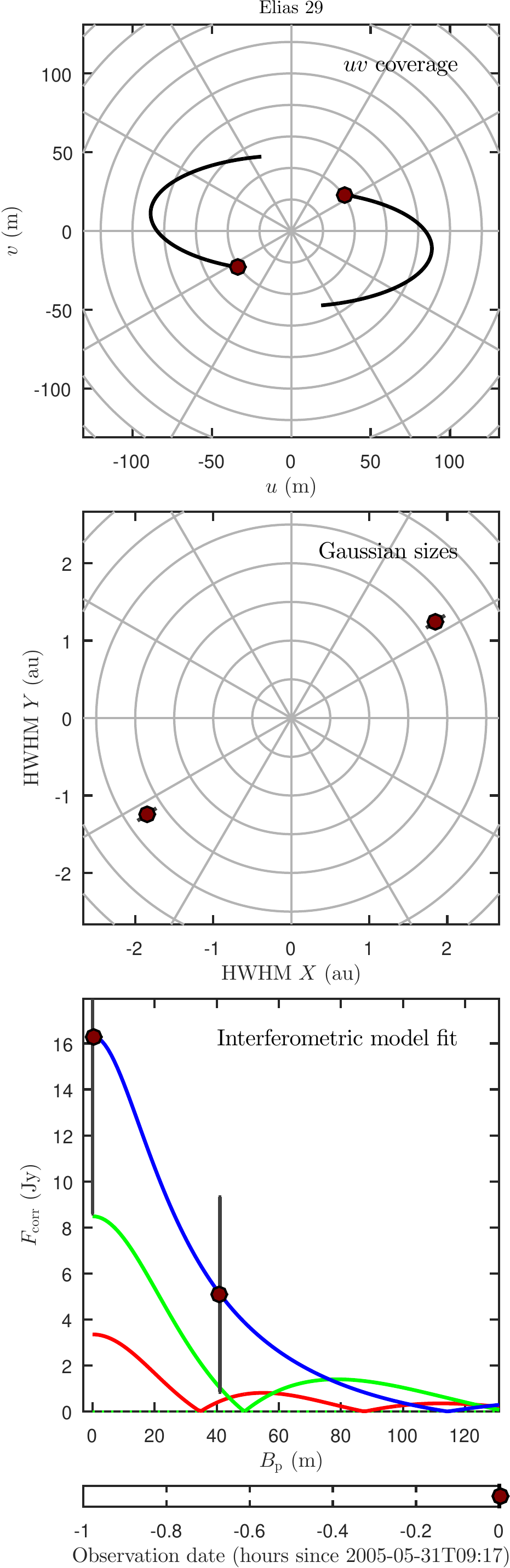}
			\includegraphics[width = 0.21\linewidth]{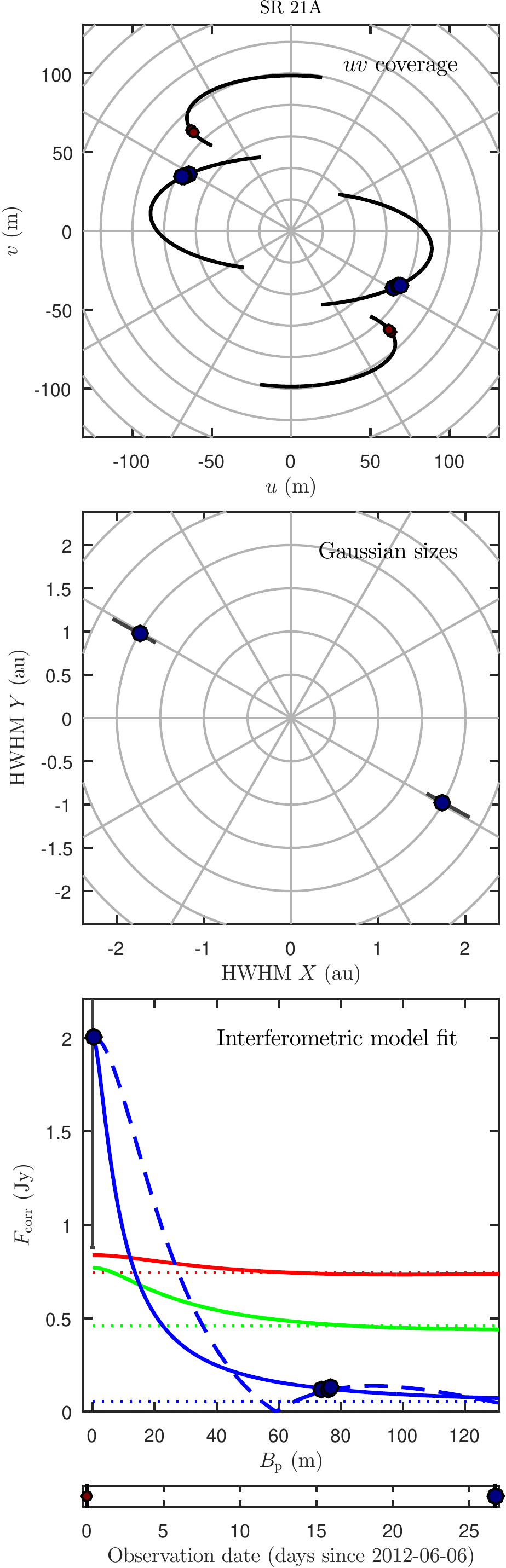}
			\includegraphics[width = 0.21\linewidth]{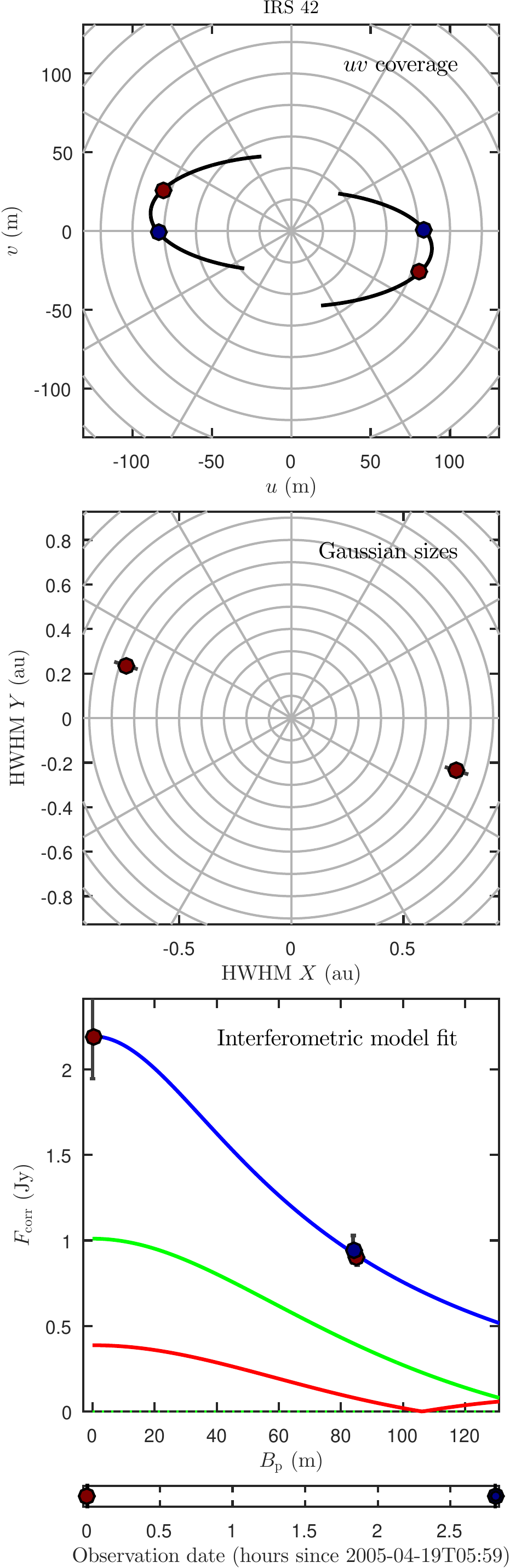}
			\includegraphics[width = 0.21\linewidth]{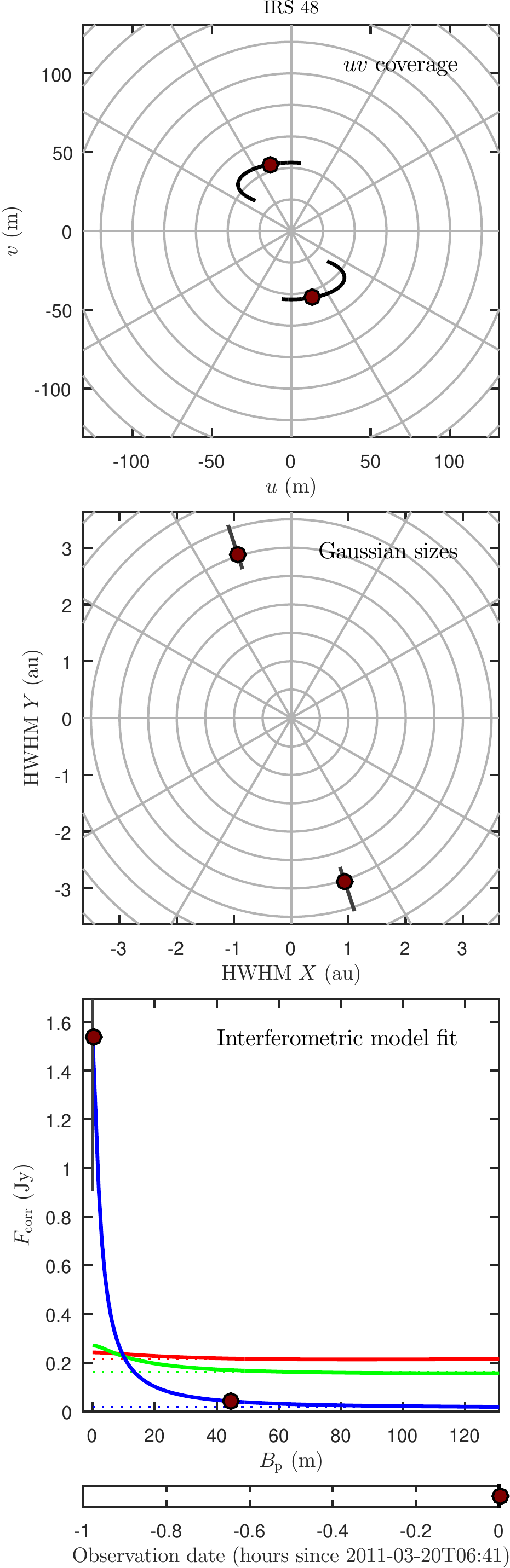}
			
		\end{figure*}
		
		\begin{figure*}[h!]
			\centering
			\includegraphics[width = 0.21\linewidth]{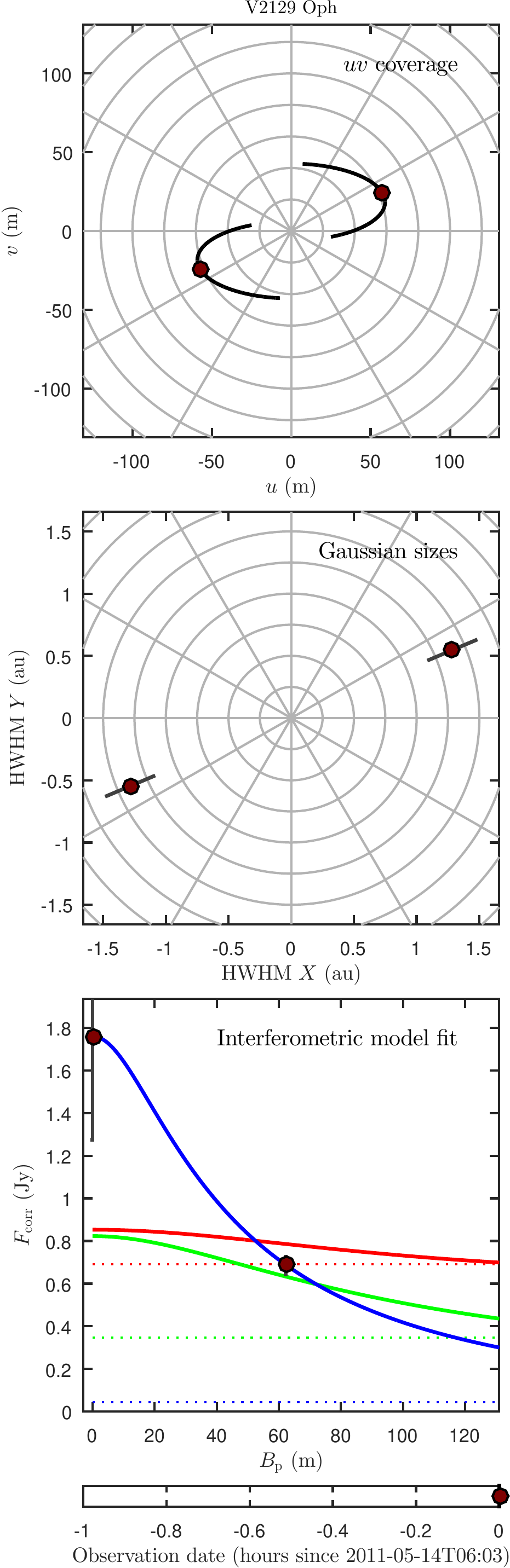}
			\includegraphics[width = 0.21\linewidth]{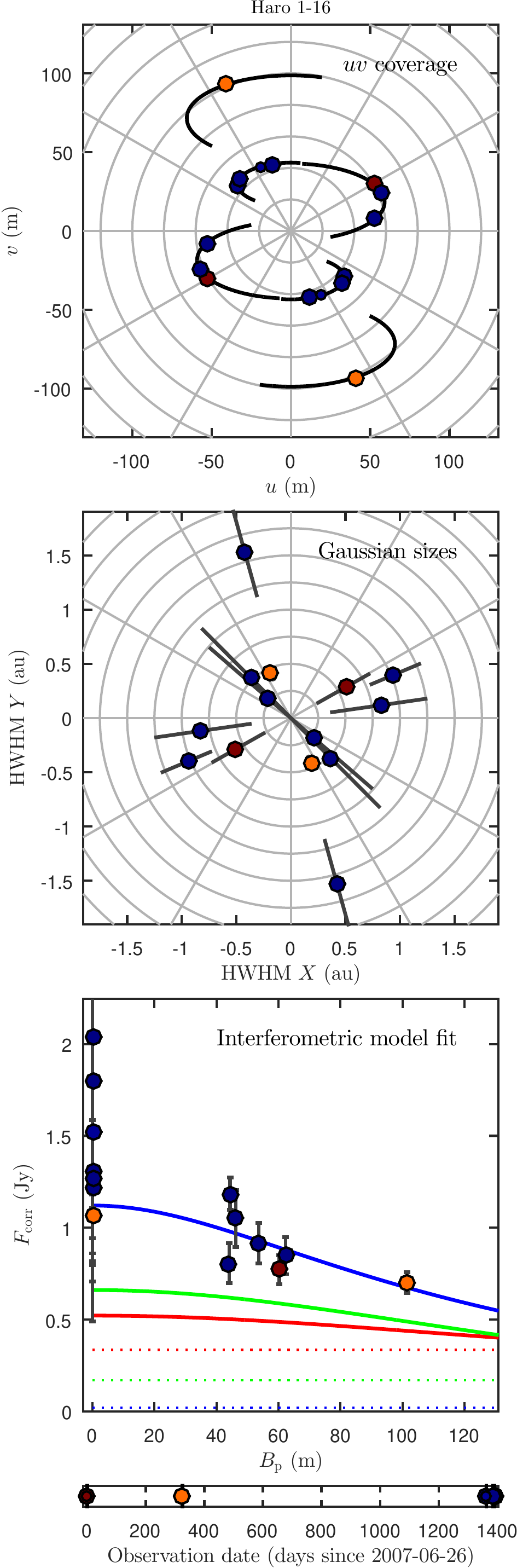}
			\includegraphics[width = 0.21\linewidth]{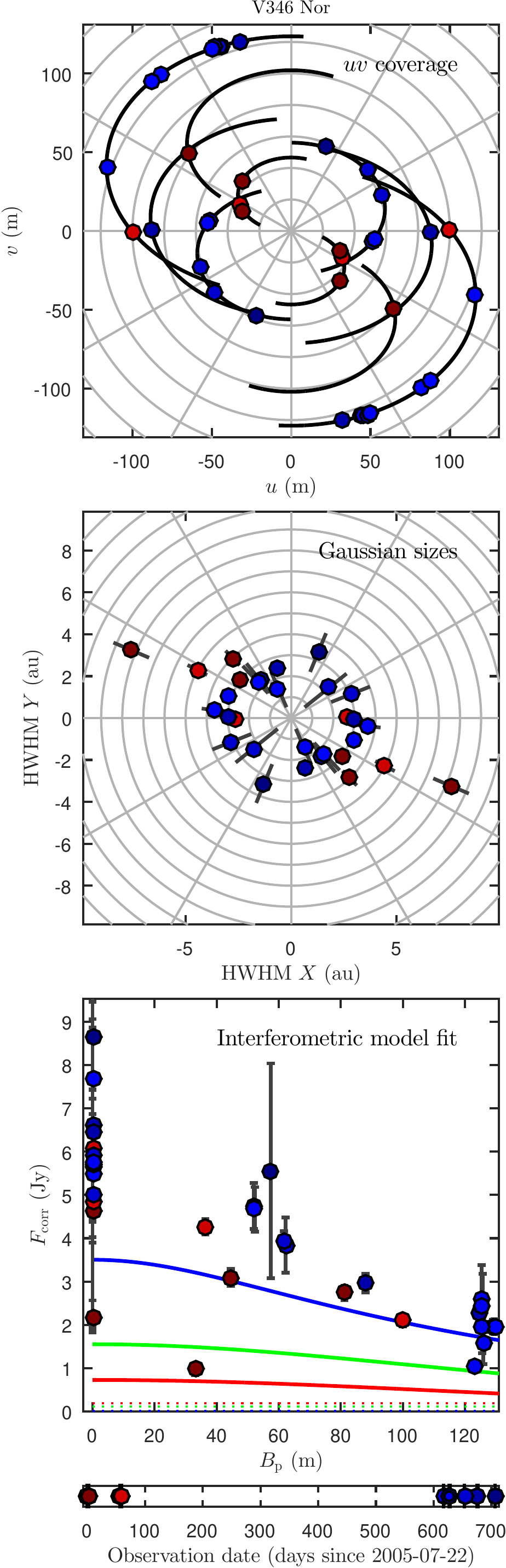}
			\includegraphics[width = 0.21\linewidth]{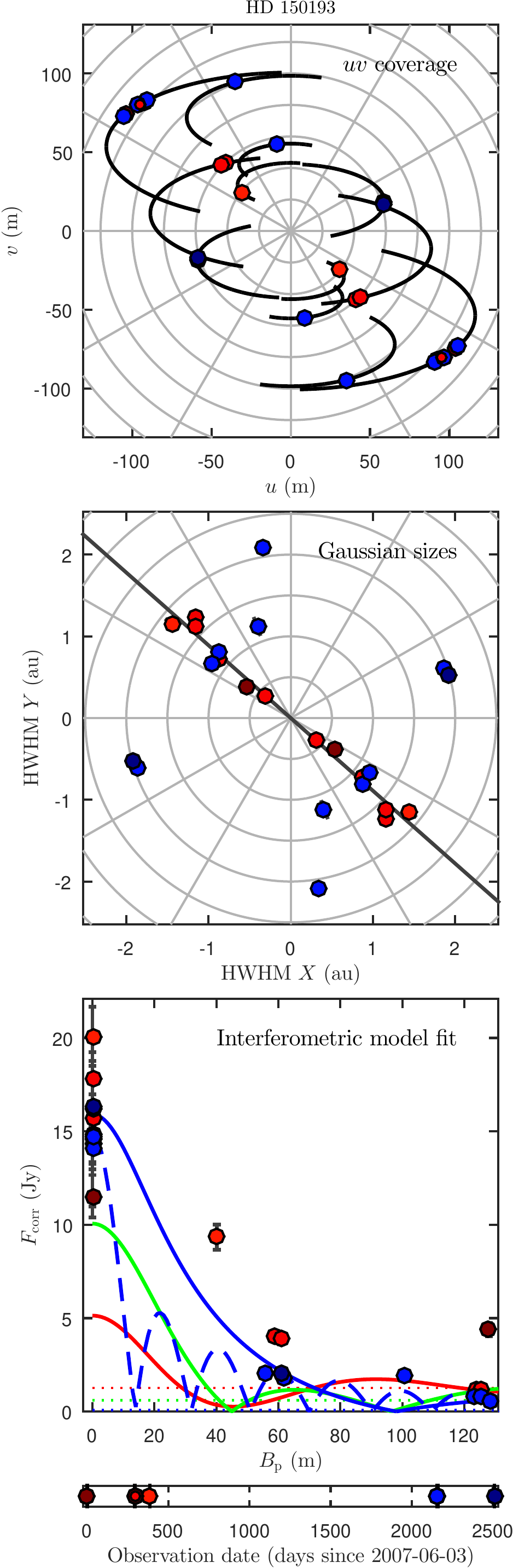}
			\includegraphics[width = 0.21\linewidth]{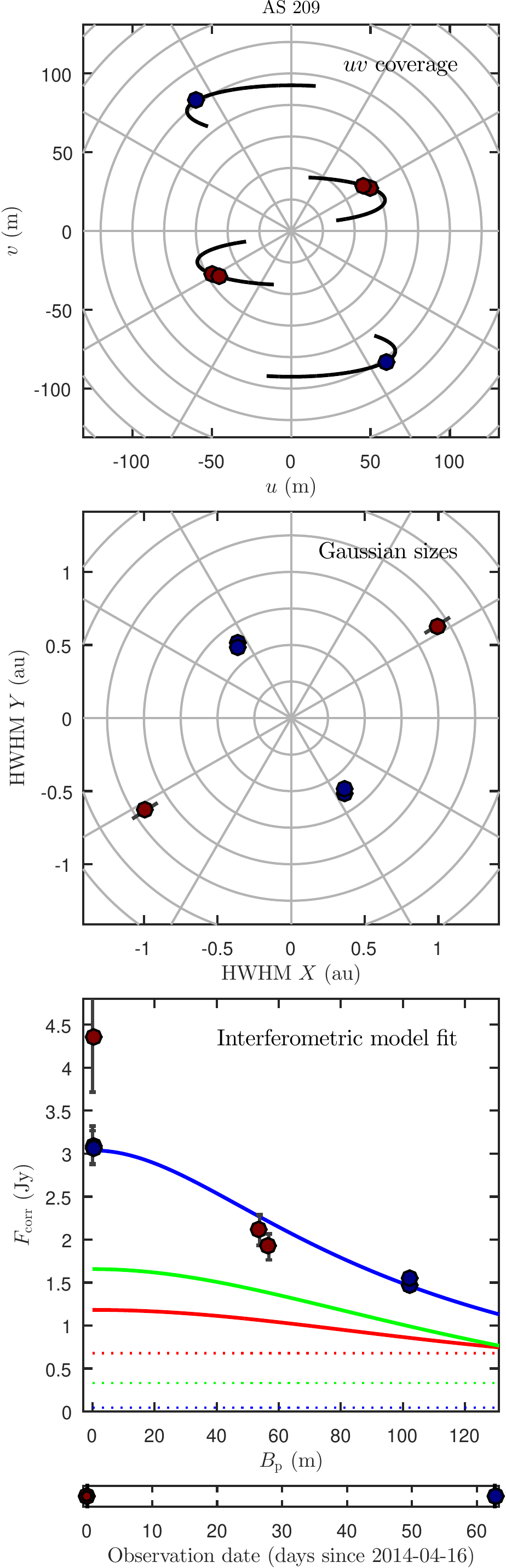}
			\includegraphics[width = 0.21\linewidth]{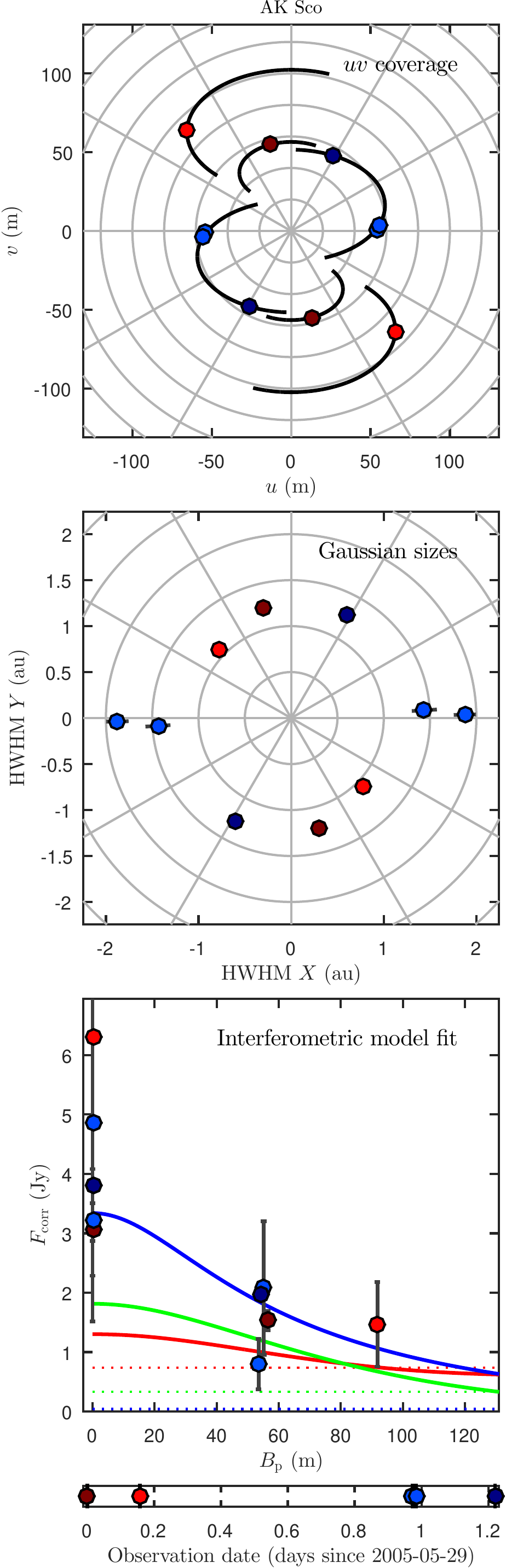}
			\includegraphics[width = 0.21\linewidth]{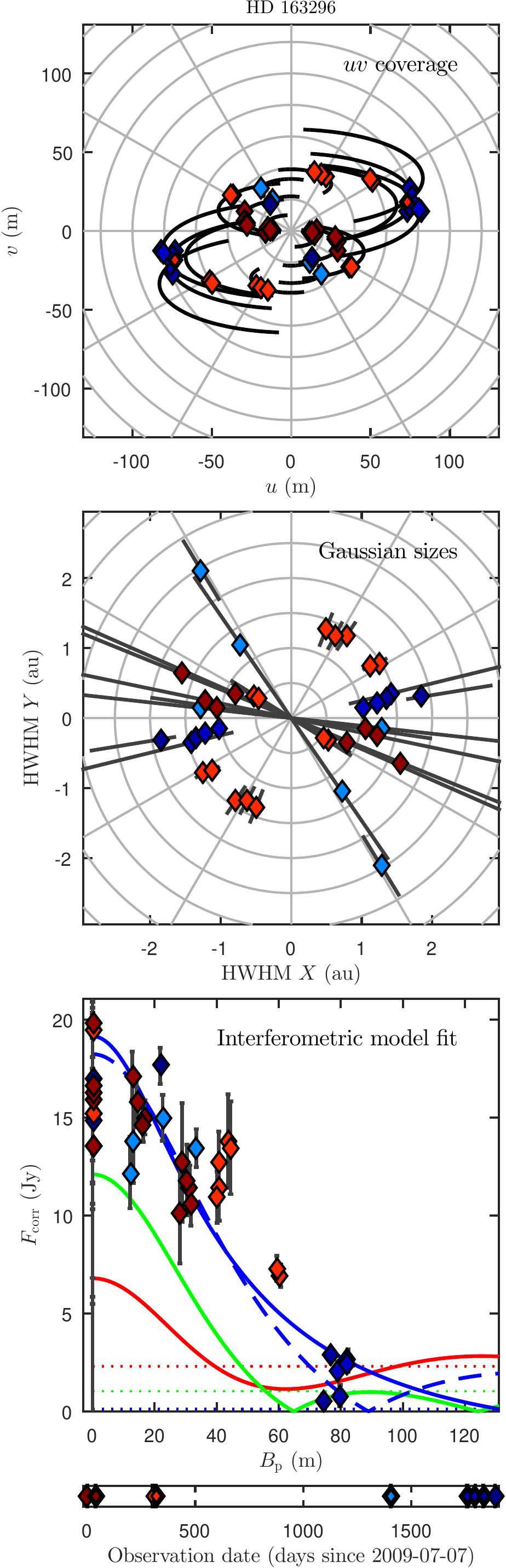}
			\includegraphics[width = 0.21\linewidth]{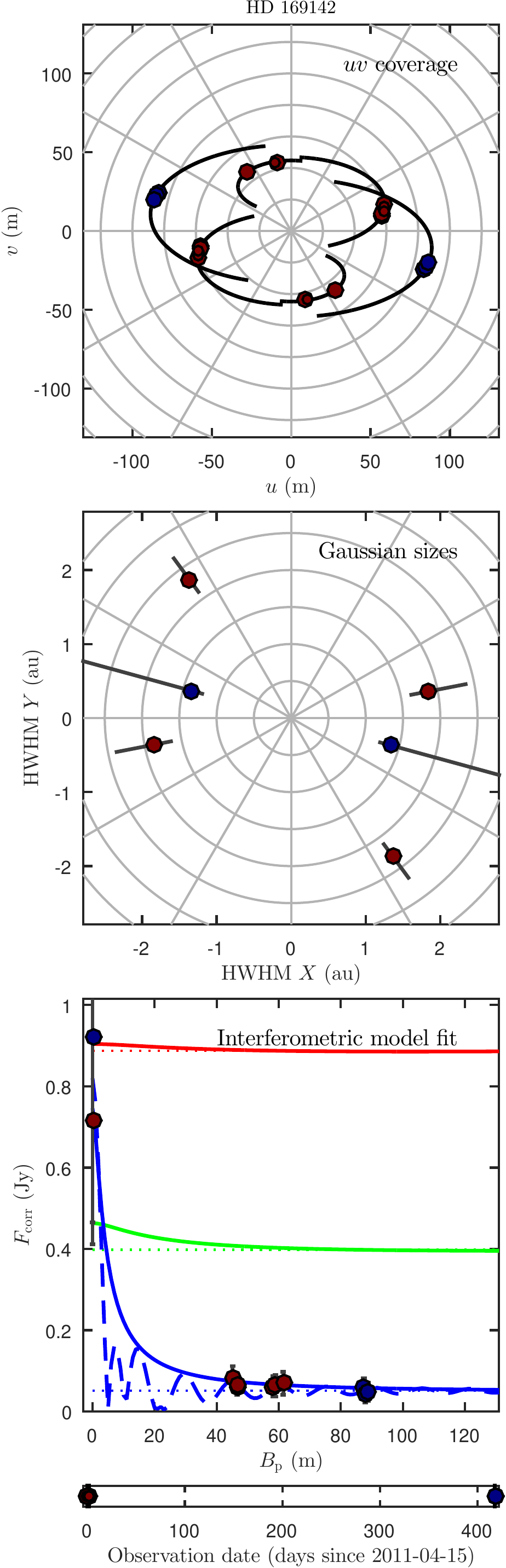}
			
		\end{figure*}
		
		\begin{figure*}[h!]
			\centering
			\includegraphics[width = 0.21\linewidth]{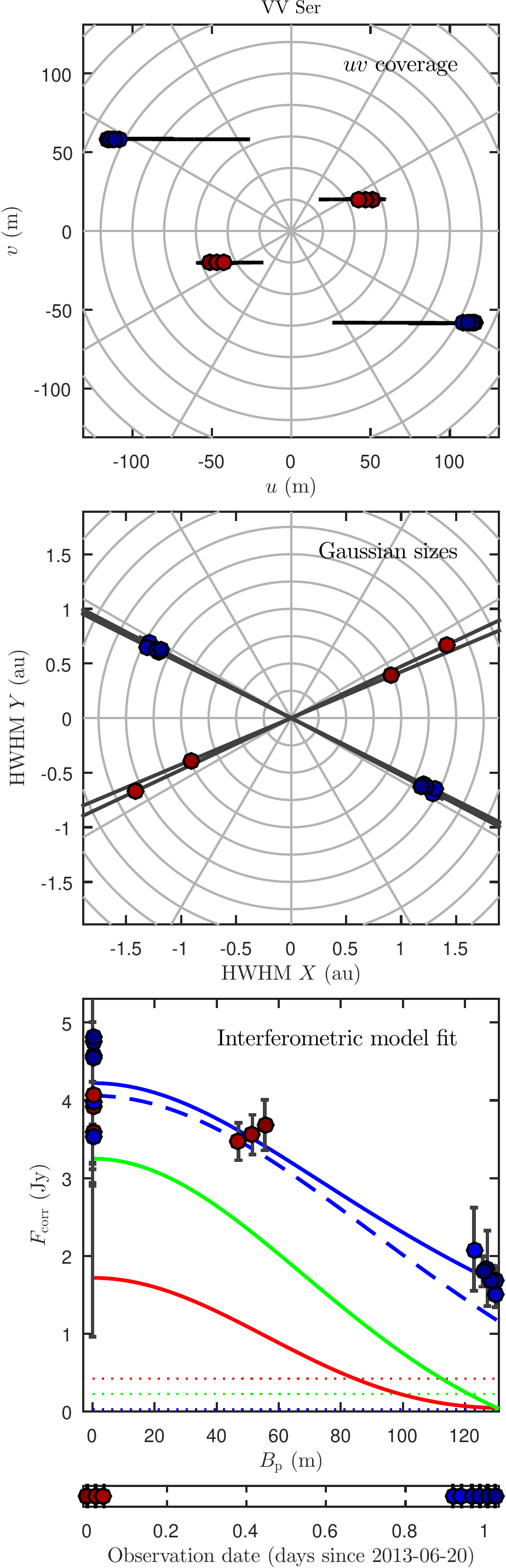}
			\includegraphics[width = 0.21\linewidth]{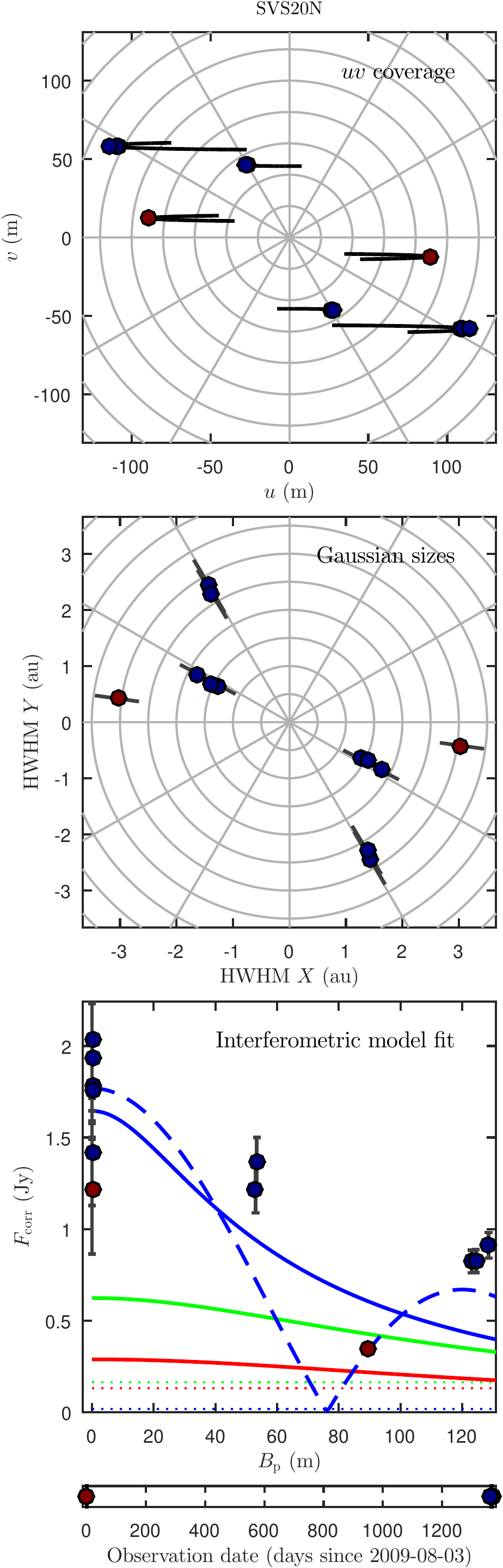}
			\includegraphics[width = 0.21\linewidth]{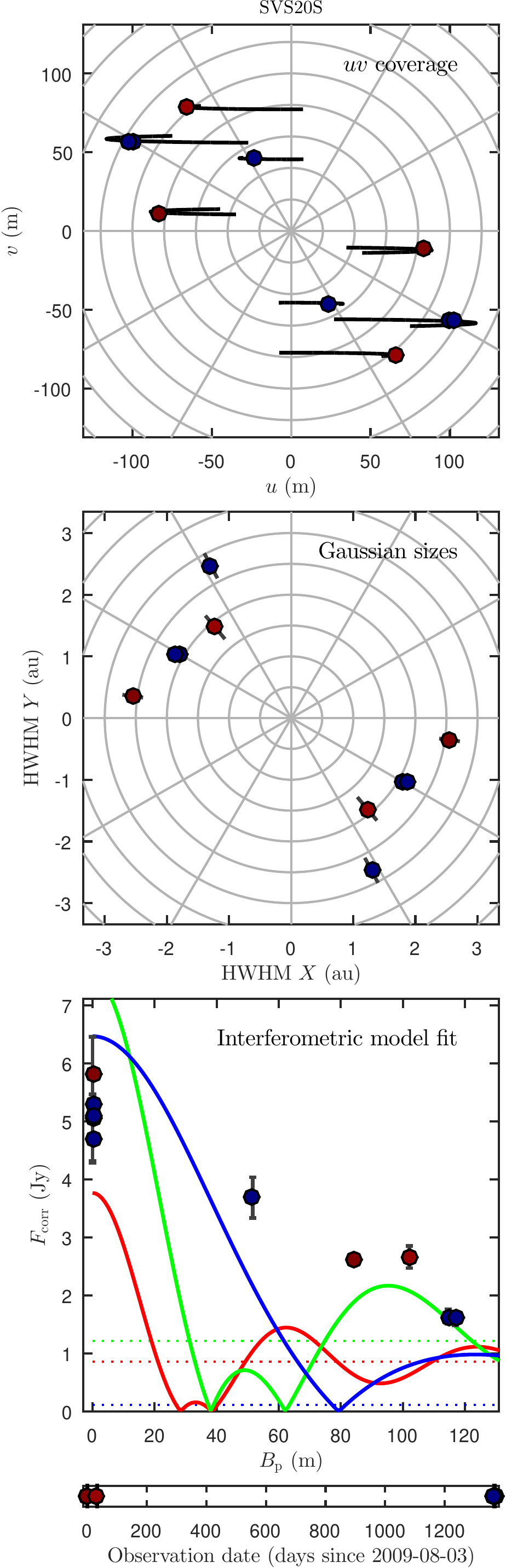}
			\includegraphics[width = 0.21\linewidth]{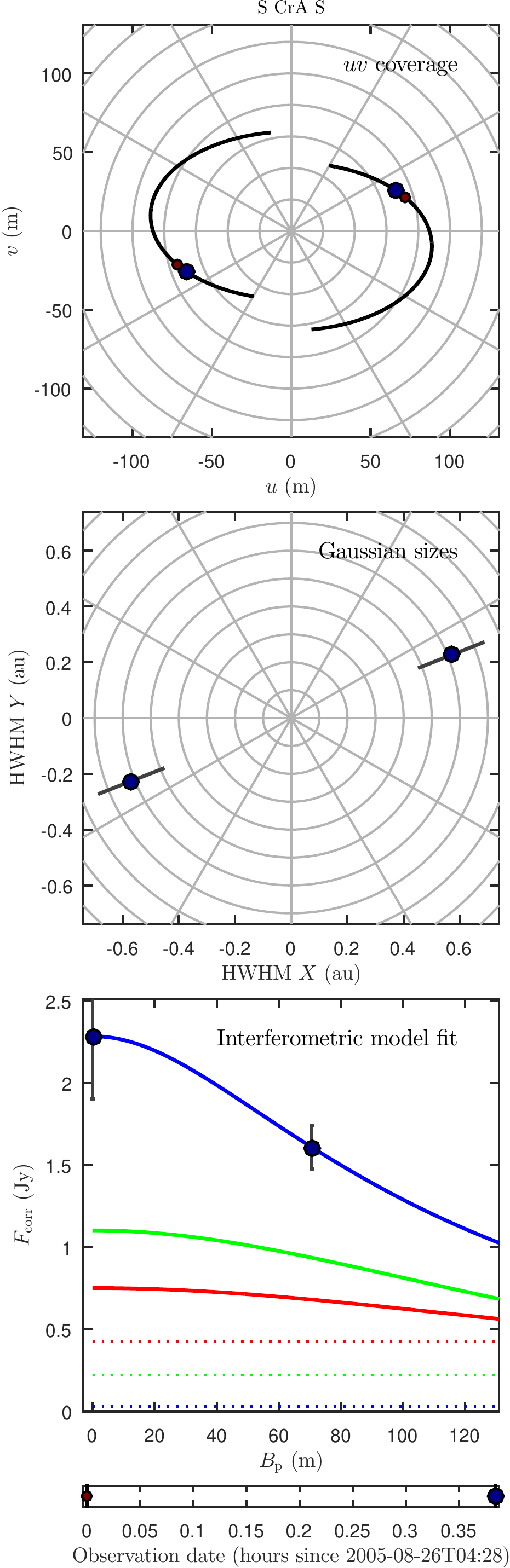}
			\includegraphics[width = 0.21\linewidth]{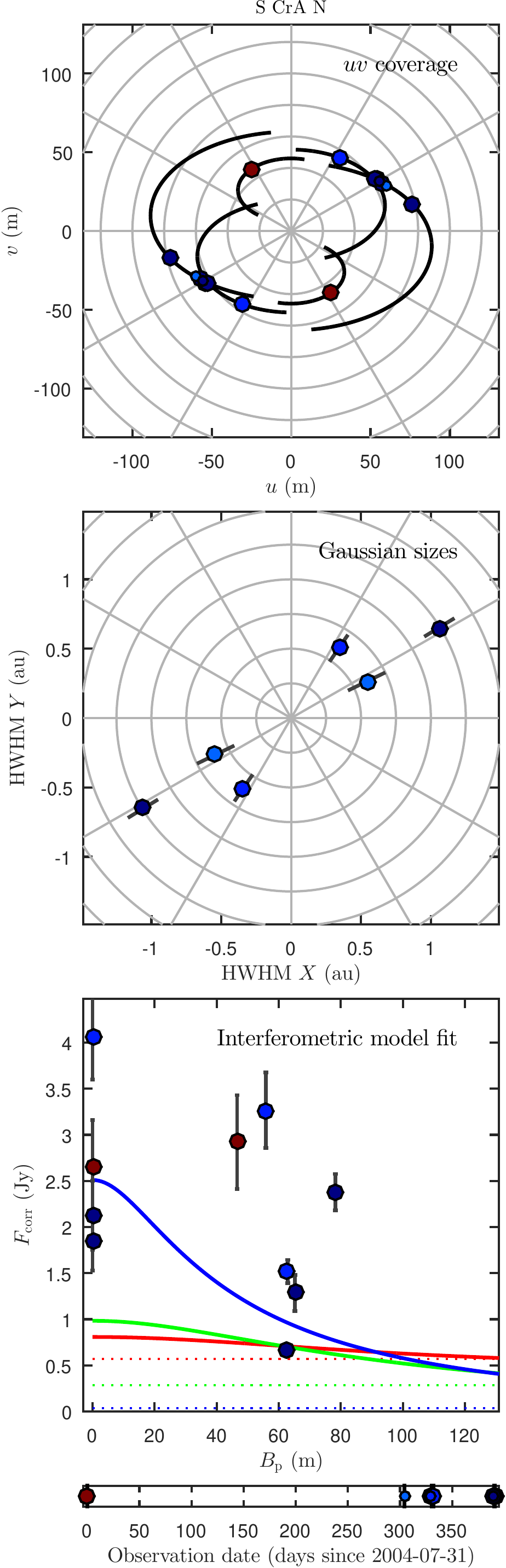}
			\includegraphics[width = 0.21\linewidth]{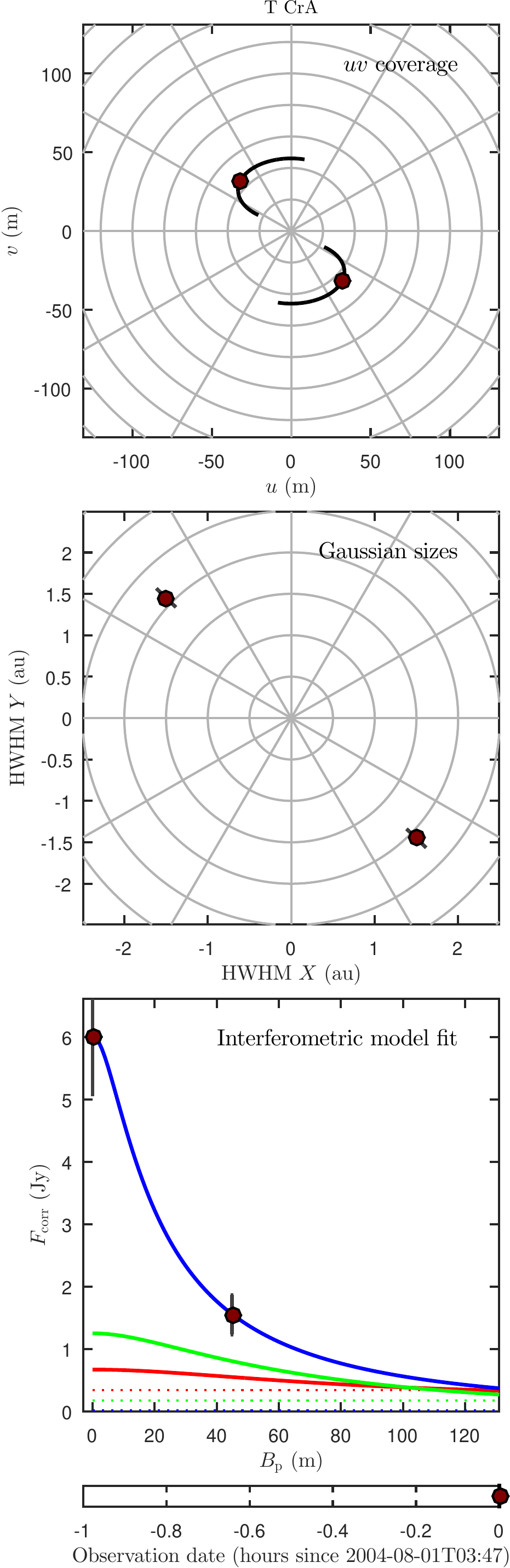}
			\includegraphics[width = 0.21\linewidth]{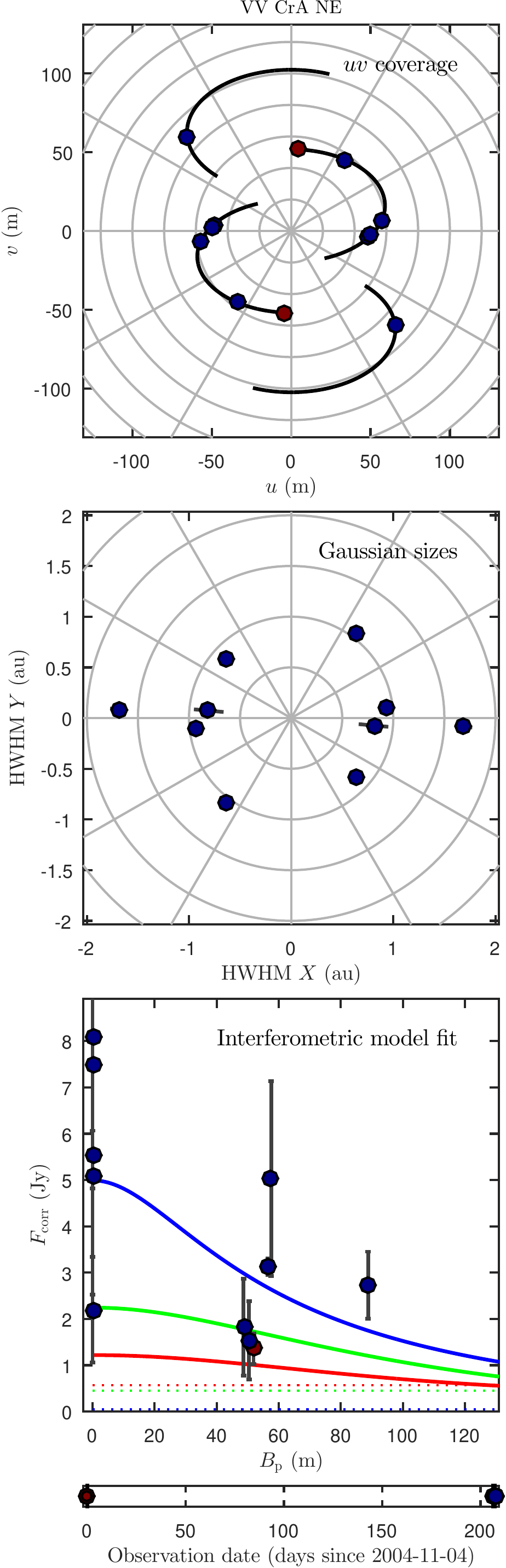}
			\includegraphics[width = 0.21\linewidth]{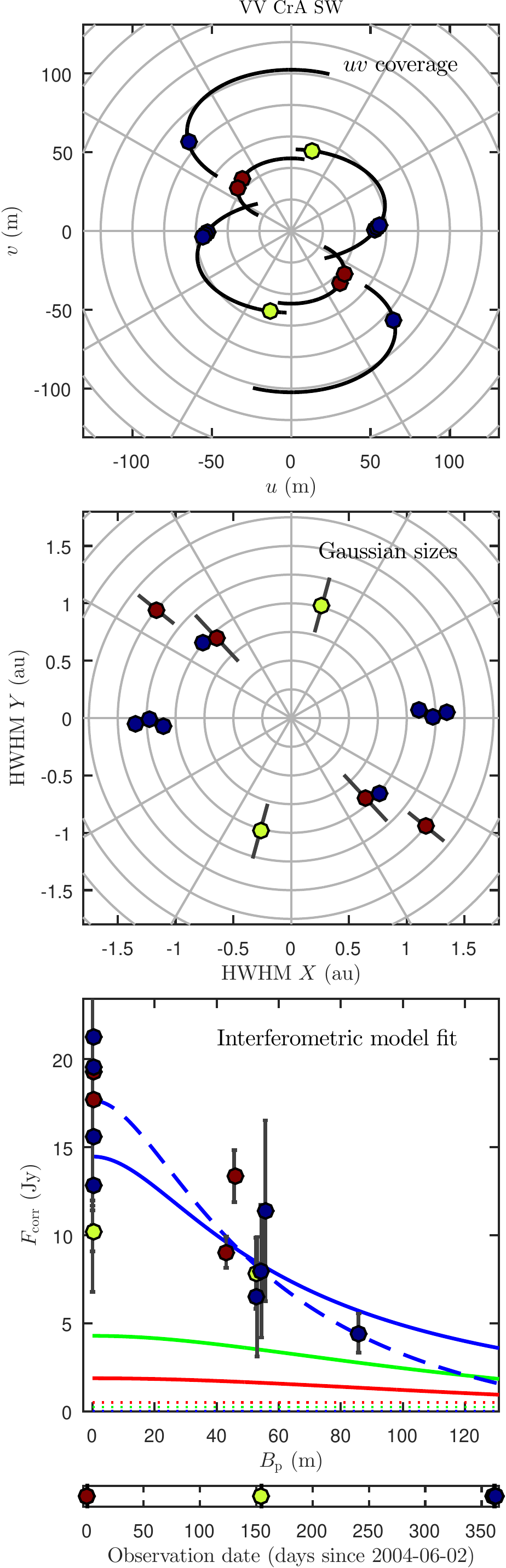}
			
		\end{figure*}
		
		\begin{figure*}[h!]
			\centering
			\includegraphics[width = 0.21\linewidth]{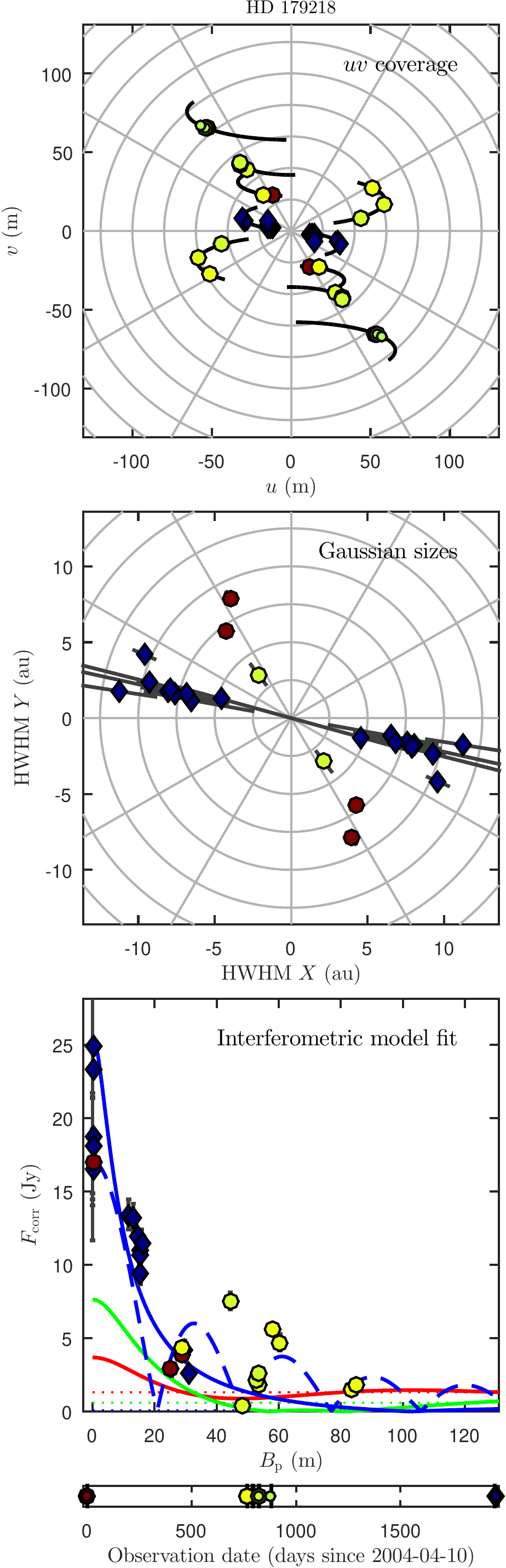}
			
		\end{figure*}

		\twocolumn
		\clearpage
		\section{Analysis of the spectra}
		\label{app:spec_class}
		
		We performed an automated analysis of the MIDI spectra to determine the basic emission characteristics, the amplitude of the silicate emission feature, and the $11.3/9.8~\mu$m flux ratio. The steps of the data processing are as follows:
		
		\begin{itemize}
			
			\item First, we fit a linear baseline to the spectra to evaluate the spectral continuum. At this step we do not know whether the feature is in emission or absorption. Hence we determine two alternative baselines, one which better fits an emission spectrum, the other which is better for an absorption spectrum. The baseline is given by two points, one from the $7.5-8.5\ \mu$m, the other from the $12-13\ \mu$m spectral window. At these wavelengths the continuum dominates over the feature.   
			The points defining the line are the lower (for emission) or upper (for absorption) $25\%$ percentile of the fluxes in each window. We use percentiles instead of minimum or maximum for robustness.
			
			\item In the next step we smooth the spectra with a third order Savitzky-Golay filter (using a window width of $1.8~\mu$m). As many of the original spectra are noisy, this step is essential to get robust parameter values.
			
			\item Now we determine the random and systematic uncertainties of the spectra. We detrend the spectra by subtracting the smoothed spectra from the original ones. Then we calculate the random errors ($\sigma_\mathrm{R}$) as the moving standard deviation of the detrended data with a window width of $0.6~\mu$m. The window width roughly corresponds to the real spectral resolution of MIDI in the low resolution mode ($R \approx 30$). Systematic errors ($\sigma_\mathrm{S}$) are then calculated as $\sigma_\mathrm{S}^2 = \sigma_\mathrm{T}^2 - \sigma_\mathrm{R}^2  $, where $\sigma_\mathrm{T}$ is the total measurement error. In the following we use the random errors to estimate the uncertainties of the derived parameters.
			
			\item Then we calculate the normalized spectra ($F_{\nu\mathrm{,\,norm}}$) as $F_{\nu\mathrm{,\,norm}} = \left(F_{\nu\mathrm{,\,smooth}} - F_{\nu\mathrm{,\,cont}}\right) / \langle F_{\nu\mathrm{,\,cont}} \rangle + 1 $, where $F_{\nu\mathrm{,\,smooth}}$ is the smoothed spectrum, $F_{\nu\mathrm{,\,cont}}$ is the fitted continuum, and $\langle F_{\nu\mathrm{,\,cont}} \rangle$ is the mean of the continuum. This normalization has the advantage of preserving the feature shape, and was earlier used by other authors \citep[e.g.,][]{vanBoekel2003}. We also compute the feature spectra ($F_{\nu\mathrm{,\,subt}}$) by simply subtracting the continuum from the smoothed spectra. We calculate two sets of $F_{\nu\mathrm{,\,norm}}$ and $F_{\nu\mathrm{,\,subt}}$ using either the upper or lower continuum fits.
			
			\item We determine the equivalent width ($W$)  of the feature, expressed in frequency units. Equivalent width is defined as  
			\begin{equation}
			W = \int\limits_{\lambda_1}^{\lambda_2} \left(1-F_{\nu\mathrm{,\ norm}} \right) \mathrm{d}\lambda,
			\end{equation}
			with $\lambda_1 = 8\,\mu$m and $\lambda_2 = 12\,\mu$m. Equivalent width also has alternative values, $W^\mathrm{u}$ and $W^\mathrm{l}$ using the upper and lower continuum fits, respectively.
			
			\item Now we determine whether the spectra show emission or absorption feature. The spectra are categorized into three groups using $W$: a) If both $W^\mathrm{u}$ and $W^\mathrm{l}$ are positive, and $W^\mathrm{l} > 0.33\ \mu$m, then we classify the spectrum as absorption; b) If both $W^\mathrm{u}$ and $W^\mathrm{l}$ are negative, and $W^\mathrm{l} < -0.33\ \mu$m, we classify it as emission; c) Finally, when $W^\mathrm{u}$ and $W^\mathrm{l}$ have different signs or $\left|W \right| < 0.33\ \mu$m, we designate the spectrum as having a weak feature.
			This value was chosen in order to be able to robustly discriminate between absorption and emission features with the typical uncertainties of MIDI spectra. Our categories are indicated in Table~\ref{tab:obs} for all correlated spectra. In the following, we use the lower continuum fit for the spectra classified as emission or weak feature, and the upper continuum fit for spectra classified as absorption.
			
			\item Then we calculate the peak amplitude of $F_{\nu\mathrm{,\ norm}}$, where $F_\mathrm{peak,\ norm} = \mathrm{max}\left(F_{\nu\mathrm{,\ norm}}\right)$ for spectra with emission or weak features, and $F_\mathrm{peak,\ norm} = \mathrm{min}\left(F_{\nu\mathrm{,\ norm}}\right)$ for spectra with absorption. We also determine the value of $F_{\nu\mathrm{,\,subt}}$ at $9.8\ \mu$m ($F_{\nu,\,\mathrm{\,subt,}\,9.8}$) and $11.3\ \mu$m ($F_{\nu,\,\mathrm{\,subt,}\,11.3}$). In the case of total spectra, the $9.4-10~\mu$m region, affected by the telluric ozone feature, is excluded from the calculation of the peak amplitude. Finally, we compute the flux ratio $F_{\nu,\,\mathrm{\,subt,}\,11.3}/F_{\nu,\,\mathrm{\,subt,}\,9.8}$, which we denote as $F_{11.3}/F_{9.8}$. 
			
		\end{itemize}    
		
		\clearpage
		\onecolumn
		\section{Observability plots}
		\label{app:matisse_obs}
		\begin{figure*}[h!]
			\centering
			\caption{Observability plots for MATISSE: $L$ versus $N$ band (top row), $K$ versus $N$ band (middle row), and $K$ versus $L$ band (bottom row) total (left column) and correlated fluxes (right column), estimated for our sample, from the continuous disk modeling. Correlated fluxes are calculated at 15~m baseline. The color of the symbols indicates the object type: Herbig Ae (blue), T Tauri (red), and  eruptive systems (orange). Vertical solid and dashed lines indicate the expected MATISSE $N$ or $L$ band sensitivity limits with or without using an external fringe tracker, respectively. Blue and red lines represent the sensitivity limits with ATs and UTs, respectively. Horizontal dashed lines indicate the $K$ sensitivity limits of the external fringe tracker with ATs or UTs (in the middle and bottom rows). In the top row, horizontal lines indicate the $L$ band sensitivity limits. Sources in the green shaded area can be observed either with the UTs and ATs, objects in the orange shaded region can only be observed with UTs. Flux limits were taken from \citet{Matter2016b}.} 
			\label{fig:obs_plot_app}
			\includegraphics[width=0.49\textwidth]{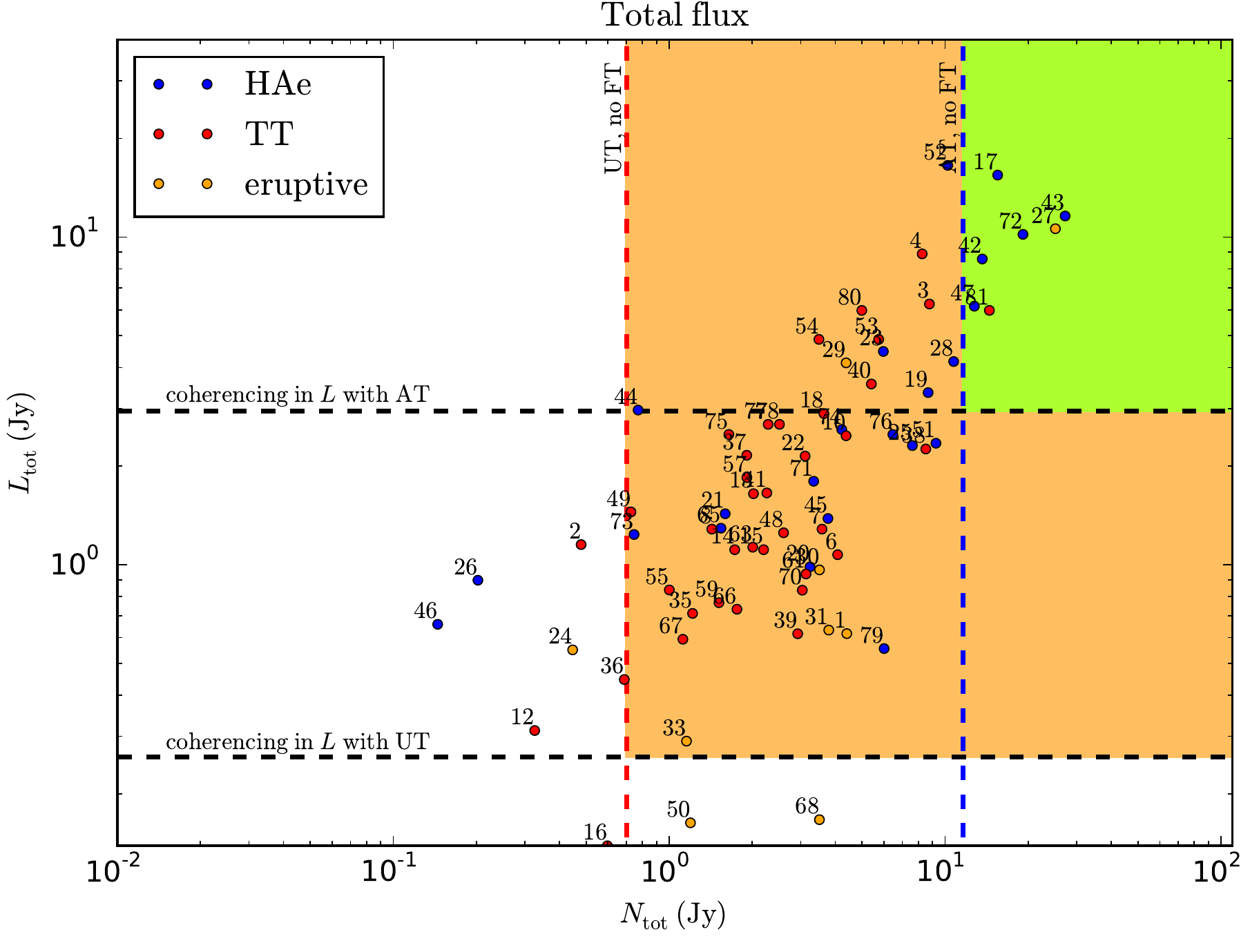}
			\includegraphics[width=0.49\textwidth]{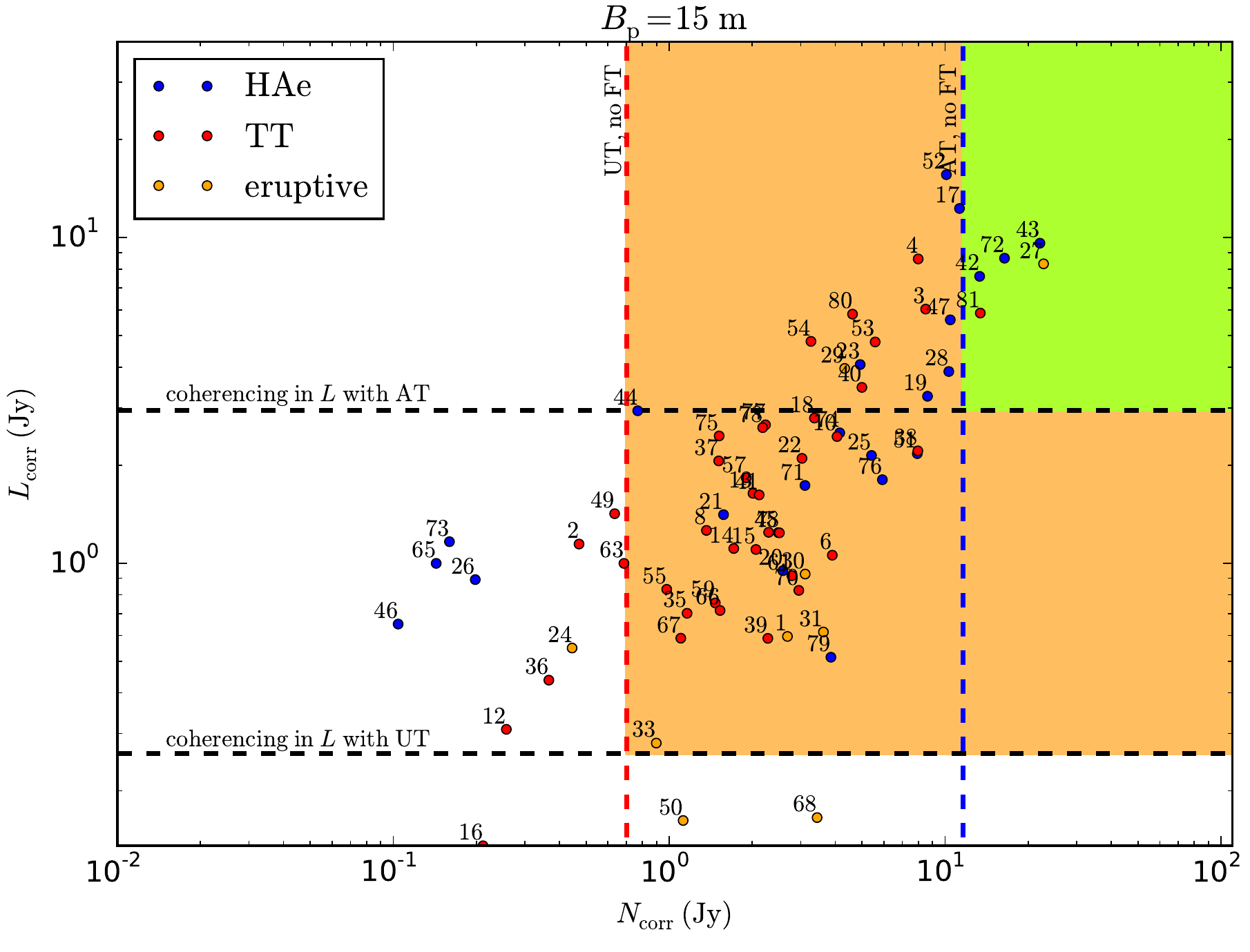}
			\includegraphics[width=0.49\textwidth]{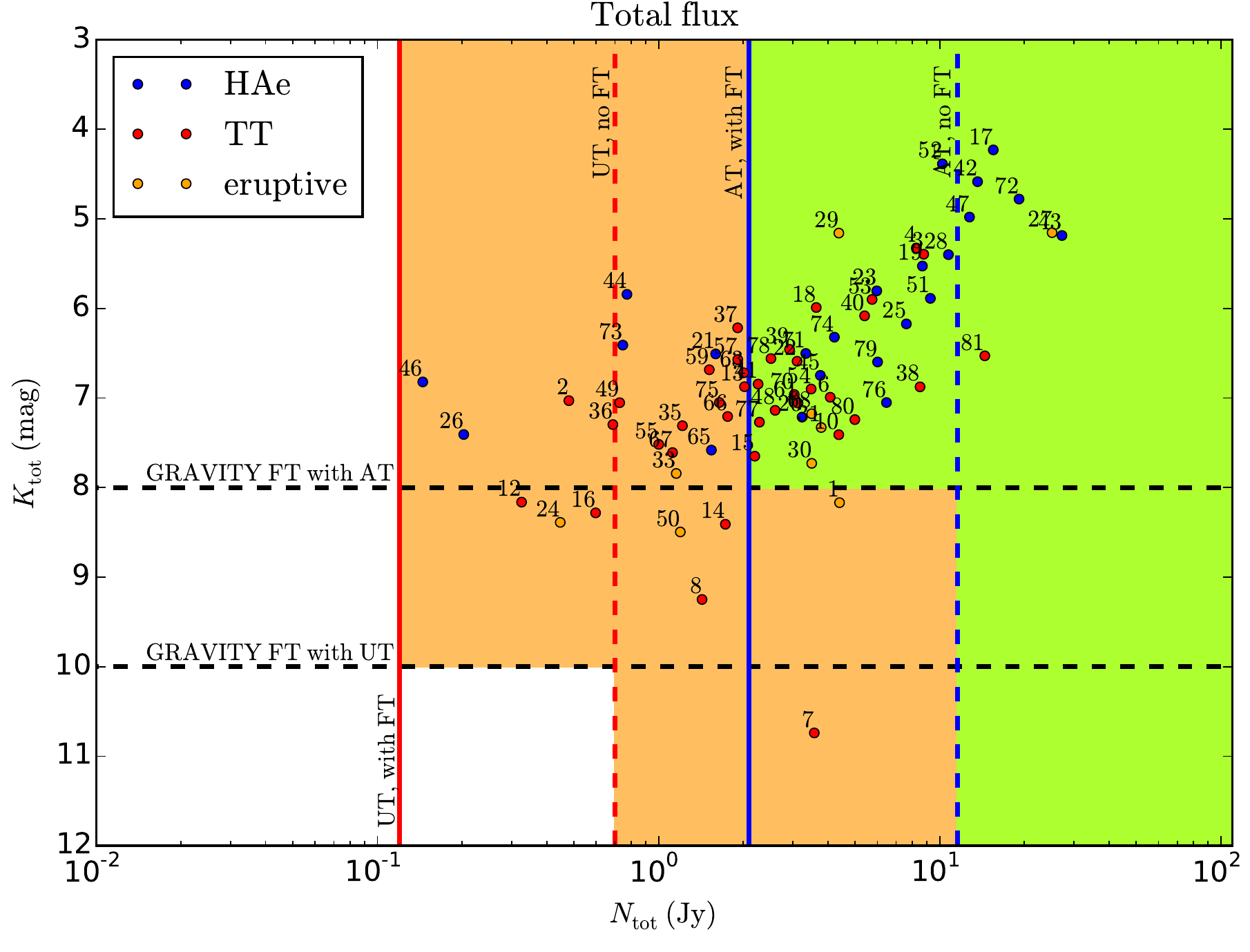}
			\includegraphics[width=0.49\textwidth]{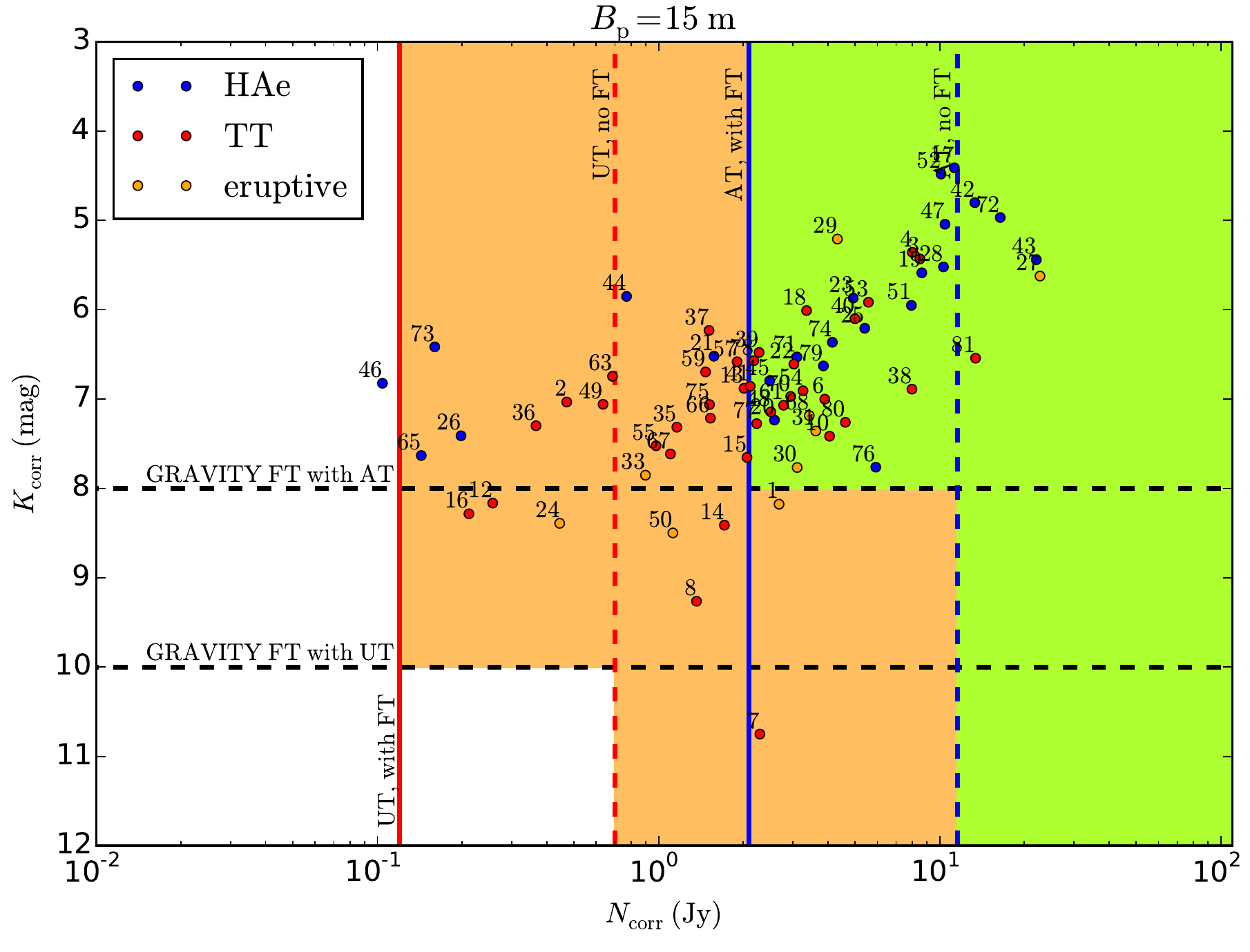}
			\includegraphics[width=0.49\textwidth]{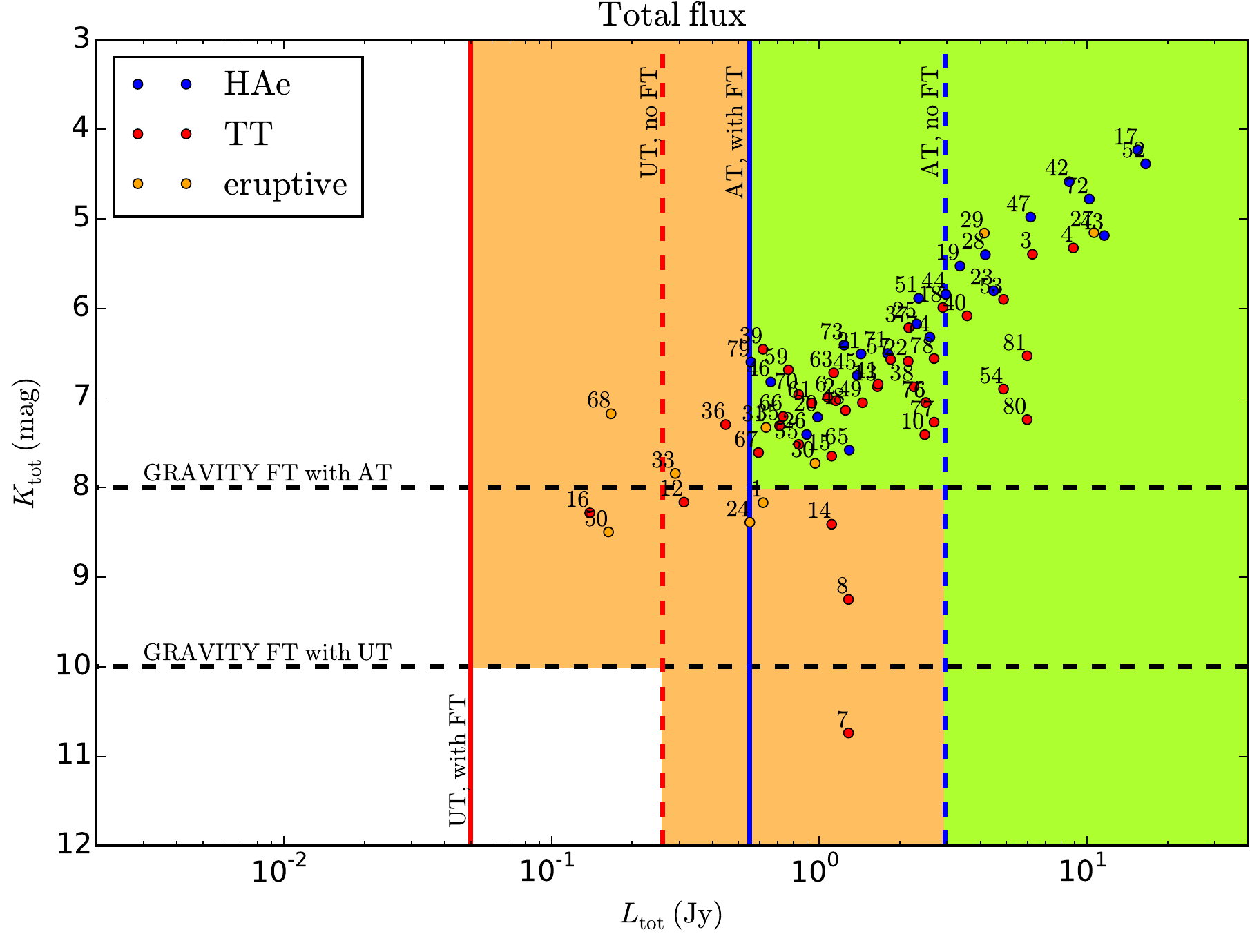}
			\includegraphics[width=0.49\textwidth]{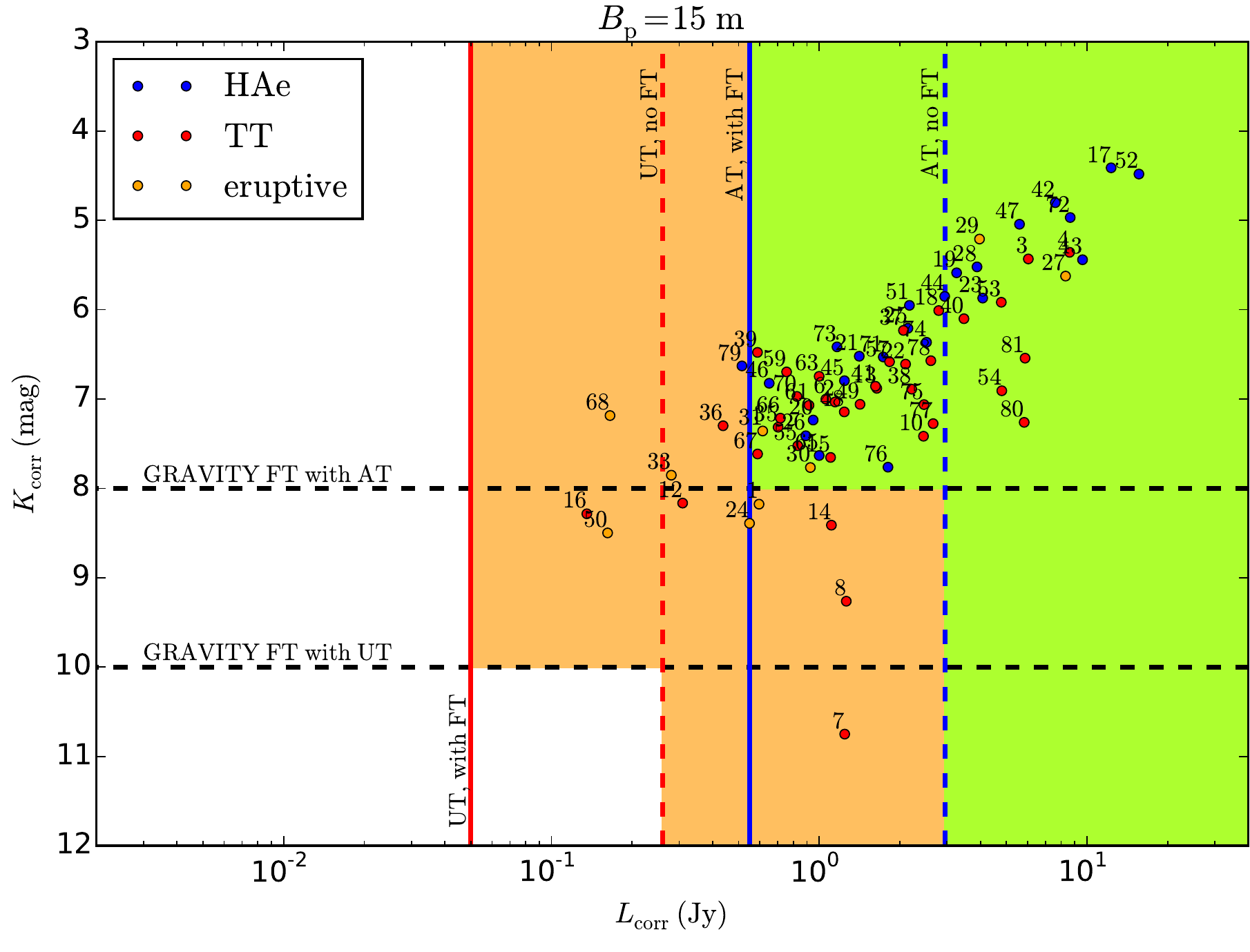}
		\end{figure*}
		
		\begin{figure*}[h!]
			\centering
			\includegraphics[width=0.49\textwidth]{matisse_observability_LN_10um7_30m-eps-converted-to.pdf}
			\includegraphics[width=0.49\textwidth]{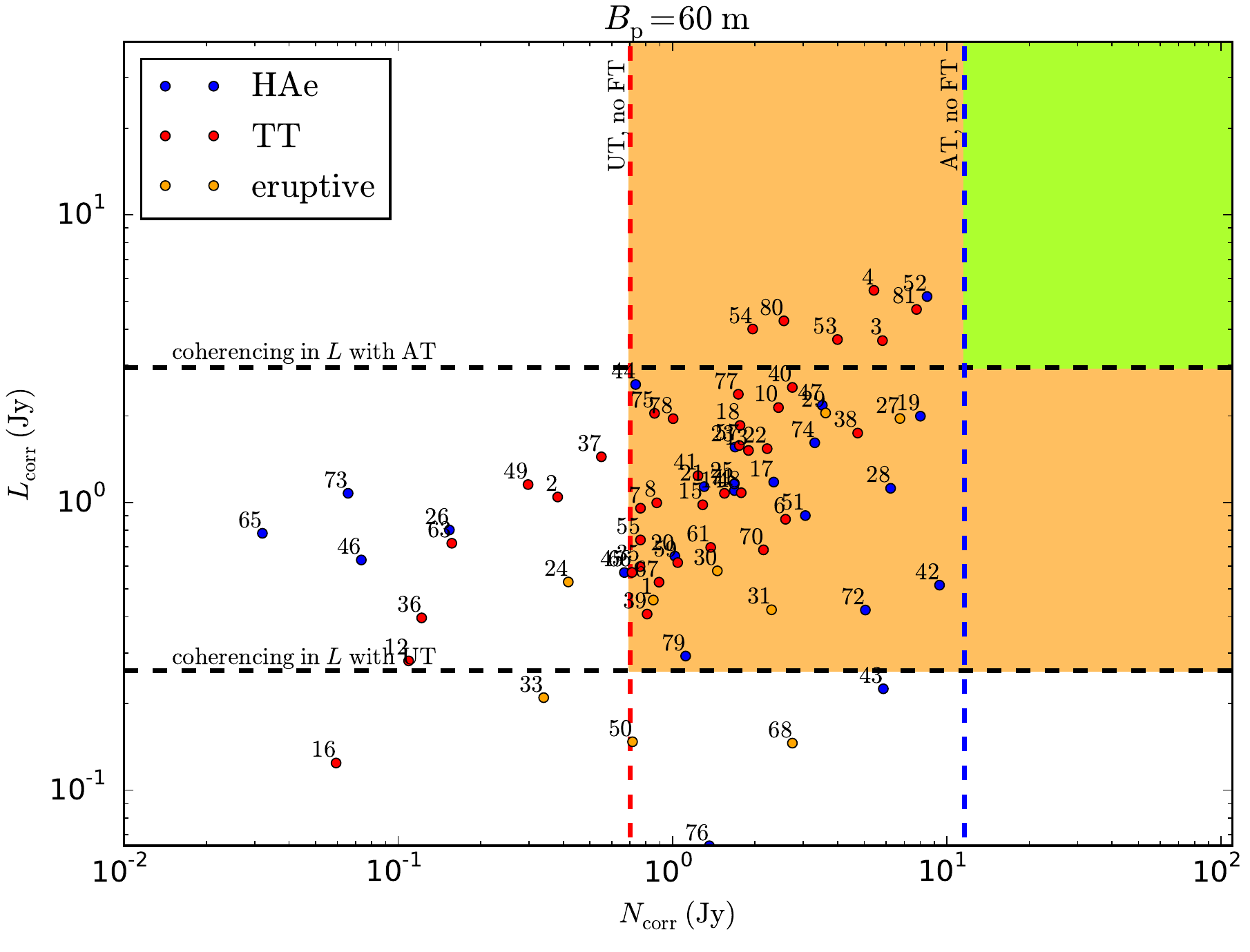}
			\includegraphics[width=0.49\textwidth]{matisse_observability_N_10um7_30m-eps-converted-to.pdf}
			\includegraphics[width=0.49\textwidth]{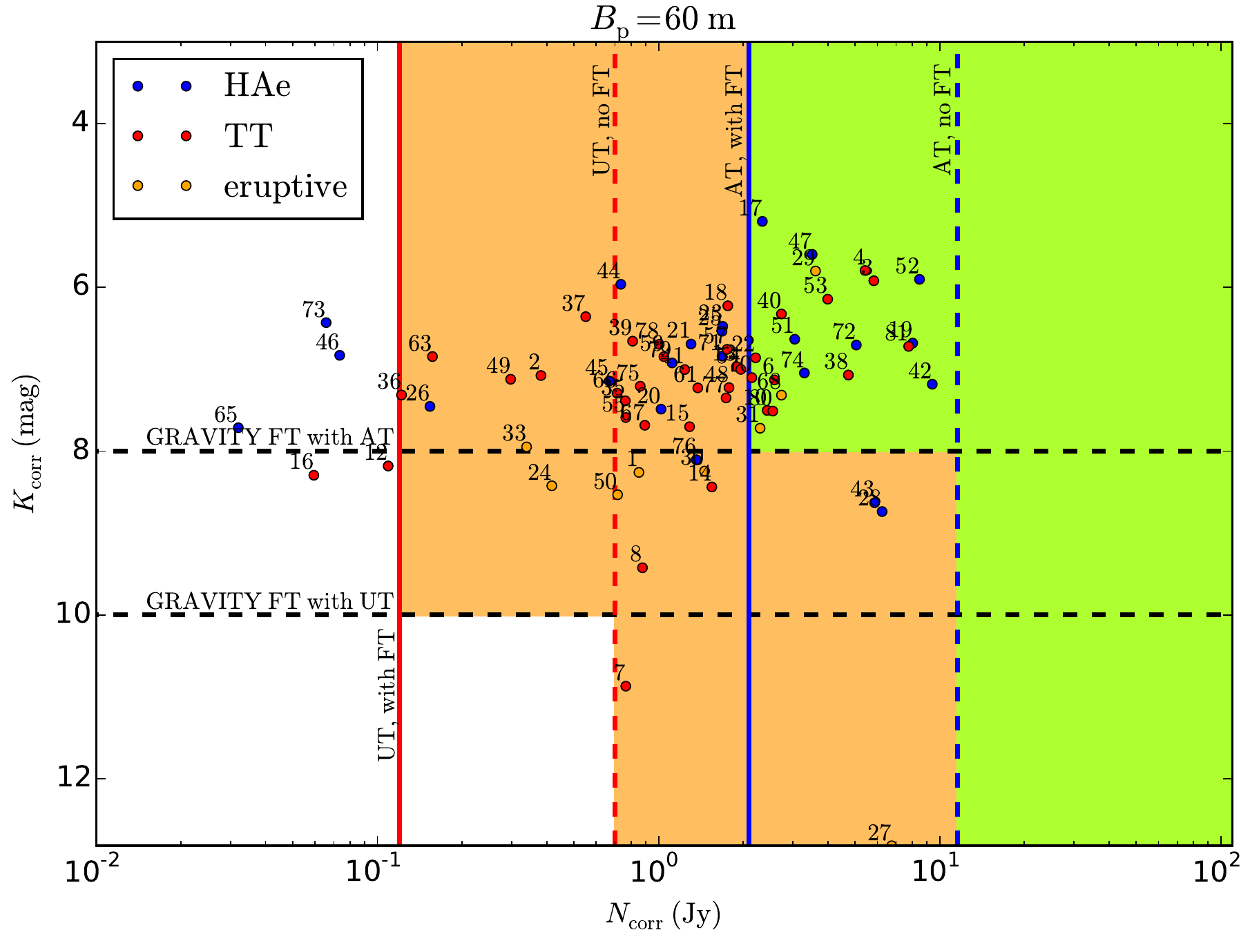}
			\includegraphics[width=0.49\textwidth]{matisse_observability_L_10um7_30m-eps-converted-to.pdf}     
			\includegraphics[width=0.49\textwidth]{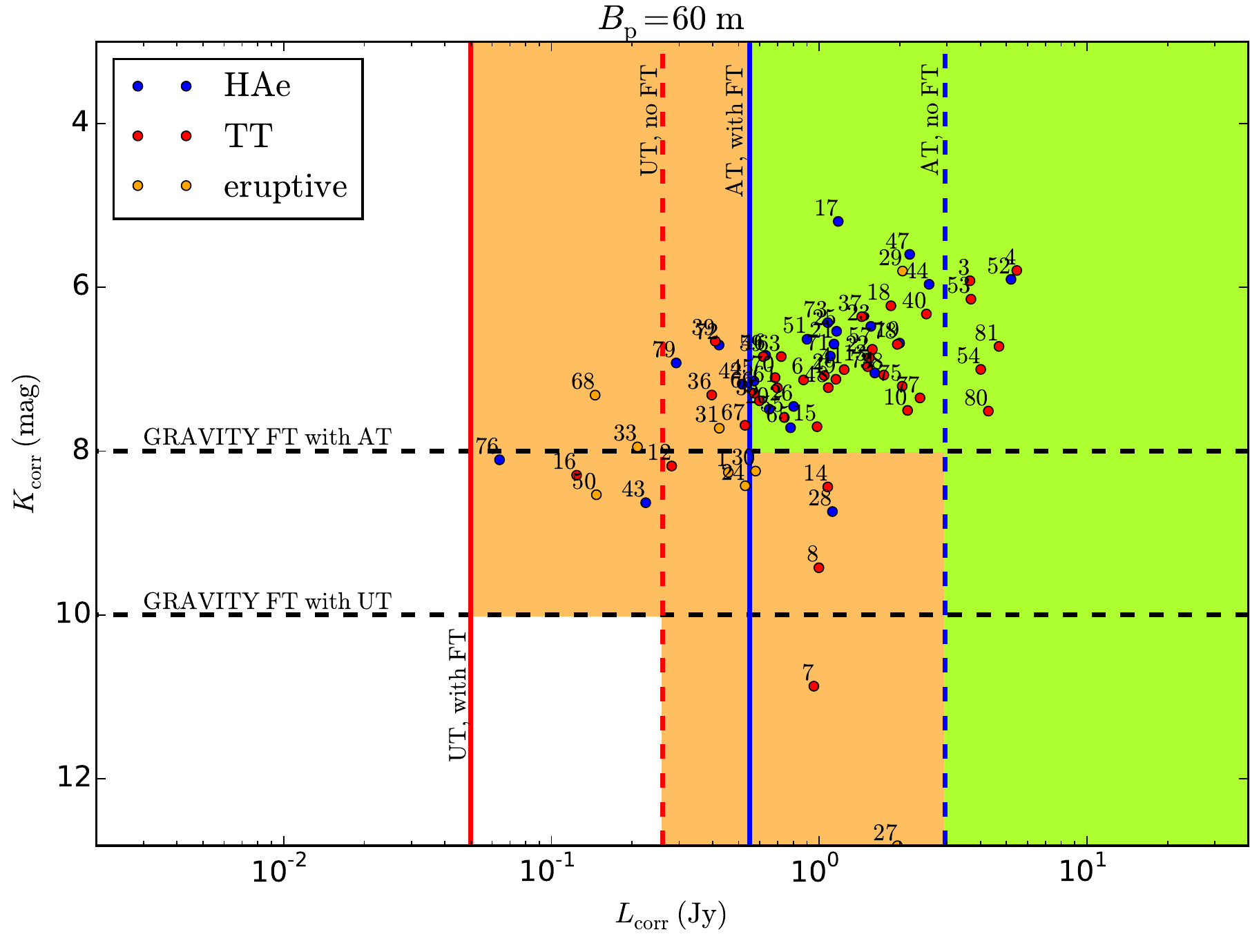}
			\caption{Same as Fig.~\ref{fig:obs_plot_app}, but for 30 and 60~m baseline lengths.}
		\end{figure*}   
		
		\begin{figure*}[h!]
			\centering
			\includegraphics[width=0.49\textwidth]{matisse_observability_LN_10um7_100m-eps-converted-to.pdf}
			\includegraphics[width=0.49\textwidth]{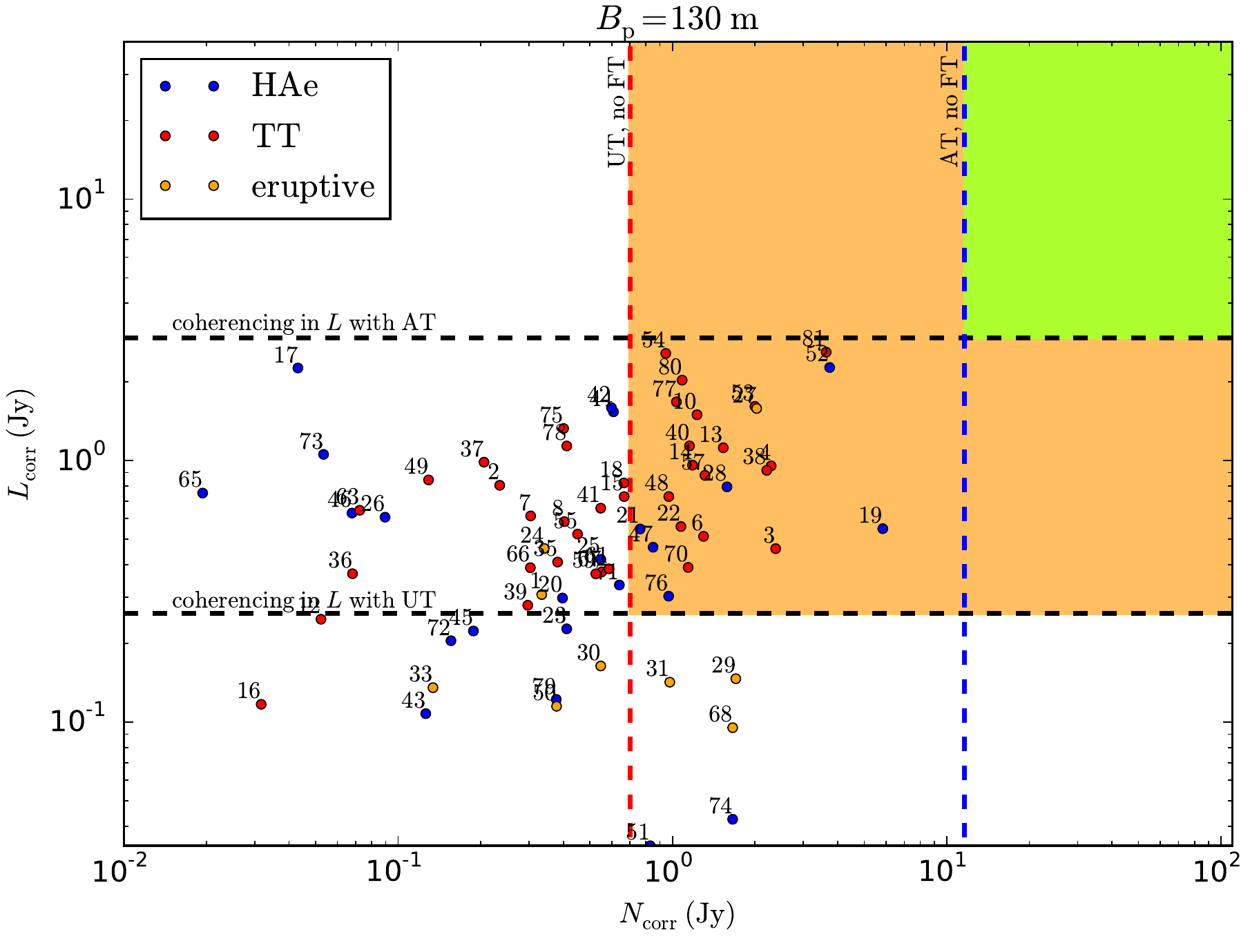}
			\includegraphics[width=0.49\textwidth]{matisse_observability_N_10um7_100m-eps-converted-to.pdf}
			\includegraphics[width=0.49\textwidth]{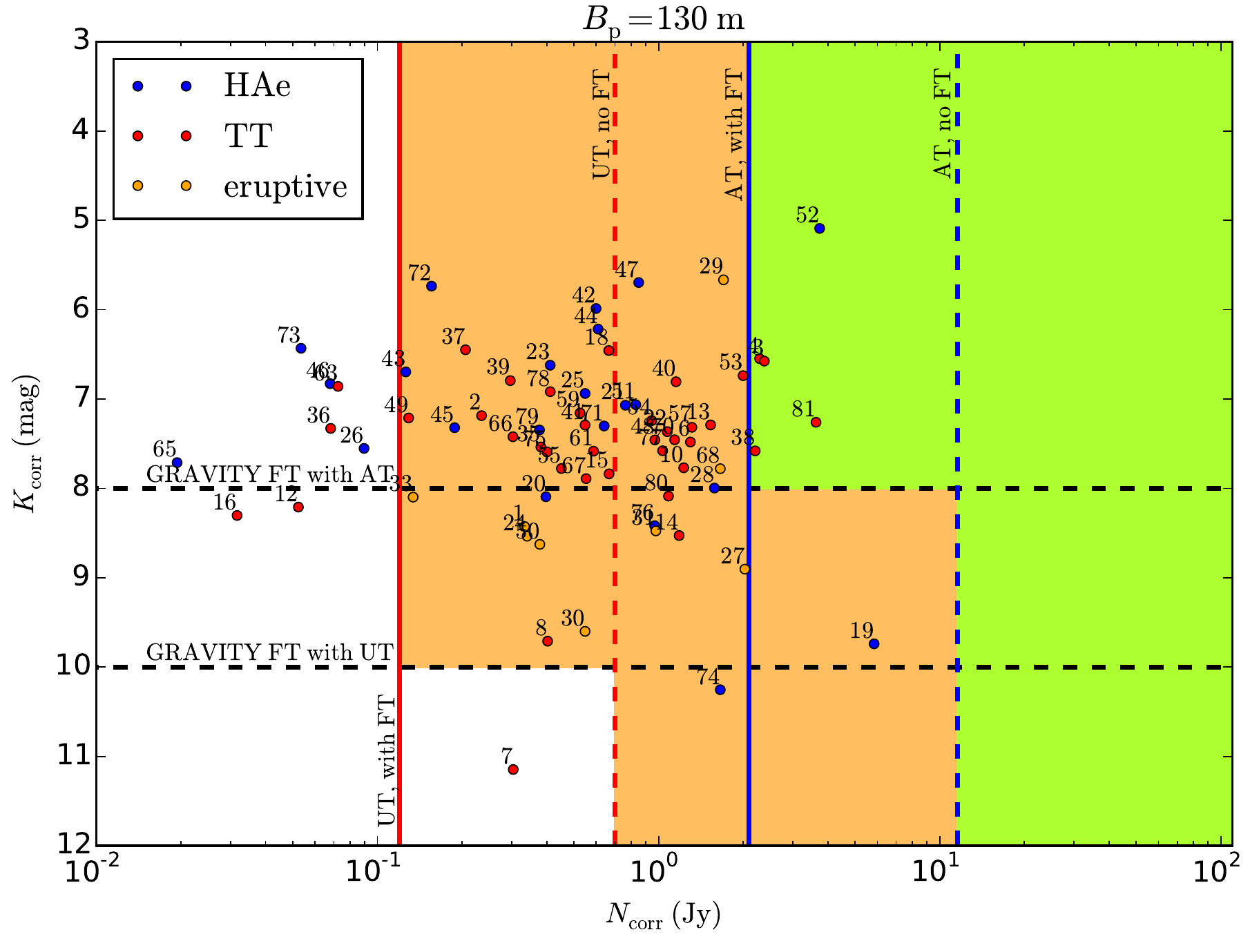}
			\includegraphics[width=0.49\textwidth]{matisse_observability_L_10um7_100m-eps-converted-to.pdf}
			\includegraphics[width=0.49\textwidth]{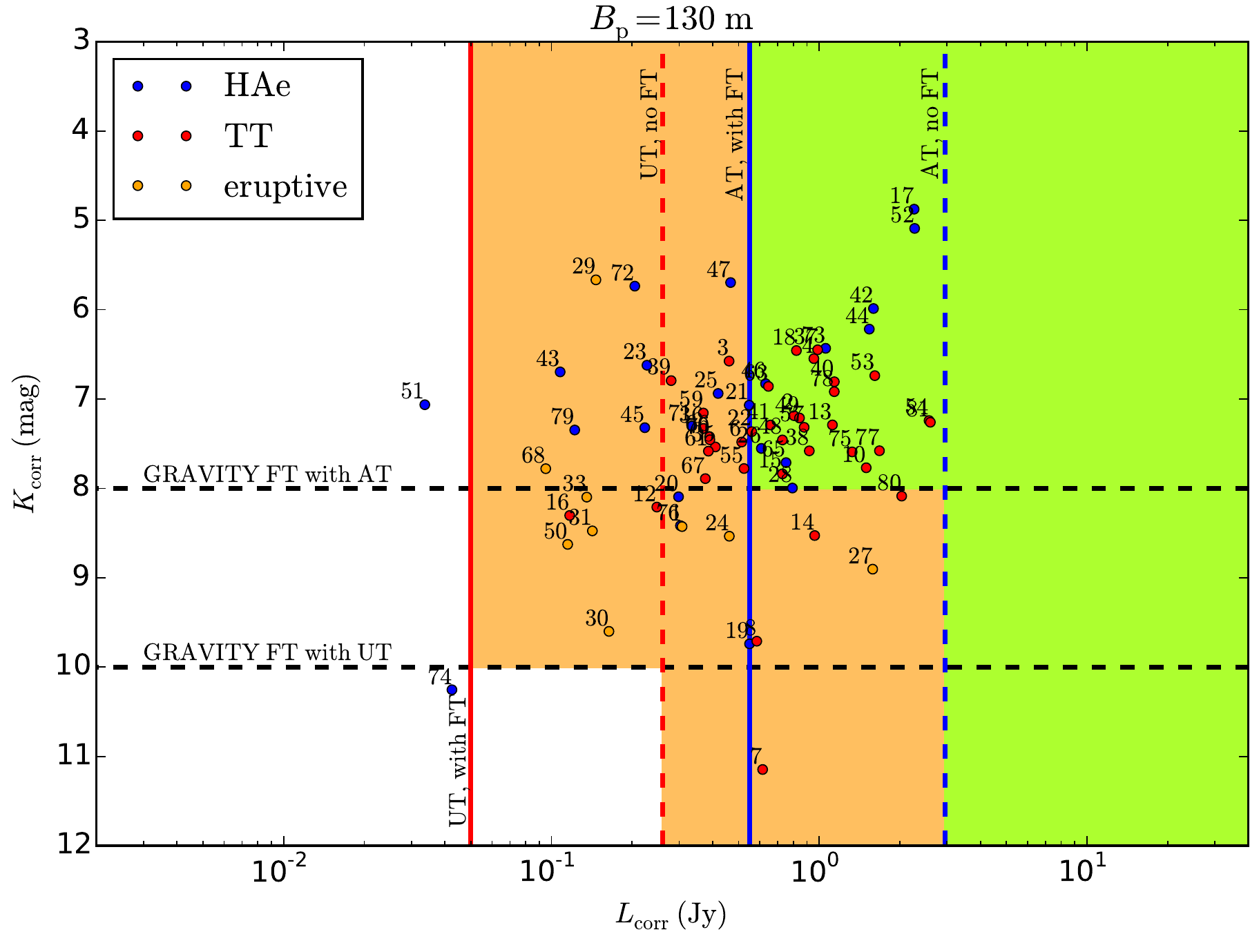}
			\caption{Same as Fig.~\ref{fig:obs_plot_app}, but for 100 and 130~m baseline lengths.}
		\end{figure*}           
		
		\begin{figure*}[h!]
			\centering
			\includegraphics[width=0.49\textwidth]{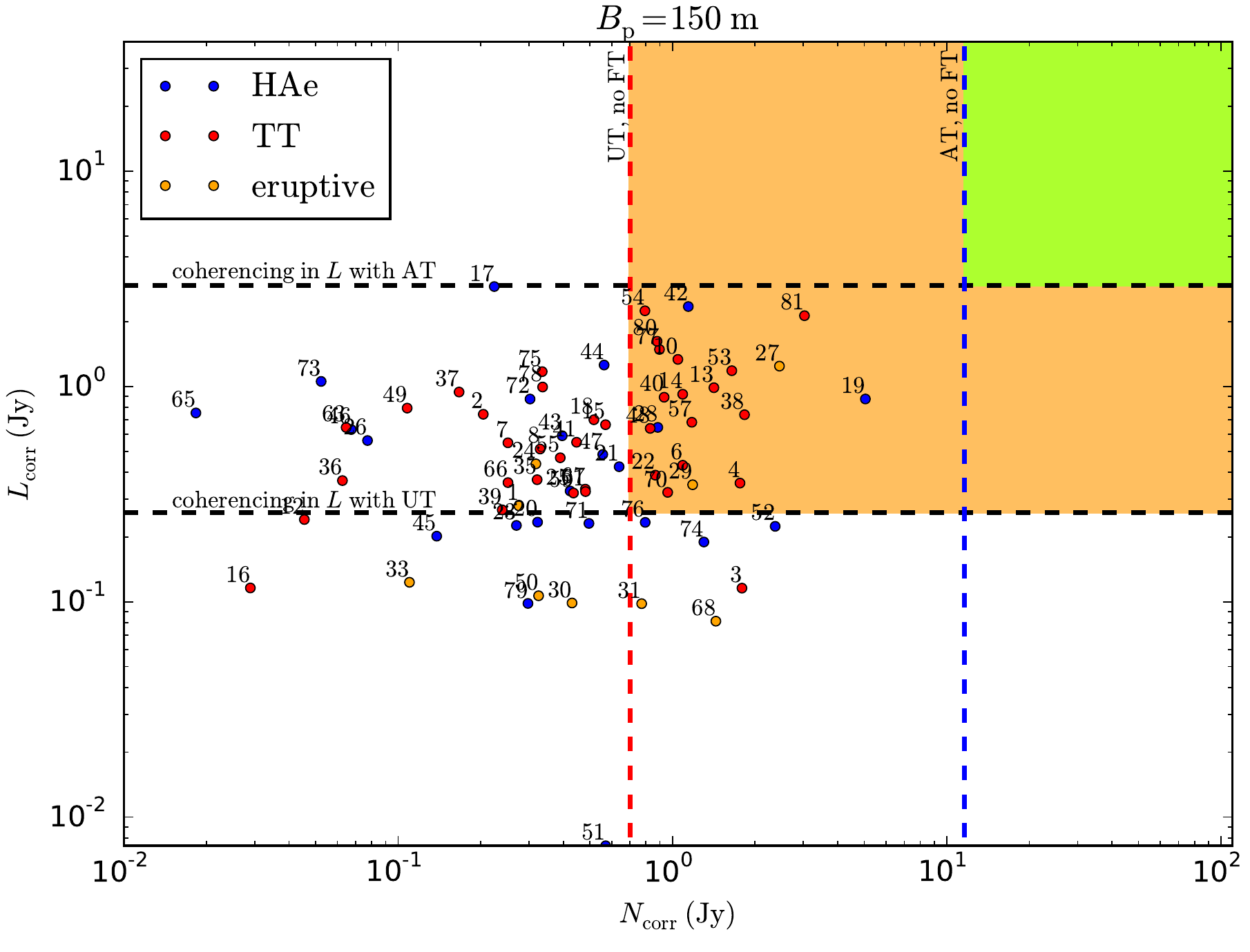}\\
			\includegraphics[width=0.49\textwidth]{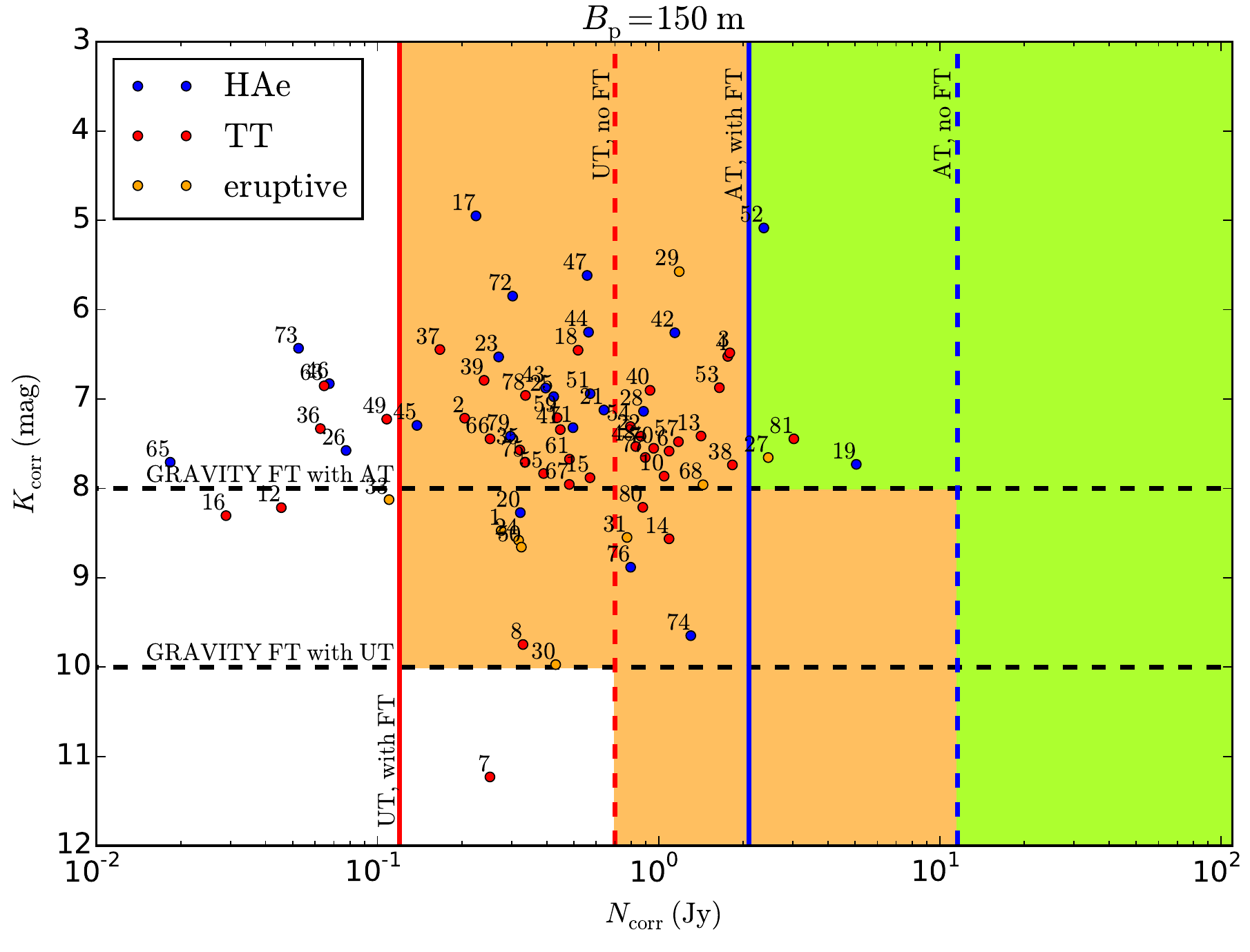}\\
			\includegraphics[width=0.49\textwidth]{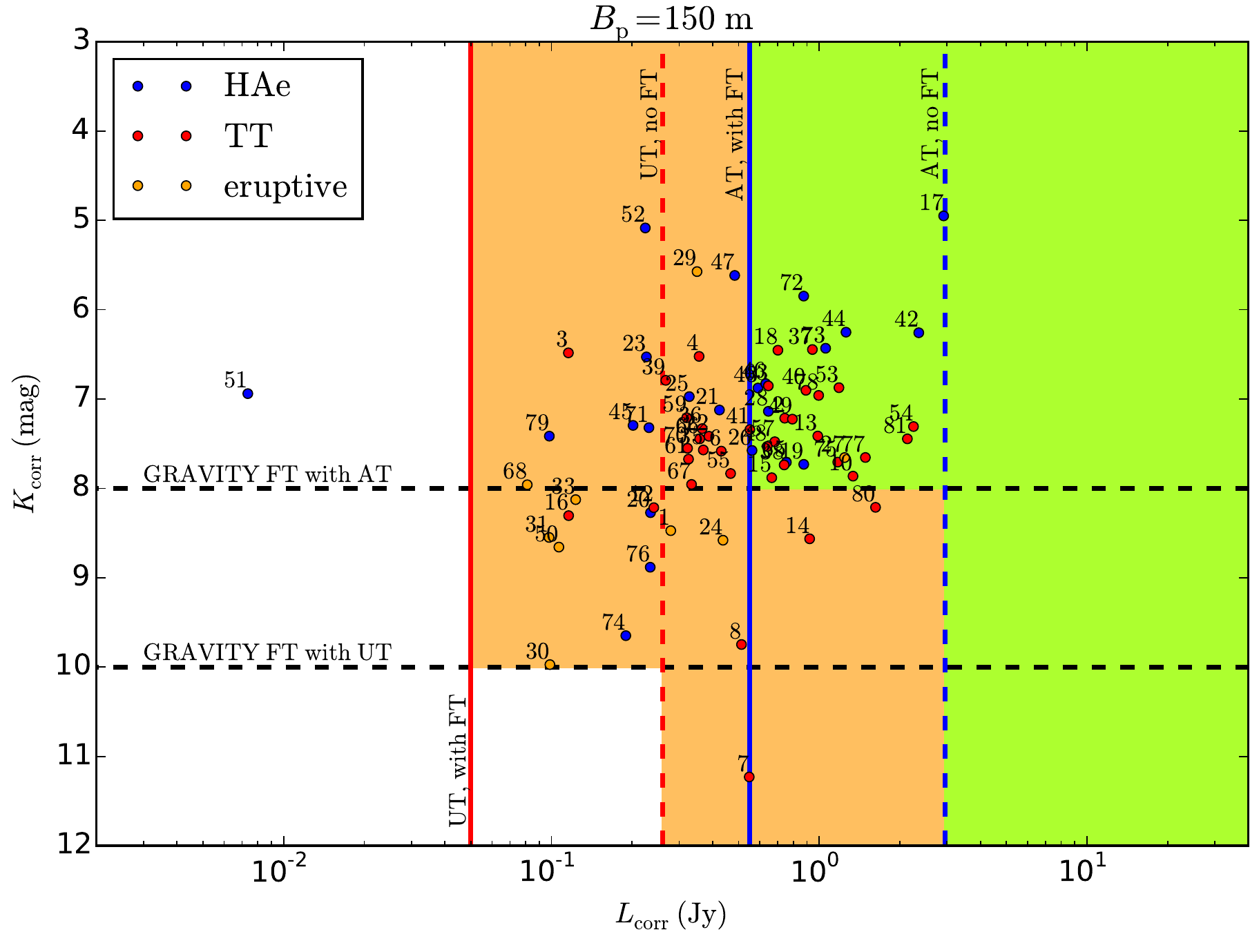}
			\caption{Same as Fig.~\ref{fig:obs_plot_app}, but for 150~m baseline length.}
		\end{figure*}
		
		\clearpage
		\section{List of MIDI observations}
		\label{app:obs}
		
		%
		%-------------------------------------------------------------
		%          For the appendices, table longer than a single page
		%-------------------------------------------------------------
		
		% Table will be print automatically at the end of the document, 
		% after the whole appendices
		
		% In the appendices do not forget to put the counter of the table 
		% as an option
		%compilation fails when trying to input the table

		%\tabcolsep=0.11cm
		%\longtab[1]
		{
			{\tiny
\begin{longtable}{l|ccrrrccc}
					\caption{Overview of the MIDI observations. The projected baseline length ($B_p$) is given in meters, and $\phi_B$ is the projected position angle of the baseline (measured from north through east). The resolution ($\theta$)
						is given as the FWHM of a fringe $\theta = \lambda / (2 B_p )$  at $\lambda = 10.7\ \mu$m, converted to physical scale. An observation flag marks unreliable data. Feature type indicates silicate emission (e), absorption (a), and weak feature or flat spectra (w).}
					\label{tab:obs}\\
					\hline\hline
					Name & Date & Base & $B_p$ & $\phi_B$ & $\theta$ & Program ID & Obs. flag & Feature\\
					&(UTC)&&(m)&($^\circ$)& (au) & & & type\\
					\hline
					\endfirsthead
					\caption{continued.}\\
					\hline
					Name & Date & Base & $B_p$ & $\phi_B$ & $\theta$ & Program ID & Obs. flag & Feature\\
					&(UTC)&&(m)&($^\circ$)& (au) & & & type\\
					\hline
					\endhead
					\hline
					\endfoot
					\hline
					\endlastfoot
					
					SVS 13A1 & 2004-10-31, 05:17 & U3U2 & 34.8 & -123.3 & 7.9 & 074.C-0649(A) &  & a \\ 
					& 2004-10-31, 05:26 & U3U2 & 35.7 & -123.6 & 7.7 & 074.C-0649(A) &  & a \\ 
					& 2004-11-04, 06:17 & U3U4 & 53.9 & 94.8 & 5.1 & 074.C-0649(A) &  & a \\ 
					\\ 
					
					LkH$\alpha$ 330 & 2011-12-11, 04:08 & U2U3 & 40.5 & 54.1 & 6.8 & 088.C-0763(A) &  & e \\ 
					& 2011-12-11, 04:46 & U2U3 & 42.9 & 51.4 & 6.4 & 088.C-0763(A) &  & e \\ 
					& 2011-12-12, 05:12 & U1U4 & 130.2 & 57.9 & 2.1 & 088.C-0763(B) &  1  & w \\ 
					& 2011-12-12, 05:24 & U1U4 & 130.1 & 56.3 & 2.1 & 088.C-0763(B) &  & a \\
\end{longtable}
			}
		}% End longtab

	\end{appendix}
	
\end{document}